\documentclass[aps,pre,twocolumn,a4paper,superscriptaddress,showpacs,showkeys,amsmath,amssymb,10pt]{revtex4-1} 

\usepackage{graphicx} 
\usepackage{dcolumn} 
\usepackage{bm} 
\usepackage{amssymb} 
\usepackage{amsmath} 
\usepackage{mathrsfs} 
\usepackage{color,soul} 
\usepackage{enumitem} 
\usepackage{makecell} 
\usepackage{hyperref} 
\hypersetup{colorlinks = true, urlcolor = blue, linkcolor = blue, citecolor = blue} 

\begin{document}

\title
{
Transfer-matrix calculations of the effects of tension and torque constraints on DNA-protein interactions
}

\author{Artem K. Efremov}
\email[Corresponding author e-mail: ]{mbiay@nus.edu.sg}
\affiliation{Mechanobiology Institute, National University of Singapore, Singapore 117411}
\affiliation{Centre for Bioimaging Sciences, National University of Singapore, Singapore 117546}

\author{Jie Yan}
\email[Corresponding author e-mail: ]{phyyj@nus.edu.sg}
\affiliation{Mechanobiology Institute, National University of Singapore, Singapore 117411}
\affiliation{Centre for Bioimaging Sciences, National University of Singapore, Singapore 117546}
\affiliation{Department of Physics, National University of Singapore, Singapore 117542}

\date{\today}

\begin{abstract}
Organization and maintenance of the chromosomal DNA in living cells strongly depends on the DNA interactions with a plethora of DNA-binding proteins. Single-molecule studies show that formation of nucleoprotein complexes on DNA by such proteins is frequently subject to force and torque constraints applied to the DNA. Although the existing experimental techniques allow to exert these type of mechanical constraints on individual DNA biopolymers, their exact effects in regulation of DNA-protein interactions are still not completely understood due to the lack of systematic theoretical methods able to efficiently interpret complex experimental observations. To fill this gap, we have developed a general theoretical framework based on the transfer-matrix calculations that can be used to accurately describe behaviour of DNA-protein interactions under force and torque constraints. Potential applications of the constructed theoretical approach are demonstrated by predicting how these constraints affect the DNA-binding properties of different types of architectural proteins. Obtained results provide important insights into potential physiological functions of mechanical forces in the chromosomal DNA organization by architectural proteins as well as into single-DNA manipulation studies of DNA-protein interactions.
\end{abstract}



\maketitle

\section{Introduction}
\label{sec:intro}

DNA-architectural proteins play a major role in the genome structural organization and maintenance of its functionality in living cells, regulating a delicate balance between the chromosomal DNA condensation level and its accessibility to various DNA-binding proteins. By synergistically cooperating or antagonizing each other's action on the chromosomal DNA, architectural proteins can adjust its mechanical properties, compaction level and supercoiling state on a local as well as the global genome scales, affecting the transcription level of numerous genes in living cells. Thus, by regulating the DNA-binding properties of architectural proteins, cells can dynamically change organization of the chromosomal DNA and rapidly switch between different gene expression patterns in response to environmental cues \cite{Luijsterburg_2008, Dame_2010, Bannister_2011}.

While DNA-architectural proteins are the key components determining the chromosomal DNA organization, it should be noted that they perform their function in the context of numerous mechanical constraints imposed on the DNA by various factors, such as multiple DNA motor proteins (topoisomerases, helicases, RNA/DNA polymerases, etc. \cite{Champoux_2001, Hubscher_2002, Rothwell_2005, Singleton_2007, Buc_2009, Spies_2013}), that generate stretching and twisting forces on the chromosomal DNA \cite{Wang_1998, Strick_2000, Wuite_2000, Gore_2006, Johnson_2007, Ma_2013}. It is also known that chromosomes form extensive adhesion contacts with a number of nuclear membrane proteins, establishing force-transmitting links between the chromosomal DNA and cytoplasmic cytoskeleton, which frequently carries strong mechanical loads \cite{Wang_2009, Shivashankar_2011, Uhler_2017}. As a result, the chromosomal DNA is a subject to the combined action of both DNA-architectural proteins and the mechanical constraints applied to it. Together, these factors not only determine the physical organization of the chromosomal DNA, but also play the major role in gene transcription regulation inside living cells. 

Indeed, it has been revealed in recent experiments that cells not only use various mechanical constraints to shape the chromosomal DNA, but actually can sense and process mechanical forces applied to the nucleus, changing the level of genes' transcription in response to their action \cite{Shivashankar_2011, Iyer_2012, Mammoto_2012, Tajik_2016, Uhler_2017}.  While the exact molecular processes responsible for such mechanosensing of living cells remain unclear, recent experimental studies suggest that this may be the result of force- and torque-dependent interactions between different groups of DNA-architectural proteins and chromosomal DNA. 

Namely, crystallographic and single-molecule experiments show that upon binding to DNA proteins frequently prompt various conformational changes in the DNA structure, which can be coupled to force and torque constraints applied to the DNA, affecting the DNA-binding properties of proteins \cite{Rice_1996, Luger_1997, Harp_2000, Swinger_2003, Skoko_2004, Noort_2004, Swinger_2004, Liu_2010, Stella_2010, Laurens_2012, Le_2013, Efremov_2015, Ahmad_2015}. What is even more interesting, existing experimental data indicate that different groups of DNA-architectural proteins frequently produce very distinct responses to the applied mechanical constraints. Indeed, according to their mechanism of interaction with DNA all architectural proteins can be divided into four major groups \cite{Luijsterburg_2008}: 1) DNA-wrapping proteins, which fold DNA into compact nucleoprotein complexes (such as eukaryotic/archaeal histones) \cite{Luger_1997, Harp_2000, Talbert_2010}; 2) DNA-bending proteins, which sharply curve DNA at the protein binding site (like bacterial HU, IHF and Fis) \cite{Rice_1996, Swinger_2003, Skoko_2004, Noort_2004, Swinger_2004, Stella_2010, Botao_2010, Le_2013}; 3) DNA-bridging proteins that cross-link DNA duplexes (for example, bacterial H-NS, human HMGA2, or any other protein that mediates DNA loops) \cite{Dame_2000, Liu_2010, Zhao_2017, Yan_2018}, and 4) DNA-stiffening proteins forming rigid nucleoprotein filaments along DNA (like archaeal TrmBL2 and Alba) \cite{Jelinska_2005, Laurens_2012, Efremov_2015}. Thus, the four major groups of DNA-architectural proteins form nucleoprotein complexes, which have very different 3D structures, leading to diverse responses of these proteins to force and torque constraints applied to DNA. 

For example, previous studies have shown that while suppressing formation of nucleoprotein complexes by DNA-bending and DNA-wrapping proteins, mechanical stretching of DNA promotes its interaction with DNA-stiffening proteins \cite{Brower_2002, Yan_2003, Mihardja_2006, Botao_2010, Le_2013, Meng_2015}. Likewise, torque exerted on DNA can either enhance or weaken binding of DNA-wrapping proteins depending on the chirality of the resulting nucleoprotein complexes and the direction of the applied torque \cite{Sheinin_2013}. Such a differential response of proteins to mechanical constraints applied to DNA suggests that it is possible to shift balance between nucleoprotein complexes formed by different groups of proteins in favour of one or the other protein group by changing the applied constraints \cite{Efremov_2015}.

Indeed, as experimental data show, this mechanism is frequently used by living cells to organize their chromosomal DNA. For example, topoisomerases I and II relax positive (right-handed) torsion accumulated in DNA during chromosome condensation by architectural proteins (histones) or due to DNA replication/transcription processes, allowing continuous assembly of left-handed nucleosome complexes that would not otherwise form on positively supercoiled DNA \cite{Pfaffle_1990, Garinther_1997, Salceda_2006}. This type of DNA organization control even more pronounced in bacterial cells, which use gyrases to maintain negatively supercoiled state of their circular chromosomal DNA to promote its interaction on a local and the global scales with various DNA-architectural proteins, such as H-NS and HU \cite{Shindo_1992, Kobryn_1999, Lal_2016}.  

To better understand potential roles of mechanical constraints in regulation of DNA interactions with architectural proteins, a number of single-DNA manipulation experimental methods have been recently developed, allowing one to control the supercoiling state of individual DNA molecules as well as to apply force and torque constraints to them \cite{Strick_1996, Cluzel_1996, Smith_1996, Wang_1997, Bryant_2003, Yan_2004, Deufel_2007, Lipfert_2010, Janssen_2012, Forth_2013}. While such experiments may provide important information regarding the effects of mechanical constraints onto the DNA-binding properties of architectural proteins, it should be noted that typical observables measured in these experiments, such as the DNA extension and linking number change, frequently have highly complex dependence on the force and torque constraints applied to the DNA, especially in the presence of DNA-binding proteins in solution. As a result, interpretation of the collected experimental data poses a challenging task that requires development of a general theoretical framework aimed at description of DNA-binding behaviour of architectural proteins in a wide range of force and torque constraints applied to DNA. 

So far, most of the previous theoretical studies have been mainly focused on understanding of the effects of stretching force on protein binding to a torsionally relaxed DNA, proposing several different approaches to investigate this question \cite{Katritch_2000, Yan_2003, Aumann_2006, Zhang_2010, Kepper_2011, Ettig_2011, Guevara_2011, Guevara_2012, Zhang_2012, Dobrovolskaia_2012, Nam_2014, Elbel_2015, Meng_2015, Dahlke_2018}. Among the proposed methods, the transfer-matrix theory developed based on a discretized semi-flexible polymer chain model of DNA has several unique advantages by providing very fast semi-analytical calculations of equilibrium conformations of DNA that allow one to easily incorporate DNA heterogeneity into the computations \cite{Yan_2003, Yan_2005, Zhang_2010, Zhang_2012}.

Furthermore, by using several famous results from the group theory, it has been recently shown that the transfer-matrix formalism can be further extended to take into consideration not only force, but also torque constraints, considerably increasing the scope of its potential applications, including but not limited to description of local DNA structural transitions and sequence-dependent response of DNA to stretching and torsional strains \cite{Efremov_2016, Efremov_2017}. What is even more important, this advancement in the transfer-matrix calculations opens a completely new way to development of a general theoretical framework aimed at description of DNA-protein interactions under both force and torque constraints. 

In this study, we show in details how such theoretical framework can be constructed based on the mathematical formalism described in ref.~\cite{Efremov_2016, Efremov_2017} and demonstrate how the developed theoretical approach can be used to obtain insights into potential roles of force and torque constraints in regulation of DNA interaction with different types of DNA-binding proteins found in living cells.

\section{General theory}
\label{sec:theory}
\subsection{Brief outline of the theoretical framework}
\label{sec:theory_outline}

In our previous work, it has been shown that DNA behaviour under mechanical constraints can be accurately described by a semiflexible polymer model in which DNA is represented by a polygonal chain consisting of straight segments whose 3D orientations in space are characterized by the three Euler rotation angles, see Figures~\ref{fig1}(a,b). Introducing transfer-matrices defined on each of the vertices joining neighbouring DNA segments, it is then possible to calculate the DNA partition function and obtain detailed information regarding the DNA conformation and DNA structural fluctuations under force and torque constraints \cite{Efremov_2016, Efremov_2017}. In this study, we describe how the previously developed transfer-matrix formalism can be further expanded to integrate DNA-protein interactions into the model. While all of the details can be found in Appendices~\ref{Appendix-Lk}-\ref{Appendix-F}, in this section we will mainly focus on the central ideas and assumptions underlying the transfer-matrix calculations for DNA behaviour under force and torque constraints in the presence of DNA-protein interactions.

As before, DNA will be represented by a discretized polygonal chain consisting of short segments, which are treated as rigid bodies with a local coordinate system $( \textbf{x}^{}_j,\textbf{y}_j,\textbf{z}^{}_j )$ attached to each of the DNA segments, see schematic Figure~\ref{fig1}(a). Here $j$ is the index enumerating all of the DNA segments from $1$ to $N$, where $N$ is the total number of segments in the discretized polymer chain representing DNA molecule. 3D orientation of each of the coordinate systems, and thus each of the DNA segments, is then can be described by the Euler rotation matrix $\textbf{R}_j = \textbf{R}_{\alpha_j}\textbf{R}_{\beta_j} \textbf{R}_{\gamma_j}$ resulting from the composition of three successive revolutions through Euler angles $\alpha_j$, $\beta_j$ and $\gamma_j$ about the fixed lab coordinate frame $( \textbf{x}^{}_0,\textbf{y}_0,\textbf{z}^{}_0 )$, see Figure~\ref{fig1}(b).

Besides the 3D orientation, DNA segments in addition are characterized by their physical state. Namely, existing experimental data show that depending on the force and torque constraints applied to DNA it may exist in several different structural states known as B-, L-, P-DNA, etc. \cite{Allemand_1998, Leger_1999,  Bryant_2003, Sheinin_2009, Sheinin_2011, Oberstrass_2012, Marko_2013}. For the sake of simplicity, in this study we consider only the following structural states of DNA, which are the most relevant to the physiological ranges of forces and torques: 1) B-DNA state, which is typical for relaxed DNA polymer; 2) L-DNA, which is favoured at negative torques, and 3) P-DNA, which is favoured at positive torques, see more detailed description of these DNA forms in ref.~\cite{Efremov_2016, Efremov_2017}. Thus, in the absence of protein binding, the DNA conformation is completely determined by the two sets of parameters: 1) rotation matrices $( \textbf{R}_1,...,\textbf{R}_N )$ describing orientations of all DNA segments, and 2) indexes $( k_1,...,k_N )$ representing the structural states of these segments, such that for each segment $j = 1,...,N$ we put $k_j = 0$ for B-DNA segments, $k_j = -1$ -- for L-DNA segments, and $k_j = -2$ -- for P-DNA segments. 

Incorporation of DNA-protein interactions into the model results in appearance of additional DNA segment states. Indeed, besides indexes $k_j = -2$, $-1$ and $0$ ($j = 1,...,N$) that indicate the structural states of bare DNA segments, we also need to have a mean to describe the states of DNA segments residing inside nucleoprotein complexes formed on DNA. Namely, to mark the positions of DNA segments in each of the nucleoprotein complexes, we will use positive values for indexes $k_j$ that will designate the sequence number of each DNA segment with respect to the DNA entry point into the complex. I.e., assuming that the protein of interest occupies $K$ DNA segments upon binding to DNA, one can assign $K$ DNA binding sites on the protein surface -- from $1$ (the first DNA binding site on the protein surface) to $K$ (the last DNA binding site on the protein surface). Correspondingly, for each DNA segment bound to the protein we put the value of $k_j$ equal to the index of the respective binding site on the surface of the protein -- from $k_j = 1$ (if the DNA segment is bound to the first binding site on the protein surface) to $k_j = K$ (if the DNA segment is bound to the last binding site on the protein surface). Thus, in the presence of DNA-protein interactions, indexes $k_j$ ($j = 1,...,N$) take integer values in the range from $-2$ to $K$, with $k_j = -2$, $-1$, $0$ representing bare DNA segments being in P-, L- or B-DNA states, respectively; and $k_j = 1,...,K$ corresponding to protein-bound DNA segments. In the latter case, for a given DNA segment, $j$, parameter $k_j$ equals to the index of the DNA binding site on the protein surface to which this DNA segment is bound. As an example, see schematic figure Figure~\ref{fig1}(c) for the case of $K = 12$.   

In the general case, the total conformational energy of DNA interacting with proteins, $E_\textrm{tot}$, can be written as a sum of the following energy terms:
\begin{equation} \label{DNA-total-energy-intro}
\!\!\! E_\textrm{tot} \!\left( k_1...k_N, \textbf{R}_1...\textbf{R}_N \right) = 
	E_\textrm{DNA} + E_\textrm{protein} + \Phi_f + \Phi_\tau
\end{equation}

\noindent
Here $E_\textrm{DNA}$ is the sum of the bending and twisting deformation energies of all protein-unbound bare DNA segments, and $E_\textrm{protein}$ is the sum of the energies associated with nucleoprotein complexes formed on the DNA. Furthermore, $\Phi_f = - ( \textbf{f} \cdot \textbf{d} )$ is the potential energy related to the stretching force $\textbf{f}$ applied to the DNA, where $\textbf{d}$ denotes the DNA end-to-end vector; and $\Phi_\tau = - 2 \pi \tau \Delta \textrm{Lk}$ is the potential energy associated with the torque $\tau$ applied to the DNA, where $\Delta \textrm{Lk}$ denotes the DNA linking number change with respect to the torsionally relaxed B-DNA state, which is used in this study as a reference state for the energy calculations. For the sake of simplicity, all of the energies in this study are presented in $k_\textrm{B} T$ units, where $k_\textrm{B}$ is Boltzmann constant and $T$ is temperature of the surrounding environment. For this reason, the force $\textbf{f}$ and torque $\tau$ are scaled by $k_\textrm{B} T$; thus, $\textbf{f}$ has a dimension of 1/length and $\tau$ is dimensionless. 

While the above energy terms will be discussed in details in the next section, here we only would like to stress that under very general assumptions it is possible to represent the total conformational energy of DNA, $E_\textrm{tot}$, as a sum of local DNA segment contributions [see Appendix~\ref{Appendix-A}]:
\begin{multline} \label{DNA-total-energy-intro-2}
E_\textrm{tot} \!\left( k_1...k_N,\textbf{R}_1...\textbf{R}_N \right) = \\
	\sum_{j=1}^{N-1} E_{k_j k_{j+1}} \!\left( \textbf{R}_j, \textbf{R}_{j+1} \right) 
	+ E_{k_N k_1} \!\left( \textbf{R}_N, \textbf{R}_1 \right)
\end{multline}

\noindent
Where $E_{k_j k_{j+1}} \!\left( \textbf{R}_j, \textbf{R}_{j+1} \right)$ is the local energy contribution by the $j^\textrm{th}$ vertex in the polygonal chain representing DNA that joins the $j^\textrm{th}$ and the $(j+1)^\textrm{th}$ DNA segments. $E_{k_j k_{j+1}} \!\left( \textbf{R}_j, \textbf{R}_{j+1} \right)$ in the general case depends on the states $k_j$ and $k_{j+1}$ of the $j^\textrm{th}$ and the $(j+1)^\textrm{th}$ DNA segments as well as their orientations, $\textbf{R}_j$ and $\textbf{R}_{j+1}$. The last term in Eq.~\eqref{DNA-total-energy-intro-2} describes the contribution of the DNA end segments, which may be considered as a part of boundary conditions imposed on the DNA. 
  
Knowing the total conformational energy of DNA, it is then straightforward to find its partition function, $Z_{f,\tau}$, which can be calculated as:
\begin{multline} \label{DNA-protein-partition-function-intro-1}
Z_{f,\tau} =
    \sum_{k_1...k_N=-2}^K \int \textrm{d} \textbf{R}_1 ... \textrm{d} \textbf{R}_N \,
    \textrm{d} \!\left[ \eta_\textrm{in} \right] \,
    \xi \!\left( \textbf{R}_N, \textbf{R}_1 \right) \times \\
    \times e^{-E_\textrm{tot} (k_1...k_N, \textbf{R}_1...\textbf{R}_N)}
\end{multline} 

\noindent
Where $\xi ( \textbf{R}_N, \textbf{R}_1 )$ is a function that imposes specific boundary conditions on the orientations of the DNA end segments. In the above formula, integrations are carried out over all of the DNA segment orientations, $( \textbf{R}_1,..., \textbf{R}_N )$. Furthermore, in order to take into account orientational freedom of nucleoprotein complexes, we in addition perform integration $\int \textrm{d} [ \eta_\textrm{in} ]$ over all possible rotations of these complexes with respect to the axes of the DNA segments entering them, for more details see comments after Eq.~\eqref{all-local-energies-DNA-wrapping} in Appendix~\ref{Appendix-A}.

Substituting Eq.~\eqref{DNA-total-energy-intro-2} into Eq.~\eqref{DNA-protein-partition-function-intro-1}, it can be shown that the exponent in Eq.~\eqref{DNA-protein-partition-function-intro-1} can be re-written as a product of local transfer-functions, $T_{k_j k_{j+1}} ( \textbf{R}_j, \textbf{R}_{j+1} )$, defined on the vertices joining neighbouring DNA segments, where $T_{k_j k_{j+1}} ( \textbf{R}_j, \textbf{R}_{j+1} ) = \int \textrm{d} \eta_\textrm{in} \, e^{- E_{k_j k_{j+1}} ( \textbf{R}_j, \textbf{R}_{j+1} )}$ if the $j^\textrm{th}$ and $(j+1)^\textrm{th}$ DNA segments are located at the interface between bare DNA and one of the nulceoprotein complexes, such that $(k_j, k_{j+1}) = (0, 1)$, $(-1,1)$ or $(-2,1)$; and $T_{k_j k_{j+1}} ( \textbf{R}_j, \textbf{R}_{j+1} ) = e^{- E_{k_j k_{j+1}} ( \textbf{R}_j, \textbf{R}_{j+1} )}$ in all other cases [for more details see Appendices~\ref{Appendix-B} and \ref{Appendix-F}]. Indeed, from Eq.~\eqref{DNA-total-energy-intro-2}-\eqref{DNA-protein-partition-function-intro-1} and the above definition of local DNA transfer-functions, it is not hard to see that the partition function, $Z_{f,\tau}$, turns into:
\begin{multline} \label{DNA-protein-partition-function-intro-2}
Z_{f,\tau} = 
    \sum_{k_1...k_N=-2}^K \int \textrm{d} \textbf{R}_1 ... \textrm{d} \textbf{R}_N
    \prod_{j=1}^{N-1} T_{k_j k_{j+1}} \!\left( \textbf{R}_j, \textbf{R}_{j+1} \right) \\
    \times \sigma_{k_N k_1} \!\left( \textbf{R}_N, \textbf{R}_1 \right)
\end{multline}

\noindent
Here all of the $\int \textrm{d} [ \eta_\textrm{in} ]$ integrals from Eq.~\eqref{DNA-protein-partition-function-intro-1} are adsorbed into $T_{k_j1} ( \textbf{R}_j, \textbf{R}_{j+1} )$ transfer-functions corresponding to the DNA segments entering nucleoprotein complexes. As for $\sigma_{k_N k_1} ( \textbf{R}_N, \textbf{R}_1 )$ functions, they describe the boundary conditions imposed on the DNA end segments and have the following simple form:
\begin{equation} \label{Sigma-function-intro}
\sigma_{k_N k_1} \!\left( \textbf{R}_N, \textbf{R}_1 \right) = 
    \xi \!\left( \textbf{R}_N, \textbf{R}_1 \right) 
	\, e^{ -E_{k_N k_1} \!\left( \textbf{R}_N, \textbf{R}_1 \right) }
\end{equation}

\noindent
Where $E_{k_N k_1} \!\left( \textbf{R}_N, \textbf{R}_1 \right)$ depends on the states of the first and the last DNA segments, and, in addition, on the potential energy of the last segment due to the force $f = |\textbf{f}|$ applied to the DNA, see Eq.~\eqref{DNA-protein-partition-function-3}-\eqref{sigma-function} in Appendix~\ref{Appendix-B}.

To calculate all of the $\int \textrm{d} \textbf{R}_j$ integrals in Eq.~\eqref{DNA-protein-partition-function-intro-2}, it is convenient to expand $T_{k_j k_{j+1}} \!\left( \textbf{R}_j, \textbf{R}_{j+1} \right)$ and $\sigma_{k_N k_1} \!\left( \textbf{R}_N, \textbf{R}_1\right)$ elements into the series of orthogonal D-functions, $D^s_{p,q} (\textbf{R})$, that form basis in the Hilbert space of square-integrable functions defined on SO(3) group of 3D rotation matrices \cite{Gelfand_1963}. Then by using orthogonality of $D^s_{p,q} (\textbf{R})$ basis, it can be shown that $\int \textrm{d} \textbf{R}_1 ... \textrm{d} \textbf{R}_N$ integrals in Eq.~\eqref{DNA-protein-partition-function-intro-2} reduce to a mere multiplication of matrices composed of the expansion coefficients of $T_{k_j k_{j+1}} \!\left( \textbf{R}_j, \textbf{R}_{j+1} \right)$ and $\sigma_{k_N k_1} \!\left( \textbf{R}_N, \textbf{R}_1\right)$ functions [see Appendix~\ref{Appendix-B}]:
\begin{equation} \label{DNA-protein-partition-function-intro-4}
Z_{f,\tau} = \textrm{Tr} \!\left( \textbf{U} \textbf{L}^{N\!-\!1} \textbf{Y} \right)
\end{equation}

\noindent
Here the entries of matrix $\textbf{L}$ are the expansion coefficients of $T_{k_j k_{j+1}} \!\left( \textbf{R}_j, \textbf{R}_{j+1} \right)$ transfer-functions; and matrices $\textbf{Y}$ and $\textbf{U}$ are composed of the expansion coefficients of $\sigma_{k_N k_1} \!\left( \textbf{R}_N, \textbf{R}_1\right)$ functions, which for convenience reasons are split into two parts [see Appendices~\ref{Appendix-B} and \ref{Appendix-E} for more details].

Knowing the DNA partition function, $Z_{f,\tau}$, it is then rather straightforward to calculate the DNA extension ($z$) and linking number change ($\Delta \textrm{Lk}$) as well as the total number of protein-bound ($N_\textrm{pr}$) and bare ($N_u$) DNA segments in each of the states, $u =$ L- or P-DNA, by differentiating $Z_{f,\tau}$ with respect to force ($f$), torque ($\tau$), protein binding energy ($\mu_\textrm{pr}$) or DNA base-pairing energy in the corresponding state ($\mu_n$, $n = -1$ or $-2$), accordingly [see Eq.~\eqref{observables} in Appendix~\ref{Appendix-F}]. From these observables it is then easy to find the DNA superhelical density ($\sigma$) and the DNA occupancy fraction by DNA-bound proteins ($O$) as: $\sigma = \Delta \textrm{Lk} / \textrm{Lk}_0$ and $O = N_\textrm{pr} / N$. Here $\textrm{Lk}_0$ is the linking number of a torsion-free B-DNA, which in the case of DNA comprised of $N_\textrm{bp}$ base-pairs equals to $\textrm{Lk}_0 = N_\textrm{bp} / h_0$, where $h_0$ is the helical repeat of B-DNA.
 
Evaluation of the above parameters based on the transfer-matrix computations of the DNA partition function provides a simple and fast way to predict changes in the DNA conformation as well as in DNA-protein interactions in response to mechanical constraints applied to the DNA, making it possible to compare theoretical results presented here to direct measurements performed in single-molecule experiments.

\subsection{DNA energy terms}
\label{sec:energy_terms}

As can be seen from the previous section, by having at hand a mathematical expression for the total conformational energy of DNA, it is possible to calculate the DNA partition function and predict the equilibrium behaviour of DNA under various force and torque constraints applied to it. To provide insights into the energy terms contributing to the total conformational energy of DNA, which were briefly mentioned in Eq.~\eqref{DNA-total-energy-intro}, here we present their detailed mathematical description with references to Appendices sections, where interested readers can find more additional information.

While $E_\textrm{DNA}$ energy term in Eq.~\eqref{DNA-total-energy-intro} has been previously discussed in details in ref.~\cite{Efremov_2016}, we would like to briefly remind that in the general case it has the following form:
\begin{align} \label{bare-DNA-sum-intro}
E_\textrm{DNA} = &
	\sum_{j=1}^{N-1} \sum_{n,m=-2}^0 \delta_{k_j n} \delta_{k_{j+1} m} \Big\{ \frac{a_n}{2} \!\left( \textbf{R}_j
	\textbf{z}_0 \!-\! \textbf{R}_{j+1} \textbf{z}_0 \right)^2 \nonumber \\
	& + \frac{c_n}{2} \!\left[ 2\pi\Delta \textrm{Tw}_j \!\left( \textbf{R}_j,\textbf{R}_{j+1} \right) \right]^2 
	+ J \left( 1 - \delta_{nm} \right) \Big\} 
	\nonumber \\
	& + q \sum_{j=1}^N \sum_{n=-2}^0 \mu_n \delta_{k_j n}
\end{align}

\noindent
Where $\delta_{n m}$ is the the Kronecker delta ($\delta_{n m} = 1$ if $n = m$ and $\delta_{n m} = 0$, otherwise). $a_n = A_n / b_n$ and $c_n = C_n / b_n$ are dimensionless parameters describing the bending and twisting rigidies of bare DNA segments being state $n$ ($n = 0$, $-1$ and $-2$ for B-, L- and P-DNA, respectively), where $A_n$, $C_n$ and $b_n$ are the bending and twisting persistence lengths of DNA, and the size of DNA segments in the respective state, accordingly (see Table~\ref{tab:DNA-parameters}). $q$ is the number of base-pairs in each of the DNA segments, which is a fixed constant having the same value for all DNA segment states. $\mu_n$ is the base-paring energy of DNA in state $n$ with respect to B-DNA form (see Table~\ref{tab:DNA-parameters}). $J$ is the domain wall penalty that accounts for the cooperativity of DNA structural transitions, describing the molecule preference for structural uniformity \cite{Sarkar_2001, Oberstrass_2012}. Finally, $\Delta \textrm{Tw}_j (\textbf{R}_{j}, \textbf{R}_{j+1})\approx \frac{1}{2\pi} \textbf{R}_j \textbf{z}_0 \cdot [ \textbf{R}_j \textbf{x}_0 \times \textbf{R}_{j+1} \textbf{x}_0 ]$ is the local DNA twist between the $j^\textrm{th}$ and $(j\!+\!1)^\textrm{th}$ DNA segments.

\begin{table*}
\resizebox{\textwidth}{!}{\begin{minipage}{\textwidth}
\caption{\label{tab:DNA-parameters} Bare DNA parameters.}
\begin{ruledtabular}
\begin{tabular}{cccccccc}
\noalign{\smallskip}
$\substack{\textbf{DNA}_{} \\ \textbf{form}}$ & 
$\substack{\textbf{Bending persistence}_{} \\ \textbf{length, } \bm{A_n} \textbf{(nm)}}$ &
$\substack{\textbf{Twisting persistence}_{} \\ \textbf{length, } \bm{C_n} \textbf{(nm)}}$ &
$\substack{\textbf{Contour length relative}^{}_{} \\ \textbf{to B-DNA form}}$ &
$\substack{\textbf{DNA helical}_{} \\ \textbf{repeat, } \bm{h_n} \textbf{(bp)}}$ &
$\substack{\textbf{Base-pairing energy relative}_{} \\ \textbf{to B-DNA form, } \bm{\mu_n} \textbf{(} \bm{k_\textrm{B} T} \textbf{)}}$ &
$\bm{\lambda_n}$ \footnote{To account for the cooperativity of the DNA structural transitions, the domain wall penalty, $J = 9.0$ $k_\textrm{B} T$ \cite{Sarkar_2001, Oberstrass_2012}, characterizing the DNA preference for structural uniformity was introduced into the transfer-matrix calculations in addition to the model parameters listed in the above table.} \\
\noalign{\smallskip}
\hline
\noalign{\smallskip}
B-DNA	& 50, \cite{Bustamante_1994, Wang_1997}	& 95, \cite{Bryant_2003, Forth_2008, Mosconi_2009}	& 1	& 10.4, \cite{Wang_1979}	& 0	& 4.3, \cite{Efremov_2016,Efremov_2017}\\

L-DNA	& 7, \cite{Sheinin_2011, Marko_2013}	& 15, \cite{Sheinin_2011, Oberstrass_2012, Marko_2013}	& 1.35, \cite{Sheinin_2011, Marko_2013}	& 16, \cite{Bryant_2003, Sheinin_2011, Oberstrass_2012}	& 5.0, \cite{Efremov_2016,Efremov_2017}	& 4.3, \cite{Efremov_2016,Efremov_2017}\\

P-DNA	& 15, \cite{Marko_2013}	& 25, \cite{Marko_2013}	& 1.7, \cite{Allemand_1998, Leger_1999, Marko_2013}	& 3, \cite{Allemand_1998, Leger_1999, Bryant_2003, Sheinin_2009, Marko_2013}	& 17.8, \cite{Efremov_2016,Efremov_2017}	& --0.5, \cite{Efremov_2016,Efremov_2017}\\[1pt]

\end{tabular}
\end{ruledtabular}
\end{minipage} }
\end{table*}

From now on we will focus our attention on the last three energy terms, $E_\textrm{protein}$, $\Phi_f$ and $\Phi_\tau$, in Eq.~\eqref{DNA-total-energy-intro} that describe the elastic deformation energy of DNA caused by DNA-protein interactions and potential energies associated with the force and torque constraints applied to DNA. To calculate them, we generally need to know the DNA conformation inside nucleoprotein complexes formed on DNA. One of the main reasons for this is dependence of the DNA linking number change, $\Delta \textrm{Lk}$, on the global DNA conformation, which is determined by the relative orientations of all of the DNA segments, including those contributing to formation of nucleoprotein complexes. As a result, $\Phi_\tau$ term generally depends on the nature of nucleoprotein complexes formed on DNA.

\begin{figure}[!htb]
\includegraphics[width=\textwidth]{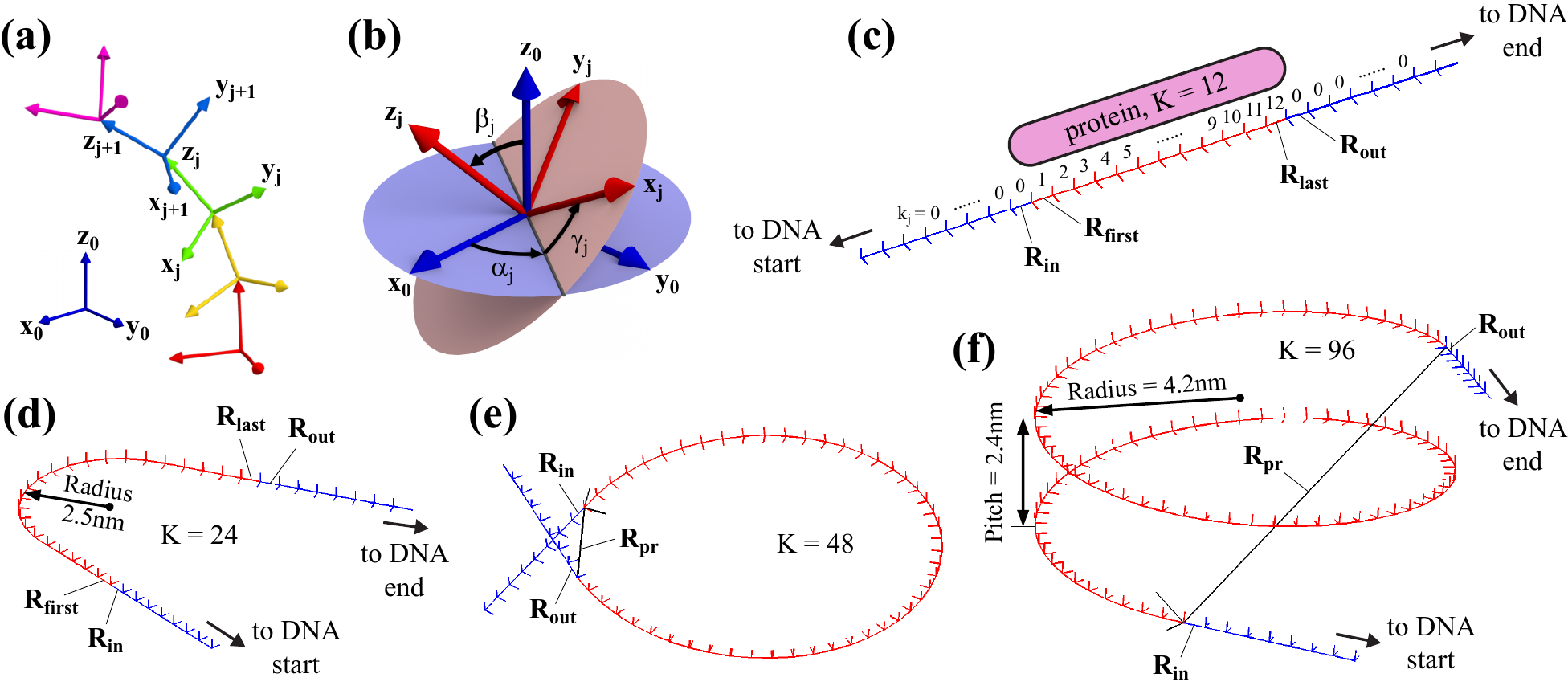}
\caption{ 
\textbf{Semiflexible polymer chain model of DNA.} \textbf{(a)} In the model, DNA is represented by a polygonal chain comprised of straight segments. The latter are considered as rigid bodies with attached local Cartesian coordinate frames, $(\textbf{x}^{}_j, \textbf{y}_j, \textbf{z}^{}_j)$, whose 3D-orientations in space with respect to the fixed global coordinate system $(\textbf{x}^{_{}}_0, \textbf{y}_0, \textbf{z}^{_{}}_0)$ are described by the Euler rotation matrices, $\textbf{R}_j$. \textbf{(b)} Each rotation matrix, $\textbf{R}_j$, results from the composition of three successive revolutions of the coordinate frame $(\textbf{x}^{}_j, \textbf{y}_j, \textbf{z}^{}_j)$ relative to the fixed coordinate system $(\textbf{x}^{_{}}_0, \textbf{y}_0, \textbf{z}^{_{}}_0)$ through Euler angles $\alpha_j$, $\beta_j$ and $\gamma_j$ shown on the graph. \textbf{(c-f)} Proteins binding to DNA results in formation of nucleoprotein complexes that constrain protein-bound DNA segments in a specific 3D conformation: DNA-stiffening proteins typically form straight nucleoprotein filaments along DNA \textbf{(c)}, while DNA-bending proteins kink DNA at the binding site \textbf{(d)}; as for DNA-wrapping proteins, such as histone tetramers and octamers, their interaction with DNA results in formation of solenoid-like nucleoprotein complexes \textbf{(e-f)}. On panels \textbf{(c-f)}, bare DNA segments are shown in blue color and protein-bound DNA segments forming the respective nucleoprotein complexes are presented in red color. On panel \textbf{(c)}, indexes $k_j$, which are displayed above the DNA segments, indicate the physical states of the corresponding segments. 
}
\label{fig1}
\end{figure}

In the case of DNA-bending proteins, such as the one schematically shown on Figure~\ref{fig1}(d), the DNA linking number change associated with the formation of nucleoprotein complexes may vary in a wide range depending on the orientations of these complexes with respect to the rest of the DNA. Hence, one cannot assign a fixed linking number change to nucleoprotein complexes formed by DNA-bending proteins, and the relative orientations of all DNA segments inside such complexes must be known in order to calculate the above energy terms, which can be done, for example, by using existing X-ray crystallographic data for nucleoprotein complexes.

In contrast, nucleoprotein complexes formed by DNA-wrapping proteins [Figures~\ref{fig1}(e,f)] make a well-defined fixed contribution, $\Delta \textrm{Lk}_\textrm{pr}$, to the DNA linking number change. Thus, one does not need to have exact information regarding the DNA conformation inside each of the nucleoprotein complexes to calculate the DNA linking number change. As a result, any such nucleoprotein complex can be replaced by a straight line connecting the entry and exit points of DNA, see Figures~\ref{fig1}(e,f). In this case, the DNA linking number change can be estimated by first calculating the contribution from all protein-unbound DNA segments, and then adding to it $\Delta \textrm{Lk}_\textrm{pr} \times M$ term, where $M$ is total number of nucleoprotein complexes formed by DNA-wrapping proteins. Such approach greatly simplifies the final expression for the DNA partition function, making its computation much more easier in comparison to the DNA-bending proteins scenario described above. 

However, it should be noted that while in the case of DNA-wrapping proteins the replaced DNA segments do not make any contribution to the formula for the DNA total conformational energy, we still need to keep track of these segments by making a corresponding register shift by $K$ DNA segments each time upon encountering one of the nucleoprotein complexes formed on DNA. One way to do this is to split the line connecting the entry and exit points of each nucleoprotein complex into $K$ smaller subintervals, assigning each of these intervals to one of the replaced DNA segments. Thus, for example, if DNA segments with indexes $j$, $j\!+\!1$, ..., $j+K-1$ are bound to one of the DNA-wrapping proteins (such that $k_j=1$, $k_{j+1}=2$, ..., $k_{j+K-1}=K$) then we simply put: $\textbf{R}_j = \textbf{R}_{j+1} = ... = \textbf{R}_{j+K-1} = \textbf{R}_\textrm{pr,j}$, where $\textbf{R}_\textrm{pr,j}$ is the rotation matrix describing the orientation of the line connecting the entry and exit points of the given nucleoprotein complex. In other words, all of the DNA segments taking part in the formation of a DNA-wrapping nucleoprotein complex can be assumed to have the same orientations, being aligned along a straight line connecting the entry and exit points of the complex, see Figures~\ref{fig1}(e-f).

Finally, we would like to note that in the special case of DNA-stiffening proteins that form straight filaments along DNA both of the above approaches lead to identical description of the resulting nucleoprotein complexes.

Following the above notes, it is not hard to obtain expressions for the DNA linking number change, $\Delta \textrm{Lk}$, as well as $\Phi_\tau$ energy term by using a combination of the famous C\u{a}lug\u{a}reanu-White's theorem \cite{Calugareanu_1961, White_1969} and the Fuller's approximate formula for the DNA writhe number \cite{Fuller_1978}. While the first states that $\Delta \textrm{Lk}$ can be expressed as a sum of two components: $\Delta \textrm{Lk} = \Delta \textrm{Tw} + \textrm{Wr}$, where $\Delta \textrm{Tw} = \sum_{j=1}^{N-1} \Delta \textrm{Tw}_j ( \textbf{R}_j, \textbf{R}_{j+1} )$ is the DNA total twist and $\textrm{Wr}$ is the DNA writhe number; the second allows to express the DNA writhe number as a sum of local DNA segments' contributions, $\textrm{Wr}^\textrm{F} = \sum_{j=1}^{N-1} \textrm{Wr}^\textrm{F}_j ( \textbf{R}_j, \textbf{R}_{j+1} )$, where $\textrm{Wr}^\textrm{F}_j ( \textbf{R}_j, \textbf{R}_{j+1} ) = \frac{1}{2\pi} ( \tilde\alpha_{j+1} \!-\! \tilde\alpha_j ) ( 1 - \cos \beta_j )$, and $\tilde\alpha_{j+1}$ and $\tilde\alpha_j$ are the azimuthal Euler angles of the $j^\textrm{th}$ and $(j\!+\!1)^\textrm{th}$ DNA segments from the extended range of $(-\infty,+\infty)$ \cite{Fain_1997, Bouchiat_2000, Efremov_2016}. The superscript $\textrm{F}$ in the above equations indicates that the DNA writhe number calculation is based on the Fuller's formula approximation. The resulting mathematical expressions for the DNA linking number change in the presence of DNA interaction with different types of DNA-binding proteins can be found in Appendix~\ref{Appendix-Lk}.

Here we would like only to stress that the Fuller's formula provides correct estimations of the DNA linking number change for those DNA conformations which can be obtained by a continuous deformation of DNA initially extended along $\textbf{z}_0$-axis direction in such a way that none of the DNA sections face the negative direction of $\textbf{z}_0$-axis for any of the intermediate DNA configurations \cite{Fuller_1978, Fain_1997, Bouchiat_2000}. A nearly straight DNA or DNA folded into a helical solenoid conformation are examples satisfying this criterion \cite{Fain_1997, Moroz_1998, Bouchiat_1998, Bouchiat_2000}. In other cases, however, the DNA writhe number $\textrm{Wr}$ does not necessarily equal to $\textrm{Wr}^\textrm{F}$. For this reason, the Fuller's formula works well only for DNA conformations that do not contain supercoiled plectoneme structures. Indeed, previous theoretical studies show that the Fuller's formula can be used to accurately predict the behaviour of DNA under a wide range of mechanical constraints up to the onset of the torque-induced buckling transition when DNA starts to develop supercoiled plectonemes \cite{Fain_1997, Moroz_1998, Bouchiat_1998, Bouchiat_2000, Yan_2003, Efremov_2016, Efremov_2017}. 

More importantly, by utilizing the Fuller's approximation, it is possible to observe collapsing of bare DNA into compact conformations upon application of sufficiently large torques, which is accompanied by increase in the absolute value of the DNA linking number \cite{Efremov_2016, Efremov_2017}. Although the resulting conformations are not necessarily the same as supercoiled DNA plectonemes, the predicted force-extension curves of DNA subjected to torque constraints resemble those observed in single-molecule experiments, exhibiting very similar behaviour near the DNA buckling transition point, see ref.~\cite{Efremov_2016, Efremov_2017} and \cite{Strick_1996}. Therefore, it is still possible to use the Fuller's approximation to describe the DNA supercoiling transition. 

The only side-effect of such approach is that it leads to a slight shift of the predicted DNA supercoiling transition boundary relative to the experimentally measured position, which, however, can be easily corrected by adding a new term, $\delta \Phi_\tau$, to $\Phi_\tau$ energy (i.e., $\Phi_\tau =  - 2 \pi \tau \Delta \textrm{Lk}^\textrm{F} + \delta \Phi_\tau$). In the case of a structurally uniform DNA, this term simply equals to $\delta \Phi_\tau = \tau \lambda \textrm{Wr}^\textrm{F}$ with $\lambda$ being a fixed scaling factor, see ref.~\cite{Efremov_2016, Efremov_2017}. Whereas in a more realistic scenario when the DNA segments are allowed to make transitions between different structural states, the correction term takes somewhat sophisticated form as each of the DNA structures (B-, L- or P-DNA) is characterized by its own value of the scaling parameter, $\lambda_n$, see ref.~\cite{Efremov_2016, Efremov_2017} and Table~\ref{tab:DNA-parameters}:
\begin{multline} \label{correction-term-intro}
\delta \Phi_\tau = 
	\tau \sum_{j=1}^{N-1} \left[ \sum_{n=-2}^0 \delta_{k_j n} \lambda_n 
	+ \lambda_\textrm{pr} \sum_{n = 1}^K \delta_{k_j n} \right] \times \\
	\times \textrm{Wr}^\textrm{F}_j \!\left( \textbf{R}_j, \textbf{R}_{j+1} \right)
\end{multline}
  
\noindent
Here $\lambda_n$ and $\lambda_\textrm{pr}$ are scaling parameters associated with different DNA structures and DNA segments residing inside nucleoprotein complexes, respectively.

In this study, we consider only the proteins that bind to B-form DNA. As a result, in all our calculations we simply put $\lambda_\textrm{pr}=\lambda_0$. Thus, in the above formula, $\lambda_0$ value is used for all of the vertices connecting neighbouring DNA segments inside nucleoprotein complexes.

The next energy term from Eq.~\eqref{DNA-total-energy-intro}, $\Phi_f$, has a very simple mathematical expression, which can be obtained by assuming that the global coordinate system $( \textbf{x}^{}_0,\textbf{y}_0,\textbf{z}^{}_0 )$ is aligned in such a way that its $\textbf{z}_0$-axis faces in the direction of force $\textbf{f}$ applied to the DNA. Then it is not hard to show that in this case $\Phi_f$ equals to:
\begin{equation} \label{Phi-f-intro}
\Phi_f =
	- \sum_{j=0}^N \sum_{n=-2}^K \delta_{k_j n} b_n f \left( \textbf{z}_0 \cdot  \textbf{R}_j \textbf{z}_0 \right)
\end{equation}

\noindent
Here $f = |\textbf{f}|$ is the force magnitude, and $b_n$ is the size of DNA segments being in state $n$. Since in this study we consider only the proteins that bind to B-form DNA, the DNA segments constrained inside nucleoprotein complexes formed by DNA-bending or DNA-stiffening proteins should have approximately the same size as protein-unbound B-form DNA segments: $b_1 = ... = b_K = b_0$. As for DNA-wrapping proteins, since all of the DNA segments bound to such proteins are replaced by the lines connecting the entry and exist points of the resulting nucleoprotein complexes, with each line being subdivided into $K$ equal intervals, we have: $b_1 = ... = b_K = r_\textrm{pr} /K$, where $r_\textrm{pr}$ is the distance between the entry and exit points of the nucleoprotein complexes. 

Finally, $E_\textrm{protein}$ energy term from Eq.~\eqref{DNA-total-energy-intro} equals to the sum of individual nucleoprotein complexes' energies, which include: 1) the protein binding energy to DNA, $\mu_\textrm{pr}$, and 2) the DNA elastic deformation energies at the entry and exit points of the nucleoprotein complex, $E_\textrm{in}$ and $E_\textrm{out}$, respectively. Thus, denoting the orientations of the DNA segments sitting next to the entry and exit points of a nucleoprotein complex by rotation matrices $\textbf{R}_\textrm{in}$ and $\textbf{R}_\textrm{out}$, and orientations of the first and the last DNA segments in the nucleoprotein complex by rotation matrices $\textbf{R}_\textrm{first}$ and $\textbf{R}_\textrm{last}$ [see Figures~\ref{fig1}(c,d)], the energy of each nucleoprotein complex can be written in the following form:
\begin{multline} \label{protein-energy-intro}
E_\textrm{pr} = 
	- \mu_\textrm{pr}
	+ E_\textrm{in} \!\left( \textbf{R}_\textrm{in}, \textbf{R}_\textrm{first} \right) 
	+ E_\textrm{out} \!\left( \textbf{R}_\textrm{last}, \textbf{R}_\textrm{out} \right)
\end{multline}

\noindent
Where in the case of DNA-wrapping proteins $\textbf{R}_\textrm{first} = \textbf{R}_\textrm{last} = \textbf{R}_\textrm{pr}$, see Figures~\ref{fig1}(e,f). As for $E_\textrm{in}$ and $E_\textrm{out}$ terms describing the DNA elastic deformation energies at the entry and exit points of a nucleoprotein complex, in the general case they equal to:
\begin{multline} \label{protein-energy-in-intro}
E_\textrm{in} \!\left( \textbf{R}_\textrm{in}, \textbf{R}_\textrm{first} \right) = 
	\frac{a_\textrm{pr}}{2} \left( \textbf{R}_\textrm{in} \textbf{A}_\textrm{in} \textbf{z}_0 - \textbf{R}_\textrm{first} 	
	\textbf{z}_0 \right)^2 \\
	+ \frac{c_\textrm{pr}}{2} \left[ 2\pi \Delta \textrm{Tw} \!\left( \textbf{R}_\textrm{in} \textbf{A}_\textrm{in}, 
	\textbf{R}_\textrm{first} \right) \right]^2
\end{multline}

\noindent
and
\begin{multline} \label{protein-energy-out-intro}
E_\textrm{out} \!\left( \textbf{R}_\textrm{last}, \textbf{R}_\textrm{out} \right) =
	\frac{a_\textrm{pr}}{2} \left( \textbf{R}_\textrm{last} \textbf{A}_\textrm{out} \textbf{z}_0 - \textbf{R}_\textrm{out} 
	\textbf{z}_0 \right)^2 \\
	+ \frac{c_\textrm{pr}}{2} \left[ 2\pi \Delta \textrm{Tw} \!\left( \textbf{R}_\textrm{last} \textbf{A}_\textrm{out},
	\textbf{R}_\textrm{out} \right) \right]^2
\end{multline}

\noindent
Here $a_\textrm{pr}$ and $c_\textrm{pr}$ are dimensionless bending and twisting elasticities of the entry and exit DNA segments of the nucleoprotein complex; $\textbf{A}_\textrm{in}$ and $\textbf{A}_\textrm{out}$ are two rotation matrices that determine the equilibrium orientations of the entry and exit DNA segments relative to the core part of the nucleoprotein complex such that in mechanical equilibrium we have: $\textbf{R}^\textrm{(eq)}_\textrm{in} \textbf{A}_\textrm{in} = \textbf{R}^\textrm{(eq)}_\textrm{first}$ and $\textbf{R}^\textrm{(eq)}_\textrm{last} \textbf{A}_\textrm{out} = \textbf{R}^\textrm{(eq)}_\textrm{out}$. Finally, $2\pi \Delta \textrm{Tw} ( \textbf{R}_\textrm{in} \textbf{A}_\textrm{in}, \textbf{R}_\textrm{first} )$ and $2\pi \Delta \textrm{Tw} ( \textbf{R}_\textrm{last} \textbf{A}_\textrm{out}, \textbf{R}_\textrm{out} )$ are the twist angles of the entry and exit DNA segments with respect to their equilibrium orientations. 

Specifically, in the case of a DNA-stiffening protein that forms straight nucleoprotein filaments along the DNA, we have: $a_\textrm{pr} = A_\textrm{pr} / (b_0 K)$ and $c_\textrm{pr} = C_\textrm{pr} / (b_0 K)$, where $A_\textrm{pr}$ and $C_\textrm{pr}$ are the bending and twisting persistence lengths of protein-covered DNA, and $K$ is the number of DNA segments bound to a single protein. Furthermore, for such a protein $\textbf{A}_\textrm{in} = \textbf{A}_\textrm{out} = \textbf{I}$, where $\textbf{I}$ is the identity matrix. As a result, in mechanical equilibrium all of the rotation matrices describing orientations of the protein-bound DNA segments have identical values: $\textbf{R}^\textrm{(eq)}_\textrm{in} = \textbf{R}^\textrm{(eq)}_\textrm{first} = ... = \textbf{R}^\textrm{(eq)}_\textrm{last} = \textbf{R}^\textrm{(eq)}_\textrm{out}$ (i.e., protein forms straight filaments).

Additional details regarding the mathematical description of the nucleoprotein complexes contribution into the total conformational energy of DNA can be found in Appendix~\ref{Appendix-A}.

Finally, it should be noted that in all of the calculations presented below, the size of the DNA segments was set to be equal to $q = 1.5$ bp for all of the DNA structural states and the DNA length was $\sim 4.7$ kbp (a total of $N = 3073$ segments in the discretized polymer chain representing DNA). The values of the rest of the model parameters are listed in Table~\ref{tab:DNA-parameters} (for bare DNA segments) and Table~\ref{tab:Protein-parameters} (for different types of nucleoprotein complexes).

The source code of the programs that have been used to obtain the results presented below can be downloaded from the personal web-sites of the authors: \href{http://www.artem-efremov.org}{AKE} and \href{https://www.physics.nus.edu.sg/~biosmm/}{YJ}.

\begin{table*}
\resizebox{\textwidth}{!}{\begin{minipage}{\textwidth}
\caption{\label{tab:Protein-parameters} Values of the model parameters for different nucleoprotein complexes studied in this work, which were used in the transfer-matrix calculations.}
\begin{ruledtabular}
\begin{tabular}{ccccccc}
\noalign{\smallskip}
$\substack{ \textbf{Protein} }$ & 
$\substack{ \textbf{Bending rigidity,}_{} \\ \bm{a_\textrm{pr}}^{ } }$ &
$\substack{ \textbf{Twisting rigidity,}_{} \\ \bm{c_\textrm{pr}}^{ } }$ &
$\substack{ \textbf{Binding energy}_{} \\ \textbf{to DNA, } \bm{\mu_\textrm{pr}} \textbf{(} \bm{k_\textrm{B} T} \textbf{)} }$ &
$\substack{ \textbf{Cooperative binding}_{} \\ \textbf{energy, } \bm{J_\textrm{pr}} \textbf{(} \bm{k_\textrm{B} T} \textbf{)} }$ &
$\substack{ \textbf{Linking number}_{} \\ \textbf{change, } \bm{\Delta \textrm{Lk}_\textrm{pr}} }$ &
$\substack{ \bm{A_\textrm{in}} \textbf{, } \bm{A_\textrm{out}} \textbf{, } \bm{A_\textrm{ht}} \textbf{ and } 
\bm{A_j} \\ \textbf{Euler rotation matrices} \footnote{Matrices $\textbf{A}_j$ describe the relative orientations of DNA segments inside the nucleoprotein complexes, see Appendix~\ref{Appendix-A} and Appendices~\ref{sec:DNA-stiffening}-\ref{sec:DNA-wrapping} for more details.}_{} }$ \\
\noalign{\smallskip}
\hline
\noalign{\medskip}

DNA-stiffening & 33.3 & 33.3 & 3.0 & 2.0 & N/A & \makecell[t]{ $\textbf{A}_\textrm{in} = \textbf{A}_\textrm{out} = 
\textbf{A}_\textrm{ht} = $ \\ $ = \textbf{A}_j = \textbf{I} (0,0,0)$ } \\[17pt]

DNA-bending	& 33.3 & 33.3 & 2.0 & 0.0 & N/A & \makecell[t]{ $\textbf{A}_\textrm{in} = \textbf{A}_\textrm{out} =
\textbf{A}_1 = $ \\ $= \textbf{I} (0,0,0)$ \\[2pt] $\textbf{A}_2 (\pi,0.2,\pi)$} \\[32pt]

Nucleosomes & 33.3 & 33.3 & 40.0 & 0.0 & --1.2 & \makecell[t]{ $\textbf{A}_\textrm{in} (0,2.12,-0.79)$ \\[2pt] $\textbf{A}_\textrm{out} (-0.79,2.12,0)$ \\[2pt] $\textbf{A}_j = \textbf{I} (0,0,0)$ } \\[32pt]

L-tetrasomes & 33.3 & 33.3 & 26.3 & 0.0 & --0.73 & \makecell[t]{ $\textbf{A}_\textrm{in,L} (0,2.26,-1.11)$ \\[2pt] $\textbf{A}_\textrm{out,L} (-1.11,2.26,0)$ \\[2pt] $\textbf{A}_j = \textbf{I} (0,0,0)$ } \\[32pt]

R-tetrasomes & 33.3 & 33.3 & 24.0 & 0.0 & +1.0 & \makecell[t]{ $\textbf{A}_\textrm{in,R} (0,2.26,1.11)$ \\[2pt] $\textbf{A}_\textrm{out,R} (1.11,2.26,0)$ \\[2pt] $\textbf{A}_j = \textbf{I} (0,0,0)$ } \\[1pt]

\end{tabular}
\end{ruledtabular}
\end{minipage} }
\end{table*}

\subsection{Main assumptions of the theory}
\label{sec:theory_assumptions}

In this section we would like to summarize all of the main assumptions used to derived Eq.~\eqref{DNA-protein-partition-function-intro-4} for the partition function of DNA interacting with proteins, which is important for understanding of potential applications that can be solved using the transfer-matrix formalism described above.

First of all, in order to derive mathematical formulas for the elements of the DNA transfer-matrix, $\textbf{L}$, in this study it was assumed that nucleoprotein complexes have fixed 3D structures, see Appendices~\ref{Appendix-A}-\ref{Appendix-E}. Therefore, application of the current theoretical framework should be restricted mainly to DNA-protein assemblies that have a well-defined conformation. While this assumption serves as a good first level of approximation to the description of DNA interaction with many different types of proteins, it should be noted that some nucleoprotein complexes may be very flexible, possessing more than one stable conformation. In this case, the formulas presented in this study should be accordingly modified to accurately depict force- and torque-dependent behaviour of such complexes.

Furthermore, the above assumption of a fixed nucleoprotein complex structure implies that the current model does not take into consideration cases of partial proteins binding to DNA, which may take place under sufficiently strong forces and torques applied to DNA. For instance, existing experimental data show that at $2-3$ pN force, the outer turn of DNA interacting with histone octamers can be unwrapped from nucleosome complexes, while the inner turn remains stably attached to the protein core, resulting in a partially bound state of histone octamers to a mechanically stretched DNA \cite{Mihardja_2006}.

However, despite the above limitations, it is very easy to make necessary modifications to the theory in order to incorporate into the model partial binding of proteins to DNA and multiple conformations of nucleoprotein complexes formed by flexible proteins. This can be achieved simply by adding new DNA segment states and/or new elements into the DNA transfer-matrix in the same way as it has been done in the case of DNA interaction with histone tetramers that can flip between the two alternative conformations, see more details in Section~\ref{sec:results-DNA-wrap}, Appendix~\ref{sec:DNA-wrapping} and at the end of Appendix~\ref{Appendix-F}.

The next assumption that has been used in our derivations is the propensity of DNA-binding proteins to form nucleoprotein complexes only on B-form DNA. While there is not much information regarding the proteins' abilities to bind to alternative DNA structures, such as L- or P-DNA, it should be noted that it will be rather straightforward to include newly discovered protein-L-DNA and protein-P-DNA complexes into the transfer-matrix calculations again by introducing additional DNA segment states into the model.

Finally, to minimize the formulas' complexity, in this study we have not considered in detail the DNA and proteins' volume exclusion effect. As a result, the current theory cannot be applied to scenarios in which the volume exclusion plays a dominant role in determining the global DNA conformation. However, in principle, it is still possible to include such an effect in a mathematically rigorous way into the transfer-matrix formalism by making use of Hubbard-Stratonovich transformation that results in addition of an auxiliary fluctuating field to the DNA total conformational energy, see ref.~\cite{Kleinert_2009} for details.

In addition to the above assumptions, we also used in this work the Fuller's approximate formula for the calculation of the DNA writhe number, see Section~\ref{sec:energy_terms}. From the existing theoretical studies, it is known that by utilizing this formula it is possible to obtain rather accurate estimations of the DNA linking number change for the most of DNA conformations up to the buckling transition point when DNA starts to develop supercoiled plectonemes \cite{Fuller_1978, Fain_1997, Bouchiat_2000, Efremov_2016, Efremov_2017}. However, as soon as plectonemes start to appear in DNA, the Fuller's formula fails to provide correct values for the DNA writhe number, which restricts application of the transfer-matrix formalism up to the buckling transition point.

Nevertheless, as has been shown in our previous studies \cite{Efremov_2016, Efremov_2017}, it is still possible to use the transfer-matrix calculations to predict transition boundaries between different structural states of DNA, including the torque-induced change between the extended and supercoiled DNA conformations. Furthermore, since binding of DNA-bending and DNA-wrapping proteins to DNA results in formation of solenoid-like complexes for which the Fuller's formula works rather well \cite{Fain_1997, Bouchiat_2000}, it is likely that the transfer-matrix formalism also can be used to obtain accurate predictions regarding the behaviour of DNA compacted by these types of proteins under force and torque constraints. This broadens application of the transfer-matrix theory to many interesting DNA-protein interaction scenarios, which are frequently studied in single-molecule experiments.

\section{Results}
\subsection{Mechanical response of bare DNA to force and torque constraints}
\label{sec:results-DNA-bare}

Using the above transfer-matrix approach, we first investigated the effects of force and torque constraints on the conformation of bare DNA and its transition between different structural states, such as B-, L- and P-DNA, in the absence of DNA-binding proteins in solution. 

It should be noted that although the case of bare DNA has been discussed in detail in our previous studies \cite{Efremov_2016, Efremov_2017}, it is used in this work as a control against which all other scenarios describing DNA interactions with proteins are compared. For this reason, we briefly recall in this section what is known about behaviour of a mechanically stretched and twisted bare DNA.

\begin{figure}[!htb]
\includegraphics[width=0.9\textwidth]{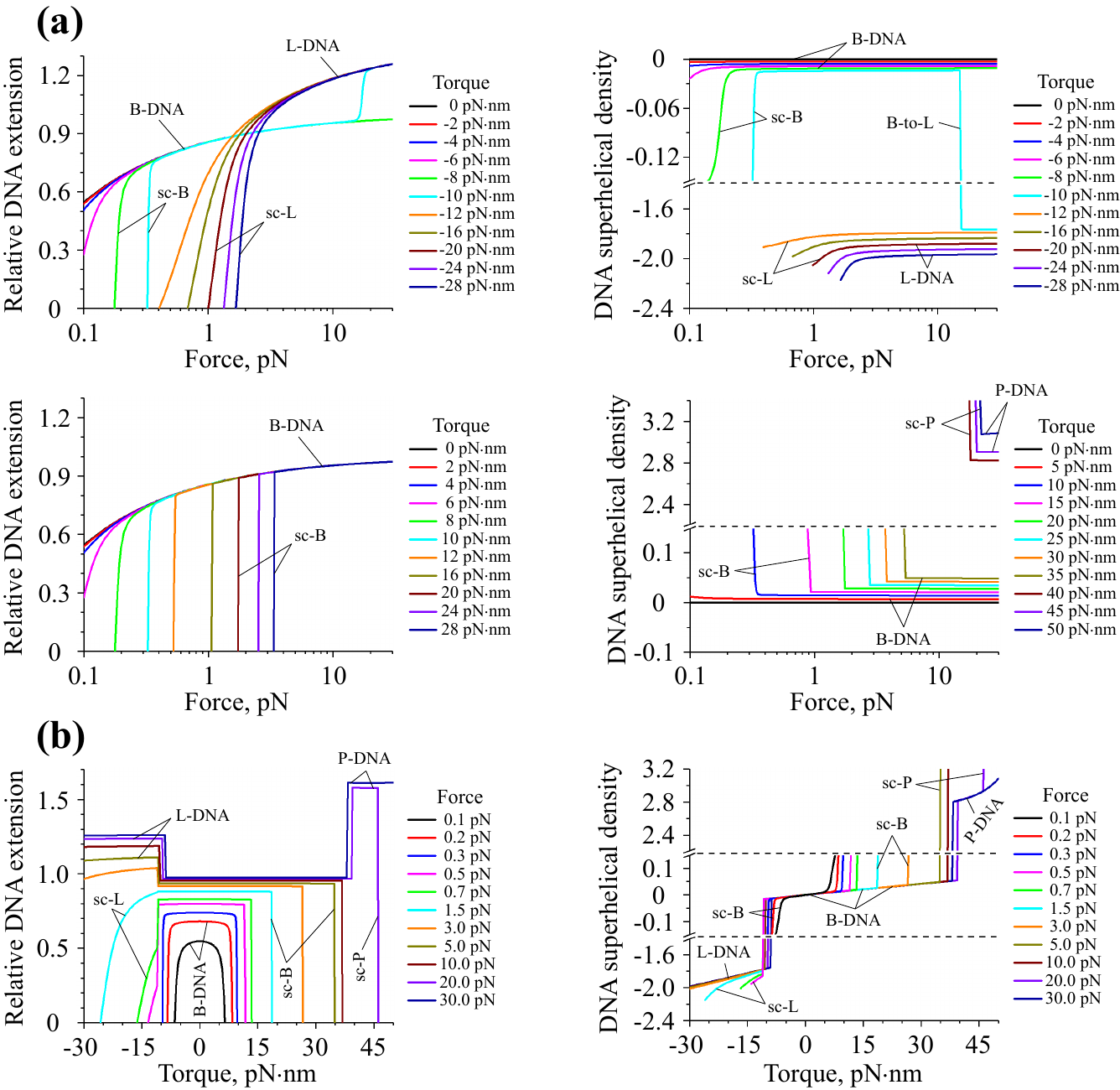}
\caption{
\textbf{Mechanical response of bare DNA to the applied force and torque constraints in the absence of DNA-binding proteins.} The figure shows \textbf{(a)} force-extension [$z (f)|_{\tau = \tau_0}$] and force-superhelical density curves [$\sigma (f)|_{\tau = \tau_0}$] as well as \textbf{(b)} torque-extension [$z (\tau)|_{f = f_0}$] and torque-superhelical density curves [$\sigma (\tau)|_{f = f_0}$] obtained at different values of the force, $f$, and torque, $\tau$, exerted to the DNA. From panel \textbf{(a)}, it can be seen that application of a sufficiently large torque ($|\tau| \geq 6$ pN$\cdot$nm) leads to collapsing of bare DNA, which is accompanied by development of supercoiled DNA structures. Panel \textbf{(b)} provides additional details, showing that at forces $f < 0.5$ pN all of the torque-extension curves have symmetric profiles with respect to both positive and negative torques, while at larger forces of $f \sim 0.5-0.7$ pN this symmetry breaks due to B-DNA switching into alternative L- and P-DNA structures, which results in the respective change of the DNA superhelical density. In all panels, the DNA extension is normalized to the total contour length of DNA in B-form. Abbreviations sc-B, sc-L and sc-P are used to indicate supercoiled states of B-, L- and P-DNA, respectively.
}
\label{fig2}
\end{figure}

By substituting the values of the model parameters listed in Table~\ref{tab:DNA-parameters} that describe the physical properties of bare DNA into Eq.~\eqref{DNA-protein-partition-function-intro-4}, it is not hard to obtain the DNA force-extension curves, $z (f) |_{\tau = \tau_0}$, and force-superhelical density curves, $\sigma (f) |_{\tau = \tau_0}$, at various torque constraints ($\tau = \tau_0$), which are shown in Figure~\ref{fig2}(a). The top and the bottom panels of Figure~\ref{fig2}(a) demonstrate the force-extension and force-superhelical density curves for the case of negative ($\tau < 0$ pN$\cdot$nm) and positive torques ($\tau > 0$ pN$\cdot$nm), respectively. 

From the graphs, it can be seen that the mechanical response of bare DNA to the applied force and torque constraints is highly non-linear. While at small torques ($-5 \leq \tau \leq 5$ pN$\cdot$nm) the DNA force-extension curves do not deviate much from the one corresponding to a torsionally relaxed DNA ($\tau = 0$ pN$\cdot$nm), application of stronger torsional stress ($|\tau| > 5$ pN$\cdot$nm) results in rapid decrease of the DNA extension as soon as the stretching forces, $f$, drops below a certain threshold value, see Figure~\ref{fig2}(a), left top and bottom panels. Calculations of the DNA superhelical density, $\sigma = \Delta \textrm{Lk} / \textrm{Lk}_0$, as a function of the applied force and torque constraints show that such torque-induced DNA collapsing is accompanied by a simultaneous steep change of the DNA superhelical density [Figure~\ref{fig2}(a), right top and bottom panels], resembling typical behaviour of strongly twisted DNA that undergoes transition into a compact supercoiled conformation, which is typically observed in single-DNA manipulation experiments \cite{Strick_1996}.

Furthermore, from the left panel of Figure~\ref{fig2}(b) showing the DNA torque-extension curves, $z (\tau) |_{f = f_0}$, calculated at various force constraints ($f=f_0$), it can be seen that the DNA folding into the supercoiled conformation occurs both at positive and negative torques in a symmetric manner at low stretching forces ($f < 0.5$ pN). However, at larger forces ($f \geq 0.5-0.7$ pN) this symmetry breaks as stronger stretching makes it harder for DNA to form compact supercoiled structures; thus, preventing release of the accumulated DNA elastic twist energy via the DNA supercoiling process. As a result, transition of DNA from B-form into alternative L- and P-DNA structures becomes a more energetically favourable way for the DNA twist elastic energy relaxation at large stretching forces ($f \geq 0.5$ pN). 

It is not hard to see the effects of these DNA structural transitions on the left panel of Figure~\ref{fig2}(b) as they manifest themselves in an abrupt change of the twist-extension curves' behaviour. For example, at forces $f \geq 5$ pN and high negative torques ($\tau < -11$ pN$\cdot$nm) the DNA extension increases by $\sim 1.1-1.3$ times comparing to the case of a torsionally relaxed B-DNA ($\tau = 0$ pN$\cdot$nm), indicating DNA transition into alternative L-DNA form, which is accompanied by a simultaneous DNA superhelical density drop to the value of $\sigma \sim -2.0$ -- see the right panel on Figure~\ref{fig2}(b) showing the DNA torque-superhelical density curves, $\sigma (\tau) |_{f = f_0}$, calculated at various force constraints. Likewise, at high positive torques ($\tau > 35$ pN$\cdot$nm) the DNA extension becomes $\sim 1.6$ times longer than that of a torsionally relaxed DNA ($\tau = 0$ pN$\cdot$nm), designating the DNA transition into P-DNA state, which is accompanied by a simultaneous large DNA superhelical density increase to the value of $\sigma \sim 3.0$, see the right panel on Figure~\ref{fig2}(b).

Similarly to B-DNA, both L- and P-DNA experience buckling transition from the extended to a compact supercoiled conformation, which is indicated on the left panel of Figure~\ref{fig2}(b) by steep decrease of the DNA extension at large negative and positive torques as soon as the applied force drops below a certain threshold, whose value is slightly larger for L-DNA ($\sim 1.5$ pN) as compared to the B-DNA case and even more higher for P-DNA ($\sim 20$ pN) due to higher elasticities of L- and P-DNA forms.

Altogether, the above results demonstrate that the global conformation and structure of bare DNA are highly sensitive to mechanical constraints applied to it, in good agreement with the exisiting experimental data previously reported in multiple single-molecule  studies \cite{Allemand_1998, Bryant_2003, Deufel_2007, Forth_2008, Sheinin_2009, Sheinin_2011, Oberstrass_2012}.

\subsection{Effects of DNA-stiffening proteins on the DNA mechanical response to force and torque constraints}
\label{sec:results-DNA-stiff}

Next, we used the transfer-matrix formalism to investigate the effects of force and torque constraints on DNA interaction with DNA-stiffening proteins, which upon binding to DNA form rigid nucleoprotein filaments that increase the DNA bending persistence lengths, and presumably the DNA twisting rigidity \cite{Noort_2004, Liu_2010, Lim_2012, Winardhi_2012, Laurens_2012, Lim_2013, Qu_2013, Winardhi_2014, Efremov_2015}. For this purpose, we carried out calculations in which the bending and twisting persistence lengths of protein-covered DNA were set to $A_\textrm{pr} = 200$ nm and $C_\textrm{pr} = 200$ nm, respectively, with the value of the bending persistence length, $A_\textrm{pr}$, falling in the range of $100$ nm $< A_\textrm{pr} < 500$ nm previously reported for different types of DNA-stiffening proteins \cite{Noort_2004, Liu_2010, Lim_2012, Winardhi_2012, Laurens_2012, Lim_2013, Qu_2013, Winardhi_2014, Efremov_2015}.

In the calculations, the proteins were allowed to bind to any place on the DNA as soon as the corresponding DNA section was in B-form (i.e., proteins interact only with B-form DNA), and each DNA-bound protein was assumed to occupy $K = 12$ DNA segments [$\sim 18$ base-pairs, see schematic Figure~\ref{fig1}(c)], which is a typical DNA binding site size for many known DNA-stiffening proteins. Having at hand the bending and twisting persistence lengths of protein-covered DNA, and the binding site size of the proteins, it is then straightforward to find the values of dimensionless bending and twisting elasticities of the entry and exit DNA segments of nucleoprotein complexes: $a_\textrm{pr} = A_\textrm{pr} / (K b_0) = c_\textrm{pr} = C_\textrm{pr} / (K b_0) = 33.3$ (see Table~\ref{tab:Protein-parameters}), which were used in all of the computations presented below.

Finally, formation of nucleoprotein complexes on DNA was associated with the DNA-protein interaction energy of $\mu_\textrm{pr} = 3.0$ $k_\textrm{B}T$. In addition, since it is known that DNA-stiffening proteins often assemble into nucleoprotein filaments on DNA through cooperative interaction with each other \cite{Noort_2004, Liu_2010, Lim_2012, Winardhi_2012, Laurens_2012, Lim_2013, Qu_2013, Winardhi_2014, Efremov_2015}, a cooperative binding energy of $J_\textrm{pr} = 2.0$ $k_\textrm{B}T$ between proteins occupying neighbouring DNA binding sites was introduced into the transfer-matrix calculations.

\begin{figure}[!htb]
\includegraphics[width=0.9\textwidth]{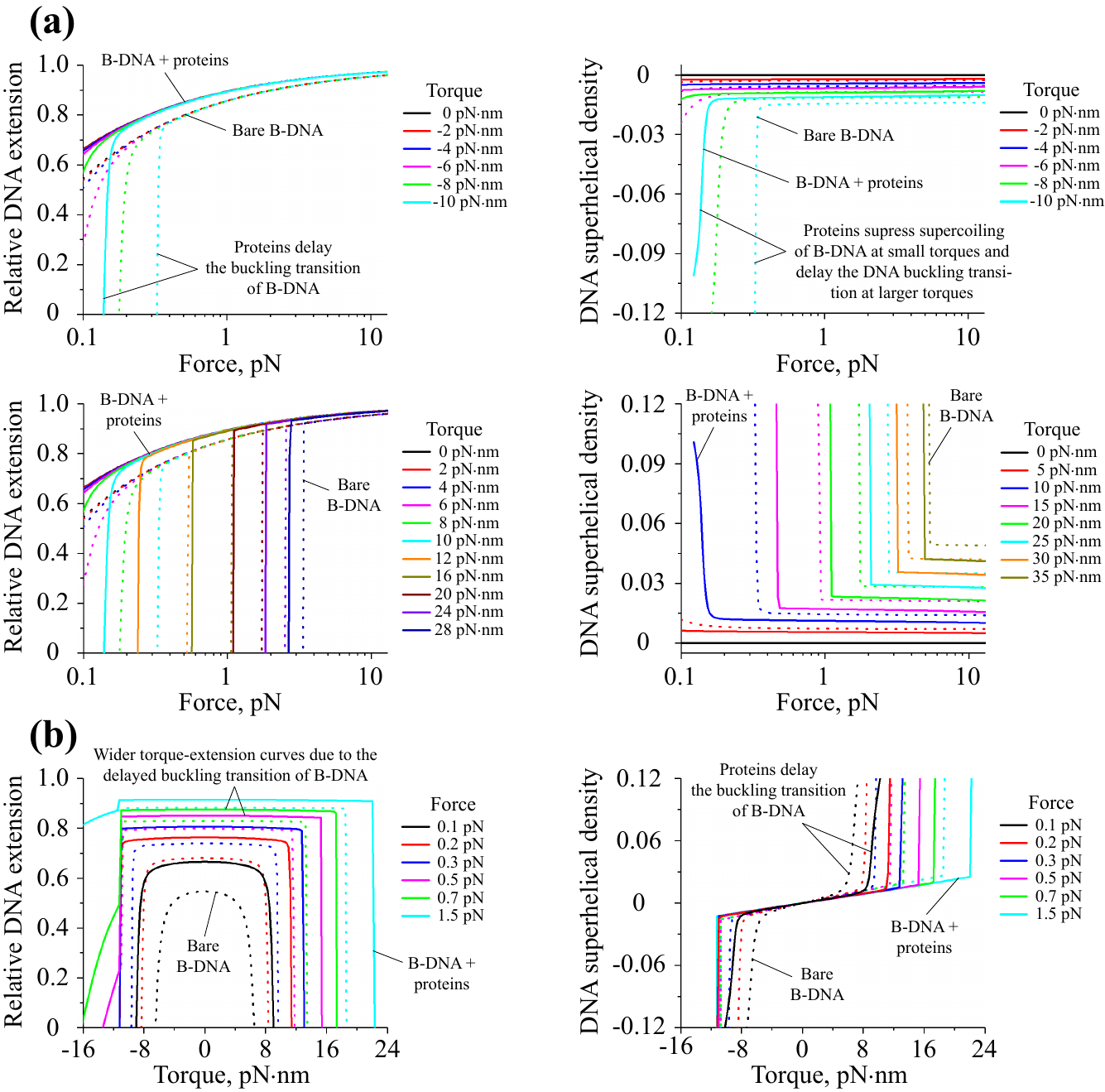}
\caption{
\textbf{Mechanical response of DNA to the applied force and torque constraints in the presence of DNA interactions with DNA-stiffening proteins.} The figure shows \textbf{(a)} force-extension and force-superhelical density curves obtained at different values of the torque, $\tau$, as well as \textbf{(b)} torque-extension and torque-superhelical density curves obtained at different values of the force, $f$, exerted to the DNA. Solid curves demonstrate the behaviour of DNA in the presence of nucleoprotein complexes formation by DNA-stiffening proteins; whereas, dotted curves indicate mechanical response of bare DNA under the same force and torque constraints. As can be seen from comparison between the force-extension and force-superhelical density curves calculated for protein-covered and bare DNA, formation of rigid nucleoprotein filaments by DNA-stiffening proteins results in either complete disappearance or leftward shift of the DNA buckling transition point to smaller values of the applied force, indicating delay in the formation of supercoiled DNA structures. Such protein-induced suppression of the DNA supercoiling can be also clearly observed from the widening of the torque-extension and torque-superhelical density curves in the presence of DNA interactions with DNA-stiffening proteins in comparison to the case of bare DNA. In all panels, the DNA extension is normalized to the total contour length of DNA in B-form.
}
\label{fig3}
\end{figure}

After substituting the above model parameters into the transfer-matrix, $\textbf{L}$, and boundary condition matrix, $\textbf{Y}$, describing DNA interaction with DNA-stiffening proteins [see Eq.~\eqref{Block-matrices-DNA-stiffening} in Appendix~\ref{sec:DNA-stiffening}], we found the values of the observables, such as the DNA extension and superhelical density, in order to investigate a potential role of DNA-stiffening proteins in modulation of the DNA conformation under force and torque constraints. The final results of the computations are shown on Figures~\ref{fig3} and \ref{figS1}. 

From the direct comparison between the force-extension curves calculated for bare DNA (dotted lines) and protein-covered DNA (solid lines) displayed on the left panels of Figure~\ref{fig3}(a), it can be seen that formation of rigid nucleoprotein filaments on DNA, as expected, results in increased extension of a torsionally relaxed DNA at low forces ($f \sim 0.1$ pN) due to the higher bending persistence length of the protein-covered DNA. In addition, the force-extension curves of protein-bound DNA demonstrate rather substantial shift in their buckling transition point at which DNA starts to collapse into a compact conformation towards lower values of the applied stretching force. This result indicates that nucleoprotein filaments assembled on DNA can delay or even completely inhibit development of supercoiled DNA structures. Indeed, the force-superhelical density curves of protein-covered DNA exhibit very similar shifts towards the lower values of the stretching force, validating that DNA interaction with DNA-stiffening proteins has an adverse effect on the formation of supercoiled DNA structures, see the right panels on Figure~\ref{fig3}(a).

Such DNA behaviour can be easily understood by recalling that DNA folding into compact supercoiled structures is initiated by DNA buckling -- formation of initial DNA loops, which eventually develop into supercoiled DNA plectonemes. Since this process requires DNA bending at the buckling site, it is clear that DNA-stiffening nucleoprotein filaments will be preventing formation of such DNA loops unless the applied torsional stress is sufficiently high to overcome the nucleoprotein filaments' resistance to the bending. As a result, onset of the DNA supercoiling transition will be delayed in the presence of DNA-stiffening proteins in solution.

The torque-extension curves shown on the left panels of Figures~\ref{fig3}(b) and \ref{figS1}(b) provide further details regarding the effect of stiff nucleoprotein filaments onto the global conformation of DNA, demonstrating that the most significant changes, such as delay in the DNA buckling transition that results in widening of the torque-extension curves, take place mainly at low forces ($f \leq 3$ pN); whereas at higher forces the mechanical response of the protein-covered DNA to force and torque constraints is practically identical to that of a bare DNA in the absence of proteins in solution [compare the left panels of Figures~\ref{fig2}(b) and \ref{figS1}(b)].

Interestingly, from the left and right panels of Figures~\ref{fig2}(b) and \ref{figS1}(b) it can be seen that binding of DNA-stiffening proteins to DNA has practically negligible suppressing effect on the DNA transitions from B- to L- or P-DNA forms. The main reason for this is that the average binding energy of the proteins to DNA per single base-pair ($< 1$ $k_\textrm{B} T$) is much lower than the free energies $\mu_u$ ($u = $ L or P) associated with the DNA transitions between different structural states ($\mu_u \sim 3-20$ $k_\textrm{B} T$ per base-pair, see Table~\ref{tab:DNA-parameters}). As a result, this does not allow proteins to efficiently interfere with the DNA structural transitions unless the protein binding energy to DNA is very high.

In addition to the above curves characterizing the DNA behaviour under mechanical constraints in the presence of DNA-stiffening protein, we also calculated the average DNA occupancy fraction by proteins as a function of the force and torque applied to the DNA, see Figure~\ref{figS5}(a). As expected for $\mu_\textrm{pr} = 3.0$ $k_\textrm{B}T$ binding energy and $J_\textrm{pr} = 2.0$ $k_\textrm{B}T$ cooperative binding energy of proteins to DNA used in the calculations, a large part of the DNA is occupied by  nucleoprotein complexes. Nevertheless, previously reported phenomenon of enhancement of the protein binding to DNA with increase in the stretching force exerted on the DNA \cite{Yan_2003} still can be clearly seen on all of the panels in Figure~\ref{figS5}(a). In contrast to the stretching force, application of stronger torsional stress to DNA promotes proteins dissociation from it, see the middle and right panels of Figure~\ref{figS5}(a). Development of supercoiled DNA structures at low forces ($f \leq 3$ pN) and high torques ($\tau > 10$ pN$\cdot$nm) further destabilizes nucleoprotein complexes formed on DNA by DNA-stiffening proteins, resulting in dramatic decrease of the DNA occupancy fraction.

Such unusual behaviour of DNA-stiffening proteins is tightly related to the changes in the DNA entropic elasticity taking place upon proteins interaction with DNA. Namely, formation of stiff nucleoprotein filaments leads to restriction of available conformations that can be taken by protein-covered DNA. As a result, there exists an entropic penalty for the binding of DNA-stiffening proteins to DNA at low forces at which DNA tends to assume more coiled conformations. On the other hand, application of stronger tension  to DNA leads to a more extended DNA conformation, resulting in reduction of the entropic penalty associated with the proteins' DNA-stiffening effect. Thus, in general, mechanical stretching of DNA promotes formation of nucleoprotein filaments by DNA-stiffening proteins.

As for the role of torque in regulation of the DNA-stiffening proteins' affinity to DNA, it is clear that rigid nucleoprotein filaments have smaller propensity to twist under applied torsional stress. This leads to a smaller change in the total DNA linking number in the case of protein-covered DNA comparing to the case of bare DNA. Thus, the potential energy associated with the DNA twisting will be smaller for bare DNA than for protein-covered DNA, suggesting that nucleoprotein complexes will be losing their stability under the applied torque. Eventually, this will result in partial dissociation of DNA-stiffening proteins from DNA.

At the buckling transition point, proteins interaction with DNA is further compromised by the DNA bending into loops that prevent formation of extended nucleoprotein filaments by DNA-stiffening proteins. This leads to apparent reduction of the proteins' binding affinity to DNA, which is manifested by the drop in the DNA occupancy fraction curves shown on the middle and right panels of Figure~\ref{figS5}(a) at the DNA buckling transition point.

\subsection{Effects of DNA-bending proteins on the DNA mechanical response to force and torque constraints}
\label{sec:results-DNA-bend}

We further investigated the effects of force and torque constraints on the DNA-binding properties of DNA-bending proteins and explored the role of this type of proteins in regulation of the global DNA conformation. As a classical example of a DNA-bending protein, we used \textit{E. coli} integration host factor (IHF) as a model DNA-architectural protein in the transfer-matrix calculations, which is known to introduce sharp DNA bending at its binding site \cite{Rice_1996}.

Following the existing structural and single-molecule data for IHF-DNA nucleoprotein complexes, the binding site size of IHF was set to $36$ bp (i.e., $K=24$ DNA segments) in all our computations, with the bending angle of DNA due to formation of the nucleoprotein complex being $150^\circ$, see schematic Figure~\ref{fig1}(d) and ref.~\cite{Rice_1996, Shimin_2013}. As in the case of DNA-stiffening proteins, in this section we assumed that IHF binds only to B-form DNA. To reproduce the experimentally measured detachment force at which IHF dissociates from DNA ($\sim 0.8$ pN \cite{Shimin_2013}), the IHF binding energy to DNA was put equal to $\mu_\textrm{pr} = 2.0$ $k_\textrm{B}T$. For the simplicity of calculations, in this study we did not consider the sequence-dependent affinity of IHF to DNA. As for the effective bending and twisting rigidities of IHF-DNA nucleoprotein complexes, $a_\textrm{pr}$ and $c_\textrm{pr}$, we used the same values for these model parameters as in the case of DNA-stiffening proteins described in the previous section, see Table~\ref{tab:Protein-parameters}.

\begin{figure}[!htb]
\includegraphics[width=0.9\textwidth]{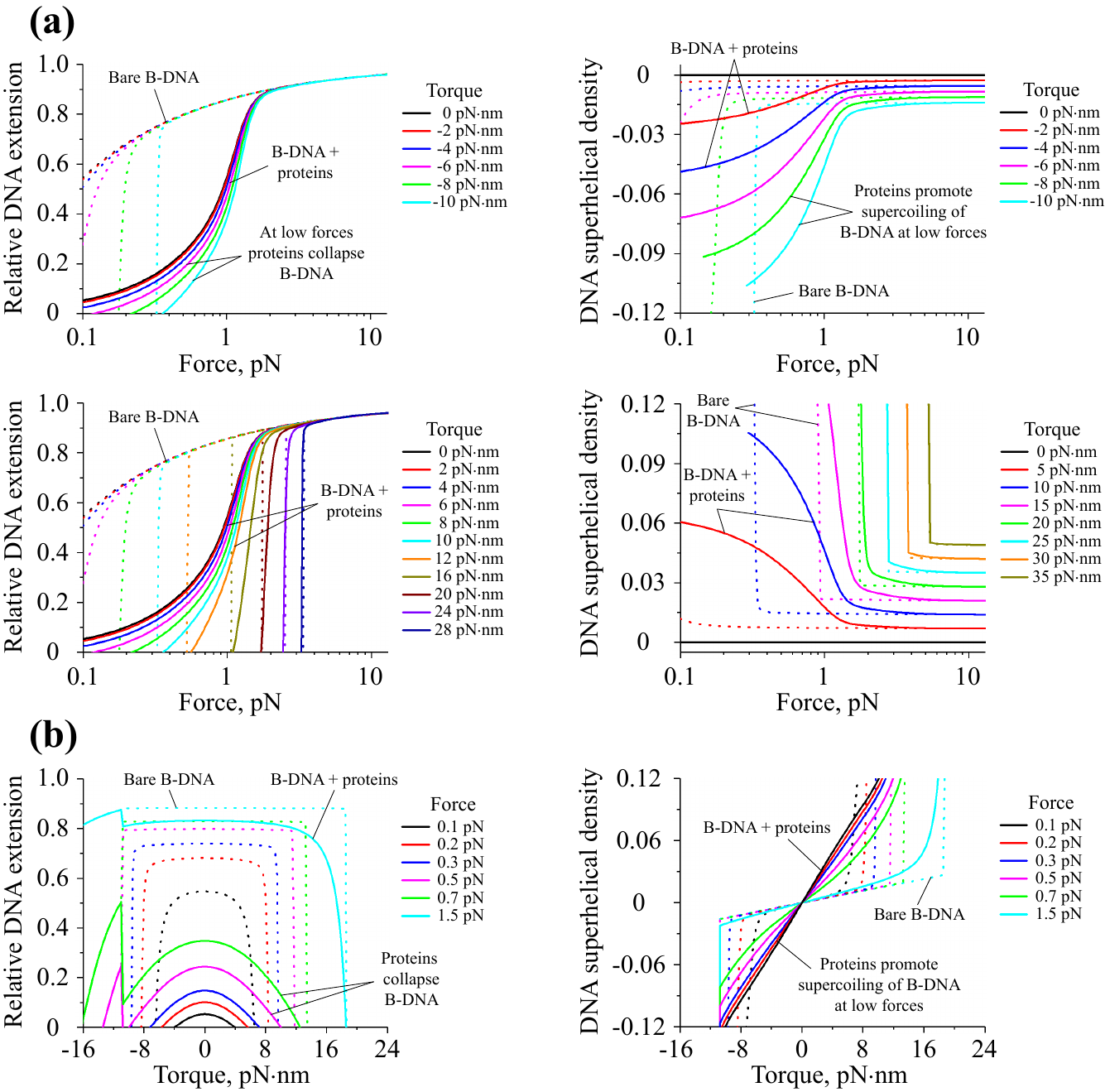}
\caption{
\textbf{Mechanical response of DNA to the applied force and torque constraints in the presence of DNA interactions with DNA-bending proteins.} The figure shows \textbf{(a)} force-extension and force-superhelical density curves obtained at different values of the torque, $\tau$, as well as \textbf{(b)} torque-extension and torque-superhelical density curves obtained at different values of the force, $f$, exerted to the DNA. Solid curves demonstrate the behaviour of DNA in the presence of nucleoprotein complexes formation by DNA-bending protein; whereas, dotted curves indicate mechanical response of bare DNA under the same force and torque constraints. From panel \textbf{(a)}, it can be seen that formation of nucleoprotein complexes by DNA-bending proteins results in DNA compaction at small forces ($f < 1$ pN), which is accompanied by a gradual increase in the magnitude of the DNA superhelical density that assumes either negative or positive sign depending on the direction of the applied torque. The left panel \textbf{(b)} provides further details, demonstrating that while having more compact shapes, the DNA torque-extension curves maintain their symmetry with respect to the torque sign up to the point where DNA experiences transition into alternative L-DNA structure at $\tau \sim -11$ pN$\cdot$nm, indicating that formed nucleoprotein complexes do not discriminate between positive or negative torques applied to the DNA. In all panels, the DNA extension is normalized to the total contour length of DNA in B-form.
}
\label{fig4}
\end{figure}

Substituting the above parameters into Eq.~\eqref{Block-matrices-DNA-bending}-\eqref{Block-matrices-DNA-bending-2} in Appendix~\ref{sec:DNA-bending} that describe the transfer-matrix, $\textbf{L}$, and boundary condition matrices, $\textbf{Y}$ and $\textbf{U}$, of DNA interacting with IHF proteins, we plotted the force- and torque-extension curves [$z (f) |_{\tau = \tau_0}$ and $z (\tau) |_{f = f_0}$] as well as the force- and torque-superhelical density curves of DNA [$\sigma (f) |_{\tau = \tau_0}$ and $\sigma (\tau) |_{f = f_0}$] at various force ($f = f_0$) and torque ($\tau = \tau_0$) constraints, see Figures~\ref{fig4} and \ref{figS2}.

The first obvious change in the conformation of DNA, which can be clearly seen from the force-extension curves calculated for protein-covered DNA (solid lines) shown on the left top and bottom panels of Figure~\ref{fig4}(a), is collapsing of DNA into a compact conformation  due to its interaction with IHF proteins that takes place at forces below $1$ pN in a wide range of the applied torque constraints ($-11 \leq \tau \leq 12$ pN$\cdot$nm). This is in stark contrast to the behaviour of bare DNA (dotted lines), which either stays in the extended conformation (at $-6 \leq \tau \leq 6$ pN$\cdot$nm torques) or undergoes supercoiling (at $| \tau | \geq 6$ pN$\cdot$nm torques), but only at considerably smaller forces than in the case of IHF-covered DNA.

Interestingly, the force-superhelical density curves of DNA interacting with IHF proteins reveal that application of even small torsional stress to the DNA ($| \tau | \leq 5$ pN$\cdot$nm) leads to development of supercoiled DNA conformations of the same sign as the applied torque [solid lines on the right top and bottom panels of Figure~\ref{fig4}(a)], which is again in sharp contrast to the bare DNA case where the superhelical density remains near zero in the same torque range (dotted lines on the same panels). This result indicates that although the IHF-mediated DNA bending does not have a preferential chirality at zero torque, it readily assumes left-handed / right-handed conformation in response to negative / positive torques applied to the DNA, suggesting that nucleoprotein complexes formed by IHF can easily flip between left- and right-handed structures. 

Furthermore, as can be seen from the right panels of Figure~\ref{fig4}(a), the magnitude of the  superhelical density of IHF-covered DNA experiences rather moderate increase with reducing stretching force in $\tau \in [-11, -6]$ and $\tau \in [6, 11]$ pN$\cdot$nm torque ranges. At the same time, bare DNA rapidly develops supercoils at these conditions, which result in the steep DNA superhelical density change. Thus, it can be concluded that IHF remains stably bound to DNA in this torque range, suppressing formation of supercoiled bare DNA structures that otherwise would form at forces $f < 1$ pN. Indeed, the DNA occupancy fraction curves shown on the left and middle panels of Figure~\ref{figS5}(b) demonstrate that the amount of DNA-bound IHF proteins stay at a constant level at low forces ($f < 1$ pN) in the broad range of the applied torque constraints ($-11 \leq \tau \leq 12$ pN$\cdot$nm).

Application of stronger positive torques ($\tau \geq 12$ pN$\cdot$nm) leads to the shift of the DNA occupancy fraction curves to higher force values, suggesting torque-induced stabilization of nucleoprotein complexes formed by IHF proteins, see the middle panel of Figure~\ref{figS5}(b). However, due to the failure of the Fuller's formula to describe the DNA writhe number beyond the buckling transition point, which results in potentially inaccurate prediction of the DNA occupancy fraction by the transfer-matrix calculations for strongly supercoiled DNA, it is not clear whether or not the DNA occupancy fraction curves eventually reach the same maximum level at $\tau \geq 12$ pN$\cdot$nm torques as in the case of lower torque values ($-11 \leq \tau \leq 12$ pN$\cdot$nm). Although, resemblance of the force-extension and force-superhelical density curves of IHF-covered DNA to those of bare DNA [Figure~\ref{fig4}(a)] suggests that IHF may partially dissociate from DNA at $\tau \geq 12$ pN$\cdot$nm torques due to the formation of supercolied bare DNA structures, similarly to the case of DNA-stiffening proteins described in the previous section.

Torque-extension curves shown on the left panels of Figure~\ref{fig4}(b) and \ref{figS2}(b) provide further details regarding the role of IHF proteins in force- and torque-dependent regulation of the DNA conformation. Namely, by comparing the results presented on the left panels of Figures~\ref{fig2}(b) and \ref{figS2}(b), it can be seen that at high forces ($f \geq 3$ pN) the torque-extension curves of DNA interacting with IHF proteins are identical to those of bare DNA, indicating that IHF binding to DNA is inhibited in this force range regardless of the magnitude of the applied torque, in full accordance with the torque-DNA occupancy fraction graphs plotted on the right panel of Figure~\ref{figS5}(b). At lower forces ($f \leq 1.5$ pN), however, formation of nucleoprotein complexes on DNA by IHF proteins leads to a very drastic change in the DNA conformation -- the DNA extension becomes significantly shorter than that of bare DNA due to the DNA bending by IHF proteins, -- see the left panels of Figures~\ref{fig2}(b) and \ref{figS2}(b), and also the left panel of Figure~\ref{fig4}(b) that displays the torque-extension curves of protein-covered DNA (solid lines) and bare DNA (dotted lines) on the same graph.

Application of torques from $\tau \in [-11, 12]$ pN$\cdot$nm range leads to further DNA extension drop with the rising torque magnitude, indicating increase in the IHF binding affinity to DNA and formation of more compact DNA-protein structures at stronger torques, see the left panel of Figure~\ref{fig4}(b). This result is in good agreement with the torque-DNA occupancy fraction curves shown on the right panel of Figure~\ref{figS5}(b) that demonstrate torque-induced promotion of the DNA interaction with IHF proteins at these conditions. Furthermore, from the left panel of Figure~\ref{fig4}(b) it can be seen that the shapes of the torque-extension curves in the case of IHF-covered DNA are much smoother than in the case of bare DNA, suggesting that in the former situation torque-induced decrease of the DNA extension is mainly caused by stronger DNA bending by IHF proteins rather than by formation of supercoiled structures typical for bare DNA. 

At larger positive torques ($\tau \geq 15$ pN$\cdot$nm), however, the torque-extension curves of DNA interacting with IHF proteins become practically identical to those obtained for bare DNA [compare the left panels of Figures~\ref{fig2}(b) and \ref{figS2}(b)], indicating IHF dissociation from the DNA due to formation of supercoiled bare DNA structures. Similarly, application of strong negative torques ($\tau < -11$ pN$\cdot$nm) also results in destabilization of nucleoprotein complexes formed by IHF proteins, but this time this happens due to the DNA transition into alternative L-DNA structural state, which is manifested by the increase in the DNA extension and the large drop in the DNA superhelical density.

Indeed, the right panel of Figure~\ref{figS2}(b) demonstrating the DNA superhelical density curves versus the applied torsional stress shows that at extreme negative ($\tau < -11$ pN$\cdot$nm) and positive ($\tau > 35$ pN$\cdot$nm) torques, where DNA experiences transitions into L- and P-DNA states, the curves look identical to those obtained in the case of bare DNA [Figure~\ref{fig2}(b), right panel]. This result suggests that similarly to DNA-stiffening proteins, IHF binding to B-DNA does not have a strong effect on the DNA transitions into alternative structural states, such as L- and P-DNA, as the protein binding energy to DNA measured per single DNA base-pair ($\mu_\textrm{pr} = 2.0$ $k_\textrm{B} T$ / $36$ bp $\approx 0.06$ $k_\textrm{B} T$ per bp) is much smaller than the free energy associated with the DNA transitions between different structural states ($\mu_u \sim 3-20$ $k_\textrm{B} T$ per base-pair, where $u = $ L or P, see Table~\ref{tab:DNA-parameters}).

Thus, it can be concluded that the most prominent changes in the conformation of DNA due to its interaction with DNA-bending proteins mostly take place in a narrow range of torques ($-11 \leq \tau \leq 16$ pN$\cdot$nm) and only at sufficiently low forces applied to DNA ($f < 1.5$ pN). Indeed, as Figure~\ref{figS5}(b) shows, only in this range the IHF density on the DNA becomes sufficiently high to alter its spatial organization. 

Finally, from the torque-superhelical density curves presented on the right panel of Figure~\ref{fig4}(b) it can be seen that the superhelical density of IHF-covered DNA switches from a negative value at negative torques to a positive value at positive torques, once again demonstrating that nucleoprotein complexes formed by DNA-bending proteins can easily flip between left- and right-handed conformations depending on the sign of the applied torque.

\subsection{Effects of DNA-wrapping proteins on the DNA mechanical response to force and torque constraints}
\label{sec:results-DNA-wrap}

The final group of architectural proteins, which we studied in this work, were DNA-wrapping proteins that not only promote formation of compact nucleoprotein complexes upon binding to DNA, but also make a well-defined fixed contribution to the total DNA linking number. In this section, we explore two famous examples of DNA-wrapping proteins: 1) histone octamers that wrap $\sim 147$ bp of DNA into a left-handed solenoidal structures known as a nucleosomes \cite{Luger_1997, Harp_2000, Talbert_2010}, and 2) histone (H3-H4)$_2$ tetramers that wrap $\sim 73$ bp of DNA into tetrasomes -- half nucleosome complexes that do not possess significant chiral preference, flipping between left- and right-handed conformations \cite{Hamiche_1996, Vlijm_2015}. Here we show how the effects of force and torque constraints applied to DNA influence on its interaction with these two protein complexes, which serve as specific examples of chiral and achiral DNA-wrapping proteins.

Since the X-ray crystal structure of nucleosomes has been previously solved \cite{Luger_1997, Harp_2000}, we used it as a template for constructing the model of nucleosome complexes, which is demonstrated on schematic Figure~\ref{fig1}(f). As for tetrasomes, their exact structure is not known yet. For this reason, we modelled them simply as a half (left-handed tetrasomes) or a mirrored half (right-handed tetrasomes) of nucleosome complexes that wrap $\sim 73$ bp of DNA \cite{Dong_1991}, see Figure~\ref{fig1}(e). 

Furthermore, due to the absence of experimental data regarding the elastic properties of nucleosomes and tetrasomes, the bending and twisting rigidities of the entry and exit DNA segments of these nucleoprotein complexes for simplicity were set equal to the same values as in the case of DNA-stiffening and DNA-bending proteins considered in the previous sections: $a_\textrm{pr} = c_\textrm{pr} = 33.3$, see Table~\ref{tab:Protein-parameters}. Although we would like to emphasize that in contrast to the case of  DNA-stiffening proteins, bending and twist rigidities of the entry and exit DNA segments of DNA-wrapping proteins play less significant roles in determining the mechanical response of protein-covered DNA to force and torque constraints, assuming that the protein binding energy to DNA, $\mu_\textrm{pr}$, is fixed at a constant value. Thus, $a_\textrm{pr}$ and $c_\textrm{pr}$ parameters have rather negligible impact on the results presented in this section.

In contrast, the binding energies of histone tetramers and octamers to DNA play the major roles in determining stabilities of tetrasome and nucleosome complexes under force and torque constraints applied to the DNA. While the exact values of these energies are not yet known, estimations based on single-molecule experimental data indicate that the value of the DNA-binding energy of histone octamers is likely to be of the order of $\sim 40$ $k_\textrm{B} T$ \cite{Yan_2007}.

In addition, single-DNA manipulation assays show that the energy associated with the unwrapping of the first DNA turn (known as outer nucleosome turn) from histone octamers equals to $12.0$ $k_\textrm{B} T$ \cite{Mihardja_2006}, and while there is no similar data for the remaining part of the nucleosome-bound DNA (inner nucleosome turn), the same experiments indicate that its affinity to histone octamers may approximately be twice as big \cite{Mihardja_2006, Kruithof_2008}. Hence, taken together, both outer and inner nucleosome turns add up to $\sim 40$ $k_\textrm{B} T$ of the nucleosome protein core binding energy to DNA, in good agreement with the chromatin stretching experiments reported in ref.~\cite{Yan_2007}. For this reason, in all our nucleosome calculations the DNA-binding energy of histone octamers to DNA was set equal to $\mu_\textrm{pr} = 40.0$ $k_\textrm{B} T$.

Furthermore, existing single-molecule data suggest that the inner nucleosome turn is formed by H3/H4-DNA interactions \cite{Sheinin_2013}. Thus, the energy associated with the unwrapping of the inner nucleosome turn may be regarded as the binding energy of (H3-H4)$_2$ histone tetramers to DNA. On top of that, experimental measurements reveal that left-handed tetrasomes have $2.3$ $k_\textrm{B}T$ energy preference over right-handed tetrasomes \cite{Vlijm_2015}. Based on these observations the DNA-binding energies for the left- and right-handed tetrasomes were put equal to $\mu_\textrm{pr}^\textrm{left} = 26.3$ $k_\textrm{B}T$ and $\mu_\textrm{pr}^\textrm{right} = 24.0$ $k_\textrm{B}T$ in all of the computations presented below.

Finally, as mentioned at the beginning of this section, formation of nucleosome and tetrasome complexes on DNA is accompanied by the change in the total DNA linking number by a well-defined amount, $\Delta \textrm{Lk}_\textrm{pr}$, per each nucleoprotein complex. From the existing experimental data it is known that the DNA linking number change due to the DNA wrapping around the nucleosome core is $\Delta \textrm{Lk}_\textrm{pr} \sim -1.2$ \cite{Prunell_1998, Vlijm_2015}; whereas, in the case of tetrasomes, experimentally measured DNA linking number changes associated with the left- and right-handed tetrasome conformations are equal to $\Delta \textrm{Lk}_\textrm{pr}^\textrm{left} = -0.73$ and $\Delta \textrm{Lk}_\textrm{pr}^\textrm{right} = +1.0$, respectively \cite{Vlijm_2015}. Thus, in all of the transfer-matrix calculations, assembly of nucleosome and tetrasome complexes on DNA was associated with the respective DNA linking number changes, see Table~\ref{tab:Protein-parameters}.

\begin{figure}[!htb]
\includegraphics[width=0.9\textwidth]{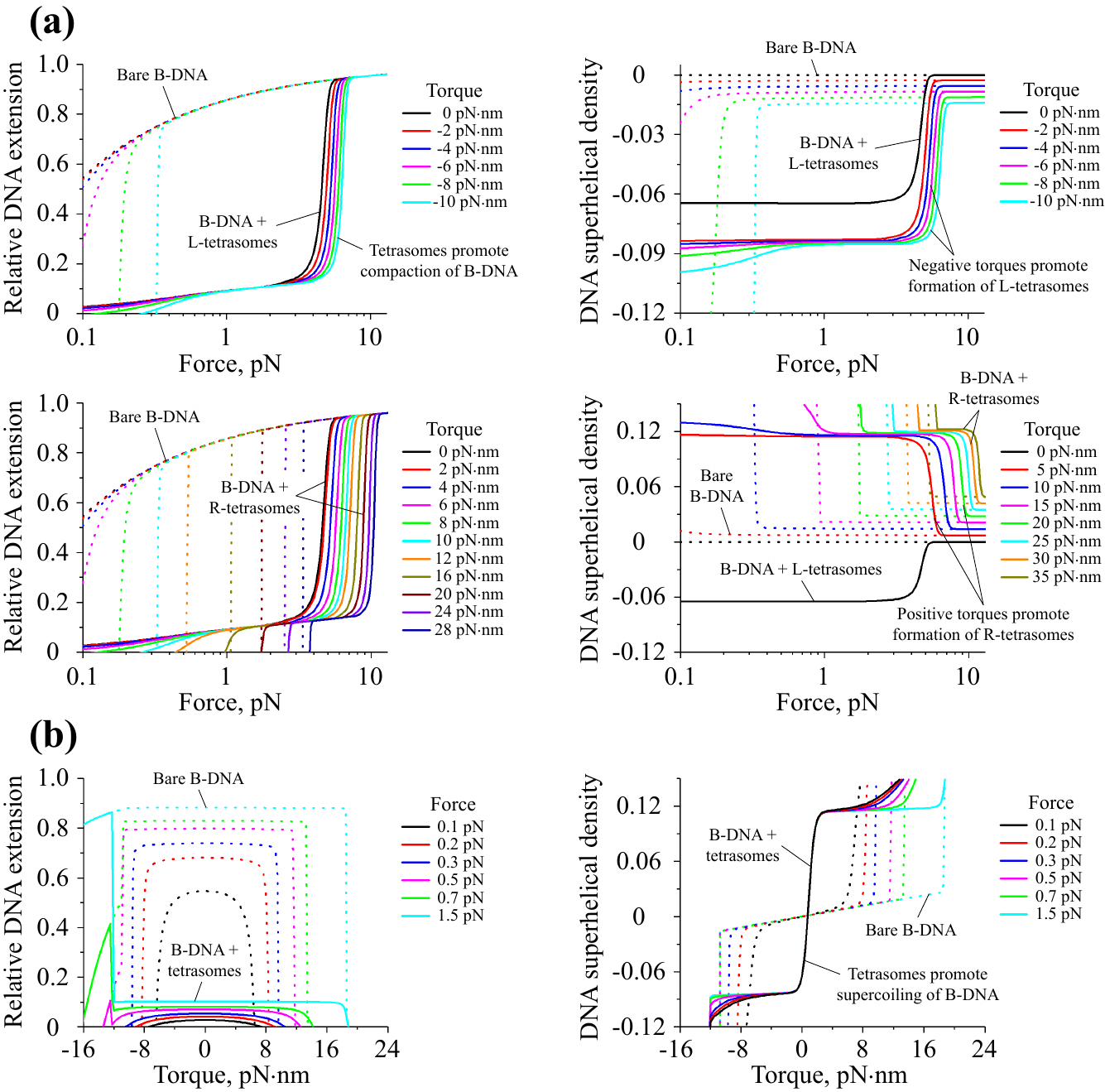}
\caption{
\textbf{Mechanical response of DNA to the applied force and torque constraints in the presence of DNA interactions with histone tetramers.} The figure shows \textbf{(a)} force-extension and force-superhelical density curves obtained at different values of the torque, $\tau$, as well as \textbf{(b)} torque-extension and torque-superhelical density curves obtained at different values of the force, $f$, exerted to the DNA. Solid curves demonstrate the behaviour of DNA in the presence of histone tetramers; whereas, dotted curves indicate mechanical response of bare DNA under the same force and torque constraints. As can be seen from the top and bottom plots on panel \textbf{(a)}, formation of tetrasome complexes on DNA leads to the molecule collapsing into a compact conformation, which is accompanied by the change in the DNA superhelical density, whose sign depends on the magnitude and direction of the applied torque. While tetrasomes can easily switch between the left- and right-handed structures, their slight preference to assume the left-handed conformation results in somewhat asymmetric behaviour of the DNA force-superhelical density curves with respect to positive and negative torques, as can be seen from the right graphs of panel \textbf{(a)}. Nevertheless, torque-extension and torque-superhelical density curves displayed on panel \textbf{(b)} still demonstrate rather symmetric shapes up to the point when DNA experiences transition into alternative L-DNA state at $\tau \sim -11$ pN$\cdot$nm torque. In all panels, the DNA extension is normalized to the total contour length of DNA in B-form. Abbreviations L-tetrasomes and R-tetrasomes are used to indicate left- and right-handed tetrasome complexes, respectively.
}
\label{fig5}
\end{figure}

\begin{figure}[!htb]
\includegraphics[width=0.9\textwidth]{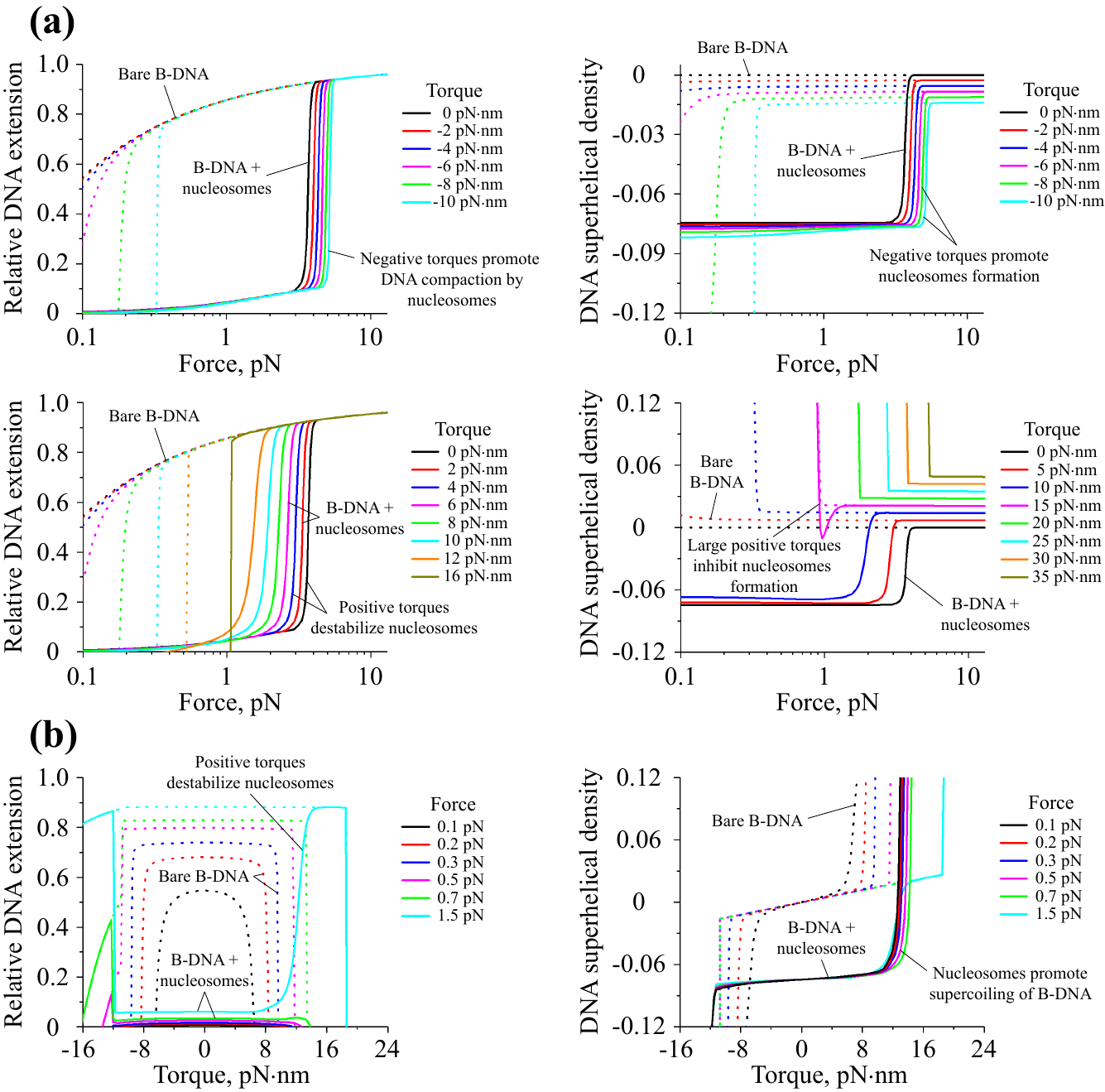}
\caption{
\textbf{Mechanical response of DNA to the applied force and torque constraints in the presence of DNA interactions with histone octamers.} The figure shows \textbf{(a)} force-extension and force-superhelical density curves obtained at different values of the torque, $\tau$, as well as \textbf{(b)} torque-extension and torque-superhelical density curves obtained at different values of the force, $f$, exerted to the DNA. Solid curves demonstrate the behaviour of DNA in the presence of histone octamers that upon binding to DNA form nucleosome complexes; whereas, dotted curves indicate mechanical response of bare DNA under the same force and torque constraints. In contrast to histone tetrasomes, nucleosomes always assume the left-handed conformation and, as a result, form on DNA only at negative ($\tau < 0$ pN$\cdot$nm) or moderate positive torques ($0 < \tau < 15$ pN$\cdot$nm). Indeed, it can be seen from panels \textbf{(a)} and \textbf{(b)} that upon binding to DNA, histone octamers collapse it into a compact conformation in $-11 \leq \tau < 15$ pN$\cdot$nm torque range; whereas, application of a higher positive torsional stress to DNA ($\tau \geq 15$ pN$\cdot$nm) leads to destabilization of nucleosome complexes, which eventually give a way to formation of supercoiled bare DNA structures. As for large negative torques ($\tau < -11$ pN$\cdot$nm), under these conditions DNA experiences transition into alternative L-DNA form, which drives dissociation of histone octamers from the DNA. In all panels, the DNA extension is normalized to the total contour length of DNA in B-form.
}
\label{fig6}
\end{figure}

Substituting the values of the above model parameters into Eq.~\eqref{Block-matrices-nucleosomes}-\eqref{Block-matrices-tetrasomes} in Appendix~\ref{sec:DNA-wrapping} and using the resulting DNA transfer-matrices to calculate the DNA partition function, we plotted the force- and torque-extension curves [$z (f) |_{\tau = \tau_0}$ and $z (\tau) |_{f = f_0}$] as well as the force- and torque-superhelical density curves of DNA [$\sigma (f) |_{\tau = \tau_0}$ and $\sigma (\tau) |_{f = f_0}$] at fixed force ($f = f_0$) and torque ($\tau = \tau_0$) constraints in the presence of tetrasome and nucleosome complexes formation on DNA. The final results of the computations are shown on Figures~\ref{fig5}, \ref{fig6}, \ref{figS3} and \ref{figS4}.

From the left top and bottom panels of Figure~\ref{fig5}(a) it can be seen that histone tetramers bind to DNA and promote its collapsing into a compact conformation in a wide range of the applied force and torque constraints. Interestingly, shift of the force-extension curves calculated for DNA interacting with histone tetramers [solid lines on Figure~\ref{fig5}(a)] towards higher force values with the increasing magnitude of the applied torque suggests that torsional stress of both positive and negative sign facilitates tetrasomes formation, resulting in a more stable compaction of the DNA. This torque-induced effect can be even more clearly observed on the left and middle panels of Figures~\ref{figS6}(a,b) demonstrating the change of the average DNA occupancy fraction by tetrasome complexes as a function of the applied force and torque constraints. 

One of the most prominent feature that stands out in Figures~\ref{figS6}(a,b) is that both positive and negative torques promote formation of tetrasomes with correspondingly right- and left-handed complex chiralities, resulting in the respective jump of the DNA superhelical density to $\pm (0.06-0.13)$, where the sign of the change is determined by the chirality of the formed nucleoprotein complexes [see the right panels of Figure~\ref{fig5}(a)]. These results are in good agreement with the previously published experimental data \cite{Vlijm_2015}, suggesting that transfer-matrix calculations correctly reproduce behaviour of tetrasome complexes revealed in single-molecule experiments.

Furthermore, from Figure~\ref{figS6}(a,b) it can be seen that while being stable at low and moderate tensions ($f < 5-9$ pN), tetrasomes quickly become destabilized by forces $f > 6-10$ pN, resulting in complete  dissociation of histone tetramers from DNA. Thus, it can be concluded that tetrasomes respond to the force and torque constraints in a completely opposite way than DNA-stiffening proteins -- while the latter prefer torsionally relaxed DNA stretched by a mechanical force, tetrasomes mostly bind to twisted DNA being under sufficiently low tension. 

Such a distinct behaviour of the two types of DNA-binding proteins stems from the large difference in the geometric and topological characteristics of their nucleoprotein complexes. Namely, DNA wrapping by histone tetramers results in $-0.73$ / $+1.0$ DNA linking number change, which leads to a strong stabilization effect of the left- and right-handed tetrasomes at high torques of the corresponding sign caused by the significant decrease of the terasomes' torque-dependent potential energy. On the other hand, DNA-stiffening proteins predominantly form straight rigid nucleoprotein filaments on DNA that rather easily lose their stability when either positive or negative torque is applied to the DNA, see Section~\ref{sec:results-DNA-stiff}. 

Furthermore, DNA compaction by tetrasomes results in a situation when mechanical stretching of DNA works against formation of tetrasome complexes, which eventually leads to destabilization of tetrasomes by the applied force. In contrast, sufficiently strong tension exerted on DNA promotes its interaction with DNA-stiffening proteins due to purely entropic reasons discussed in Section~\ref{sec:results-DNA-stiff}.

Finally, it should be noted that besides having different response to force and torque constraints, tetrasomes and DNA-stiffening complexes also have very distinct effects on the global DNA conformation, which are not only can be clearly seen from the DNA force-extension and force-superhelical density curves shown on Figures~\ref{fig3}(a) and \ref{fig5}(a), but also strongly pronounced in the behaviour of the DNA torque-extension and torque-superhelical density curves presented on Figures~\ref{fig3}(b) and \ref{fig5}(b). Indeed, direct comparison between the left panels of Figures~\ref{fig3}(b) and \ref{fig5}(b) demonstrates that while DNA interaction with DNA-stiffening proteins results in widening of the DNA torque-extension curves in the force range of $0 \leq f \leq 3$ pN due to formation of rigid nucleoprotein filaments delaying the DNA buckling transition into a supercoiled conformation, binding of histone tetramers to DNA leads to almost complete collapsing of the torque-extension curves as a result of assembly of compact tetrasome complexes on the DNA.

It is also interesting to note from the right panels of Figures~\ref{fig4}(b) and \ref{fig5}(b) that while nucleoprotein complexes formed by DNA-bending proteins and histone tetramers both can easily flip between the left- and right-handed conformations, the DNA torque-superhelical density curves corresponding to these complexes exhibit very different behaviours. In the case of tetrasomes, these curves reach two plateaus: $\sim -0.09$ at negative torques ($-10 \leq \tau < 0$ pN$\cdot$nm) and $\sim 0.12$ at positive torques ($0 < \tau \leq 30$ pN$\cdot$nm) applied to the DNA, see the right panels of Figures~\ref{fig5}(b) and \ref{figS3}(b); whereas, in the case of DNA-bending protein, IHF, no such plateaus can be observed, see the right panels of Figures~\ref{fig4}(b) and \ref{figS2}(b). The main reason for such distinct behaviour of the two proteins is previously mentioned fact that histone tetramers make a well-defined contribution to the DNA linking number change upon formation of tetrasome complexes on the DNA. At the same time, the contribution of DNA-bending proteins, such as IHF, to the DNA linking number mainly depends on the relative orientations of the resulting nucleoprotein complexes with the respect to the rest of the DNA, which can be changed by modulating the magnitude and sign of the torque applied to the DNA.

As for nucleosomes, their behaviour is practically identical to that of left-handed tetrasomes. Namely, from the top left panel of Figure~\ref{fig6}(a) it can be seen that nucleosomes promote collapsing of DNA into a compact conformation with the resulting effect being enhanced by negative torques applied to DNA. Indeed, the left panel of Figure~\ref{figS6}(c) shows that larger negative torques facilitate formation of nucleosome complexes on DNA. On the other hand, the bottom left panel of Figure~\ref{fig6}(a) and the middle panel Figure~\ref{figS6}(c) indicate that application of large positive torques to DNA results in strong destabilization of nucleosomes, causing DNA unwrapping from histone octamers with their subsequent dissociation from the DNA.

Such asymmetric response of nucleosome complexes to the applied torque constraints can be also clearly seen on the left panel of Figure~\ref{fig6}(b) demonstrating the DNA torque-extension curves in the presence of DNA interaction with histone octamers (solid lines). The figure shows that while DNA is compacted by nucleosome complexes in the torque range of $-11 \leq \tau \leq 15$ pN$\cdot$nm, at large positive torques ($\tau > 15$ pN$\cdot$nm) it behaves in the same way as in the absence of histone octamers in solution, suggesting that histone octamers dissociate from DNA at these conditions [for more details compare the left panels of Figures~\ref{fig2}(b) and \ref{figS4}(b)].

The above observations result from the fact that due to the negative linking number change of DNA upon formation of nucleosome complexes ($\Delta \textrm{Lk}_\textrm{pr} = -1.2$), negative torsional stresses applied to the DNA decrease the torque-dependent potential energy of nucleosome complexes, enhancing their stability and promoting their formation on DNA; whereas, positive torques result in the nucleosomes' potential energy increase, which eventually drives dissociation of histone octamers from DNA. 

Furthermore, from the right panels of Figure~\ref{fig6} that show the force- and torque-superhelical density curves of DNA interacting with histone octamers it can be seen that inability of nucleosome complexes to change their chirality by flipping from the left-handed to a right-handed conformation results in the negative superhelical density of DNA ($\sim -0.08$), which is covered by nucleosome complexes. In addition, this leads to appearance of only one, negative plateau ($\sim -0.08$), in torque-superhelical density curves at $-11 \leq \tau \leq 15$ pN$\cdot$nm torques, in sharp contrast to the the case of tetrasome complexes, whose capability to switch between the left- and right-handed conformations causes formation of the two plateaus (negative and positive) in the torque-superhelical density curves, see the right panel of Figure~\ref{fig5}(b).

\subsection{Force-torque phase diagrams of DNA structures and DNA-protein complexes}
\label{sec:results-DNA-phase}

\begin{figure}[!htb]
\includegraphics[width=\textwidth]{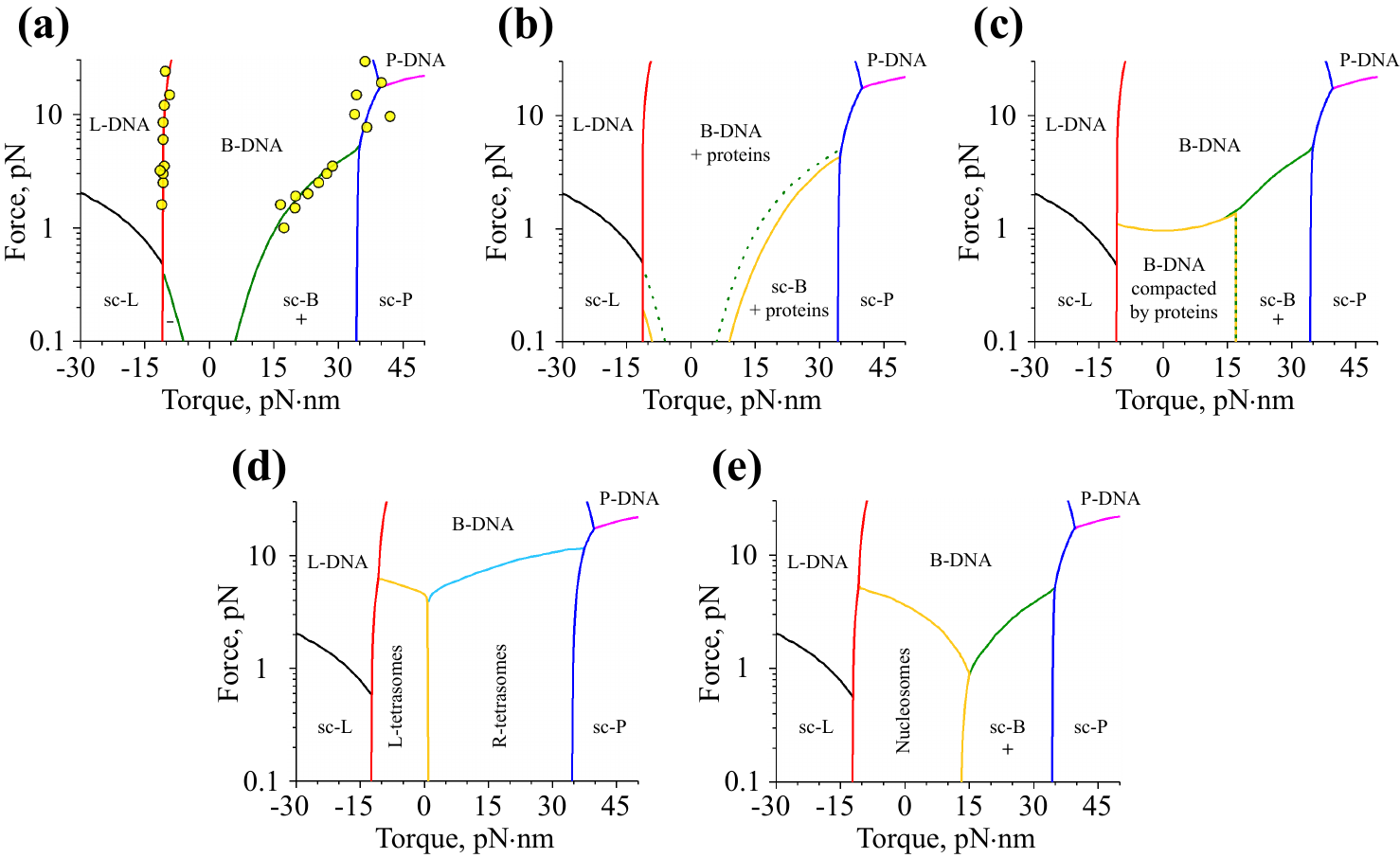}
\caption{
\textbf{DNA phase diagrams.} The figure shows force-torque phase diagrams for: bare DNA \textbf{(a)}, DNA interacting with DNA-stiffening \textbf{(b)} and DNA-bending proteins \textbf{(c)} as well as for DNA in the presence of tetrasome \textbf{(d)} and nucleosome \textbf{(e)} complexes formation. Solid curves predicted by the transfer-matrix calculations indicate transition boundaries between extended (B, L and P) and supercoiled (sc-B, sc-L and sc-P) states of DNA as well as between various DNA-protein conformations. Presented phase diagrams summarize all of the theoretical results plotted on Figures~\ref{fig2}-\ref{fig6}. From the figure, it can be seen that while DNA-stiffening proteins delay formation of supercoiled DNA structures, forcing the DNA to stay in the extended conformation (i.e., the boundary on panel \textbf{(b)} between B-DNA and sc-B states recedes to higher values of the applied torque), DNA-bending and wrapping proteins promote the DNA compaction via assembly of nucleoprotein complexes inducing DNA supercoiling. Circles on the phase diagram of bare DNA  \textbf{(a)} indicate experimental data points, which were digitized from ref. \cite{Allemand_1998, Bryant_2003, Deufel_2007, Forth_2008, Sheinin_2009, Sheinin_2011, Oberstrass_2012}. Dotted lines on panel \textbf{(b)} demonstrate position of the boundary between extended and supercoiled B-DNA states in the absence of DNA-stiffening proteins in solution (i.e., in the case of bare DNA). 
}
\label{fig7}
\end{figure}

Using the obtained theoretical results, we have plotted force-torque phase diagrams that show the transition boundaries between different structural states of DNA and/or DNA-protein complexes for the five scenarios considered in the above sections, including the bare DNA case and DNA interacting with the four different types of DNA-architectural proteins, see Figure~\ref{fig7}.

The boundaries between B- and L-DNA as well as between B- and P-DNA structural states were defined as the set of points $(f,\tau)$ at which $\sim \!50\%$ of the DNA segments are in L- or P-DNA forms, respectively. Furthermore, the boundary between extended and supercoiled conformations of DNA in a particular structural state was determined as a set of points at which DNA extension experiences $\sim \!50\%$ drop with respect to the value predicted by the worm-like chain model for the corresponding form of DNA being in a torsionally relaxed state. 

Finally, the boundaries between bare DNA and protein-covered DNA states were assumed to pass through the points at which half of the maximum DNA-binding sites are occupied by the studied protein. Here we would like to note that the total number of DNA-binding sites is not necessarily equivalent to the total number of DNA segments, see, for example, Figure~\ref{figS6}(c) showing that the maximum occupancy fraction of DNA by nucleosomes never goes above $\sim 90 \%$. The main reason for this is the existence of bare DNA gaps between nucleoprotein complexes that correspond to DNA linkers connecting neighbouring protein-DNA complexes. In the case of reconstituted nucleosome arrays or densily packed yeast chromatin, the minimal length of such DNA linkers was found to be of the order of $\sim 10-20$ bp \cite{Lohr_1977, Shimamura_1988}. For this reason, the minimal possible spacing between neighbouring nucleosomes was set to $18$ bp (i.e., $12$ DNA segments) in all of the transfer-matrix calculations. The same minimal length of the DNA linkers was also used in the computations of DNA interacting with histone tetramers and IHF proteins, as previously reported structural data suggest that such linkers likely exist in-between nucleoprotein complexes formed by IHF proteins as well \cite{Rice_1996}, see Appendices~\ref{sec:DNA-bending}-\ref{sec:DNA-wrapping} for details. 

The resulting phase diagrams plotted using the above definitions for the DNA transition boundaries for the cases of bare DNA and DNA interacting with DNA-stiffening, DNA-bending (IHF) and DNA-wrapping proteins (hitone tetramers and octamers) are depicted on Figure~\ref{fig7}.

While the case of bare DNA has been previously discussed in detailes in our earlier publications \cite{Efremov_2016, Efremov_2017}, here we will mainly focus on the description of the rest of the phase diagrams using the bare DNA graph shown on Figure~\ref{fig7}(a) as a reference point to identify main changes in the DNA behaviour upon addition of different DNA-binding proteins into solution.

The next panel, [Figure~\ref{fig7}(b)], demonstrates the phase diagram of DNA in the presence of nucleoprotein filaments formation by the DNA-stiffening protein that was described in Section~\ref{sec:results-DNA-stiff}. As can be seen from the figure, proteins binding to DNA leads to the leftward and rightward shifts of the boundaries between extended and supercoiled B-DNA conformations at negative and positive torques, respectively, comparing to the case of bare DNA. Such receding of the DNA supercoiling transition boundaries results from the delay in the DNA buckling transition due to the DNA-stiffening effect produced by rigid nucleoprotein filaments, which polymerize on DNA as a result of DNA-protein interactions, see Section~\ref{sec:results-DNA-stiff} for more details.

In the case of DNA interaction with the DNA-bending protein (IHF) described in Section~\ref{sec:results-DNA-bend}, the most prominent effect that can be seen from the phase diagram displayed on Figure~\ref{fig7}(c) is appearance of a new DNA-protein state in $-11 \leq \tau \leq 17$ pN$\cdot$nm torque range and at forces  $f < 1.0-1.5$ pN that corresponds to the formation of compact nucleoprotein complexes by IHF proteins on DNA. As the transfer matrix calculations show, these complexes assume left-handed chirality at negative torques ($-11 \leq \tau < 0$ pN$\cdot$nm) and, more importantly, have free energy, which is smaller than the energy of supercoiled bare B-DNA, see Section~\ref{sec:results-DNA-bend}. This results in complete disappearance of the latter state from the phase diagram of IHF-bound DNA at negative torques. At positive torques, however, the DNA behaviour is slightly more complicated. While at $0 < \tau \leq 17$ pN$\cdot$nm torques IHF binding to DNA leads to formation of compact nucleoprotein complexes with right-handed chirality, further increase of the torque causes dissociation of IHF proteins from DNA, which give a way to formation of positively supercoiled bare B-DNA structures, see Section~\ref{sec:results-DNA-bend} for more details.

The final two panels shown on Figures~\ref{fig7}(d,e) demonstrate the phase diagrams of DNA in the presence of tetrasome (d) and nucleosome (e) complexes formation. From Figure~\ref{fig7}(d) it can be seen that in the case of tetrasomes, the most prominent changes emerging on the phase diagram of DNA is appearance of the two new DNA states corresponding to assembly of the left-handed tetrasomes at negative torques and right-handed tetrasomes at positive torques. Interestingly, in contrast to DNA-bending proteins, strong drop in the DNA free energy associated with the formation of tetrasome complexes not only leads to complete disappearance of the supercoiled bare B-DNA state at negative torques, but at positive torques as well.

In the case of nucleosomes, Figure~\ref{fig7}(e) demonstrates that they form in a more narrow torque range ($-11 \leq \tau \leq 15$ pN$\cdot$nm) comparing to tetrasome complexes. Indeed, as the transfer-matrix calculations discussed in Section~\ref{sec:results-DNA-wrap} show, nucleosomes become highly destabilized at large positive torques due to their left-handed chirality. As a result, while nucleosomes assembly on DNA leads to disappearance of supercoiled bare B-DNA state at negative torques, at high positive torques ($\tau \geq 15$ pN$\cdot$nm) DNA keeps developing supercoiled structures that drive dissociation of histone octamers from the DNA. Another interesting feature that can be seen on Figure~\ref{fig7}(e) is a rather steep boundary between the nucleosome-covered and extended bare B-DNA states, indicating that nucleosomes formation on B-DNA is more sensitive to the applied torque constraints than in the case of other nucleoprotein complexes discussed in this work -- an effect which may be employed by living cells in regulation of the chromatin structure and its spatial organization.

\subsection{Application of the transfer-matrix theory for processing of experimental data}
\label{sec:results-data-fitting}

To demonstrate practical utility of the transfer-matrix formalism, in this section we describe how to exploit it in order to extract valuable information about DNA-protein interactions from experimentally measured force-extension curves of DNA. For this purpose, we use experimental data obtained on a torsionally relaxed 48,502 bp $\lambda$-DNA incubated in the presence of different amounts of TrmBL2 protein in solution \cite{Efremov_2015}.

\begin{figure}[!htb]
\includegraphics[width=0.4\textwidth]{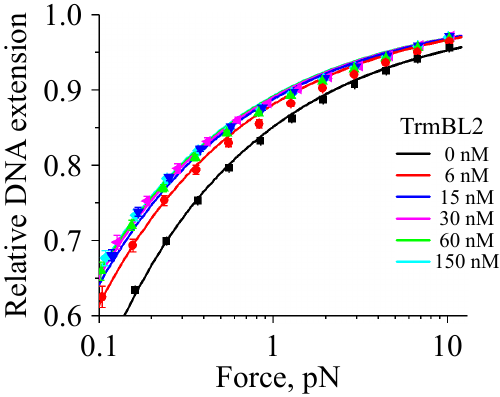}
\caption{
\textbf{Force-extension curves of DNA in the presence of different amounts of TrmBL2 protein in solution.} The figure shows fitting of the experimentally measured force-extension curves of DNA obtained at different concentrations of DNA-stiffening protein, TrmBL2, in solution to the theoretical results predicted by the transfer-matrix theory. Solid symbols on the plot represent the experimental data points collected during stretching cycles of $\lambda$-DNA; whereas, solid curves demonstrate theoretical data fitting based on the transfer-matrix calculations described in the main text. Error bars show experimental SEM values of the corresponding data points.
}
\label{fig8}
\end{figure}

It has been shown in our previous study that TrmBL2 is a DNA-stiffening protein, which binds to DNA in a cooperative manner, resulting in polymerization of rigid nucleoprotein filaments \cite{Efremov_2015}. Furthermore, it has been found that TrmBL2 has two different binding modes to DNA, which manifest themselves in a protein concentration-dependend manner \cite{Efremov_2015}. While it is not hard to introduce both of these modes into the transfer-matrix calculations [see comments in Section~\ref{sec:theory_assumptions}], here we deal only with the experimental data obtained at $0-150$ nM protein concentrations, at which TrmBL2 interaction with DNA can be described by a single binding mode \cite{Efremov_2015}. This makes it possible to directly use Eq.~\eqref{Block-matrices-DNA-stiffening} and \eqref{DNA-protein-partition-function-intro-4} in order to fit experimentally measured force-extension curves of DNA (solid symbols on Figure~\ref{fig8}) in the presence of TrmBL2 protein in solution to the theoretical graphs predicted by the transfer-matrix theory (solid lines on Figure~\ref{fig8}).

For the fitting procedure we used the Nelder-Mead simplex algorithm \cite{Lagarias_1998}, which enables to search for the optimal values of the model parameters at which the total deviation between the experimental data points and the theoretical curves is minimal. To fit the data, the following three model parameters were varied in the calculations: 1) the bending persistence length of protein-covered DNA [$A_\textrm{pr}$], 2) equilibrium dissociation constant of the protein from DNA [$K_\textrm{d}$], and 3) the cooperative binding energy of proteins to DNA [$J_\textrm{pr}$]. At each algorithm step, the proteins' binding energy to DNA at a given concentration, $c$, of TrmBL2 in solution was calculated using the following classical formula: $\mu_\textrm{pr} = \ln ( c / K_\textrm{d} )$. The final results in the form of DNA force-extension curves predicted by the transfer-matrix theory for the optimum values of the model parameters are shown on Figure~\ref{fig8}.

As can be seen from the figure, the theoretical graphs demonstrate very good agreement with the experimental data. Furthermore, the obtained optimal values of the model parameters: $A_\textrm{pr} = 88$ nm, $K_\textrm{d} = 3.8$ nM and $J_\textrm{pr} = 4.26$ $k_\textrm{B} T$ are very close to those previously reported in ref.~\cite{Efremov_2015}, which were acquired by an independent method via fitting the experimental data to the Marko-Siggia formula and Hill equation. This consistency indicates that the transfer-matrix theory presented in this study accurately describes DNA-protein interactions and can be easily implemented for extraction of important information regarding the DNA-binding affinities of studied proteins and physical properties of nucleoprotein complexes from single-molecule experiments performed on individual DNA molecules.

\section{Discussion}
\label{sec:discussion}

In this study, we have developed a new theoretical approach based on the transfer-matrix calculations for investigation of DNA-protein interactions under force and torque constraints, which makes it possible to evaluate changes in the DNA conformation due to formation of nucleoprotein complexes by DNA-binding proteins in a wide range of mechanical forces applied to the DNA. As a result, the constructed theoretical framework may be used in future to provide better understanding of the potential role of such constraints in regulation of the DNA-binding properties of different types of DNA-architectural proteins. 

It should be noted that although in this study the transfer-matrix approach has been demonstrated using examples of proteins which equally well bind to all of the DNA segments, the nature of the transfer-matrix formalism easily allows one to include sequence-dependent behaviour of DNA-binding proteins into the calculations. Indeed, according to Eq.~\eqref{DNA-protein-partition-function-intro-4}, the DNA partition function is determined by the product of transfer matrices, which are defined locally on the vertices connecting neighbouring DNA segments in the polygonal chain representing the DNA polymer. Thus, proteins sequence-specific binding to DNA can be straightforwardly implemented by introduction of site-dependent DNA transfer-matrices, $\textbf{L}_j$ ($j = 1,...,N\!-\!1$), and replacement of $\textbf{L}^{N-1}$ matrices product with $\prod_{j=1}^{N-1} \textbf{L}_j$ in Eq.~\eqref{DNA-protein-partition-function-intro-4}.

Furthermore, flexibility of the developed transfer-matrix approach makes it possible not only use it to study formation of nucleoprotein complexes by a single type of DNA-architectural proteins at a time, but, more importantly, to investigate competitive binding of different types of proteins to the same DNA and its potential regulation by mechanical constraints applied to the DNA. Indeed, calculations presented in this study demonstrate that force and torque constraints imposed on DNA frequently have a strong effect on the proteins' DNA-binding affinity, whose strength may either increase or drop depending on the architecture of the nucleoprotein complexes as well as the magnitude and direction of the applied mechanical forces. These results immediately imply that by changing the mechanical constraints it may be  possible to modulate the balance between nucleoprotein complexes formed on DNA by different groups of DNA-binding proteins, warranting future study.

Considering mounting experimental evidences showing that the chromosomal DNA in living cells is subject to a large number of various mechanical constraints, and taking into account that there exist many different types of DNA-binding proteins involved in regulation of the DNA organization inside living cells, this kind of research may help to gain better understanding of how the force- and torque-dependent interaction of DNA-architectural proteins and transcription factors with DNA results in experimentally observed activation or suppression of a number of specific genes in response to mechanical forces applied to the nucleus and/or chromosomal DNA in living cells \cite{Shivashankar_2011, Iyer_2012, Mammoto_2012, Tajik_2016, Uhler_2017}. 

Finally, it should be noted that the transfer-matrix calculations developed in this study appear to be much faster than the existing Brownian / molecular dynamics simulation (MD) and Metropolis-Monte Carlo (MC) computation algorithms, which are frequently used to model DNA behaviour under mechanical constraints in the presence or absence of DNA-binding proteins \cite{Katritch_2000, Aumann_2006, Kepper_2011, Ettig_2011, Guevara_2011, Guevara_2012, Dobrovolskaia_2012, Nam_2014}. For example, computation of torque-extension curves of a micrometer size DNA could be done in several seconds by running transfer-matrix calculations on a laptop, while for MC algorithm it takes several days of intensive calculations on a computer cluster to obtain similar results (data not shown). This gives the transfer-matrix approach a strong advantage in interpretation of experimental data obtained in single-DNA manipulation assays.

Indeed, as demonstrated in the example of DNA interaction with TrmBL2 proteins in Section~\ref{sec:results-data-fitting}, fast transfer-matrix calculations described in our study allow one to vary parameters to achieve best fitting to the experimentally measured force- and torque-extension curves as well as force- and torque-superhelical density curves of DNA in a sufficiently short amount of time. By doing so, it is possible to obtain accurate and detailed information about the DNA-binding affinities of studied proteins and physical properties of nucleoprotein complexes formed on DNA from the experimental data, providing important information about the role of force and torque constraints in regulation of DNA-protein interactions. 

For this reason, we believe that the transfer-matrix formalism presented in our work may be used in future to quickly estimate potential changes in the DNA conformation under various mechanical constraints imposed on DNA in the presence or absence of DNA-binding proteins in surrounding environment, pinpointing the most important questions and problems that can be later studied in detail by utilizing the classical MD and MC simulation methods. By utilizing such a combination of the transfer-matrix calculations and MD / MC algorithms, it will be then possible to gain deep insights into the role of force and torque constraints in modulation of DNA-protein interactions, which will be important for better understanding of multiple experimental findings suggesting a major role of mechanical forces in regulation of the cell genome organization.

In summary, the transfer-matrix formalism developed in this study allows one to gain valuable insights into physical processes governing formation of nucleoprotein complexes by DNA-binding proteins under force and torque constraints applied to the DNA. The flexibility and advantages of this method make it a powerful tool for a broad range of future applications, including but not limited to investigation of the DNA organization by multiple DNA-binding proteins as well as processing and interpretation of single-molecule experimental data obtained in single-DNA manipulation assays. 

\section{Acknowledgements}
We would like to thank Irina Belyanskaya, Ladislav Hovan and Yang Kaiyuan for their invaluable help in translation of the transfer-matrix calculation programs into C++ language and uploading of the program source files to the internet. This research was funded by Ministry of Education (Singapore) Academic Research Fund Tier 3 (Grant No. MOE2012-T3-1-001), the National Research Foundation (NRF), Prime Minister's Office, Singapore under its NRF Investigatorship Programme (NRF Investigatorship Award No. NRF-NRFI2016-03) and the National Research Foundation (Singapore) through the Mechanobiology Institute Singapore to J.Y.

\onecolumngrid
\appendix

\section{Calculation of the DNA linking number change}
\label{Appendix-Lk}

As has been mentioned in the main text, to calculate the DNA linking number change, we use in this study two famous results from the knot theory. The first one is the C\u{a}lug\u{a}reanu-White's theorem \cite{Calugareanu_1961, White_1969} stating that $\Delta \textrm{Lk}$ can be expressed as a sum of two components: $\Delta \textrm{Lk} = \Delta \textrm{Tw} + \textrm{Wr}$, where $\Delta \textrm{Tw} = \sum_{j=1}^{N-1} \Delta \textrm{Tw}_j ( \textbf{R}_j, \textbf{R}_{j+1} )$ is the DNA total twist and $\textrm{Wr}$ is the DNA writhe number. The second is the Fuller's approximate formula that allows to express the writhe number of DNA as a sum of local DNA segments contributions, $\textrm{Wr}^\textrm{F} = \sum_{j=1}^{N-1} \textrm{Wr}^\textrm{F}_j ( \textbf{R}_j, \textbf{R}_{j+1} )$, where $\textrm{Wr}^\textrm{F}_j ( \textbf{R}_j, \textbf{R}_{j+1} ) = \frac{1}{2\pi} ( \tilde\alpha_{j+1} \!-\! \tilde\alpha_j ) ( 1 - \cos \beta_j )$, and $\tilde\alpha_{j+1}$ and $\tilde\alpha_j$ are the azimuthal Euler angles of the $j^\textrm{th}$ and $(j\!+\!1)^\textrm{th}$ DNA segments from the extended range of $(-\infty,+\infty)$ \cite{Fain_1997, Bouchiat_2000, Efremov_2016}. The superscript $\textrm{F}$ in the above equations indicates that the DNA writhe number calculation is based on the Fuller's formula approximation.

Assuming for a moment that DNA does not transit between alternative structural states, always staying in B-DNA form, and combining together the above mathematical expressions, it is not hard to see that the DNA linking number change can be represented as a sum of local DNA segments contributions:
\begin{equation} \label{delta-Lk-intro-1}
\Delta \textrm{Lk}^\textrm{F} = \Delta \textrm{Tw} + \textrm{Wr}^\textrm{F} =
 \sum_{j=1}^{N-1} \Delta \textrm{Lk}^\textrm{F}_j \!\left( \textbf{R}_j, \textbf{R}_{j+1} \right)
\end{equation}

\noindent 
Where 
\begin{equation} \label{delta-Lk-intro-2}
\Delta \textrm{Lk}^\textrm{F}_j \!\left( \textbf{R}_j, \textbf{R}_{j+1} \right) = 
	\Delta \textrm{Tw}_j \!\left( \textbf{R}_j, \textbf{R}_{j+1} \right)
	+\textrm{Wr}^\textrm{F}_j \!\left( \textbf{R}_j, \textbf{R}_{j+1} \right)
= \frac{1}{2\pi} \!\left( \tilde\alpha_{j+1} + \tilde\gamma_{j+1} - \tilde\alpha_j - \tilde\gamma_j
	\right)
\end{equation}

\noindent
Here $\Delta \textrm{Tw}_j (\textbf{R}_j, \textbf{R}_{j+1}) = \frac{1}{2\pi} ( \tilde\alpha_{j+1} \!-\! \tilde\alpha_j )  \cos \beta_j + \frac{1}{2\pi} ( \tilde\gamma_{j+1} \!-\! \tilde\gamma_j )$, where $\tilde\alpha_j$ and $\tilde\gamma_j$ are the Euler angles of the $j^\textrm{th}$ DNA segment from the extended range of $(-\infty; +\infty)$ \cite{Fain_1997, Bouchiat_2000}.

Now, in order to take into account contribution of various DNA structures into the linker number change, all we need to do is to add an additional term to the above Eq.~\eqref{delta-Lk-intro-1}:
\begin{equation} \label{delta-Lk-intro-3}
\Delta \textrm{Lk}^\textrm{F} = 
	\sum_{j=1}^{N-1} \Delta \textrm{Lk}^\textrm{F}_j \!\left( \textbf{R}_j, \textbf{R}_{j+1} \right)
	+ q \sum_{j=1}^N \sum_{n=-2}^0 \delta_{k_j n} \Delta lk^{\left( n \right)}_0
\end{equation}

\noindent
Where $\delta_{n m}$ is the the Kronecker delta ($\delta_{n m} = 1$ if $n = m$ and $\delta_{n m} = 0$, otherwise); $q$ is the number of base-pairs in each of the DNA segments; and $\Delta lk^{(n)}_0 = lk_{0,n} - lk_{0,0}$ is the linking number change per single base-pair during the DNA structural transition from B-DNA state to the state corresponding to index $n$ ($n = 0$, $-1$ and $-2$ for B-, L- and P-DNA, respectively). In the last formula, $lk_{0,n} = \pm h_n^{-1}$ is the relaxed linking number per single base-pair of DNA in state $n$, which is assigned to be positive for right-handed DNA helical structures (like B- or P-DNA) and negative for left-handed structures (L-DNA). Here $h_n$ is the helical repeat of DNA in the respective state (see Table~\ref{tab:DNA-parameters}).

While Eq.~\eqref{delta-Lk-intro-3} can be directly used to calculate the DNA linking number change in the case of DNA interactions with DNA-bending or DNA-stiffening proteins, it needs to be slightly modified in order to apply it to the case of DNA-wrapping proteins. Indeed, as has been mentioned in the main text (see Section~\ref{sec:energy_terms}), nucleoprotein complexes formed by such proteins make a fixed contribution, $\Delta \textrm{Lk}_\textrm{pr}$, to the total DNA linking number change. As a result, in the case of DNA interactions with DNA-wrapping proteins, Eq.~\eqref{delta-Lk-intro-3} takes the following form:
\begin{equation} \label{delta-Lk-intro-4}
\Delta \textrm{Lk}^\textrm{F} = 
	\sum_{j=1}^{N-1} \Delta \textrm{Lk}^\textrm{F}_j \!\left( \textbf{R}_j, \textbf{R}_{j+1} \right) 
	+ q \sum_{j=1}^N \sum_{n=-2}^0 \delta_{k_j n} \Delta lk^{\left( n \right)}_0
	+ \frac{ \Delta \textrm{Lk}_\textrm{pr} }{ K } \sum_{j=1}^N \sum_{n=1}^K \delta_{k_j n} 
\end{equation}
 
\noindent
Where the last sum describes the contribution of nucleoprotein complexes to the DNA linking number change. In the above expression, $K$ is the number of DNA segments bound to a single protein, see Section~\ref{sec:theory_outline} for more details.

Applying Eq.~\eqref{delta-Lk-intro-3}-\eqref{delta-Lk-intro-4} based on the Fuller's approximation, it is then straightforward to obtain a formula for the potential energy $\Phi_\tau = - 2 \pi \tau \Delta \textrm{Lk}^\textrm{F} + \delta \Phi_\tau$ associated with the torque $\tau$ applied to the DNA, where $\delta \Phi_\tau$ is the correction term described by Eq.~\eqref{correction-term-intro}.

\section{Conformational energy of DNA interacting with proteins}
\label{Appendix-A}

In this and the next Appendix section, we are going to derive the exact formula for the DNA total conformational energy and to prove Eq.~\eqref{DNA-protein-partition-function-intro-4} for the DNA partition function, which will be then used in Appendices~\ref{Appendix-D}-\ref{Appendix-E} to find expressions for all of the DNA transfer-matrix elements.

Let's focus first on writing down the total conformational energy of DNA interacting with DNA-binding proteins under force and torque constraints. To this aim, we will start with a simple scenario when DNA does not change its structural state, always staying in B-form, in addition making assumption that the protein binding site size, $K$, spans only three DNA segments ($K = 3$). Then after finding formulas for the total energy and partition function of DNA for such a hypothetical case, we will generalize the obtained results to the case when DNA can transit between alternative structural states and proteins that have an arbitrary large binding site size on DNA.

As has been mentioned in Section~\ref{sec:theory_outline}, the global conformation of DNA in the general case is completely determined by the two sets of parameters: 1) rotation matrices $( \textbf{R}_1,...,\textbf{R}_N )$ describing orientations of all DNA segments, and 2) indexes $( k_1,...,k_N )$ designating the physical states of these segments, which in the case considered here take integer values from $0$ to $3$, where DNA segments with $k_j = 0$ correspond to bare B-DNA state, and segments with $k_j = 1$, $2$ and $3$ to protein-bound states. Here $j$ is the index enumerating all of the DNA segments from $1$ to $N$, where $N$ is the total number of segments in the discretized polymer chain representing DNA.

Let's now consider one by one the energy terms from Eq.~\eqref{DNA-total-energy-intro} taking into account that bare DNA segments always stay in B-form. As can be seen from Eq.~\eqref{bare-DNA-sum-intro}, in this case the first energy term from Eq.~\eqref{DNA-total-energy-intro} describing elastic deformations of bare DNA parts takes the following form:
\begin{equation} \label{bare-DNA-energy-1}
E_\textrm{DNA} =  
	\sum_{j=1}^{N-1} \delta_{k_j 0} \delta_{k_{j+1} 0} E_\textrm{bare} \!\left( \textbf{R}_j, \textbf{R}_{j+1} \right)
\end{equation}

\noindent
Where $E_\textrm{bare} ( \textbf{R}_j, \textbf{R}_{j+1} )$ is the local elastic deformation energy of DNA corresponding to the vertex joining the $j^\textrm{th}$ and $(j\!+\!1)^\textrm{th}$ segments of the polygonal chain representing the polymer:
\begin{equation} \label{bare-DNA-energy-2}
E_\textrm{bare} \!\left( \textbf{R}_j, \textbf{R}_{j+1} \right) =  
	\frac{a_0}{2} \!\left( \textbf{R}_j \textbf{z}_0 \!-\! \textbf{R}_{j+1} \textbf{z}_0 \right)^2 
	+ \frac{c_0}{2} \!\left[ 2\pi\Delta \textrm{Tw}_j \!\left( \textbf{R}_j,\textbf{R}_{j+1} \right) \right]^2
\end{equation}

\noindent
Here $a_0 = A_0 / b_0$ and $c_0 = C_0 / b_0$ are dimensionless parameters designating the bending and twisting elasticities of bare B-DNA segments in the semiflexible polymer chain model of DNA, where $A_0$ and $C_0$ are the bending and twisting persistence lengths of B-DNA (see Table~\ref{tab:DNA-parameters}), and $b_0$ is the size of bare B-DNA segments in the model. The latter equals to the the number of base-pairs in a single DNA segment, $q$, multiplied by the $0.34$ nm rise of each base-pair in B-DNA form (since in all our calculations $q = 1.5$ base-pairs, we have $b_0 = 0.5$ nm for B-DNA segments). Finally, $\Delta \textrm{Tw}_j ( \textbf{R}_j, \textbf{R}_{j+1} )$ is the local DNA twist between the $j^\textrm{th}$ and $(j\!+\!1)^\textrm{th}$ DNA segments, which equals to the twisting angle between the $j^\textrm{th}$ and $(j\!+\!1)^\textrm{th}$ DNA segments normalized to $2\pi$.

To derive the next formula for $E_\textrm{protein}$ energy term from Eq.~\eqref{DNA-total-energy-intro}, we need first to provide several additional details regarding the mathematical treatment of nucleoprotein complexes in this study. 

In all of the calculations, nucleoprotein complexes are considered as rigid bodies that may freely rotate in space. To describe the orientations of DNA segments constrained inside such complexes, we will still use Euler rotation matrices, $\textbf{R}_j$. However, since the 3D structure of nucleoprotein complexes is fixed, it is clear that DNA segments residing inside these complexes also must have fixed orientations relative to one another. Indeed, let $( \textbf{R}^0_\textrm{first}, \textbf{R}^0_\textrm{second}, \textbf{R}^0_\textrm{third} )$ be a set of rotation matrices describing orientations of the first ($k_j = 1$), second ($k_j = 2$) and the third ($k_j = 3$) DNA segments in one of the nucleoprotein complexes with respect to the global coordinate system. Then it can be easily seen that the relative orientations of the protein-bound DNA segments are characterized by the following two matrices:
\begin{equation} \label{A1-A2-matrices-1}
\textbf{A}_1 = \left( \textbf{R}^0_\textrm{first} \right)^{-1} \textbf{R}^0_\textrm{second}
\quad \textrm{and} \quad
\textbf{A}_2 = \left( \textbf{R}^0_\textrm{second} \right)^{-1} \textbf{R}^0_\textrm{third}
\end{equation}

It is not hard to check that matrices $\textbf{A}_1$ and $\textbf{A}_2$ do not change upon rotation of the nucleoprotein complex as a rigid body, and thus can be used to represent the relative orientations of the protein-bound DNA segments. More importantly, by knowing matrices $\textbf{A}_1$ and $\textbf{A}_2$ as well as the orientation of one of the DNA segments inside a nucleoprotein complex, it is straightforward to find the orientations of the rest of the DNA segments in the same complex. For example, given the orientation $\textbf{R}_\textrm{first}$ of the first DNA segment in a nucleoprotein complex, one can calculate the orientations of the second and the third DNA segments as: 
\begin{equation} \label{A1-A2-matrices-2}
\textbf{R}_\textrm{second} = \textbf{R}_\textrm{first} \textbf{A}_1
\quad \textrm{and} \quad
\textbf{R}_\textrm{third} = \textbf{R}_\textrm{second} \textbf{A}_2 = \textbf{R}_\textrm{first} \textbf{A}_1 \textbf{A}_2
\end{equation}

While here we consider the case of a protein with the binding site size of three DNA segments ($K = 3$), it is clear that very similar approach works equally well for proteins that have an arbitrarily large binding site on DNA. For example, in the case of a DNA-bending protein that has an arbitrary  binding site size $K$, the relative orientations of DNA segments in the resulting nucleoprotein complexes will be described by $K\!-\!1$ rotation matrices $\textbf{A}_1,...,\textbf{A}_{K-1}$. The same is true for DNA-stiffening and DNA-wrapping proteins, for which we in addition have the following set of equations: $\textbf{A}_1 = \textbf{A}_2 = ... = \textbf{A}_{K-1} = \textbf{I}$ (where $\textbf{I}$ is the identity rotation matrix), -- as DNA-stiffening proteins form straight filaments and since protein-bound DNA segments in DNA-wrapping complexes are represented by small intervals aligned along the line connecting the entry and exit points of DNA, see Section~\ref{sec:energy_terms} for more details.

Before moving to the next step, it should be noted that matrices $\textbf{A}_\textrm{in}$ and $\textbf{A}_\textrm{out}$ in Eq.~\eqref{protein-energy-in-intro} and Eq.~\eqref{protein-energy-out-intro} have very similar geometric interpretations as matrices $\textbf{A}_1$ and $\textbf{A}_2$. Specifically, matrices $\textbf{A}_\textrm{in}$ and $\textbf{A}_\textrm{out}$ describe the equilibrium orientations of the two DNA segments entering a nucleoprotein complex with respect to the first and the last segments of the complex in the same way as matrices $\textbf{A}_1$ and $\textbf{A}_2$ describe the relative orientations of neighbouring DNA segments inside the nucleoprotein complex.

Having at hand rotation matrices characterizing the 3D structure of nucleoprotein complexes, it is then rather straightforward to find the exact expression for the second energy term, $E_\textrm{protein}$, from Eq.~\eqref{DNA-total-energy-intro}. As has been mentioned in Section~\ref{sec:energy_terms}, it includes both the DNA-binding energies of proteins that form nucleoprotein complexes on DNA and elastic deformation energies of the DNA segments entering these complexes, see Eq.~\eqref{protein-energy-intro}-\eqref{protein-energy-out-intro}. On top of that, in order to correctly represent the structure of nucleoprotein complexes in the DNA partition function calculations, we are going to add two additional terms to $E_\textrm{protein}$.

First, to impose the matrix constraints shown in Eq.~\eqref{A1-A2-matrices-2}, we will utilize Dirac $\delta$-functions defined on SO(3) group of Euler rotation matrices. Namely, let's assume that we have a protein bound to DNA segments with indexes $j$, $j\!+\!1$ and $j\!+\!2$. Then the relative orientations of these segments will be described by Eq.~\eqref{A1-A2-matrices-2}, which can be enforced in the DNA partition function calculations by using the two Dirac $\delta$-functions: $\delta ( \textbf{R}_j \textbf{A}_1 \! - \! \textbf{R}_{j+1} )$ and $\delta ( \textbf{R}_{j+1} \textbf{A}_2 \! - \! \textbf{R}_{j+2} )$. For the sake of formulas simplicity, it is convenient to add these two functions to $E_\textrm{protein}$ term in a form of the Dirac $\delta$-function logarithms, $-\ln [ \delta ( \textbf{R}_j \textbf{A}_1 \! - \! \textbf{R}_{j+1} ) ]$ and $-\ln [ \delta ( \textbf{R}_{j+1} \textbf{A}_2 \! - \! \textbf{R}_{j+2} ) ]$, instead of inserting them directly under the integral sign into the DNA partition function. These logarithms are defined as generalized functions, which after exponentiation result in the Dirac $\delta$-functions: $ \exp [ \ln \delta ( \textbf{R} - \textbf{R}' ) ] = \delta ( \textbf{R} - \textbf{R}' )$, where $\textbf{R}$ and $\textbf{R}'$ are some rotation matrices. Since such generalized functions have to be added to $E_\textrm{protein}$ for each of the DNA segment bound to a protein, it is clear that $E_\textrm{protein}$ must be modified by the following sum: $- \sum_{j=1}^{N-1} \sum_{n=1,2} \delta_{k_j n} \ln{ \delta ( \textbf{R}_j \textbf{A}_n - \textbf{R}_{j+1} ) }$.

It should be noted that such approach has a small drawback -- by using Dirac $\delta$-functions to impose predefined relative orientations on the protein-bound DNA segments, we implicitly offset the free energies of the corresponding nucleoprotein complexes by a constant term, $\mu_\textrm{off}$, whose value can be easily found using the transfer-matrix calculations, see comments in Appendix~\ref{Appendix-F}. Thus, to accurately describe proteins interaction with DNA, the energy of each nucleoprotein complex must be decreased by the same amount of $\mu_\textrm{off}$.

Next, we would like to note that for the sake of formulas simplicity in this study we only consider scenario when proteins form complete nucleoprotein complexes upon binding to DNA and never assemble into partially unfolded structures. To this aim, we set the energy of all of the DNA-protein conformations that contain one or more partially unfolded nucleoprotein complexes equal to infinity. This way DNA-protein conformations containing improperly formed nucleoprotein complexes do not make any contribution to the DNA partition function.

To distinguish correct DNA-protein states from those corresponding to partially unfolded nucleoprotein complexes, we use the following approach. It is clear that in the case of properly formed nucleoprotein complexes each pair $(k_j, k_{j+1})$ of the neighbouring DNA segments' states can have only one of the following values: $(0,0)$, $(0,1)$, $(1,2)$, $(2,3)$, $(3,0)$, $(3,1)$ since the protein considered in this section binds only to $K = 3$ DNA segments. All other combinations of states $(k_j,k_{j+1})$, such as $(1,1)$, $(2,1)$, $(0,2)$, etc., correspond to the situation when there is one or more partially unfolded nucleoprotein complexes formed on DNA. Thus, to set the energy of such DNA-protein conformations to infinity, all we need to do is to add the following sum $\sum_{j=1}^{N-1} \sum_{( n, m ) \notin G} \delta_{k_j n} \delta_{k_{j+1} m} \times \infty$ to $E_\textrm{protein}$ energy term, where $G = \{ (0,0),(0,1),(1,2),(2,3),(3,0),(3,1) \}$ is the set of correct combinations of the neighbouring DNA segments' states corresponding to properly folded nucleoprotein complexes. Here we use a typical mathematical convention that $0 \times \infty = 0$.

Collecting together all of the above energy terms, we finally obtain the following formula for $E_\textrm{protein}$ energy:
\begin{align} \label{DNA-protein-energy}
E_\textrm{protein} = & 
	- \frac{ \mu_\textrm{pr} + \mu_\textrm{off} }{ K } \sum_{j=1}^N \left( 1 - \delta_{k_j 0} \right)
	+ \sum_{j=1}^{N-1} \delta_{k_j 0} \delta_{k_{j+1} 1} E_\textrm{in} \!\left( \textbf{R}_j, \textbf{R}_{j+1} \right)
	+ \sum_{j=1}^{N-1} \delta_{k_j K} \delta_{k_{j+1} 0} E_\textrm{out} \!\left( \textbf{R}_j, \textbf{R}_{j+1} \right)
	\nonumber\\
	& - J_\textrm{pr} \sum_{j=1}^{N-1} \delta_{k_j K} \delta_{k_{j+1} 1}
	+ \sum_{j=1}^{N-1} \delta_{k_j K} \delta_{k_{j+1} 1} E_\textrm{ht} \!\left( \textbf{R}_j, \textbf{R}_{j+1} \right)
	- \sum_{j=1}^{N-1} \sum_{n=1}^{K-1} \delta_{k_j n} \ln{ \delta \!\left( \textbf{R}_j \textbf{A}_n - \textbf{R}_{j+1}
	\right) } \nonumber\\
	& + \sum_{j=1}^{N-1} \sum_{\left( n,m \right) \notin G} \delta_{k_j n} \delta_{k_{j+1} m} \times \infty
	+ \left( 1 - \delta_{k_1 0} \right) \times \infty + \left( 1 - \delta_{k_N 0} \right) \times \infty
\end{align}

\noindent
Where we have included a few additional terms into $E_\textrm{protein}$ energy, such as $( 1 - \delta_{k_1 0} ) \times \infty$ and $( 1 - \delta_{k_N 0} ) \times \infty$, which are introduced to prohibit formation of partially unfolded nucleoprotein complexes on the DNA end segments, and the sum of $E_\textrm{ht} ( \textbf{R}_j, \textbf{R}_{j+1} )$ energies that describe the elastic deformations of proteins bound to neighbouring DNA sites in a head-to-tail configuration. The latter takes place only in the case when proteins interact with DNA in a cooperative manner, forming continuous nucleoprotein filaments along the DNA, which is typical for DNA-stiffening proteins. Each local energy contribution, $E_\textrm{ht} ( \textbf{R}_j, \textbf{R}_{j+1} )$, comprises the same bending and twisting deformation energy terms as in Eq.~\eqref{protein-energy-out-intro} with the only difference being that matrix $\textbf{A}_\textrm{out}$ is replaced by matrix $\textbf{A}_\textrm{ht}$, which represents the relative equilibrium orientations of neighbouring nucleoprotein complexes in the head-to-tail configuration. Finally, $J_\textrm{pr}$ is the proteins' cooperative binding energy to DNA. 

As for the last two energy terms from Eq.~\eqref{DNA-total-energy-intro}, $\Phi_f$ and $\Phi_\tau$, they undergo only minor changes under the previously mentioned assumptions. Namely, recalling that at the moment we consider a hypothetical scenario when bare DNA segments always stay in B-DNA form, it is easy to find from Eq.~\eqref{Phi-f-intro} that $\Phi_f$ term corresponding to the DNA potential energy associated with the stretching force, $f$, takes the following form: 
\begin{equation} \label{Phi-f}
\Phi_f =
	- \sum_{j=0}^N \sum_{n=0}^K \delta_{k_j n} b_n f \left( \textbf{z}_0 \cdot  \textbf{R}_j \textbf{z}_0 \right)
\end{equation}

In the case if the protein considered in this section belongs either to DNA-bending or DNA-stiffening type, both protein-bound and bare DNA segments in the above equation have the same size: $b_1 = b_2 = b_3 = b_0$; otherwise, if it is a DNA-wrapping protein, for protein-bound segments we have: $b_1 = b_2 = b_3 = r_\textrm{pr} / 3$, see comments after Eq.~\eqref{Phi-f-intro} in Section~\ref{sec:energy_terms} for more details. Here, as before, $r_\textrm{pr}$ is the distance between the entry and exit points of DNA in nucleoprotein complexes.

As for the potential energy $\Phi_\tau$ associated with the torque $\tau$ exerted to the DNA, from Eq.~\eqref{correction-term-intro} and $\Phi_\tau =  - 2 \pi \tau \Delta \textrm{Lk}^\textrm{F} + \delta \Phi_\tau$ formula it follows that:
\begin{equation} \label{Phi-tau}
\Phi_\tau = 
	- 2 \pi \tau \Delta \textrm{Lk}^\textrm{F}
	+ \tau \sum_{j=1}^{N-1} \left[ \delta_{k_j 0} \lambda_0 \\
	+ \lambda_\textrm{pr} \sum_{n = 1}^K \delta_{k_j n} \right]  
	\textrm{Wr}^\textrm{F}_j \!\left( \textbf{R}_j, \textbf{R}_{j+1} \right)
\end{equation}

\noindent
Where in the case of a DNA-bending or DNA-stiffening protein, the DNA linking number change, $\Delta \textrm{Lk}^\textrm{F}$, is defined by Eq.~\eqref{delta-Lk-intro-1}-\eqref{delta-Lk-intro-2}; and in the case of a DNA-wrapping protein we have [see Eq.~\eqref{delta-Lk-intro-4}]:
\begin{equation} \label{delta-Lk-wrapping-proteins}
\Delta \textrm{Lk}^\textrm{F} = 
	\sum_{j=1}^{N-1} \Delta \textrm{Lk}^\textrm{F}_j \!\left( \textbf{R}_j, 
	\textbf{R}_{j+1} \right) 
	+ \frac{ \Delta \textrm{Lk}_\textrm{pr} }{ K } \sum_{j=1}^N \sum_{n=1}^K \delta_{k_j n} 
\end{equation}

Substituting Eq.~\eqref{bare-DNA-energy-1}, \eqref{DNA-protein-energy}, \eqref{Phi-f} and \eqref{Phi-tau} into Eq.~\eqref{DNA-total-energy-intro}, it is not hard to see that the total conformational energy of DNA can be represented as a sum of local energy contributions, $E_{nm}$, by neighbouring DNA segments:
\begin{align} \label{DNA-total-energy}
E_\textrm{tot} \!\left( k_1...k_N,\textbf{R}_1...\textbf{R}_N \right) = &
	\sum_{j=1}^{N-1} \sum_{n,m = 0}^K \delta_{k_j n}\delta_{k_{j+1} m} E_{nm} \!\left( \textbf{R}_j, \textbf{R}_{j+1} 
    \right) 
    - b_0 f \left( \textbf{z}_0 \cdot \textbf{R}_N \textbf{z}_0 \right) \nonumber\\
    & + \left( 1 - \delta_{k_1 0} \right) \times \infty
	+ \left( 1 - \delta_{k_N 0} \right) \times \infty
\end{align}

\noindent
Where indexes $n$ and $m$ correspond to the states of neighbouring DNA segments. 

As in this study nucleoprotein complexes formed by DNA-bending and DNA-wrapping proteins are treated in slightly different ways, the exact form of $E_{nm}$ energy terms generally depends on the nature of nucleoprotein complexes formed on DNA. Namely, in the case of DNA interaction with a DNA-bending or DNA-stiffening protein, it is not hard to find from Eq.~\eqref{DNA-total-energy-intro}, \eqref{delta-Lk-intro-1} and \eqref{bare-DNA-energy-1}-\eqref{Phi-tau} that $E_{nm}$ energy terms take the following shapes:
\begin{multline} \label{all-local-energies-DNA-bending}
\begin{cases}
E_{00} \!\left( \textbf{R}_j, \textbf{R}_{j+1} \right) = 
 		\frac{a_0}{2} \left( \textbf{R}_j \textbf{z}_0 - \textbf{R}_{j+1} \textbf{z}_0 \right)^2
 		+ \frac{c_0}{2} \left[ 2\pi \Delta \textrm{Tw}_j \!\left( \textbf{R}_j, \textbf{R}_{j+1} \right) \right]^2
 		- b_0 f \left( \textbf{z}_0 \cdot \textbf{R}_j \textbf{z}_0 \right) \\
 		\hspace{207pt} - \tau \left( 2\pi - \lambda_0 \right) \Delta \textrm{Lk}_j^\textrm{F} \!\left( \textbf{R}_j, 		
 		\textbf{R}_{j+1} \right) 
 		- \tau \lambda_0 \Delta \textrm{Tw}_j \!\left( \textbf{R}_j, \textbf{R}_{j+1} \right) \\[10pt]
E_{01} \!\left( \textbf{R}_j, \textbf{R}_{j+1} \right) =
 		\frac{a_\textrm{pr}}{2} \left( \textbf{R}_j \textbf{A}_{\textrm{in}} \textbf{z}_0 - \textbf{R}_{j+1} 
 		\textbf{z}_0\right)^2 
 		+ \frac{c_\textrm{pr}}{2} \left[ 2\pi \Delta \textrm{Tw}_j \left( \textbf{R}_j \textbf{A}_{\textrm{in}} ,
 		\textbf{R}_{j+1} \right) \right]^2
 		- b_0 f \left( \textbf{z}_0 \cdot \textbf{R}_j \textbf{z}_0 \right) \\
 		\hspace{200pt} - \tau \left( 2\pi - \lambda_0 \right) \Delta \textrm{Lk}_j^\textrm{F} \!\left( 
 		\textbf{R}_j , \textbf{R}_{j+1} \right) 
 		- \tau \lambda_0 \Delta \textrm{Tw}_j \!\left( \textbf{R}_j , \textbf{R}_{j+1} \right) \\[10pt]
E_{12} \!\left( \textbf{R}_j, \textbf{R}_{j+1} \right) = 
 		- \textstyle{ \frac{ \mu_\textrm{pr} + \mu_\textrm{off} }{ K } } 
 		- b_0 f \left( \textbf{z}_0 \cdot \textbf{R}_j \textbf{z}_0 \right) 
		- \ln{ \delta \!\left( \textbf{R}_j \textbf{A}_1 - \textbf{R}_{j+1} \right) } \\
		\hspace{200pt} - \tau \left( 2\pi - \lambda_\textrm{pr} \right) \Delta \textrm{Lk}_j^\textrm{F} \!\left( 
		\textbf{R}_j, \textbf{R}_{j+1} \right) 
 		- \tau \lambda_\textrm{pr} \Delta \textrm{Tw}_j \!\left( \textbf{R}_j, \textbf{R}_{j+1} \right) \\[10pt]
E_{23} \!\left( \textbf{R}_j, \textbf{R}_{j+1} \right) = 
 		- \textstyle{ \frac{ \mu_\textrm{pr} + \mu_\textrm{off} }{ K } } 
 		- b_0 f \left( \textbf{z}_0 \cdot \textbf{R}_j \textbf{z}_0 \right) 
		- \ln{ \delta \!\left( \textbf{R}_j \textbf{A}_2 - \textbf{R}_{j+1} \right) } \\
		\hspace{200pt} - \tau \left( 2\pi - \lambda_\textrm{pr} \right) \Delta \textrm{Lk}_j^\textrm{F} \!\left( 
		\textbf{R}_j, \textbf{R}_{j+1} \right) 
 		- \tau \lambda_\textrm{pr} \Delta \textrm{Tw}_j \!\left( \textbf{R}_j, \textbf{R}_{j+1} \right) \\[10pt]
E_{30} \!\left( \textbf{R}_j, \textbf{R}_{j+1} \right) = 
		- \textstyle{ \frac{ \mu_\textrm{pr} + \mu_\textrm{off} }{ K } } 
 		+ \frac{a_{\textrm{pr}}}{2} \left( \textbf{R}_j \textbf{A}_{\textrm{out}} \textbf{z}_0 - \textbf{R}_{j+1} 
 		\textbf{z}_0\right)^2
 		+ \frac{c_{\textrm{pr}}}{2} \left[ 2\pi \Delta \textrm{Tw}_j \left( \textbf{R}_j \textbf{A}_{\textrm{out}},
 		\textbf{R}_{j+1} \right) \right]^2 \\
 		\hspace{125pt} - b_0 f \left( \textbf{z}_0 \cdot \textbf{R}_j \textbf{z}_0 \right)
 		- \tau \left( 2\pi - \lambda_\textrm{pr} \right) \Delta \textrm{Lk}_j^\textrm{F} \!\left( \textbf{R}_j, 		
 		\textbf{R}_{j+1} \right) 
 		- \tau \lambda_\textrm{pr} \Delta \textrm{Tw}_j \!\left( \textbf{R}_j, \textbf{R}_{j+1} \right) \\[10pt]
E_{31} \!\left( \textbf{R}_j, \textbf{R}_{j+1} \right) = 
		- \textstyle{ \frac{ \mu_\textrm{pr} + \mu_\textrm{off} }{ K } } - J_\textrm{pr}
 		+ \frac{a_{\textrm{pr}}}{2} \left( \textbf{R}_j \textbf{A}_{\textrm{ht}} \textbf{z}_0 - \textbf{R}_{j+1} 
 		\textbf{z}_0\right)^2
 		+ \frac{c_{\textrm{pr}}}{2} \left[ 2\pi \Delta \textrm{Tw}_j \left( \textbf{R}_j \textbf{A}_{\textrm{ht}},
 		\textbf{R}_{j+1} \right) \right]^2 \\
 		\hspace{125pt} - b_0 f \left( \textbf{z}_0 \cdot \textbf{R}_j \textbf{z}_0 \right)
 		- \tau \left( 2\pi - \lambda_\textrm{pr} \right) \Delta \textrm{Lk}_j^\textrm{F} \!\left( \textbf{R}_j, 		
 		\textbf{R}_{j+1} \right) 
 		- \tau \lambda_\textrm{pr} \Delta \textrm{Tw}_j \!\left( \textbf{R}_j, \textbf{R}_{j+1} \right) \\[10pt]
E_{02} = E_{03} = E_{10} = E_{11} = E_{13} = E_{20} = E_{21} = E_{22} = E_{32} = E_{33} = \infty	
\end{cases}
\end{multline}

Similarly, in the case of DNA interaction with a DNA-wrapping protein, we have:
\begin{multline} \label{all-local-energies-DNA-wrapping}
\begin{cases}
E_{00} \!\left( \textbf{R}_j, \textbf{R}_{j+1} \right) = 
 		\frac{a_0}{2} \left( \textbf{R}_j \textbf{z}_0 - \textbf{R}_{j+1} \textbf{z}_0 \right)^2
 		+ \frac{c_0}{2} \left[ 2\pi \Delta \textrm{Tw}_j \!\left( \textbf{R}_j, \textbf{R}_{j+1} \right) \right]^2
 		- b_0 f \left( \textbf{z}_0 \cdot \textbf{R}_j \textbf{z}_0 \right) \\
 		\hspace{207pt} - \tau \left( 2\pi - \lambda_0 \right) \Delta \textrm{Lk}_j^\textrm{F} \!\left( \textbf{R}_j, 		
 		\textbf{R}_{j+1} \right) 
 		- \tau \lambda_0 \Delta \textrm{Tw}_j \!\left( \textbf{R}_j, \textbf{R}_{j+1} \right) \\[10pt]
E_{01} \!\left( \textbf{R}_j, \textbf{R}_{j+1} \right) =
 		\frac{a_\textrm{pr}}{2} \left( \textbf{R}_j \textbf{A}_{\textrm{in}} \textbf{z}_0 - \textbf{R}_{j+1} 
 		\textbf{z}_0\right)^2 
 		+ \frac{c_\textrm{pr}}{2} \left[ 2\pi \Delta \textrm{Tw}_j \left( \textbf{R}_j \textbf{A}_{\textrm{in}} ,
 		\textbf{R}_{j+1} \right) \right]^2
 		- b_0 f \left( \textbf{z}_0 \cdot \textbf{R}_j \textbf{z}_0 \right) \\
 		\hspace{200pt} - \tau \left( 2\pi - \lambda_0 \right) \Delta \textrm{Lk}_j^\textrm{F} \!\left( 
 		\textbf{R}_j , \textbf{R}_{j+1} \right) 
 		- \tau \lambda_0 \Delta \textrm{Tw}_j \!\left( \textbf{R}_j , \textbf{R}_{j+1} \right) \\[10pt]
E_{12} \!\left( \textbf{R}_j, \textbf{R}_{j+1} \right) = E_{23} \!\left( \textbf{R}_j, \textbf{R}_{j+1} \right) =
 		- \textstyle{ \frac{ \mu_\textrm{pr} + \mu_\textrm{off} + 2 \pi \tau \Delta \textrm{Lk}_\textrm{pr} }{ K } } 
 		- \textstyle{ \frac{ 1 }{ K } } r_\textrm{pr} f \left( \textbf{z}_0 \cdot \textbf{R}_j \textbf{z}_0 \right) 
		- \ln{ \delta \!\left( \textbf{R}_j - \textbf{R}_{j+1} \right) } \\[10pt]
E_{30} \!\left( \textbf{R}_j, \textbf{R}_{j+1} \right) = 
		- \textstyle{ \frac{ \mu_\textrm{pr} + \mu_\textrm{off} + 2 \pi \tau \Delta \textrm{Lk}_\textrm{pr} }{ K } } 
 		+ \frac{a_{\textrm{pr}}}{2} \left( \textbf{R}_j \textbf{A}_{\textrm{out}} \textbf{z}_0 - \textbf{R}_{j+1} 
 		\textbf{z}_0\right)^2 
 		+ \frac{c_{\textrm{pr}}}{2} \left[ 2\pi \Delta \textrm{Tw}_j \left( \textbf{R}_j \textbf{A}_{\textrm{out}},
 		\textbf{R}_{j+1} \right) \right]^2 \\
 		\hspace{114pt} - \textstyle{ \frac{ 1 }{ K } } r_\textrm{pr} f \left( \textbf{z}_0 \cdot \textbf{R}_j 
 		\textbf{z}_0 \right)
 		- \tau \left( 2\pi - \lambda_\textrm{pr} \right) \Delta \textrm{Lk}_j^\textrm{F} \!\left( \textbf{R}_j, 		
 		\textbf{R}_{j+1} \right) 
 		- \tau \lambda_\textrm{pr} \Delta \textrm{Tw}_j \!\left( \textbf{R}_j, \textbf{R}_{j+1} \right) \\[10pt]
E_{31} \!\left( \textbf{R}_j, \textbf{R}_{j+1} \right) = 
		- \textstyle{ \frac{ \mu_\textrm{pr} + \mu_\textrm{off} + 2 \pi \tau \Delta \textrm{Lk}_\textrm{pr} }{ K } } 
		- J_\textrm{pr}
 		+ \frac{a_{\textrm{pr}}}{2} \left( \textbf{R}_j \textbf{A}_{\textrm{ht}} \textbf{z}_0 - \textbf{R}_{j+1} 
 		\textbf{z}_0\right)^2 
 		+ \frac{c_{\textrm{pr}}}{2} \left[ 2\pi \Delta \textrm{Tw}_j \left( \textbf{R}_j \textbf{A}_{\textrm{ht}},
 		\textbf{R}_{j+1} \right) \right]^2 \\
 		\hspace{114pt} - \textstyle{ \frac{ 1 }{ K } } r_\textrm{pr} f \left( \textbf{z}_0 \cdot \textbf{R}_j 
 		\textbf{z}_0 \right)
 		- \tau \left( 2\pi - \lambda_\textrm{pr} \right) \Delta \textrm{Lk}_j^\textrm{F} \!\left( \textbf{R}_j, 		
 		\textbf{R}_{j+1} \right) 
 		- \tau \lambda_\textrm{pr} \Delta \textrm{Tw}_j \!\left( \textbf{R}_j, \textbf{R}_{j+1} \right) \\[10pt]
E_{02} = E_{03} = E_{10} = E_{11} = E_{13} = E_{20} = E_{21} = E_{22} = E_{32} = E_{33} = \infty	
\end{cases}
\end{multline}

\noindent
Where we used $\textrm{Wr}^\textrm{F}_j ( \textbf{R}_j, \textbf{R}_{j+1} ) = \Delta \textrm{Lk}^\textrm{F}_j ( \textbf{R}_j, \textbf{R}_{j+1} ) - \Delta \textrm{Tw}_j ( \textbf{R}_j, \textbf{R}_{j+1} )$ formula to express the local DNA segment contributions to the DNA writhe number, $\textrm{Wr}^\textrm{F}_j ( \textbf{R}_j, \textbf{R}_{j+1} )$, as a function of the local DNA linking number change and DNA twist number, $\Delta \textrm{Lk}^\textrm{F}_j ( \textbf{R}_j, \textbf{R}_{j+1} )$ and $\Delta \textrm{Tw}_j ( \textbf{R}_j, \textbf{R}_{j+1} )$, respectively. In addition, in Eq.~\eqref{all-local-energies-DNA-wrapping} we have taken into account that the local DNA linking number changes, $\Delta \textrm{Lk}^\textrm{F}_j ( \textbf{R}_j, \textbf{R}_{j+1} ) = \frac{1}{2\pi} (\tilde\alpha_{j+1} \!+\! \tilde\gamma_{j+1} \!-\! \tilde\alpha_j \!-\! \tilde\gamma_j)$, and local DNA twist numbers, $\Delta \textrm{Tw}_j ( \textbf{R}_j, \textbf{R}_{j+1} )$, in $E_{12}$ and $E_{23}$ energy terms equal to zero since all protein-bound segments in DNA-wrapping complexes are represented by intervals that have identical 3D orientations (i.e., $\textbf{R}_{j+1} = \textbf{R}_j$, and thus $\tilde\alpha_{j+1} = \tilde\alpha_j$ and $\tilde\gamma_{j+1} = \tilde\gamma_j$). Here, as before, $\tilde\alpha_j$ and $\tilde\gamma_j$ are the Euler angles of the $j^\textrm{th}$ DNA segment from the extended range of $(-\infty; +\infty)$.

Despite the daunting look of the above equations, it can be seen that the most of $E_{nm}$ energy terms have very similar functional forms, which only slightly vary from one line of the equation to another. This makes it easy to obtain formulas for all of the elements of the DNA transfer-matrix, $\textbf{L}$, described in the next Appendix section, as soon as we know a mathematical expression only for one of them.

But before proceeding to the description of the transfer-matrix formalism, we need to make the last important note in this section. As was briefly mentioned in Section~\ref{sec:theory_outline}, all of the nucleoprotein complexes formed on DNA have a certain orientational freedom -- upon binding to DNA proteins may form nucleoprotein complexes on either side of the DNA duplex due to its double-stranded helical structure. This introduces a new degree of freedom into the model, which we have not considered so far.

From a physical point of view, such positional freedom means that in Eq.~\eqref{all-local-energies-DNA-bending} and Eq.~\eqref{all-local-energies-DNA-wrapping} we need to replace rotation matrix $\textbf{R}_j$ in the formula for $E_{01}$ term with the matrices product $\textbf{R}_j \textbf{B}$, where $\textbf{B} = \textbf{B} (\eta_\textrm{in}, 0, 0)$. Here angle $\eta_\textrm{in} \in [0, 2\pi]$ describes the relative position of the nucleoprotein complex with respect to the axis of the DNA segment entering it, which basically tells on which side of the DNA the nucleoprotein complex is formed. Indeed, since $\textbf{R}_j \textbf{B} = \textbf{R}_{\alpha_j} \textbf{R}_{\beta_j} \textbf{R}_{\gamma_j} \textbf{R}_{\eta_\textrm{in}} = \textbf{R}_{\alpha_j} \textbf{R}_{\beta_j} \textbf{R}_{\gamma_j+\eta_\textrm{in}}$, it is clear that angle $\eta_\textrm{in}$ simply introduces rotation of the nucleoprotein complex with respect to the axis of the DNA segment entering it. Here $(\alpha_j, \beta_j, \gamma_j)$ are the Euler angles corresponding to the rotation matrix $\textbf{R}_j$; and $\textbf{R}_{\alpha_j}$, $\textbf{R}_{\beta_j}$, $\textbf{R}_{\gamma_j}$ and $\textbf{R}_{\eta_\textrm{in}}$ are rotation matrices describing the respective coordinate system revolutions through the angles $\alpha_j$, $\beta_j$, $\gamma_j$ and $\eta_\textrm{in}$.

It should be noted that the above matrices product, $\textbf{R}_j \textbf{B}$, has to be used only in $E_{01}$ energy term, without making similar changes in other energy terms corresponding to the downstream DNA segments, as their orientations will be completely defined by the orientation of the first DNA segment bound to the protein. Analogously, in the case of $E_{31}$ term describing nucleoprotein complexes in the head-to-tail configuration, the equilibrium orientation of the downstream nucleoprotein complex is completely determined by the orientation of the one at front of it, resulting in a lack of orientational freedom of the downstream nucleoprotein complex. Thus, no changes are required to $E_{31}$ term either.

As we have now all of the equations necessary for introduction of the transfer-matrix formalism, let's proceed to its description, applying it to calculate the DNA partition function.

\section{DNA partition function}
\label{Appendix-B}

Knowing the DNA total conformational energy, $E_\textrm{tot}$, the partition function of DNA that always stays in B-form can be calculated as [see Eq.~\eqref{DNA-protein-partition-function-intro-1} in Section~\ref{sec:theory_outline}]:
\begin{equation} \label{DNA-protein-partition-function}
Z_{f,\tau} = 
    \sum_{k_1...k_N=0}^K \int \textrm{d} \textbf{R}_1 ... \textrm{d} \textbf{R}_N \,
    \textrm{d} \!\left[ \eta_\textrm{in} \right] 
    e^{-E_\textrm{tot} \left( k_1...k_N,\textbf{R}_1...\textbf{R}_N \right)}
    \xi \!\left( \textbf{R}_N, \textbf{R}_1 \right)
\end{equation} 

\noindent
Where $\xi ( \textbf{R}_N, \textbf{R}_1 )$ is a function that imposes specific boundary conditions on the orientations of the DNA ends. Integrations in the above mathematical expression are carried out over all of the DNA segment orientations (i.e., $\int \! \textrm{d} \textbf{R}_1 ... \textrm{d} \textbf{R}_N  = \int_0^{2\pi} \! \textrm{d} \alpha_1 ... \textrm{d} \alpha_N \int_0^{2\pi} \! \textrm{d} \gamma_1 ... \textrm{d} \gamma_N \int_0^{\pi} \! \sin \beta_1 \, \textrm{d} \beta_1 ... \sin \beta_N \, \textrm{d} \beta_N$) as well as over the set $[ \eta_\textrm{in} ]$ of angles $\eta_{\textrm{in},j}$ describing the relative orientations of nucleoprotein complexes with respect to the DNA segments entering them. Here subscript $j$ is used to enumerate angles $\eta_\textrm{in}$ corresponding to different nucleoprotein complexes according to their positions on the DNA, with angle $\eta_{\textrm{in},j}$ referring to a nucleoprotein complex occupying DNA segments with indexes $j\!+\!1$, $j\!+\!2$ and $j\!+\!3$.

Applying the vector-valued integration technique described in ref.~\cite{Efremov_2016}, it is not hard to show that the DNA partition function defined by Eq.~\eqref{DNA-protein-partition-function} obeys a number of recurrence relations, which are very similar to those derived in ref.~\cite{Efremov_2016}. Using these relations, it is then possible to greatly simplify the expression for the DNA partition function by utilizing the transfer-matrix formalism. To demonstrate it, we will use the same example of a protein with a binding site size of $K = 3$ DNA segments, which has been discussed in the previous Appendix section.

First, from the comments after Eq.~\eqref{all-local-energies-DNA-wrapping}, it can be seen that integrals $\int \textrm{d} [ \eta_\textrm{in} ]$ are only relevant for the DNA segments being in state $k_j = 0$, which are followed by a segment in state $k_{j+1} = 1$, as the angle $\eta_\textrm{in}$ appears only in $E_{01}$ energy term in the form of the rotation matrix $\textbf{B}$. This makes it possible to re-write the DNA partition function in a slightly different form:
\begin{equation} \label{DNA-protein-partition-function-2}
Z_{f,\tau} = 
    \sum_{k_1...k_N=0}^K \int \textrm{d} \textbf{R}_1 ... \textrm{d} \textbf{R}_N \, \textrm{d} \theta_1 ... 
    \textrm{d} \theta_N \,
    e^{-E_\textrm{tot} \left( k_1...k_N,\textbf{R}_1...\textbf{R}_N \right)}
    \xi \!\left( \textbf{R}_N, \textbf{R}_1 \right)
\end{equation} 

\noindent
Here $\int \textrm{d} \theta_1 ... \textrm{d} \theta_N$ is a shorthand notation for $\int \textrm{d} [ \eta_\textrm{in} ]$ integrals, which are calculated only over DNA segments being in the respective states:
\begin{equation} \label{Nucleoprotein-integrals}
\textrm{d} \theta_j = 
\begin{cases}
	\textrm{d} \eta_{\textrm{in},j}, \textrm{ if } k_j = 0 \textrm{ and } k_{j+1} = 1 \\
	\quad 1 \quad , \textrm{ otherwise} \\
\end{cases}
\end{equation}

To further simplify Eq.~\eqref{DNA-protein-partition-function-2}, it is convenient to introduce transfer-functions $T_{nm}$ defined as:
\begin{equation} \label{Transfer-matrix-elements}
T_{nm} \!\left( \textbf{R}, \textbf{R}' \right) = 
\begin{cases}
		\displaystyle \int_0^{2\pi} \!\textrm{d} \eta_\textrm{in} 
		\, e^{-E_{nm} \!\left( \textbf{R}, \textbf{R}' \right)} 
		, \textrm{ if } \left( n, m \right) = \left( 0, 1 \right)  \\[10pt]
		\hspace{23pt} e^{-E_{nm} \!\left( \textbf{R}, \textbf{R}' \right)} 
		\hspace{20pt} , \textrm{ otherwise} 
\end{cases}
\end{equation}  

\noindent
Where $E_{nm} ( \textbf{R}, \textbf{R}' )$ are the local energy terms defined on neighbouring DNA segments, whose structural states are denoted by indexes $n$ and $m$, and whose orientations are described by rotation matrices $\textbf{R}$ and $\textbf{R}'$, see Eq.~\eqref{all-local-energies-DNA-bending} and \eqref{all-local-energies-DNA-wrapping}.

Substituting Eq.~\eqref{DNA-total-energy} into Eq.~\eqref{DNA-protein-partition-function-2} and using Eq.~\eqref{Nucleoprotein-integrals} and \eqref{Transfer-matrix-elements}, we get:
\begin{equation} \label{DNA-protein-partition-function-3}
Z_{f,\tau} = 
    \sum_{k_1...k_N=0}^K \!\!\! \delta_{k_1 0} \times \int \textrm{d} \textbf{R}_1 ... \textrm{d} \textbf{R}_N
    \prod_{j=1}^{N-1} T_{k_j k_{j+1}} \!\left( \textbf{R}_j, \textbf{R}_{j+1} \right) 
    \times \sigma_{k_N} \!\left( \textbf{R}_N, \textbf{R}_1 \right)
\end{equation}

\noindent
Where
\begin{equation} \label{sigma-function}
\sigma_{k_N} \!\left( \textbf{R}_N, \textbf{R}_1 \right) = 
\delta_{k_N 0} \, e^{ b_0 f \left( \textbf{z}_0 \cdot \textbf{R}_N \textbf{z}_0 \right)} \, \xi \!\left( \textbf{R}_N, \textbf{R}_1 \right)
\end{equation}

Using Eq.~\eqref{DNA-protein-partition-function-3}, it is then not very hard to show that the DNA partition function obeys a number of important recurrence relations, which can be used to further simplify it. To derive these relations, let's first define intermediary partition functions as:
\begin{equation} \label{partition-function-intermediates}
Z_s \!\left( k_s, \textbf{R}_s, \textbf{R}_1 \right) = 
    \!\!\! \sum_{k_{s+1}...k_N=0}^K \int \textrm{d} \textbf{R}_{s+1} ... \textrm{d} \textbf{R}_N
    \prod_{j=s}^{N-1} T_{k_j k_{j+1}} \!\left( \textbf{R}_j, \textbf{R}_{j+1} \right) 
    \times \sigma_{k_N} \!\left( \textbf{R}_N, \textbf{R}_1 \right)
\end{equation} 

\noindent
Here $1 \leq s \leq N\!-\!1$. Then from Eq.~\eqref{DNA-protein-partition-function-3} and \eqref{partition-function-intermediates} it is not hard to see that the DNA partition function, $Z_{f,\tau}$, equals to:
\begin{equation} \label{DNA-protein-partition-function-4}
Z_{f,\tau} = \sum_{k_1=0}^K \delta_{k_1 0} \times \int \!\! \textrm{d} \textbf{R}_1 \,  Z_1\!\left( k_1, \textbf{R}_1, \textbf{R}_1 \right) =
	\int \!\! \textrm{d} \textbf{R}_1 \,
	\left(\begin{array}{cccc}
		\!1 & 0 & 0 & 0\!
    \end{array}\right)
    \times
    \left(\begin{array}{c}
		\!Z_1\!\left( 0, \textbf{R}_1,\textbf{R}_1 \right) \\[2pt]
		\!Z_1\!\left( 1, \textbf{R}_1,\textbf{R}_1 \right) \\[2pt]
		\!Z_1\!\left( 2, \textbf{R}_1,\textbf{R}_1 \right) \\[2pt]
		\!Z_1\!\left( 3, \textbf{R}_1,\textbf{R}_1 \right)
    \end{array}\right)
\end{equation}

What is even more important, from the definition of the intermediary partition functions it follows that obey the following recurrence relation:
\begin{equation} \label{partition-function-intermediates-relation-1}
Z_{s-1} \!\left( k_{s-1},\textbf{R}_{s-1},\textbf{R}_1 \right) = \int \!\! \textrm{d} \textbf{R}_s \,
	\left(\begin{array}{cccc}
		\!T_{k_{s-1} 0} & T_{k_{s-1} 1} & T_{k_{s-1} 2} & T_{k_{s-1} 3} \!
    \end{array}\right)
    \times
    \left(\begin{array}{c}
		\!Z_s\!\left( 0,\textbf{R}_s,\textbf{R}_1 \right) \\[2pt]
		\!Z_s\!\left( 1,\textbf{R}_s,\textbf{R}_1 \right) \\[2pt]
		\!Z_s\!\left( 2,\textbf{R}_s,\textbf{R}_1 \right) \\[2pt]
		\!Z_s\!\left( 3,\textbf{R}_s,\textbf{R}_1 \right)
    \end{array}\right),
\end{equation}

\noindent
which can be conveniently re-written in a more compact form using the vector-valued integration technique (see Appendix F in ref.~\cite{Efremov_2016}):
\begin{equation} \label{partition-function-intermediates-relation-2}
    \left(\begin{array}{c}
		\!Z_{s-1}\!\left( 0,\textbf{R}_{s-1},\textbf{R}_1 \right) \\[2pt]
		\!Z_{s-1}\!\left( 1,\textbf{R}_{s-1},\textbf{R}_1 \right) \\[2pt]
		\!Z_{s-1}\!\left( 2,\textbf{R}_{s-1},\textbf{R}_1 \right) \\[2pt]
		\!Z_{s-1}\!\left( 3,\textbf{R}_{s-1},\textbf{R}_1 \right) 
    \end{array}\right)
    = \int \!\! \textrm{d} \textbf{R}_s \, \textbf{T} \!\left( \textbf{R}_{s-1}, \textbf{R}_s \right) \times
    \left(\begin{array}{c}
		\!Z_s\!\left( 0,\textbf{R}_s,\textbf{R}_1 \right) \\[2pt]
		\!Z_s\!\left( 1,\textbf{R}_s,\textbf{R}_1 \right) \\[2pt]
		\!Z_s\!\left( 2,\textbf{R}_s,\textbf{R}_1 \right) \\[2pt]
		\!Z_s\!\left( 3,\textbf{R}_s,\textbf{R}_1 \right)
    \end{array}\right)
\end{equation}

\noindent
Here $\textbf{T} ( \textbf{R}_{s-1}, \textbf{R}_s )$ is the DNA transfer-matrix, which is defined as:
\begin{equation} \label{Transfer-matrix-2}
\textbf{T} \!\left( \textbf{R}_{s-1}, \textbf{R}_s	\right) =
	    \left(\begin{array}{cccc}
		T_{00} & T_{01} & T_{02} & T_{03} \\[2pt]
		T_{10} & T_{11} & T_{12} & T_{13} \\[2pt]
		T_{20} & T_{21} & T_{22} & T_{23} \\[2pt]
		T_{30} & T_{31} & T_{32} & T_{33}
    \end{array}\!\right) =
    \left(\begin{array}{cccc}
		T_{00} & T_{01} & 0 & 0 \\[2pt]
		0 & 0 & T_{12} & 0 \\[2pt]
		0 & 0 & 0 & T_{23} \\[2pt]
		T_{30} & T_{31} & 0 & 0
    \end{array}\!\right)
\end{equation}

\noindent
Where in the right part of the above equation, we simply took into account that all of the matrix entries corresponding to partially unfolded nucleoprotein complexes become nullified due to the infinitely high energy of such DNA-protein conformations, see Eq.~\eqref{DNA-protein-energy}, \eqref{all-local-energies-DNA-bending} and \eqref{all-local-energies-DNA-wrapping}. For the sake of the formula simplicity, the arguments, $( \textbf{R}_{s-1}, \textbf{R}_s )$, of transfer-functions $T_{nm} ( \textbf{R}_{s-1}, \textbf{R}_s )$ are omitted in Eq.~\eqref{partition-function-intermediates-relation-1} and \eqref{Transfer-matrix-2}.

Combining together Eq.~\eqref{DNA-protein-partition-function-4} and \eqref{partition-function-intermediates-relation-2}, we finally obtain the formula for the DNA partition function in terms of the transfer-matrices product:
\begin{equation} \label{DNA-protein-partition-function-5}
Z_{f,\tau} = 
	\left(\begin{array}{cccc}
		\!1 & 0 & 0 & 0\!\!
    \end{array}\right)
    \times
    \int \textrm{d} \textbf{R}_1 ... \textrm{d} \textbf{R}_N 
    \prod_{j=1}^{N-1} \textbf{T} \!\left( \textbf{R}_j, \textbf{R}_{j+1} \right) 
    \times \bm{\sigma} \!\left( \textbf{R}_N, \textbf{R}_1 \right)
\end{equation} 

\noindent
Where the boundary condition vector, $\bm{\sigma} ( \textbf{R}_N, \textbf{R}_1 )$, is:
\begin{equation} \label{Sigma-matrix}
\bm{\sigma} \!\left( \textbf{R}_N, \textbf{R}_1 \right) = 
    \left(\begin{array}{c}
		\sigma_0 \!\left( \textbf{R}_N, \textbf{R}_1 \right) \\[2pt]
		0 \\[2pt]
		0 \\[2pt]
		0 \\[2pt]
    \end{array}\right)
\end{equation}

\noindent 
Here $\sigma_0 ( \textbf{R}_N, \textbf{R}_1 ) = \xi ( \textbf{R}_N, \textbf{R}_1 ) e^{ b_0 f ( \textbf{z}_0 \cdot \textbf{R}_N \textbf{z}_0 )}$.

Eq.~\eqref{DNA-protein-partition-function-5} can be further streamlined by recalling that any square-integrable function defined on SO(3) group of 3D rotation matrices parametrized by Euler angles $( \alpha, \beta, \gamma )$ can be expanded into a series of orthogonal D-functions, $D^s_{p,q} (\alpha, \beta, \gamma)$ \cite{Gelfand_1963}. Performing such an expansion with respect to the both arguments of $T_{nm} ( \textbf{R}, \textbf{R}' )$ elements of the transfer-matrix $\textbf{T} ( \textbf{R}, \textbf{R}' )$, we obtain the following series of $D^s_{p,q}$ functions (see Appendices~\ref{Appendix-C}-\ref{Appendix-D} for details):
\begin{equation} \label{Transfer-matrix-elements-expansion}
T_{nm} \!\left( \textbf{R}, \textbf{R}' \right) =
\frac{1}{8\pi^2} \!\!\! \sum_{p,p'\!\!,\,q,q'\!\!,\,s,s'\!} \!\!\!\!\!\! \sqrt{\left( 2s\!+\!1 \right) \!\left( 2s'\!\!+\!1 \right)} \, \left( T_{nm} \right)^{p'\!\!,\,q'\!\!,\,s'}_{p,\,q,\,s} D^s_{p,q} \!\left( \textbf{R} \right) \overline{D}^{s'}_{p'\!,q'} \!\left( \textbf{R}' \right)
\end{equation}

\noindent
Where $( T_{nm} )^{p'\!\!,\,q'\!\!,\,s'}_{p,\,q,\,s}$ are the expansion coefficients.

Then by taking into account the linear property of matrices, it is not hard to see that the DNA transfer-matrix $\textbf{T} \!\left( \textbf{R},\textbf{R}' \right)$ can be presented as:
\begin{equation} \label{Transfer-matrix-expansion}
\textbf{T} \!\left( \textbf{R}, \textbf{R}' \right) =
\frac{1}{8\pi^2} \!\!\! \sum_{p,p'\!\!,\,q,q'\!\!,\,s,s'\!} \!\!\!\!\!\! \sqrt{\left( 2s\!+\!1 \right) \!\left( 2s'\!\!+\!1 \right)} \; \textbf{T}^{p'\!\!,\,q'\!\!,\,s'}_{p,\,q,\,s} D^s_{p,q} \!\left( \textbf{R} \right) \overline{D}^{s'}_{p'\!,q'} \!\left( \textbf{R}' \right)
\end{equation}

\noindent
Where $\textbf{T}^{p'\!\!,\,q'\!\!,\,s'}_{p,\,q,\,s}$ denotes the matrix built of the expansion coefficients of $T_{nm} ( \textbf{R}, \textbf{R}' )$ functions:
\begin{equation} \label{Transfer-matrix-expansion-coeff}
\textbf{T}^{p'\!\!,\,q'\!\!,\,s'}_{p,\,q,\,s} =
    \left(\begin{array}{cccc}
		\! \left( T_{00} \right)^{p'\!\!,\,q'\!\!,\,s'}_{p,\,q,\,s} & \left( T_{01} \right)^{p'\!\!,\,q'\!\!,\,s'}_{p,\,q,
		\,s} & 0 & 0 \\[2pt]
		0 & 0 & \left( T_{12} \right)^{p'\!\!,\,q'\!\!,\,s'}_{p,\,q,\,s} & 0 \\[2pt]
		0 & 0 & 0 & \left( T_{23} \right)^{p'\!\!,\,q'\!\!,\,s'}_{p,\,q,\,s} \\[2pt]
		\! \left( T_{30} \right)^{p'\!\!,\,q'\!\!,\,s'}_{p,\,q,\,s} & \left( T_{31} \right)^{p'\!\!,\,q'\!\!,\,s'}_{p,\,q,
		\,s} & 0 & 0
    \end{array}\!\!\!\right)
\end{equation}

Analogously, for the boundary condition vector, $\bm{\sigma} ( \textbf{R}_N, \textbf{R}_1 )$, we have:
\begin{equation} \label{Sigma-matrix-expansion}
\pmb{\sigma} \!\left( \textbf{R}_N, \textbf{R}_1 \right) =
\frac{1}{8\pi^2} \!\!\!\! \sum_{\substack{\,p_1\!,\,q_1\!,\,s_1 \\ p_N \!,\,q_N\!,\,s_N}} \!\!\!\!\!\! \sqrt{\left( 2s_1\!+\!1 \right) \!\left( 2s_N\!+\!1 \right)} \; \pmb{\sigma}^{\;p_1\!,\,q_1\!,\,s_1}_{p_N \!,\,q_N\!,\,s_N} D^{s_N}_{p_N,q_N} \!\left( \textbf{R}_N \right) \overline{D}^{s_1}_{p_1,q_1} \!\left( \textbf{R}_1 \right)
\end{equation}

\noindent
Where $\pmb{\sigma}^{\;p_1\!,\,q_1\!,\,s_1}_{p_N \!,\,q_N\!,\,s_N}$ is the following vector of expansion coefficients:
\begin{equation} \label{Sigma-matrix-expansion-coeff}
\pmb{\sigma}^{\;p_1\!,\,q_1\!,\,s_1}_{p_N \!,\,q_N\!,\,s_N} = 
	\left(\begin{array}{c}
		\!\left( \sigma_0 \right)^{\;p_1\!,\,q_1\!,\,s_1}_{p_N \!,\,q_N\!,\,s_N} \\[2pt]
		0 \\[2pt]
		0 \\[2pt]
		0
    \end{array} \!\!\right)
\end{equation}

After substituting Eq.~\eqref{Transfer-matrix-expansion} and \eqref{Sigma-matrix-expansion} into Eq.~\eqref{DNA-protein-partition-function-5}, and using orthogonality of $D^s_{p,q}$ functions [Eq.~\eqref{D-functions-orthogonality}], it can be shown that all of the integrals in Eq.~\eqref{DNA-protein-partition-function-5} reduce to mere summations over the indexes of the expansion coefficient matrices:
\begin{equation} \label{DNA-protein-partition-function-6}
Z_{f,\tau} =  
	\left(\begin{array}{cccc}
		\!1 & 0 & 0 & 0\!\!
    \end{array}\right) \times
\!\!\!\! \sum_{\substack{p_1...p_N \\ q_1...q_N \\ s_1...s_N}} \!\! \left[  \prod_{j=1}^{N-1} \textbf{T}_{\!\!\!\!\!\!\!\quad p_j, \;q_j, \;s_j}^{p_{j+1},\,q_{j+1},\,s_{j+1}} \times \pmb{\sigma}^{\;p_1\!,\,q_1\!,\,s_1}_{p_N \!,\,q_N\!,\,s_N} \right]
\end{equation}

While the exact mathematical forms of the expansion coefficients $( T_{nm} )^{p'\!\!,\,q'\!\!,\,s'}_{p,\,q,\,s}$ that compose matrices $\textbf{T}^{p'\!\!,\,q'\!\!,\,s'}_{p,\,q,\,s}$ are derived in Appendix~\ref{Appendix-D} [see Eq.~\eqref{T00-expansion-coeff}, \eqref{T01-expansion-coeff}, \eqref{T31-expansion-coeff} and \eqref{T12-expansion-coeff}], here we would only like to note that all these coefficients contain $\delta_{pp'}$ Kronecker delta prefactor. Furthermore, if the boundary condition function $\sigma_0 ( \textbf{R}_N, \textbf{R}_1 )$ has a symmetry with respect to $\textbf{z}_0$-axis of the global coordinate system (which is frequently the case in \textit{in vitro} experiments), it can be shown that all of the expansion coefficients $( \sigma_0 )^{\;p_1\!,\,q_1\!,\,s_1}_{p_N \!,\,q_N\!,\,s_N}$ forming matrices $\pmb{\sigma}^{\;p_1\!,\,q_1\!,\,s_1}_{p_N \!,\,q_N\!,\,s_N}$ have $\delta_{p_N 0}$ prefactor, see, for example, Eq.~\eqref{sigma0-unconstrained-expansion-coeff} and Eq.~\eqref{sigma0-parallel-expansion-coeff} in Appendix~\ref{sec:expansion-boundary}. Combined together, all these Kronecker deltas lead to nullification of all $p_j$ indexes in Eq.~\eqref{DNA-protein-partition-function-6} via a domino-like effect in the same way as in the case of bare DNA scenario discussed in Appendix C of ref.~\cite{Efremov_2016}. However, in contrast to the case of bare DNA, the same nullification effect usually does not take place for $q_j$ indexes as the expansion coefficients $( T_{12} )^{p'\!\!,\,q'\!\!,\,s'}_{p,\,q,\,s}$,  $( T_{23} )^{p'\!\!,\,q'\!\!,\,s'}_{p,\,q,\,s}$ and $( T_{31} )^{p'\!\!,\,q'\!\!,\,s'}_{p,\,q,\,s}$ do not contain $\delta_{qq'}$ prefactor in the general case, see Eq.~\eqref{T31-expansion-coeff} and \eqref{T12-expansion-coeff}.

By taking into account the above notes and putting $p_j = 0$ for all $j=1,...,N$, we obtain the following expression for the DNA partition function:
\begin{equation} \label{DNA-protein-partition-function-7}
Z_{f,\tau} =  
	\left(\begin{array}{cccc}
		\!1 & 0 & 0 & 0\!\!
    \end{array}\right) \times
\!\!\!\! \sum_{\substack{q_1...q_N \\ s_1...s_N}} \!\! \left[  \prod_{j=1}^{N-1} \textbf{T}_{0, \,q_j, \,s_j}^{0, \,q_{j+1}, s_{j+1}} \times \pmb{\sigma}^{0, \,q_1, \,s_1}_{0, \,q_N \!,\, s_N} \right]
\end{equation}

To further simplify Eq.~\eqref{DNA-protein-partition-function-7}, it is convenient to slightly rearrange multidimensional arrays $\textbf{T}_{0, \,q, \,s}^{0, \,q'\!\!, \,s'}$ and $\pmb{\sigma}^{0, \,q_1, \,s_1}_{0, \,q_N \!, \,s_N}$ by recalling from the definition of $D^s_{p,q}$ functions that index $q$ varies in the range of $-s \leq q \leq s$ for any given value of index $s$, see Appendix~\ref{Appendix-C}. Similarly, for $q'$ we have: $-s' \leq q' \leq s'$. As a result, it can be shown that for any fixed pair of integers $n$ and $m$ ($0 \leq n,m \leq 3$) the expansion coefficients $( T_{nm} )^{0,\,q'\!\!,\,s'}_{0,\,q,\,s}$ can be re-organized in the form of a two-dimensional matrix, $\textbf{S}_{nm}$, whose elements are enumerated by indexes $v$ and $v'$ that relate to the old indexes $q, q', s$ and $s'$ as $v = q+s(s+1)$ and $v' = q'+s'(s'+1)$, such that $( \textbf{S}_{nm} )_{vv'} = ( T_{nm} )^{0,\,q'\!\!,\,s'}_{0,\,q,\,s}$. In a very similar way, it is possible to rearrange the expansion coefficients $( \sigma_0 )^{0, \,q_1, \,s_1}_{0, \,q_N \!, \,s_N}$ in the form of a new two-dimensional boundary condition matrix, $\textbf{V}_0$, such that $( \textbf{V}_0 )_{v_N v_1} = ( \sigma_0 )^{0, \,q_1, \,s_1}_{0, \,q_N \!, \,s_N}$, where indexes $v_1$ and $v_N$ are defined by the same equations as $v$ and $v'$: $v_1 = q_1+s_1(s_1+1)$ and $v_N = q_N+s_N(s_N+1)$.

Substituting the newly formed matrices, $\textbf{S}_{nm}$ and $\textbf{V}_0$, into Eq.~\eqref{Transfer-matrix-expansion-coeff} and \eqref{Sigma-matrix-expansion-coeff}, we obtain arrays of the expansion coefficients, $\textbf{L}_{vv'}$ and $\textbf{Y}_{v_N v_1}$, which have reduced dimensionalities comparing to $\textbf{T}_{0, \,q, \,s}^{0, \,q'\!\!, \,s'}$ and $\pmb{\sigma}^{0, \,q_1, \,s_1}_{0, \,q_N \!, \,s_N}$:
\begin{equation} \label{2D-matrix-forms}
\textbf{L}_{vv'} = 
    \left(\begin{array}{cccc}
		\! \left( \textbf{S}_{00} \right)_{vv'} & \left( \textbf{S}_{01} \right)_{vv'} & 0 & 0 \\[2pt]
		0 & 0 & \left( \textbf{S}_{12} \right)_{vv'} & 0 \\[2pt]
		0 & 0 & 0 & \left( \textbf{S}_{23} \right)_{vv'} \\[2pt]
		\! \left( \textbf{S}_{30} \right)_{vv'} & \left( \textbf{S}_{31} \right)_{vv'} & 0 & 0
    \end{array}\!\!\!\right) =
\textbf{T}_{0, \,q, \,s}^{0, \,q'\!\!, \,s'} 
\textrm{\quad and \quad} 
\textbf{Y}_{v_N v_1} = 
	\left(\begin{array}{c}
		\!\left( \textbf{V}_0 \right)_{v_N v_1} \\[2pt]
		0 \\[2pt]
		0 \\[2pt]
		0
    \end{array} \!\!\right) =
\pmb{\sigma}^{0, \,q_1, \,s_1}_{0, \,q_N \!, \,s_N}
\end{equation} 

It should be noted that in the general case matrices $\textbf{S}_{nm}$ and $\textbf{V}_0$ have infinite size. However, calculations show that the value of the DNA partition function is typically determined by several first harmonics corresponding to indexes $-s \leq q \leq s$, $-s' \leq q' \leq s'$ and $0 \leq s,s' \leq s_\textrm{max}$, where $s_\textrm{max} \sim 14-15$, see ref.~\cite{Efremov_2016}. Thus, in real computations it makes sense to use finite $( s_\textrm{max} \!+\! 1 )^2 \times ( s_\textrm{max} \!+\! 1 )^2$ square matrices $\textbf{S}_{nm}$ and $\textbf{V}_0$, which include only the first $( s_\textrm{max} \!+\! 1 )^2$ rows and columns related to the above harmonics.

Anyway, utilizing the new matrices $\textbf{L}_{vv'}$ and $\textbf{Y}_{v_N v_1}$, it is then straightforward to apply the mathematical technique based on the generalized matrix multiplication formula described in Appendix E of ref.~\cite{Efremov_2016} in order to streamline Eq.~\eqref{DNA-protein-partition-function-7}. Indeed, substituting Eq.~\eqref{2D-matrix-forms} into Eq.~\eqref{DNA-protein-partition-function-7}, we get the following expression for the DNA partition function:
\begin{equation} \label{DNA-protein-partition-function-8}
Z_{f,\tau} =  
	\left(\begin{array}{cccc}
		\!1 & 0 & 0 & 0\!\!
    \end{array}\right) \times
\!\!\!\! \sum_{v_1...v_N} \!\! \left[  \prod_{j=1}^{N-1} \textbf{L}_{v_j v_{j+1}} \times \textbf{Y}_{v_N v_1} \right]
\end{equation}

Now, by using the classical definition of the matrix product, it is not hard to check that each of the sums over indexes $v_2,..,v_{N\!-\!1}$ in Eq.~\eqref{DNA-protein-partition-function-8} reduces to a mere multiplication of the DNA transfer-matrices:
\begin{equation} \label{new-matrices-product-1}
\sum_t 
    \left.\left(\begin{array}{cccc}
		\textbf{S}_{00} & \textbf{S}_{01} & 0 & 0 \\[2pt]
		0 & 0 & \textbf{S}_{12} & 0 \\[2pt]
		0 & 0 & 0 & \textbf{S}_{23} \\[2pt]
		\textbf{S}_{30} & \textbf{S}_{31} & 0 & 0
    \end{array} \!\!\right) \right\vert_{st}
    \left.\left(\begin{array}{cccc}
		\textbf{S}_{00} & \textbf{S}_{01} & 0 & 0 \\[2pt]
		0 & 0 & \textbf{S}_{12} & 0 \\[2pt]
		0 & 0 & 0 & \textbf{S}_{23} \\[2pt]
		\textbf{S}_{30} & \textbf{S}_{31} & 0 & 0
    \end{array} \!\!\right) \right\vert_{tp} =
    \left.\left(\begin{array}{cccc}
		\textbf{S}_{00} & \textbf{S}_{01} & 0 & 0 \\[2pt]
		0 & 0 & \textbf{S}_{12} & 0 \\[2pt]
		0 & 0 & 0 & \textbf{S}_{23} \\[2pt]
		\textbf{S}_{30} & \textbf{S}_{31} & 0 & 0
    \end{array} \!\!\right)^{\!\!\!\!2} \,\right\vert_{sp}
\end{equation}

\noindent 
Where we have employed the following short-hand notation:
\begin{equation} \label{new-matrices-product-2}
    \left.\left(\begin{array}{cccc}
		\textbf{S}_{00} & \textbf{S}_{01} & 0 & 0 \\[2pt]
		0 & 0 & \textbf{S}_{12} & 0 \\[2pt]
		0 & 0 & 0 & \textbf{S}_{23} \\[2pt]
		\textbf{S}_{30} & \textbf{S}_{31} & 0 & 0
    \end{array} \!\!\right) \right\vert_{st} =
    \left(\begin{array}{cccc}
		\! \left( \textbf{S}_{00} \right)_{st} & \left( \textbf{S}_{01} \right)_{st} & 0 & 0 \\[2pt]
		0 & 0 & \left( \textbf{S}_{12} \right)_{st} & 0 \\[2pt]
		0 & 0 & 0 & \left( \textbf{S}_{23} \right)_{st} \\[2pt]
		\! \left( \textbf{S}_{30} \right)_{st} & \left( \textbf{S}_{31} \right)_{st} & 0 & 0
    \end{array}\!\!\!\right)
\end{equation}

Applying Eq.~\eqref{new-matrices-product-1} $N\!-\!2$ times to Eq.~\eqref{DNA-protein-partition-function-8}, it is not hard to see that the mathematical expression for the DNA partition function takes the following form:
\begin{equation} \label{DNA-protein-partition-function-9}
Z_{f,\tau} =  
	\left(\begin{array}{cccc}
		\!1 & 0 & 0 & 0\!\!
    \end{array}\right) \times
\!\! \sum_{v_1,v_N} \!\! \left[  \left. \textbf{L}^{N\!-\!1} \right\vert_{v_1 v_N} \times \left. \textbf{Y} \right\vert_{v_N v_1} \right]
\end{equation}

\noindent
Where block-matrices $\textbf{L}$ and $\textbf{Y}$ are:
\begin{equation} \label{SV-big-block-matrices}
\textbf{L} = 
	\left(\begin{array}{cccc}
		\textbf{S}_{00} & \textbf{S}_{01} & 0 & 0 \\[2pt]
		0 & 0 & \textbf{S}_{12} & 0 \\[2pt]
		0 & 0 & 0 & \textbf{S}_{23} \\[2pt]
		\textbf{S}_{30} & \textbf{S}_{31} & 0 & 0
    \end{array} \!\!\right)
\quad \textrm{and} \quad
\textbf{Y} = 
	\left(\begin{array}{c}
		\!\textbf{V}_0 \\[2pt]
		0 \\[2pt]
		0 \\[2pt]
		0
    \end{array} \!\right)
\end{equation}

Finally, after a few simple algebraic re-arrangements, Eq.~\eqref{DNA-protein-partition-function-9} can be presented in the form of Eq.~\eqref{DNA-protein-partition-function-intro-4}:
\begin{equation} \label{DNA-protein-partition-function-10}
Z_{f,\tau} = \textrm{Tr} \!\left( \textbf{U} \textbf{L}^{N\!-\!1} \textbf{Y} \right)
\end{equation}

\noindent
Where block-matrix $\textbf{U} = \begin{pmatrix} \textbf{I} & 0 & 0 & 0 \end{pmatrix}$, with $\textbf{I}$ being the square $( s_\textrm{max} \!+\! 1 )^2 \times ( s_\textrm{max} \!+\! 1 )^2$ identity matrix ($\textbf{I}_{nm} = \delta_{nm}$).

All that remains now is to derive mathematical expressions for the elements of matrices $\textbf{S}_{nm}$ and $\textbf{V}_0$, which is done in the next two Appendix sections~\ref{Appendix-C} and \ref{Appendix-D}, thus concluding the description of the transfer-matrix approach for the special case of proteins that have the binding site size of three DNA segments ($K = 3$). In Appendix~\ref{Appendix-E}, the obtained formulas will be further generalized for the case of DNA-binding proteins that have an arbitrary large binding site size on DNA.

\section{Orthogonal D-functions}
\label{Appendix-C}

To find out the elements of matrices $\textbf{S}_{nm}$ and $\textbf{V}_0$, we will use several famous results from the group theory, which have been described in our previous work~\cite{Efremov_2016} and which we are going to repeat in this Appendix section for the sake of convenience, as we will be using them quite extensively in our derivations.

First of all, we would like to recall that from the group theory it is known that any square-integrable function defined on SO(3) group can be expanded into a series of orthogonal functions, $D^s_{p,q}$, which have the following canonical form, see p. 101 in \cite{Gelfand_1963}:
\begin{equation} \label{D-functions}
D^s_{p,q} \!\left( \alpha,\beta,\gamma \right) = 
	e^{-ip\alpha} P^s_{p,q} \!\left(\cos \beta \right) e^{-iq\gamma}
\end{equation}

\noindent
Here $( \alpha,\beta,\gamma )$ are the three Euler rotation angles, which are usually used to parametrize SO(3) group ($\alpha,\gamma \in [0,2\pi]$ and $\beta \in [0,\pi]$); $s$, $p$, $q$ are integers such that $s \geq 0$ and $-s \leq p,q \leq s$; finally, $P^s_{p,q}$ are polynomials, which relate to the elements of so-called small Wigner d-matrix, $d^s_{p,q}$, as: $P^s_{p,q} ( \cos \beta ) = i^{p-q} d^s_{p,q} ( \beta )$.

Functions $D^s_{p,q}$ and polynomials $P^s_{p,q}$ possess a number of important properties, which will come in handy in our derivations of the formulas for the transfer-matrix elements.

First, by substituting $( \alpha,\beta,\gamma ) = ( 0,0,0 )$ into Eq.~\eqref{D-functions} and taking into account that $d^s_{p,q} ( 0 ) = \delta_{pq}$, we get:
\begin{equation} \label{D-functions-000}
D^s_{p,q} \!\left( 0,0,0 \right) = P^s_{p,q} \!\left( 1 \right) = i^{p-q} d^s_{p,q} \!\left( 0 \right) = \delta_{pq}
\end{equation}

\noindent
Here, as before, $\delta_{pq}$ is the Kronecker delta ($\delta_{pq} = 1$ if $p = q$ and $\delta_{pq} = 0$, otherwise).

Furthermore, since for any indexes $s \geq 0$ and $-s \leq p,q \leq s$: $d^s_{p,q} ( \beta )$ are real functions obeying the following symmetric relations $d^s_{p,q} ( \beta ) = ( -1 )^{q-p} d^s_{q,p} ( \beta ) = d^s_{-q,-p} ( \beta )$, it is not very hard to see that:
\begin{equation} \label{P-functions-symmetry}
\left( -1 \right)^{q-p} \overline{P}^s_{p,q} \!\left( x \right) = P^s_{p,q} \!\left( x \right) = P^s_{q,p} \!\left( x \right) = P^s_{-p,-q} \!\left( x \right)
\end{equation}

\noindent
Where the bar over the function denotes the complex conjugate.

Combining together Eq.~\eqref{D-functions} and \eqref{P-functions-symmetry}, we obtain:
\begin{equation} \label{D-functions-conjugate}
\overline{D}^s_{p,q} \!\left( \textbf{R} \right) = \left( -1 \right)^{p-q} D^n_{-p,-q} \!\left( \textbf{R} \right)
\end{equation}

\noindent
Here and below for the sake of formulas simplicity we use $D^s_{p,q} ( \textbf{R} )$ notation to address functions $D^s_{p,q} ( \alpha,\beta,\gamma )$, where $\textbf{R}$ is the Euler rotation matrix corresponding to angles $( \alpha,\beta,\gamma )$.

Using Eq.~\eqref{D-functions}, \eqref{P-functions-symmetry} and \eqref{D-functions-conjugate}, it is straightforward to show that:
\begin{equation} \label{D-functions-inverse}
D^s_{p,q} \!\left( \textbf{R}^{-1} \right) = \overline{D}^s_{q,p} \!\left( \textbf{R} \right)
\end{equation}

\noindent
Here matrix $\textbf{R}^{-1}$ corresponding to Euler angles $( \pi \!-\! \gamma,\beta,\pi \!-\! \alpha )$ is the inverse of matrix $\textbf{R}$ (i.e., $\textbf{R}^{-1} \textbf{R} = \textbf{R} \textbf{R}^{-1} = \textbf{I}$, where $\textbf{I}$ is the $3 \times 3$ identity matrix: $I_{pq} = \delta_{pq}$).

Next, functions $D^s_{p,q}$ obey the following important multiplication rules \cite{Gelfand_1963}:
\begin{equation} \label{D-functions-product-1}
D^s_{p,q} \!\left( \textbf{R}_1 \textbf{R}_2 \right) = 
	\sum_{t=-s}^s \! D^s_{p,t} \!\left( \textbf{R}_1 \right) D^s_{t,q} \!\left( \textbf{R}_2 \right)
\end{equation}

\noindent
and
\begin{equation} \label{D-functions-product-2}
D^{s_1}_{p_1 \!, q_1} \!\!\left( \textbf{R} \right) D^{s_2}_{p_2, q_2} \!\!\left( \textbf{R} \right) = 
	\sum_{s} \left\langle s_1 s_2 p_1 p_2 | s \!\left( p_1 \!+\! p_2 \right) \right\rangle \left\langle s_1 s_2 q_1 q_2 |
	s \!\left( q_1 \!+\! q_2 \right) \right\rangle D^s_{p_1\!+p_2,q_1\!+q_2} \!\!\left( \textbf{R} \right)
\end{equation}

\noindent
Where $\langle s_1 s_2 p_1 p_2 | s_3 p_3 \rangle$ are Clebsh-Gordan coefficients. For the sake of the formulas simplicity and compactness, below we will use Winger 3-j symbols instead of Clebsh-Gordan coefficients, which relate to each other as:
\begin{equation} \label{Wigner-3j}
\left(
    \begin{array}{ccc}
		s_1 & s_2 & s_3 \\
		p_1 & p_2 & p_3
	\end{array}
\!\right) 
= \frac{\left( -1 \right)^{s_1-s_2-p_3}}{\sqrt{2s_3\!+\!1}} \left\langle s_1 s_2 p_1 p_2 | s_3 \!\left( -p_3 \right) \right\rangle
\end{equation}

The final important property of $D^s_{p,q}$ functions required for the DNA transfer-matrix derivation is their orthogonality, which was mentioned in the beginning of this Appendix section. Namely, it can be shown that \cite{Gelfand_1963}:
\begin{equation} \label{D-functions-orthogonality}
\int \!\textrm{d} \textbf{R} \, \overline{D}^{s_1}_{p_1 \!, q_1}\!\!\left( \textbf{R} \right) D^{s_2}_{p_2, q_2}\!\!\left( \textbf{R} \right) = 
	\frac{8\pi^2}{2s_1\!+\!1} \, \delta_{s_1 s_2} \delta_{p_1 p_2} \delta_{q_1 q_2}
\end{equation}

\noindent
Where the integration in the above formula is carried out over all of the possible combinations of the Euler angles $( \alpha,\beta,\gamma )$:
\begin{equation} \label{dR-integral}
\int \! \textrm{d} \textbf{R} = 
	\int_0^{2\pi} \!\!\!\! \textrm{d} \alpha \int_0^{2\pi} \!\!\!\! \textrm{d} \gamma \int_0^{\pi} \!\! \sin \beta \,
	\textrm{d} \beta
\end{equation}

Orthogonality and completeness of $D^s_{p,q}$ functions make it possible to use them as a Hilbert basis in the space of square-integrable functions, $F ( \alpha,\beta,\gamma ) = F ( \textbf{R} )$, defined on SO(3) group \cite{Gelfand_1963}. Therefore, any such function, $F ( \textbf{R} )$, can be expanded into the following series:
\begin{equation} \label{D-expansion-1}
F \! \left( \textbf{R} \right) = 
	\sum_{s=0}^{\infty} \, \sum_{p,q=-s}^s \!\!\! F_{p, q, s} D^s_{p,q} \! \left( \textbf{R} \right)
\end{equation}

\noindent
Where the expansion coefficients $F_{p, q, s}$ are:
\begin{equation} \label{expansion-coeff-1}
F_{p, q, s} = 
	\frac{2s\!+\!1}{8\pi^2} \int \!\! \textrm{d} \textbf{R} \, \overline{D}^s_{p,q} \! \left( \textbf{R} \right) F \! 
	\left( \textbf{R} \right)
\end{equation}

Analogously, for any square-integrable function $F ( \textbf{R},\textbf{R}' )$, where $\textbf{R}$ and $\textbf{R}'$ are two rotation matrices, we have:
\begin{equation} \label{D-expansion-2}
F \!\left( \textbf{R},\textbf{R}' \right) = 
	\sum_{s,s'=0}^{\infty} \, \sum_{p,q=-s}^s \, \sum_{p'\!,q'=-s'}^{s'} \!\!\!\! 
	F^{p'\!\!,\,q'\!\!,\,s'}_{p,\,q,\,s}
	D^s_{p,q} \!\left( \textbf{R} \right) \overline{D}^{s'}_{p'\!,q'} \! \left( \textbf{R}' \right)
\end{equation}

\noindent
Where the expansion coefficients $F^{p'\!\!,\,q'\!\!,\,s'}_{p,\,q,\,s}$ are:
\begin{equation} \label{expansion-coeff-2}
F^{p'\!\!,\,q'\!\!,\,s'}_{p,\,q,\,s} = 
	\frac{ \left(2s\!+\!1\right)\!\left(2s'\!\!+\!1\right) }{ \left(8\pi^2\right)^2 } 
	\int \! \textrm{d} \textbf{R} \textrm{d} \textbf{R}' \,
	\overline{D}^s_{p,q} \!\left( \textbf{R} \right) F \!\left( \textbf{R}, \textbf{R}' \right)
    D^{s'}_{p'\!,q'} \!\left( \textbf{R}' \right)
\end{equation}

In order to use Eq.~\eqref{D-expansion-2} and \eqref{expansion-coeff-2} for the DNA partition function calculations, it will be more convenient to slightly re-organize these two formulas, since as it can be seen from Eq.~\eqref{D-functions-orthogonality}, functions $D^s_{p,q}$ have the $\textrm{L}_2$-norm $\| D^s_{p,q} \|_2 = \sqrt{\frac{8\pi^2}{2s+1}}$, and thus are not normalized:
\begin{equation}\label{D-functions-norm}
\| D^s_{p,q} \|^2_2 = 
	\int \!\!\textrm{d} \textbf{R} \, \overline{D}^s_{p,q} \!\left( \textbf{R} \right) 
	D^s_{p,q}\!\left( \textbf{R} \right)
= \frac{8\pi^2}{2s\!+\!1}
\end{equation}

Hence, while being orthogonal, the basis formed by $D^s_{p,q}$ functions is not orthonormal. Using Eq.~\eqref{D-functions-norm}, we can easily normalize it by switching from $D^s_{p,q}$ to $\sqrt{\frac{2s+1}{8\pi^2}} D^s_{p,q}$ functions. By doing this, Eq.~\eqref{D-expansion-2} turns into:
\begin{equation}\label{D-expansion-3}
F \!\left( \textbf{R},\textbf{R}' \right) = 
	\frac{1}{8\pi^2} \! \sum_{s,s'=0}^{\infty} \, \sum_{p,q=-s}^s \, \sum_{p'\!,q'=-s'}^{s'} \!\!\!\!
	\sqrt{\left( 2s\!+\!1 \right) \!\left( 2s'\!\!+\!1 \right)} \,
	F^{p'\!\!,\,q'\!\!,\,s'}_{p,\,q,\,s}
	D^s_{p,q} \!\left( \textbf{R} \right) \overline{D}^{s'}_{p'\!,q'} \! \left( \textbf{R}' \right)
\end{equation}

\noindent
Where the expansion coefficients $F^{p'\!\!,\,q'\!\!,\,s'}_{p,\,q,\,s}$ are:
\begin{equation} \label{expansion-coeff-3}
F^{p'\!\!,\,q'\!\!,\,s'}_{p,\,q,\,s} = 
	\frac{ \sqrt{\left( 2s\!+\!1 \right) \!\left( 2s'\!\!+\!1 \right)} }{ 8\pi^2 } 
	\int \! \textrm{d} \textbf{R} \textrm{d} \textbf{R}' \,
	\overline{D}^s_{p,q} \!\left( \textbf{R} \right) F \!\left( \textbf{R}, \textbf{R}' \right)
	 D^{s'}_{p'\!,q'} \!\left( \textbf{R}' \right)
\end{equation}

With all of the above formulas at hand, we will now deduct mathematical expressions for the transfer-matrix elements, which were discussed in Appendix~\ref{Appendix-B}.

\section{Expansion formulas}
\label{Appendix-D}

From Eq.~\eqref{Transfer-matrix-elements-expansion} and comments after Eq.~\eqref{DNA-protein-partition-function-7} it can be seen that in order to find the elements of $\textbf{S}_{nm}$ matrices that constitute the DNA transfer-matrix $\textbf{L}$, all we need to do is to derive formulas for the expansion coefficients of $T_{nm}$ transfer-functions. The simplest way to achieve this goal is to note that all of $E_{nm}$ energy terms in Eq.~\eqref{all-local-energies-DNA-bending} and \eqref{all-local-energies-DNA-wrapping} describing local contributions of neighbouring DNA segments into the DNA total conformational energy can be divided into two big groups: 1) terms that look like $\frac{a}{2} ( \textbf{R} \textbf{A} \textbf{z}_0 - \textbf{R}' \textbf{z}_0 )^2 + \frac{c}{2} [ 2\pi \Delta \textrm{Tw} ( \textbf{R} \textbf{A}, \textbf{R}' ) ]^2 - b f ( \textbf{z}_0 \cdot \textbf{R} \textbf{z}_0 ) - \tau ( 2\pi - \lambda ) \Delta \textrm{Lk}^\textrm{F} ( \textbf{R}, \textbf{R}' ) - \tau \lambda \Delta \textrm{Tw} ( \textbf{R}, \textbf{R}' ) + \textrm{const}$ (namely, $E_{00}$, $E_{01}$, $E_{30}$ and $E_{31}$); and 2) terms, which have the following mathematical form: $- b f ( \textbf{z}_0 \cdot \textbf{R} \textbf{z}_0 ) - \ln{ \delta ( \textbf{R} \textbf{A} - \textbf{R}' ) } - \tau ( 2\pi - \lambda ) \Delta \textrm{Lk}^\textrm{F} ( \textbf{R}, \textbf{R}' ) - \tau \lambda \Delta \textrm{Tw} ( \textbf{R}, \textbf{R}' ) + \textrm{const}$ (namely, $E_{12}$ and $E_{23}$). Thus, all of the expansion coefficients of $T_{nm}$ transfer-functions can be easily obtained from the expansion series of the following two functions:
\begin{align} \label{functions-F1-F2}
F_1 \!\left( \textbf{R}, \textbf{R}' \right) = &
	\; e^{ 
	- \frac{a}{2} \left( \textbf{R} \textbf{A} \textbf{z}_0 - \textbf{R}' \textbf{z}_0 \right)^2 
	- \frac{c}{2} \left[ 2\pi \Delta \textrm{Tw} \left( \textbf{R} \textbf{A}, \textbf{R}' \right) \right]^2 
	+ b f \left( \textbf{z}_0 \cdot \textbf{R} \textbf{z}_0 \right)
	+ \tau \left( 2\pi - \lambda \right) \Delta \textrm{Lk}^\textrm{F} \left( \textbf{R}, \textbf{R}' \right)
	+ \tau \lambda \Delta \textrm{Tw} \left( \textbf{R}, \textbf{R}' \right)
	} \nonumber \\
F_2 \!\left( \textbf{R}, \textbf{R}' \right) = &
	\;
	\delta \!\left( \textbf{R} \textbf{A} - \textbf{R}' \right) 
	\, e^{ 
	b f \left( \textbf{z}_0 \cdot \textbf{R} \textbf{z}_0 \right)	
	+ \tau \left( 2\pi - \lambda \right) \Delta \textrm{Lk}^\textrm{F} \left( \textbf{R}, \textbf{R}' \right)
	+ \tau \lambda \Delta \textrm{Tw} \left( \textbf{R}, \textbf{R}' \right)
	}
\end{align}

\subsection{Expansion of $F_1$ function}\label{sec:expansion-F1}

To find out the expansion series of $F_1$ function, we will follow the same steps as in Appendix B of ref.~\cite{Efremov_2016}: first, we will derive separate expansion formulas for exponential functions comprising the global, $e^{ b f ( \textbf{z}_0 \cdot \textbf{R} \textbf{z}_0 ) + \tau ( 2\pi - \lambda ) \Delta \textrm{Lk}^\textrm{F} \!( \textbf{R}, \textbf{R}' ) }$, and local energy terms, $e^{ - \frac{a}{2} ( \textbf{R} \textbf{A} \textbf{z}_0 - \textbf{R}' \textbf{z}_0 )^2 - \frac{c}{2} [ 2\pi \Delta \textrm{Tw} ( \textbf{R} \textbf{A}, \textbf{R}' ) ]^2 + \tau \lambda \Delta \textrm{Tw} ( \textbf{R}, \textbf{R}' ) }$, and then combine them together by using Eq.~\eqref{D-functions-product-2} to get the final result. Here $\textbf{R}$ and $\textbf{R}^{\prime}$ are two Euler matrices describing the orientations of neighbouring DNA segments (DNA segment corresponding to the matrix $\textbf{R}$ is followed by one corresponding to the matrix $\textbf{R}'$); $\textbf{A}$ is a rotation matrix characterizing the relative equilibrium orientations of the neighbouring DNA segments in the absence of mechanical constraints applied to the DNA (i.e., when $f = 0$ pN and $\tau = 0$ pN$\cdot$nm); and $a$, $b$, $c$ and $\lambda$ are fixed model parameters describing the physical characteristics of the DNA polymer.

Let's start with the exponential function that incorporates the global energy terms, $e^{ b f ( \textbf{z}_0 \cdot \textbf{R} \textbf{z}_0 ) + \tau ( 2\pi - \lambda ) \Delta \textrm{Lk}^\textrm{F} \!( \textbf{R}, \textbf{R}' ) }$. In this function, $\Delta \textrm{Lk}^\textrm{F} \!( \textbf{R}, \textbf{R}' ) = \frac{1}{2\pi} ( \tilde{\alpha}' \!+\! \tilde{\gamma}' \!-\! \tilde{\alpha} \!-\! \tilde{\gamma} )$ denotes the contribution of neighbouring DNA segments to the total DNA linking number change, which is calculated using the Fuller's formula [Eq.~\eqref{delta-Lk-intro-2}]. Here $\tilde{\alpha}$, $\tilde{\gamma}$, $\tilde{\alpha}'$, $\tilde{\gamma}'$ are the Euler angles from the extended range of $( -\infty, \infty )$ corresponding to matrices $\textbf{R}$ and $\textbf{R}'$, which relate to usual Euler angles $\alpha, \gamma, \alpha'$ and $\gamma'$ as:
\begin{equation} \label{Euler-angles}
\alpha = \tilde{\alpha} \bmod 2\pi, \quad \gamma = \tilde{\gamma} \bmod 2\pi, \quad \alpha' = \tilde{\alpha}' \bmod 2\pi, \quad \gamma' = \tilde{\gamma}' \bmod 2\pi
\end{equation}

By applying Eq.~\eqref{D-expansion-2} and \eqref{expansion-coeff-2}, it is not very hard to find an analytical expression for the expansion series of the above exponential function in the general case, assuming that the values of the model parameters are selected in such a way that neighbouring DNA segments in most DNA conformations are only slightly rotated relative to each other. To this aim, in this study the size of the DNA segments, $b$, is chosen to be much smaller than the bending, $A$, and twisting, $C$, persistence lengths of DNA ($b \ll A$ and $b \ll C$). Furthermore, all matrices $\textbf{A}$ describing equilibrium orientations of neighbouring DNA segments are either set equal to the unit matrix, $\textbf{I}$, or only slightly deviating from it (the special case of DNA-wrapping proteins for which $\textbf{A}_\textrm{in}$ and $\textbf{A}_\textrm{out}$ matrices strongly deviate from the unit matrix, $\textbf{I}$, is discussed separately in Appendices~\ref{sec:expansion-S01} and \ref{sec:expansion-S30-S31}). As a result of such parameters' selection, the coordinate frames attached to each pair of neighbouring DNA segments in the vast majority of physically relevant DNA conformations will be only slightly rotated relative to each other. Then Eq.~\eqref{delta-Lk-intro-2} for the local DNA linking number change can be re-written in the form of a periodic function defined on SO(3)$\times$SO(3) group:
\begin{equation} \label{LkF}
\Delta \textrm{Lk}^\textrm{F} \!\!\left( \textbf{R}, \textbf{R}' \right) = 
	\textstyle{ \frac{1}{2\pi} } \left( \tilde{\alpha}' + \tilde{\gamma}' - \tilde{\alpha} - \tilde{\gamma} \right)
	\approx
	\textstyle{ \frac{1}{2\pi} } \sin \!\left( \alpha' + \gamma' - \alpha - \gamma \right)
\end{equation}

Substituting Eq.~\eqref{LkF} into $e^{ b f ( \textbf{z}_0 \cdot \textbf{R} \textbf{z}_0 ) + \tau ( 2\pi - \lambda ) \Delta \textrm{Lk}^\textrm{F} \!( \textbf{R}, \textbf{R}' ) }$ and taking into account that $( \textbf{z}_0 \cdot \textbf{Rz}_0 ) = ( \textbf{z}_0 \cdot \textbf{z} ) = \cos \beta$, where $\beta$ is the angle between $\textbf{z}_0$-axis of the lab coordinate system and $\textbf{z}$-axis of the system generated by Euler rotations $( \alpha, \beta, \gamma )$ [Figure~\ref{fig1}(b)], it is easy to see that:
\begin{equation} \label{global-potentials-1}
e^{ b f \left( \textbf{z}_0 \cdot \textbf{Rz}_0 \right) + \tau \left( 2\pi - \lambda \right) \Delta \textrm{Lk}^\textrm{F}
    \!\left( \textbf{R}, \textbf{R}' \right) } = 
	e^{ b f \!\cos \beta + \tau \left( 1 - \frac{\lambda}{2\pi} \right) \sin \left( \alpha' \!+ \gamma'	\! - \alpha - 
	\gamma \right) }
\end{equation}

To obtain the expansion formula for the above function, it is convenient to use Jacobi-Anger equation (p. 687, \cite{Arfken_2005}):
\begin{equation} \label{Jacobi-Anger-1}
e^{iq \cos \varphi} = \sum_{n=-\infty}^{+\infty} \! i^n J_n \!\left( q \right) e^{in \varphi}
\end{equation}

\noindent
Where $J_n \!\left( x \right)$ are Bessel functions of the first kind; $i$ is imaginary unit and $q$ is an arbitrary constant. All we need to do is slightly re-write Eq.~\eqref{Jacobi-Anger-1} in an alternative form by putting $\varphi = \frac{\pi}{2} \!-\! \psi$ and $q = - i \rho$:
\begin{equation}\label{Jacobi-Anger-2}
e^{\rho \sin \psi} = \sum_{n=-\infty}^{+\infty} \! i^{-n} I_n \!\left( \rho \right) e^{in \psi}
\end{equation}

\noindent
Here $I_n ( x ) = i^{-n} J_n ( ix )$ are modified Bessel functions of the first kind, which have the following properties: $I_{-n} ( x ) = I_n ( x )$ and $I_n ( -x ) = ( -1 )^n I_n ( x )$, see p. 714 in ref.~\citep{Arfken_2005}. 

Utilizing functions $I_n ( x )$, it is also possible to re-organize Eq.~\eqref{Jacobi-Anger-1} in yet another convenient form, which we will be applying below in our derivations:
\begin{equation}\label{Jacobi-Anger-3}
e^{\rho \cos \varphi} = \sum_{n=-\infty}^{+\infty} \! I_n \!\left( \rho \right) e^{in \varphi}
\end{equation}

Anyway, by using Eq.~\eqref{expansion-coeff-2} and \eqref{Jacobi-Anger-2}, it is rather straightforward to find the expansion coefficients for the exponential function defined by Eq.~\eqref{global-potentials-1}:
\begin{align} \label{global-potentials-coeff}
& F^{p'\!\!,\,q'\!\!,\,s'}_{p,\,q,\,s} = 
	\frac{ \left( 2s \!+\! 1 \right) \!\left( 2s' \!\!+\! 1 \right) }{ \left( 8\pi^2 \right)^2 }
	\int \!\! \textrm{d} \textbf{R} \textrm{d} \textbf{R}' \, \overline{D}^s_{p,q} \!\left( \textbf{R} \right) 
	e^{ b f \left( \textbf{z}_0 \cdot \textbf{Rz}_0 \right) 
	+ \tau \left( 2\pi - \lambda \right) \Delta \textrm{Lk}^\textrm{F} \!\left( \textbf{R},\textbf{R}' \right) }
	D^{s'}_{p'\!,q'} \!\left( \textbf{R}' \right) = 
	\frac{ \left( 2s \!+\! 1 \right) \!\left(2s' \!\!+\! 1 \right) }{ \left( 8\pi^2 \right)^2 } \times \nonumber\\
	& \quad \;\; \times\!\! \int \!\! \textrm{d} \textbf{R} \textrm{d} \textbf{R}' \,
	\overline{P}^s_{p,q} \!\left( \cos \beta \right) e^{ b f \!\cos \beta} \times 
	P^{s'}_{p'\!,q'} \!\left( \cos \beta' \right) \times 
	\!\!\!\!\sum_{k=-\infty}^{+\infty} \!\! i^{-k} 
	I_k \!\left( \tau \left[ 1 \!-\! \textstyle{ \frac{ \lambda }{ 2\pi } } \right] \right)
	e^{i \left( p - k \right) \alpha + i \left( q - k \right) \gamma - i \left( p' \!- k \right) \alpha' \!- i \left(
	q' \!- k \right) \gamma' } = \nonumber\\
	& \quad \;\; = \delta_{pp'} \delta_{qq'} \delta_{pq} \!\times\! 
	\frac{1}{4} \left( 2s \!+\! 1 \right) \!\left(2s' \!\!+\! 1 \right) i^{-p} 
	I_p \!\left( \tau \left[ 1 \!-\! \textstyle{ \frac{ \lambda }{ 2\pi } } \right] \right)
	\mathscr{L}^s_p \!\left( -bf \right) \mathscr{L}^{s'}_p \!\!\left( 0 \right)
\end{align}

\noindent
Where $\mathscr{L}^s_p ( x )$ designates bilateral Laplace transform of $P^s_{p,p}$ polynomial (or, which is the same thing, diagonal element, $d^s_{p,p}$, of Wigner small d-matrix):
\begin{equation} \label{L-coeff}
\mathscr{L}^s_p \!\left( x \right) = 
\int^{1}_{-1} \!\! P^s_{p,p} \!\left( y \right) e^{-xy} \textrm{d} y = 
\int^{1}_{-1} \!\! d^s_{p,p} \!\left( \cos^{-1} y \right) e^{-xy} \textrm{d} y
\end{equation}

Substituting Eq.~\eqref{global-potentials-coeff} into Eq.~\eqref{D-expansion-2}, we finally get the desired expansion formula for the first part of function $F_1$:
\begin{equation} \label{global-potentials-expansion}
e^{ b f \left( \textbf{z}_0 \cdot \textbf{Rz}_0 \right) + \tau \left( 2\pi - \lambda \right) \Delta \textrm{Lk}^\textrm{F} \!\left( \textbf{R}, \textbf{R}' \right) } =
	\frac{1}{4} \!\sum_{s,s'\!,p} \!\left( 2s \!+\! 1 \right) \!\left(2s' \!\!+\! 1 \right) i^{-p}
	I_p \!\left( \tau \left[ 1 \!-\! \textstyle{ \frac{ \lambda }{ 2\pi } } \right] \right) 
	\mathscr{L}^s_p \!\left( -bf \right) \mathscr{L}^{s'}_p \!\!\left( 0 \right) D^s_{p,p} \!\left( \textbf{R} \right)
	\overline{D}^{s'}_{p,p} \!\left( \textbf{R}' \right)
\end{equation}

To derive the expansion series of the second part, $e^{ - \frac{a}{2} ( \textbf{R} \textbf{A} \textbf{z}_0 - \textbf{R}' \textbf{z}_0 )^2 - \frac{c}{2} [ 2\pi \Delta \textrm{Tw} ( \textbf{R} \textbf{A}, \textbf{R}' ) ]^2 + \tau \lambda \Delta \textrm{Tw} ( \textbf{R}, \textbf{R}' ) }$, we will again follow the logic described in our previous study \cite{Efremov_2016}. But first, we will note that since the rotation matrix $\textbf{A}$ only slightly deviates from the unit matrix, $\textbf{I}$, the local DNA twist between neighbouring DNA segments can be represented in the following form: $\Delta \textrm{Tw} ( \textbf{R}, \textbf{R}' ) \approx \Delta \textrm{Tw} ( \textbf{R}, \textbf{RA} ) \!+\! \Delta \textrm{Tw} ( \textbf{RA}, \textbf{R}' ) = \Delta \textrm{Tw} ( \textbf{I}, \textbf{A} ) \!+\! \Delta \textrm{Tw} ( \textbf{RA}, \textbf{R}' )$, where $\Delta \textrm{Tw} ( \textbf{I}, \textbf{A} ) = \frac{ 1 }{ 2\pi } ( \alpha_A \!+\! \gamma_A )$ is the twist between the coordinate frame corresponding to the rotation matrix $\textbf{A} = \textbf{A} (\alpha_A, \beta_A, \gamma_A )$ and the global coordinate system, whose orientation in space is described by the unit rotation matrix, $\textbf{I}$. 

Substituting the above formula for the local DNA twist into the exponential function, it becomes clear that the latter depends only on $( \textbf{RA} )^{-1} \textbf{R}'$ product of Euler rotation matrices. Indeed, it is not hard to see that $( \textbf{RAz}_0 \!-\! \textbf{R}' \textbf{z}_0 )^2 = 2 \!-\! 2 ( \textbf{RAz}_0 \cdot \textbf{R}' \textbf{z}_0 ) = 2 \!-\! 2 ( \textbf{z}_0 \cdot (\textbf{RA})^{-1} \textbf{R}' \textbf{z}_0 )$ and $\Delta \textrm{Tw} ( \textbf{RA}, \textbf{R}' ) = \Delta \textrm{Tw} ( \textbf{I}, (\textbf{RA})^{-1} \textbf{R}' )$. In other words, the twisting angle between the coordinate systems corresponding to Euler matrices $\textbf{RA}$ and $\textbf{R}'$ as well as the bending angle between their $\textbf{z}$-axes depend only on the relative orientation of the two coordinate systems and is independent from their exact alignments with respect to the lab coordinate frame $( \textbf{x}^{}_0,\textbf{y}_0,\textbf{z}^{}_0 )$. 

As a result, the expansion series of $e^{ - \frac{a}{2} ( \textbf{R} \textbf{A} \textbf{z}_0 - \textbf{R}' \textbf{z}_0 )^2 - \frac{c}{2} [ 2\pi \Delta \textrm{Tw} ( \textbf{R} \textbf{A}, \textbf{R}' ) ]^2 + \tau \lambda \Delta \textrm{Tw} ( \textbf{R}, \textbf{R}' ) }$ function can be found in two steps. First, we will consider the special case in which the coordinate system corresponding to the matrices product $\textbf{RA}$ is identical to the lab coordinate system ($\textbf{RA} = \textbf{I}$), and the coordinate frame corresponding to matrix $\textbf{R}'$ is only slightly rotated relative to it. Second, by substituting $\textbf{R}' \rightarrow (\textbf{RA})^{-1} \textbf{R}^{\prime}$ into the formula obtained for the special case and by using Eq.~\eqref{D-functions-product-1}, we will get the desired expansion series for the above exponential function in the general case.

Let $\left( \alpha',\beta',\gamma' \right)$ be the Euler angles corresponding to matrix $\textbf{R}'$. Then by taking into account the above notes, for the special case of $\textbf{RA} = \textbf{I}$ we have: $( \textbf{RAz}_0 \!-\! \textbf{R}' \textbf{z}_0 )^2 = 2 \!-\! 2 ( \textbf{z}_0 \cdot \textbf{R}' \textbf{z}_0 ) = 2 \!-\! 2 \cos{\beta'}$ and  $\Delta \textrm{Tw} ( \textbf{RA}, \textbf{R}' ) = \Delta \textrm{Tw} ( \textbf{I}, \textbf{R}' )$. Furthermore, since the coordinate frame corresponding to matrix $\textbf{R}'$ is only slightly rotated relative to the lab coordinate system, it is clear that:
\begin{equation} \label{twist-approx}
2\pi \Delta \textrm{Tw} \!\left( \textbf{I}, \textbf{R}' \right) \approx
	\alpha' + \gamma' \approx
	\sin \!\left( \alpha' + \gamma' \right) 
\quad \textrm{and} \quad 
\left[ 2\pi \Delta \textrm{Tw} \!\left( \textbf{I}, \textbf{R}' \right) \right]^2 \approx 
	2 - 2 \cos \!\left( \alpha' + \gamma' \right)
\end{equation}

Thus, in the special case of $\textbf{RA} = \textbf{I}$:
\begin{align} \label{local-potentials-1}
& e^{ - \frac{a}{2} \left( \textbf{RAz}_0 - \textbf{R}' \textbf{z}_0  \right)^2 
	- \frac{c}{2} \left[ 2\pi \Delta \textrm{Tw} \left( \textbf{RA}, \textbf{R}' \right) \right]^2 
	+ \tau \lambda \Delta \textrm{Tw} \left( \textbf{R}, \textbf{R}' \right) } =
e^{ - \frac{a}{2} \left( \textbf{z}_0 - \textbf{R}' \textbf{z}_0  \right)^2 
	- \frac{c}{2} \left[ 2\pi \Delta \textrm{Tw} \left( \textbf{I}, \textbf{R}' \right) \right]^2
	+ \tau \lambda \Delta \textrm{Tw} \left( \textbf{I}, \textbf{R}' \right)
	+ \tau \lambda \Delta \textrm{Tw} \left( \textbf{I}, \textbf{A} \right) } = \nonumber\\
& \! = e^{ - a - c + \tau\lambda \Delta \textrm{Tw} \left( \textbf{I}, \textbf{A} \right) } \,
	e^{ a \cos \beta' + c \cos \left( \alpha' + \gamma' \right) + \frac{ \tau \lambda }{ 2\pi } \sin \left( \alpha' +
	\gamma' \right) }
\end{align}

Applying Eq.~\eqref{expansion-coeff-1} and \eqref{Jacobi-Anger-3} to Eq.~\eqref{local-potentials-1}, it is easy to derive a mathematical formula for the expansion coefficients of the above exponential function:
\begin{align} \label{local-potentials-coeff}
& F_{p'\!,q'\!,s'} =
	\frac{ 2s' \!\!+\! 1 }{ 8\pi^2 } \, e^{ \tau \lambda \Delta \textrm{Tw} \left( \textbf{I}, \textbf{A} \right) }
	\int \! \textrm{d} \textbf{R}' \, 
	\overline{D}^{s'}_{p'\!,q'} \!\left( \textbf{R}' \right) 
	e^{ - \frac{a}{2} \left( \textbf{z}_0 - \textbf{R}' \textbf{z}_0  \right)^2 
	- \frac{c}{2} \left[ 2\pi \Delta \textrm{Tw} \left( \textbf{I}, \textbf{R}' \right) \right]^2
	+ \tau \lambda \Delta \textrm{Tw} \left( \textbf{I}, \textbf{R}' \right) } = \nonumber\\
& = \frac{ 2s' \!\!+\! 1 }{ 8\pi^2 } \, 
	e^{ - a - c + \tau \lambda \Delta \textrm{Tw} \left( \textbf{I}, \textbf{A} \right) }
	\int \! \textrm{d} \textbf{R}' \, 
	\overline{P}^{s'}_{p'\!,q'} \!\left( \cos \beta' \right) e^{ a \cos \beta' } \!\times\! 
	\!\!\sum_{k=-\infty}^{+\infty} \!\! I_k \!\left( c \sqrt{ 1 \!+\! \chi^2 } \right) 
	e^{i \left( p' \!+ k \right) \alpha' \!+ i \left( q' \!+ k \right) \gamma' \!- ik \omega} = \nonumber\\
& = \delta_{p'q'} \times \frac{1}{2} \left( 2s' \!\!+\! 1 \right)
	e^{ - a - c + \tau \lambda \Delta \textrm{Tw} \left( \textbf{I}, \textbf{A} \right) } \,
	e^{ i p' \omega} I_{p'} \!\left( c \sqrt{ 1 \!+\! \chi^2 } \right) \mathscr{L}^{s'}_{p'} \!\left( - a \right) 
\end{align}

\noindent
Where $\chi = \frac{ \tau \lambda }{ 2\pi c }$ and $\omega = \tan^{-1} \!\left( \chi \right)$.

Then by substituting Eq.~\eqref{local-potentials-coeff} into Eq.~\eqref{D-expansion-1}, we finally obtain the desired expansion series for the special case of $\textbf{RA} = \textbf{I}$:
\begin{multline} \label{local-potentials-expansion-1}
e^{ - \frac{a}{2} \left( \textbf{z}_0 - \textbf{R}' \textbf{z}_0  \right)^2 
	- \frac{c}{2} \left[ 2\pi \Delta \textrm{Tw} \left( \textbf{I}, \textbf{R}' \right) \right]^2
	+ \tau \lambda \Delta \textrm{Tw} \left( \textbf{I}, \textbf{R}' \right)
	+ \tau \lambda \Delta \textrm{Tw} \left( \textbf{I}, \textbf{A} \right) } = \\
= \frac{1}{2} \, e^{ - a - c + \tau \lambda \Delta \textrm{Tw} \left( \textbf{I}, \textbf{A} \right) }
	\sum_{s'\!,p'} \left( 2s' \!\!+\! 1 \right) e^{ i p' \omega } I_{p'} \!\left( c \sqrt{ 1 \!+\! \chi^2 } \right)
	\mathscr{L}^{s'}_{p'} \!\left( -a \right) D^{s'}_{p'\!,p'} \!\left( \textbf{R}' \right)
\end{multline}

To extend the above expression to the general case, we simply need to put $\textbf{R}' \rightarrow (\textbf{RA})^{-1} \textbf{R}'$ and use the previously mentioned multiplication property of $D^s_{p,q}$ functions [Eq.~\eqref{D-functions-product-1}]. By doing so, we get:
\begin{multline} \label{local-potentials-expansion-2}
e^{ - \frac{a}{2} \left( \textbf{RAz}_0 - \textbf{R}' \textbf{z}_0  \right)^2 
	- \frac{c}{2} \left[ 2\pi \Delta \textrm{Tw} \left( \textbf{RA}, \textbf{R}' \right) \right]^2 
	+ \tau \lambda \Delta \textrm{Tw} \left( \textbf{R}, \textbf{R}' \right) } = \\
= \frac{1}{2} \, e^{ - a - c + \tau \lambda \Delta \textrm{Tw} \left( \textbf{I}, \textbf{A} \right) }
	\!\!\sum_{s,p,q,v} \!\!\left( 2s \!+\! 1 \right) e^{ - i p \omega } I_p \!\left( c \sqrt{ 1 \!+\! \chi^2} \right)
	\mathscr{L}^s_p \!\left( -a \right) 
	D^s_{v,p} \!\left( \textbf{A} \right)	
	D^s_{q,v} \!\left( \textbf{R} \right) 
	\overline{D}^s_{q,p} \!\left( \textbf{R}' \right)
\end{multline}

\noindent
Here, in addition to the indexes change $s' \rightarrow s$ and $p' \rightarrow -p$, we also used Eq.~\eqref{P-functions-symmetry}-\eqref{D-functions-inverse} and \eqref{L-coeff} to re-write the expansion formula in the form of Eq.~\eqref{D-expansion-2}.

Having the expansion series for the exponential functions containing both the global and local energy terms, it is now simple enough to obtain the expansion formula for $F_1$ function defined by Eq.~ \eqref{functions-F1-F2}. Namely, by  multiplying Eq.~\eqref{global-potentials-expansion} and \eqref{local-potentials-expansion-2}, and using twice the multiplication rule for $D^s_{p,q}$ functions [Eq.~\eqref{D-functions-product-2}] together with Eq.~\eqref{Wigner-3j}, we arrive to the following result:
\begin{align} \label{F1-function-expansion-1}
& F_1 \!\left( \textbf{R}, \textbf{R}' \right) = 
	\; e^{ 
	- \frac{a}{2} \left( \textbf{R} \textbf{A} \textbf{z}_0 - \textbf{R}' \textbf{z}_0 \right)^2 
	- \frac{c}{2} \left[ 2\pi \Delta \textrm{Tw} \left( \textbf{R} \textbf{A}, \textbf{R}' \right) \right]^2 
	+ b f \left( \textbf{z}_0 \cdot \textbf{R} \textbf{z}_0 \right)
	+ \tau \left( 2\pi - \lambda \right) \Delta \textrm{Lk}^\textrm{F} \left( \textbf{R}, \textbf{R}' \right)
	+ \tau \lambda \Delta \textrm{Tw} \left( \textbf{R}, \textbf{R}' \right)
	} = \nonumber\\[15pt]
& \! = \frac{1}{8} \, e^{ - a - c + \tau \lambda \Delta \textrm{Tw} \left( \textbf{I}, \textbf{A} \right) }
	\!\!\!\!\!\!\sum_{s,p,q,v,k,k'\!,r} \!\!\!\!\!\! 
	\left( 2s \!+\! 1 \right) \!\left( 2k \!+\! 1 \right) \!\left( 2k' \!\!+\! 1 \right)
	i^{-r} e^{ -i p \omega } 
	I_p \!\left( c \sqrt{ 1 \!+\! \chi^2 } \right) I_r \!\left( \tau \left[ 1 \!-\! \textstyle{ \frac{ \lambda }{ 2\pi } }
	\right] \right) \times \nonumber\\[-5pt]
& \hspace{68pt} \times \mathscr{L}^s_p \!\left( -a \right) \mathscr{L}^k_r \!\left( - b f \right)
	\mathscr{L}^{k'}_r \!\!\left( 0 \right) D^s_{v,p} \!\left( \textbf{A} \right) \times
	D^s_{q,v} \!\left( \textbf{R} \right) D^k_{r,r} \!\left( \textbf{R} \right) \times
	\overline{D}^s_{q,p} \!\left( \textbf{R}' \right) \overline{D}^{k'}_{r,r} \!\left( \textbf{R}' \right) 
	= \nonumber\\[15pt]
& \! = \frac{1}{8} \, e^{ - a - c + \tau \lambda \Delta \textrm{Tw} \left( \textbf{I}, \textbf{A} \right) }
	\!\!\!\!\!\!\!\!\!\!\sum_{s,p,q,v,k,k'\!,r,u,u'} \!\!\!\!\!\!\!\!\!\!\! 
	\left( 2s \!+\! 1 \right) \!\left( 2k \!+\! 1 \right) \!\left( 2k' \!\!+\! 1 \right) 
	\!\left( 2u \!+\! 1 \right) \!\left( 2u' \!\!+\! 1 \right)
	\left( -1 \right)^{v+p}
	i^{-r} e^{ - i p \omega } I_p \!\left( c \sqrt{ 1 \!+\! \chi^2 } \right) \times \nonumber\\[-5pt]
& \quad \times\! I_r \!\left( \tau \left[ 1 \!-\! \textstyle{ \frac{ \lambda }{ 2\pi } } \right] \right)
	D^s_{v,p} \!\left( \textbf{A} \right) \!\times\!
	\mathscr{L}^s_p \!\left( -a \right) \mathscr{L}^k_r \!\left( - b f \right)
	\mathscr{L}^{k'}_r \!\!\left( 0 \right) \!\times\!
    \left(\begin{array}{ccc}
		s & \;\;k & \!u \\
		q & \;\;r & \!-q\!-\!r
	\end{array}\right)
	\left(\begin{array}{ccc}
		s & \;\;k & \!u \\
		v & \;\;r & \!-v\!-\!r
	\end{array}\right)
    \!\times \nonumber\\	
& \quad \times\!	
	\left(\begin{array}{ccc}
		s & \;\;k' & \!u' \\
		q & \;\;r & \!-q\!-\!r
	\end{array}\right)
    \left(\begin{array}{ccc}
		s & \;\;k' & \!u' \\
		p & \;\;r & \!-p\!-\!r
	\end{array}\right)
	\!\times\! D^{\,u}_{q+r,v+r} \!\left( \textbf{R} \right) \overline{D}^{\,u'}_{q+r,p+r} \!\left( \textbf{R}' \right)
\end{align}

By performing one more indexes change $u \rightarrow s$, $u' \rightarrow s'$, $s \rightarrow t$, $q \rightarrow p - r$, $v \rightarrow q - r$ and $p \rightarrow q' - r$, it is not very hard to re-write the above expansion formula for $F_1$ function in the form of Eq.~\eqref{D-expansion-3}:
\begin{equation} \label{F1-function-expansion-2}
F_1 \!\left( \textbf{R}, \textbf{R}' \right) =
	\frac{1}{8\pi^2} \!\!\! \sum_{p,p'\!\!,\,q,q'\!\!,\,s,s'\!} \!\!\!\!\!\! 
	\sqrt{ \left( 2s \!+\! 1 \right) \!\left( 2s' \!\!+\! 1 \right) }
	\, \left( F_1 \right)^{p'\!\!,\,q'\!\!,\,s'}_{p,\,q,\,s} D^s_{p,q} \!\left( \textbf{R} \right) 
	\overline{D}^{s'}_{p'\!,q'} \!\left( \textbf{R}' \right)
\end{equation}

\noindent
Where the expansion coefficients $( F_1 )^{p'\!\!,\,q'\!\!,\,s'}_{p,\,q,\,s}$ are:
\begin{align}\label{F1-function-expansion-coeff}
\left( F_1 \right)^{p'\!\!,\,q'\!\!,\,s'}_{p,\,q,\,s} =
	\delta_{pp'} \times \pi^2 \sqrt{ \left( 2s \!+\! 1 \right) \!\left( 2s' \!\!+\! 1 \right) } \,
	\left( -1 \right)^{q+q'} \!
	e^{ - a - c + \tau \lambda \Delta \textrm{Tw} \left( \textbf{I}, \textbf{A} \right) }
	\!\!\sum_{t,k,k'\!,r} \!\!
	\left( 2t \!+\! 1 \right) \!\left( 2k \!+\! 1 \right) \!\left( 2k' \!\!+\! 1 \right)
	i^{-r} e^{ -i \left( q' \!- r \right) \omega } \times & \nonumber \\[-2pt]
	\times \, I_{q'-r} \!\left( c \sqrt{1\!+\!\chi^2} \right) 
	I_r \!\left( \tau \left[ 1 \!-\! \textstyle{ \frac{ \lambda }{ 2\pi } } \right] \right) 
	D^{\,t}_{q-r,q'\!-r} \!\left( \textbf{A} \right) \!\times 
	\mathscr{L}^t_{q'-r} \!\left( -a \right) \mathscr{L}^k_r \!\left( - b f \right)
	\mathscr{L}^{k'}_r \!\!\left( 0 \right) \times & \nonumber \\[3pt]
	\times
    \left(\begin{array}{ccc}
		t & k & \,s \\
		p\!-\!r & r & \,-p
	\end{array}\right)
	\left(\begin{array}{ccc}
		t & k & \,s \\
		q\!-\!r & r & \,-q
	\end{array}\right)
	\left(\begin{array}{ccc}
		t & k' & \,s' \\
		p\!-\!r & r & \,-p
	\end{array}\right)
    \left(\begin{array}{ccc}
		t & k' & \,s' \\
		q'\!\!-\!r & r & \,-q'
	\end{array}\right) \hspace{6pt} &
\end{align}

\subsection{Expansion of $F_2$ function}\label{sec:expansion-F2}

In order to find the expansion series for $F_2$ function defined by Eq.~\eqref{functions-F1-F2}, we will use the same strategy as in the previous section, deriving first separate expansion formulas for the two parts of $F_2$ function, $e^{ b f ( \textbf{z}_0 \cdot \textbf{R} \textbf{z}_0 ) + \tau ( 2\pi - \lambda ) \Delta \textrm{Lk}^\textrm{F} \!( \textbf{R}, \textbf{R}' ) }$ and $\delta ( \textbf{RA} - \textbf{R}' ) \, e^{ \tau \lambda \Delta \textrm{Tw} ( \textbf{R}, \textbf{R}' ) }$, and then combining them together by  using the multiplication rule for $D^s_{p,q}$ functions [Eq.~\eqref{D-functions-product-2}] to get the final result.

The best part of such approach is that we already have the expansion series for $e^{ b f ( \textbf{z}_0 \cdot \textbf{R} \textbf{z}_0 ) + \tau ( 2\pi - \lambda ) \Delta \textrm{Lk}^\textrm{F} \!( \textbf{R}, \textbf{R}' ) }$, see Eq.~\eqref{global-potentials-expansion}. Thus, all that remains to do is to find an expansion formula for the remaining $\delta ( \textbf{RA} - \textbf{R}' ) \, e^{ \tau \lambda \Delta \textrm{Tw} ( \textbf{R}, \textbf{R}' ) }$ function, which can be done by applying Eq.~\eqref{D-functions-product-1}, \eqref{D-functions-orthogonality} and \eqref{expansion-coeff-2} in order to obtain the desired expansion coefficients:
\begin{align} \label{F2-function-delta-expansion-coeff}
F^{p'\!\!,\,q'\!\!,\,s'}_{p,\,q,\,s} & = 
	\frac{ \left(2s\!+\!1\right)\!\left(2s'\!\!+\!1\right) }{ \left(8\pi^2\right)^2 } 
	\int \! \textrm{d} \textbf{R} \textrm{d} \textbf{R}' \,
	\overline{D}^s_{p,q} \!\left( \textbf{R} \right) 
	\delta \!\left( \textbf{RA} - \textbf{R}' \right) e^{ \tau \lambda \Delta \textrm{Tw} \left( \textbf{R}, \textbf{R}'
	\right) }
	D^{s'}_{p'\!,q'} \!\left( \textbf{R}' \right) = \nonumber \\
& = \frac{ \left(2s\!+\!1\right)\!\left(2s'\!\!+\!1\right) }{ \left(8\pi^2\right)^2 } \,
	e^{ \tau \lambda \Delta \textrm{Tw} \left( \textbf{I}, \textbf{A} \right) }
	\int \! \textrm{d} \textbf{R} \,
	\overline{D}^s_{p,q} \!\left( \textbf{R} \right)
	D^{s'}_{p'\!,q'} \!\left( \textbf{RA} \right) = \nonumber \\
& = \frac{ \left(2s\!+\!1\right)\!\left(2s'\!\!+\!1\right) }{ \left(8\pi^2\right)^2 } \,
	e^{ \tau \lambda \Delta \textrm{Tw} \left( \textbf{I}, \textbf{A} \right) }
	\sum_k D^{s'}_{k,q'} \!\left( \textbf{A} \right)
	\int \! \textrm{d} \textbf{R} \,
	\overline{D}^s_{p,q} \!\left( \textbf{R} \right)
	D^{s'}_{p'\!,k} \!\left( \textbf{R} \right) = \nonumber \\
& = \delta_{ss'} \delta_{pp'} \times \frac{ 2s \!+\! 1 }{ 8\pi^2 } 
	e^{ \tau \lambda \Delta \textrm{Tw} \left( \textbf{I}, \textbf{A} \right) } 
	D^s_{q,q'} \!\left( \textbf{A} \right)
\end{align}

\noindent
Where we have taken into account that $\Delta \textrm{Tw} ( \textbf{R}, \textbf{RA} ) = \Delta \textrm{Tw} ( \textbf{I}, \textbf{A} )$. 

Substituting the above coefficients into Eq.~\eqref{D-expansion-2}, it immediately follows that:
\begin{equation} \label{F2-function-delta-expansion}
\delta \!\left( \textbf{RA} - \textbf{R}' \right) e^{ \tau \lambda \Delta \textrm{Tw} \left( \textbf{R}, \textbf{R}' \right) } = 
	\frac{ 1 }{ 8\pi^2 } \, e^{ \tau \lambda \Delta \textrm{Tw} \left( \textbf{I}, \textbf{A} \right) }
	\!\sum_{s,p,q,q'} \!\!\left( 2s \!+\! 1 \right) D^s_{q,q'} \!\left( \textbf{A} \right)
	D^s_{p,q} \!\left( \textbf{R} \right) \overline{D}^s_{p,q'} \!\left( \textbf{R}' \right)
\end{equation}

To derive the final expansion series for $F_2$ function, all we need to do now is to multiply Eq.~\eqref{global-potentials-expansion} and \eqref{F2-function-delta-expansion}, and use twice the multiplication rule for $D^s_{p,q}$ functions [Eq.~\eqref{D-functions-product-2}] together with Eq.~\eqref{Wigner-3j}. By doing so, we get the following formula for $F_2$ function:
\begin{align} \label{F2-function-expansion-1}
& F_2 \!\left( \textbf{R}, \textbf{R}' \right) = 
	\;
	\delta \!\left( \textbf{R} \textbf{A} - \textbf{R}' \right) 
	e^{ 
	b f \left( \textbf{z}_0 \cdot \textbf{R} \textbf{z}_0 \right)	
	+ \tau \left( 2\pi - \lambda \right) \Delta \textrm{Lk}^\textrm{F} \left( \textbf{R}, \textbf{R}' \right)
	+ \tau \lambda \Delta \textrm{Tw} \left( \textbf{R}, \textbf{R}' \right)
	} = \nonumber\\[15pt]
& \! = \frac{ 1 }{ 32\pi^2 } \, e^{ \tau \lambda \Delta \textrm{Tw} \left( \textbf{I}, \textbf{A} \right) }
	\!\!\!\!\!\!\sum_{s,p,q,q'\!,k,k'\!,r} \!\!\!\!\!\! 
	\left( 2s \!+\! 1 \right) \!\left( 2k \!+\! 1 \right) \!\left( 2k' \!\!+\! 1 \right)
	i^{-r} I_r \!\left( \tau \left[ 1 \!-\! \textstyle{ \frac{ \lambda }{ 2\pi } } \right] \right)
	D^s_{q,q'} \!\left( \textbf{A} \right)
	\mathscr{L}^k_r \!\left( - b f \right) \mathscr{L}^{k'}_r \!\!\left( 0 \right) \times \nonumber\\[-5pt]
& \hspace{230pt} \times
	D^s_{p,q} \!\left( \textbf{R} \right) D^k_{r,r} \!\left( \textbf{R} \right) \times
	\overline{D}^s_{p,q'} \!\left( \textbf{R}' \right) \overline{D}^{k'}_{r,r} \!\left( \textbf{R}' \right) 
	= \nonumber\\[15pt]
& \! = \frac{ 1 }{ 32\pi^2 } \, e^{ \tau \lambda \Delta \textrm{Tw} \left( \textbf{I}, \textbf{A} \right) }
	\!\!\!\!\!\!\!\!\!\!\sum_{s,p,q,q'\!,k,k'\!,r,u,u'} \!\!\!\!\!\!\!\!\!\!
	\left( 2s \!+\! 1 \right) \!\left( 2k \!+\! 1 \right) \!\left( 2k' \!\!+\! 1 \right)
	\!\left( 2u \!+\! 1 \right) \!\left( 2u' \!\!+\! 1 \right)
	\left( -1 \right)^{q+q'} i^{-r} 
	I_r \!\left( \tau \left[ 1 \!-\! \textstyle{ \frac{ \lambda }{ 2\pi } } \right] \right)
	D^s_{q,q'} \!\left( \textbf{A} \right) \times \nonumber\\[0pt]
& \quad \times \mathscr{L}^k_r \!\left( - b f \right) \mathscr{L}^{k'}_r \!\!\left( 0 \right) \times 
    \left(\begin{array}{ccc}
		s & \;\;k & \!u \\
		p & \;\;r & \!-p\!-\!r
	\end{array}\right)
	\left(\begin{array}{ccc}
		s & \;\;k & \!u \\
		q & \;\;r & \!-q\!-\!r
	\end{array}\right)
	\left(\begin{array}{ccc}
		s & \;\;k' & \!u' \\
		p & \;\;r & \!-p\!-\!r
	\end{array}\right)
    \left(\begin{array}{ccc}
		s & \;\;k' & \!u' \\
		q' & \;\;r & \!-q'\!\!-\!r
	\end{array}\right)
	\times \nonumber\\[-3pt]
& \quad \times	
	D^{\,u}_{p+r,q+r} \!\left( \textbf{R} \right) \overline{D}^{\,u'}_{p+r,q'\!+r} \!\left( \textbf{R}' \right)
\end{align}

As in the previous section, by performing indexes change $u \rightarrow s$, $u' \rightarrow s'$, $s \rightarrow t$, $p \rightarrow p - r$, $q \rightarrow q - r$ and $q' \rightarrow q' - r$, it is not very hard to obtain the expansion formula for $F_2$ function in the form of Eq.~\eqref{D-expansion-3}:
\begin{equation} \label{F2-function-expansion-2}
F_2 \!\left( \textbf{R}, \textbf{R}' \right) =
	\frac{1}{8\pi^2} \!\!\! \sum_{p,p'\!\!,\,q,q'\!\!,\,s,s'\!} \!\!\!\!\!\! 
	\sqrt{ \left( 2s \!+\! 1 \right) \!\left( 2s' \!\!+\! 1 \right) }
	\, \left( F_2 \right)^{p'\!\!,\,q'\!\!,\,s'}_{p,\,q,\,s} D^s_{p,q} \!\left( \textbf{R} \right) 
	\overline{D}^{s'}_{p'\!,q'} \!\left( \textbf{R}' \right)
\end{equation}

\noindent
Where the expansion coefficients $( F_2 )^{p'\!\!,\,q'\!\!,\,s'}_{p,\,q,\,s}$ are:
\begin{multline}\label{F2-function-expansion-coeff}
\left( F_2 \right)^{p'\!\!,\,q'\!\!,\,s'}_{p,\,q,\,s} =
	\delta_{pp'} \times \frac{1}{4} \sqrt{ \left( 2s \!+\! 1 \right) \!\left( 2s' \!\!+\! 1 \right) } \,
	\left( -1 \right)^{q+q'} \!
	e^{ \tau \lambda \Delta \textrm{Tw} \left( \textbf{I}, \textbf{A} \right) }
	\!\!\sum_{t,k,k'\!,r} \!\!
	\left( 2t \!+\! 1 \right) \!\left( 2k \!+\! 1 \right) \!\left( 2k' \!\!+\! 1 \right)
	i^{-r} I_r \!\left( \tau \left[ 1 \!-\! \textstyle{ \frac{ \lambda }{ 2\pi } } \right] \right) 
	\times \\[-2pt]
\times D^{\,t}_{q-r,q'\!-r} \!\left( \textbf{A} \right)
\mathscr{L}^k_r \!\left( - b f \right) \mathscr{L}^{k'}_r \!\!\left( 0 \right) \times
    \left(\begin{array}{ccc}
		t & k & \,s \\
		p\!-\!r & r & \,-p
	\end{array}\right)
	\left(\begin{array}{ccc}
		t & k & \,s \\
		q\!-\!r & r & \,-q
	\end{array}\right)
	\left(\begin{array}{ccc}
		t & k' & \,s' \\
		p\!-\!r & r & \,-p
	\end{array}\right)
    \left(\begin{array}{ccc}
		t & k' & \,s' \\
		q'\!\!-\!r & r & \,-q'
	\end{array}\right)
\end{multline}

While Eq.~\eqref{F2-function-expansion-coeff} holds for any matrix $\textbf{A}$ which is sufficiently close to the unit matrix, $\textbf{I}$; for the special case of $\textbf{A} = \textbf{I}$ it is possible to get even more simpler expression by noting that in this case $F_2 ( \textbf{R}, \textbf{R}' ) = \delta ( \textbf{R} - \textbf{R}' ) \, e^{ b f ( \textbf{z}_0 \cdot \textbf{R} \textbf{z}_0 ) }$ since $\Delta \textrm{Tw} ( \textbf{R}, \textbf{R} ) = \Delta \textrm{Lk}^\textrm{F} ( \textbf{R}, \textbf{R} ) = 0$. Thus, by deriving separate expansion formulas for the two parts, $\delta ( \textbf{R} - \textbf{R}' )$ and $e^{ b f ( \textbf{z}_0 \cdot \textbf{R} \textbf{z}_0 ) }$, of such $F_2$ function and combining them together in the same way as it was done in the general case, it is easy to obtain expansion series of $F_2$ function for the special case of $\textbf{A} = \textbf{I}$.

Expansion coefficients of the Dirac $\delta ( \textbf{R} - \textbf{R}' )$ function can be found rather easily by applying Eq.~\eqref{D-functions-orthogonality} and \eqref{expansion-coeff-2}:
\begin{align} \label{delta-special-expansion-coeff}
F^{p'\!\!,\,q'\!\!,\,s'}_{p,\,q,\,s} & = 
	\frac{ \left(2s\!+\!1\right)\!\left(2s'\!\!+\!1\right) }{ \left(8\pi^2\right)^2 } 
	\int \! \textrm{d} \textbf{R} \textrm{d} \textbf{R}' \,
	\overline{D}^s_{p,q} \!\left( \textbf{R} \right) 
	\delta \!\left( \textbf{R} - \textbf{R}' \right)
	D^{s'}_{p'\!,q'} \!\left( \textbf{R}' \right) = \nonumber \\
& = \frac{ \left(2s\!+\!1\right)\!\left(2s'\!\!+\!1\right) }{ \left(8\pi^2\right)^2 } \,
	\int \! \textrm{d} \textbf{R} \,
	\overline{D}^s_{p,q} \!\left( \textbf{R} \right)
	D^{s'}_{p'\!,q'} \!\left( \textbf{R} \right) = 
\delta_{ss'} \delta_{pp'} \delta_{qq'} \times \frac{ 2s \!+\! 1 }{ 8\pi^2 }
\end{align}

Hence, for this function we have:
\begin{equation} \label{delta-special-expansion}
\delta \!\left( \textbf{R} - \textbf{R}' \right) = 
	\frac{ 1 }{ 8\pi^2 }
	\sum_{s,p,q} \left( 2s \!+\! 1 \right)
	D^s_{p,q} \!\left( \textbf{R} \right) \overline{D}^s_{p,q} \!\left( \textbf{R}' \right)
\end{equation}

As for the second part of $F_2$ function, $e^{ b f ( \textbf{z}_0 \cdot \textbf{R} \textbf{z}_0 ) }$, its expansion coefficients can be simply obtained by using Eq.~\eqref{expansion-coeff-1}:
\begin{align} \label{exp-special-expansion-coeff}
F_{p,\,q,\,s} & = 
	\frac{ 2s \!+\! 1 }{ 8\pi^2 } 
	\int \! \textrm{d} \textbf{R} \,
	\overline{D}^s_{p,q} \!\left( \textbf{R} \right) 
	e^{ b f \left( \textbf{z}_0 \cdot \textbf{R} \textbf{z}_0 \right) } = 
	\frac{ 2s \!+\! 1 }{ 8\pi^2 } 
	\int \! \textrm{d} \textbf{R} \,
	\overline{P}^s_{p,q} \!\left( \cos \beta \right) e^{ b f \cos \beta } e^{ i p \alpha } e^{ i q \gamma } = \nonumber \\
& = \delta_{p0} \delta_{q0} \times \frac{ 2s \!+\! 1 }{ 2 } \mathscr{L}^s_0 \!\left( - b f \right) =
	\delta_{p0} \delta_{q0} \times \left( 2s \!+\! 1 \right) i_s \!\left( b f \right)
\end{align}

\noindent
Where we have taken into account that $\mathscr{L}^s_0 ( - x ) = 2 \, i_s ( x )$, see Eq.~(D3) from ref.~\cite{Efremov_2016}. Here $i_s ( x )$ is the modified spherical Bessel function of the first kind.

Substituting Eq.~\eqref{exp-special-expansion-coeff} into Eq.~\eqref{D-expansion-1}, we get the following expansion formula:
\begin{equation} \label{exp-special-expansion}
	e^{ b f \left( \textbf{z}_0 \cdot \textbf{R} \textbf{z}_0 \right) } = 
	\sum_s \left( 2s \!+\! 1 \right)
	i_s \!\left( b f \right) D^s_{0,0} \!\left( \textbf{R} \right)
\end{equation}

Multiplying Eq.~\eqref{delta-special-expansion} and \eqref{exp-special-expansion} and using Eq.~\eqref{D-functions-product-2} together with Eq.~\eqref{Wigner-3j}, it is not very hard to derive the following expression for $F_2$ function in the special case of $\textbf{A} = \textbf{I}$:
\begin{multline} \label{F2-special-expansion}
\delta \!\left( \textbf{R} - \textbf{R}' \right) e^{ b f \left( \textbf{z}_0 \cdot \textbf{R} \textbf{z}_0 \right) } = 
	\frac{ 1 }{ 8\pi^2 } \!\!\sum_{s,s'\!,p,q}\!\!
	\left( 2s \!+\! 1 \right) \left( 2s' \!\!+\! 1 \right)
	i_{s'} \!\left( b f \right) \times
	D^s_{p,q} \!\left( \textbf{R} \right) D^{s'}_{0,0} \!\left( \textbf{R} \right) \times
	\overline{D}^s_{p,q} \!\left( \textbf{R}' \right) = \\[0pt]
 = \frac{ 1 }{ 8\pi^2 } \!\!\!\sum_{s,s'\!,p,q,k}\!\!\!
	\left( 2s \!+\! 1 \right) \left( 2s' \!\!+\! 1 \right) \left( 2k \!+\! 1 \right)
	\left( -1 \right)^{p+q} i_{s'} \!\left( b f \right) \times 
	\left(\begin{array}{ccc}
		s & \;s' & k \\
		p & \;0 & -p
	\end{array}\right)
	\left(\begin{array}{ccc}
		s & \;s' & k \\
		q & \;0 & -q
	\end{array}\right)
	\times
	D^k_{p,q} \!\left( \textbf{R} \right) \overline{D}^s_{p,q} \!\left( \textbf{R}' \right) 
\end{multline}

By performing indexes change $k \rightarrow s$, $s \rightarrow s'$ and $s' \rightarrow k$ in Eq.~\eqref{F2-special-expansion}, it is then rather straightforward to obtain the expansion series of $F_2$ function in the special case of $\textbf{A} = \textbf{I}$ in the form of Eq.~\eqref{F2-function-expansion-2}, where the expansion coefficients $( F_2 )^{p'\!\!,\,q'\!\!,\,s'}_{p,\,q,\,s}$ are:
\begin{equation}\label{F2-function-expansion-coeff-special}
\left( F_2 \right)^{p'\!\!,\,q'\!\!,\,s'}_{p,\,q,\,s} =
	\delta_{pp'} \delta_{qq'} \times \sqrt{ \left( 2s \!+\! 1 \right) \!\left( 2s' \!\!+\! 1 \right) } \,
	\left( -1 \right)^{p+q}
	\sum_k
	\left( 2k \!+\! 1 \right) i_k \!\left( b f \right)
	\left(\begin{array}{ccc}
		s' & \;k & s \\
		p  & \;0 & -p
	\end{array}\right)
	\left(\begin{array}{ccc}
		s' & \;k & s \\
		q  & \;0 & -q
	\end{array}\right)
\end{equation}

\subsection{Elements of $\textbf{S}_{00}$ matrix}\label{sec:expansion-S00}

Having expansion series for $F_1$ and $F_2$ functions, it is now possible to derive the expansion formulas for $T_{nm}$ transfer-functions and elements of $\textbf{S}_{nm}$ matrices. Let's start with $T_{00}$ function.

Substituting the first line of Eq.~\eqref{all-local-energies-DNA-bending} [or Eq.~\eqref{all-local-energies-DNA-wrapping}]  into Eq.~\eqref{Transfer-matrix-elements} and comparing the resulting expression to Eq.~\eqref{functions-F1-F2}, it is clear that the expansion series for $T_{00}$ transfer-function can be obtained from Eq.~\eqref{F1-function-expansion-2}-\eqref{F1-function-expansion-coeff} by using the following values of the model parameters: $a = a_0$, $b = b_0$, $c = c_0$, $\lambda = \lambda_0$ and $\textbf{A} = \textbf{I}$. Then by taking into account Eq.~\eqref{D-functions-000} saying that $D^s_{p,q} ( \textbf{I} ) = \delta_{pq}$, it is easy to find the following formula for the expansion coefficients of $T_{00}$ function:
\begin{align} \label{T00-expansion-coeff}
\left( T_{00} \right)^{p'\!\!,\,q'\!\!,\,s'}_{p,\,q,\,s} =
	\delta_{pp'} \delta_{qq'} \times \pi^2 \sqrt{ \left( 2s \!+\! 1 \right) \!\left( 2s' \!\!+\! 1 \right) } \,
	e^{ - a_0 - c_0 }
	\!\!\sum_{t,k,k'\!,r} \!\!
	\left( 2t \!+\! 1 \right) \!\left( 2k \!+\! 1 \right) \!\left( 2k' \!\!+\! 1 \right)
	i^{-r} e^{ -i \left( q \!- r \right) \omega_0 } \times & \nonumber \\[-2pt]
	\times \, I_{q-r} \!\left( c_0 \sqrt{1\!+\!\chi_0^2} \right) 
	I_r \!\left( \tau \left[ 1 \!-\! \textstyle{ \frac{ \lambda_0 }{ 2\pi } } \right] \right) 
	\!\times 
	\mathscr{L}^t_{q-r} \!\left( -a_0 \right) \mathscr{L}^k_r \!\left( - b_0 f \right)
	\mathscr{L}^{k'}_r \!\!\left( 0 \right) \times & \nonumber \\[3pt]
	\times
    \left(\begin{array}{ccc}
		t & k & \,s \\
		p\!-\!r & r & \,-p
	\end{array}\right)
	\left(\begin{array}{ccc}
		t & k & \,s \\
		q\!-\!r & r & \,-q
	\end{array}\right)
	\left(\begin{array}{ccc}
		t & k' & \,s' \\
		p\!-\!r & r & \,-p
	\end{array}\right)
    \left(\begin{array}{ccc}
		t & k' & \,s' \\
		q\!-\!r & r & \,-q
	\end{array}\right) \hspace{6pt} &
\end{align}

\noindent
Where $\chi_0 = \frac{ \tau \lambda_0 }{ 2 \pi c_0 }$ and $\omega_0 = \tan^{-1} ( \chi_0 )$.

Comparing the above formula to Eq.~(C8) from ref.~\cite{Efremov_2016}, it is not hard to see that it is absolutely the same as the one derived for bare DNA in the absence of protein-DNA interactions, which was obtained in our previous study.

From Eq.~\eqref{Transfer-matrix-expansion-coeff}, \eqref{2D-matrix-forms} and \eqref{T00-expansion-coeff}, it then is easy to see that the elements of $\textbf{S}_{00}$ matrix equal to:
\begin{align} \label{S00-matrix-elements-1}
\left( \textbf{S}_{00} \right)_{vv'} =
	\delta_{qq'} \times \pi^2 \sqrt{ \left( 2s \!+\! 1 \right) \!\left( 2s' \!\!+\! 1 \right) } \,
	e^{ - a_0 - c_0 }
	\!\!\sum_{t,k,k'\!,r} \!\!
	\left( 2t \!+\! 1 \right) \!\left( 2k \!+\! 1 \right) \!\left( 2k' \!\!+\! 1 \right)
	i^{-r} e^{ -i \left( q \!- r \right) \omega_0 } \times & \nonumber \\[-2pt]
	\times \, I_{q-r} \!\left( c_0 \sqrt{1\!+\!\chi_0^2} \right) 
	I_r \!\left( \tau \left[ 1 \!-\! \textstyle{ \frac{ \lambda_0 }{ 2\pi } } \right] \right) 
	\!\times 
	\mathscr{L}^t_{q-r} \!\left( -a_0 \right) \mathscr{L}^k_r \!\left( - b_0 f \right)
	\mathscr{L}^{k'}_r \!\!\left( 0 \right) \times & \nonumber \\[3pt]
	\times
    \left(\begin{array}{ccc}
		t & k & \,s \\
	   -r & r & \,0
	\end{array}\right)
	\left(\begin{array}{ccc}
		t & k & \,s \\
		q\!-\!r & r & \,-q
	\end{array}\right)
	\left(\begin{array}{ccc}
		t & k' & \,s' \\
	   -r & r & \,0
	\end{array}\right)
    \left(\begin{array}{ccc}
		t & k' & \,s' \\
		q\!-\!r & r & \,-q
	\end{array}\right) \hspace{6pt} &
\end{align}

\noindent
Where, as before, indexes $v$ and $v'$ are defined as: $v = q+s(s+1)$ and  $v' = q'+s'(s'+1)$.

In fact, we can further simplify Eq.~\eqref{S00-matrix-elements-1} by noting that $\delta_{qq'}$ pref-factor in Eq.~\eqref{S00-matrix-elements-1} and $\delta_{q0}$ pref-factor in Eq.~\eqref{S01-matrix-elements-1} [or Eq.~\eqref{S01-matrix-elements-4} in the case of DNA-wrapping proteins] lead to a domino-like effect, which is similar to that discussed in Appendix~\ref{Appendix-B} for $p$ indexes, resulting in nullification of all of the $q$ indexes of matrices $\textbf{S}_{00}$ and $\textbf{S}_{01}$, and all of the $q'$ indexes of matrices $\textbf{S}_{00}$ and $\textbf{S}_{30}$ in the DNA partition function calculations. For this reason, we can set both indexes, $q$ and $q'$, in Eq.~\eqref{S00-matrix-elements-1} equal to zero, as a result getting the following final formula for the elements of $\textbf{S}_{00}$ matrix, which appears to be practically identical to Eq.~(C19) from ref.~\cite{Efremov_2016}:
\begin{multline} \label{S00-matrix-elements-2}
\left( \textbf{S}_{00} \right)_{vv'} =
	\delta_{q0} \delta_{q'0} \times \pi^2 \sqrt{ \left( 2s \!+\! 1 \right) \!\left( 2s' \!\!+\! 1 \right) } \,
	e^{ - a_0 - c_0 }
	\!\!\sum_{t,k,k'\!,r} \!\!
	\left( 2t \!+\! 1 \right) \!\left( 2k \!+\! 1 \right) \!\left( 2k' \!\!+\! 1 \right)
	i^{-r} e^{ i r \omega_0 } \times \\[-2pt]
	\times \, I_r \!\left( c_0 \sqrt{1\!+\!\chi_0^2} \right) 
	I_r \!\left( \tau \left[ 1 \!-\! \textstyle{ \frac{ \lambda_0 }{ 2\pi } } \right] \right) 
	\!\times 
	\mathscr{L}^t_r \!\left( -a_0 \right) \mathscr{L}^k_r \!\left( - b_0 f \right)
	\mathscr{L}^{k'}_r \!\!\left( 0 \right) \times
	\left(\begin{array}{ccc}
		t & k & \,s \\
	   -r & r & \,0
	\end{array}\right)^{\!2}
	\left(\begin{array}{ccc}
		t & k' & \,s' \\
	   -r & r & \,0
	\end{array}\right)^{\!2}
\end{multline}

\subsection{Elements of $\textbf{S}_{01}$ matrix}\label{sec:expansion-S01}

In order to derive a mathematical expression for the elements of $\textbf{S}_{01}$ matrix, we have to go pretty much through the same procedure as in the previous section. Namely, first we need to find a formula for the expansion coefficients of $T_{01}$ transfer-function, from which we can then get the elements of $\textbf{S}_{01}$ matrix. To this aim, it should be noted that from Eq.~\eqref{all-local-energies-DNA-bending}-\eqref{all-local-energies-DNA-wrapping}, \eqref{Transfer-matrix-elements}, \eqref{functions-F1-F2} and comments at the end of Appendix~\ref{Appendix-A} it is not hard to deduce that the expansion series of $T_{01}$ function can be obtained from Eq.~\eqref{F1-function-expansion-2}-\eqref{F1-function-expansion-coeff} via the following two steps: 1) by making the model parameters' substitutions $a \rightarrow a_\textrm{pr}$, $b \rightarrow b_0$, $c \rightarrow c_\textrm{pr}$, $\lambda \rightarrow \lambda_0$, $\textbf{A} \rightarrow \textbf{A}_\textrm{in}$ and $\textbf{R} \rightarrow \textbf{R} \textbf{B}$, where $\textbf{B} = \textbf{B} ( \eta_\textrm{in}, 0, 0 )$, and 2) by performing integration over the angle $\eta_\textrm{in}$. As can be seen from Eq.~\eqref{D-functions}, \eqref{D-functions-000}, \eqref{D-functions-product-1} and \eqref{F1-function-expansion-2} the latter integration results in nullification of index $q$ and multiplication of the whole expression by $2\pi$ prefactor, leading us to the following formula for the expansion coefficients of $T_{01}$ transfer-function:
\begin{align} \label{T01-expansion-coeff}
\left( T_{01} \right)^{p'\!\!,\,q'\!\!,\,s'}_{p,\,q,\,s} =
	\delta_{pp'} \delta_{q0} \times 2\pi^3 \sqrt{ \left( 2s \!+\! 1 \right) \!\left( 2s' \!\!+\! 1 \right) } \,
	\left( -1 \right)^{q'} \!
	e^{ - a_\textrm{pr} - c_\textrm{pr} + \tau \lambda_0 \Delta \textrm{Tw} \left( \textbf{I}, 
	\textbf{A}_\textrm{in} \right) }
	\!\!\sum_{t,k,k'\!,r} \!\!
	\left( 2t \!+\! 1 \right) \!\left( 2k \!+\! 1 \right) \!\left( 2k' \!\!+\! 1 \right)
	i^{-r} \times & \nonumber \\[-2pt]
	\times \, e^{ -i \left( q' \!- r \right) \omega_\textrm{pr} } I_{q'-r} \!\left( c_\textrm{pr}
	\sqrt{ 1 \!+\! \chi_\textrm{pr}^2 } \right) 
	I_r \!\left( \tau \left[ 1 \!-\! \textstyle{ \frac{ \lambda_0 }{ 2\pi } } \right] \right) 
	D^{\,t}_{-r,q'\!-r} \!\left( \textbf{A}_\textrm{in} \right) \!\times 
	\mathscr{L}^t_{q'-r} \!\left( -a_\textrm{pr} \right) \mathscr{L}^k_r \!\left( - b_0 f \right)
	\mathscr{L}^{k'}_r \!\!\left( 0 \right) \times & \nonumber \\[3pt]
	\times
    \left(\begin{array}{ccc}
		t & k & \,s \\
		p\!-\!r & r & \,-p
	\end{array}\right)
	\left(\begin{array}{ccc}
		t & k & \,s \\
		-r & r & \,0
	\end{array}\right)
	\left(\begin{array}{ccc}
		t & k' & \,s' \\
		p\!-\!r & r & \,-p
	\end{array}\right)
    \left(\begin{array}{ccc}
		t & k' & \,s' \\
		q'\!\!-\!r & r & \,-q'
	\end{array}\right) \hspace{6pt} &
\end{align}

\noindent
Where $\chi_\textrm{pr} = \frac{ \tau \lambda_0 }{ 2 \pi c_\textrm{pr} }$ and $\omega_\textrm{pr} = \tan^{-1} ( \chi_\textrm{pr} )$.

From Eq.~\eqref{Transfer-matrix-expansion-coeff}, \eqref{2D-matrix-forms} and \eqref{T01-expansion-coeff}, it is then straightforward to obtain the elements of $\textbf{S}_{01}$ matrix:
\begin{align} \label{S01-matrix-elements-1}
\left( \textbf{S}_{01} \right)_{vv'} =
	\delta_{q0} \times 2\pi^3 \sqrt{ \left( 2s \!+\! 1 \right) \!\left( 2s' \!\!+\! 1 \right) } \,
	\left( -1 \right)^{q'} \!
	e^{ - a_\textrm{pr} - c_\textrm{pr} + \tau \lambda_0 \Delta \textrm{Tw} \left( \textbf{I}, 
	\textbf{A}_\textrm{in} \right) }
	\!\!\sum_{t,k,k'\!,r} \!\!
	\left( 2t \!+\! 1 \right) \!\left( 2k \!+\! 1 \right) \!\left( 2k' \!\!+\! 1 \right)
	i^{-r} \times & \nonumber \\[-2pt]
	\times \, e^{ -i \left( q' \!- r \right) \omega_\textrm{pr} } I_{q'-r} \!\left( c_\textrm{pr}
	\sqrt{ 1 \!+\! \chi_\textrm{pr}^2 } \right) 
	I_r \!\left( \tau \left[ 1 \!-\! \textstyle{ \frac{ \lambda_0 }{ 2\pi } } \right] \right) 
	D^{\,t}_{-r,q'\!-r} \!\left( \textbf{A}_\textrm{in} \right) \!\times 
	\mathscr{L}^t_{q'-r} \!\left( -a_\textrm{pr} \right) \mathscr{L}^k_r \!\left( - b_0 f \right)
	\mathscr{L}^{k'}_r \!\!\left( 0 \right) \times & \nonumber \\[3pt]
	\times
    \left(\begin{array}{ccc}
		t & k & \,s \\
		-r & r & \,0
	\end{array}\right)^{\!\!2}
	\left(\begin{array}{ccc}
		t & k' & \,s' \\
		-r & r & \,0
	\end{array}\right)
    \left(\begin{array}{ccc}
		t & k' & \,s' \\
		q'\!\!-\!r & r & \,-q'
	\end{array}\right) \hspace{6pt} &
\end{align}

\noindent
Where $v = q+s(s+1)$ and $v' = q'+s'(s'+1)$.

Furthermore, we would like to stress that for many DNA-binding proteins it is a frequent situation that the rotation matrix $\textbf{A}_\textrm{in}$ describing the equilibrium orientations of the DNA segments entering the corresponding nucleoprotein complexes equals to the unit matrix: $\textbf{A}_\textrm{in} = \textbf{I}$. In this case, Eq.~\eqref{S01-matrix-elements-1} can be further simplified by noting that $D^{\,t}_{-r,q'\!-r} ( \textbf{I} ) = \delta_{q'0}$ [see Eq.~\eqref{D-functions-000}], which leads us to the following expression for the elements of $\textbf{S}_{01}$ matrix:
\begin{multline} \label{S01-matrix-elements-2}
\left( \textbf{S}_{01} \right)_{vv'} =
	\delta_{q0} \delta_{q'0} \times 2\pi^3 \sqrt{ \left( 2s \!+\! 1 \right) \!\left( 2s' \!\!+\! 1 \right) } \,
	e^{ - a_\textrm{pr} - c_\textrm{pr} }
	\!\!\sum_{t,k,k'\!,r} \!\!
	\left( 2t \!+\! 1 \right) \!\left( 2k \!+\! 1 \right) \!\left( 2k' \!\!+\! 1 \right)
	i^{-r} e^{ i r \omega_\textrm{pr} } \times \\[-2pt]
	\times \, I_r \!\left( c_\textrm{pr} \sqrt{ 1 \!+\! \chi_\textrm{pr}^2 } \right) 
	I_r \!\left( \tau \left[ 1 \!-\! \textstyle{ \frac{ \lambda_0 }{ 2\pi } } \right] \right) \!\times 
	\mathscr{L}^t_r \!\left( -a_\textrm{pr} \right) \mathscr{L}^k_r \!\left( - b_0 f \right)
	\mathscr{L}^{k'}_r \!\!\left( 0 \right) \times 
    \left(\begin{array}{ccc}
		t & k & \,s \\
		-r & r & \,0
	\end{array}\right)^{\!\!2}
	\left(\begin{array}{ccc}
		t & k' & \,s' \\
		-r & r & \,0
	\end{array}\right)^{\!\!2}
\end{multline}

It should be noted that Eq.~\eqref{S01-matrix-elements-1}-\eqref{S01-matrix-elements-2} were obtained for the case when the rotation matrix $\textbf{A}_\textrm{in}$ is either equal to the unit matrix, $\textbf{I}$, or only slightly deviates from it. However, as can be seen from Figures~\ref{fig1}(e,f), nucleoprotein complexes formed by DNA-wrapping proteins, such as histone tetramers and octamers, do not belong to either of these two scenarios as the equilibrium angle between the DNA segment entering the nucleoprotein complex and the line connecting the entry and exit points of the complex reaches quite a large value, indicating strong deviation of matrix $\textbf{A}_\textrm{in}$ from the unit matrix, $\textbf{I}$. In this case, the above Eq.~\eqref{S01-matrix-elements-1} cannot be used directly to find the elements of $\textbf{S}_{01}$ matrix. Nevertheless, Eq.~\eqref{S01-matrix-elements-1} still come in handy even for the case of DNA-wrapping proteins as it can be applied to find an approximate analytic expression for the elements of $\textbf{S}_{01}$ matrix.

Indeed, in the above derivations, $\textbf{A}_\textrm{in} \approx \textbf{I}$ assumption is used only to estimate the local DNA linking number change, $\Delta \textrm{Lk}^\textrm{F} ( \textbf{R}, \textbf{R}' )$, and DNA twist, $\Delta \textrm{Tw} ( \textbf{R}, \textbf{R}' )$, at the entry points of nucleoprotein complexes. In the general case, both of these quantities contribute to the total DNA linking number change and DNA writhe, influencing the global conformation of the polymer under force and torque constrains applied to it. However, in the case of DNA-wrapping proteins, we can safely neglect variations of $\Delta \textrm{Lk}^\textrm{F} ( \textbf{R}, \textbf{R}' )$ and $\Delta \textrm{Tw} ( \textbf{R}, \textbf{R}' )$ quantities due to the thermal fluctuations of DNA segments, as the resulting effect of such variations is typically much smaller than the effect created by the nucleoprotein complex itself [i.e., $\Delta \textrm{Lk}^\textrm{F} ( \textbf{R}, \textbf{R}' ) - \langle \Delta  \textrm{Lk}^\textrm{F} ( \textbf{R}, \textbf{R}' ) \rangle \ll \Delta \textrm{Lk}_\textrm{pr}$ and $\Delta \textrm{Tw} ( \textbf{R}, \textbf{R}' ) - \langle \Delta \textrm{Tw} ( \textbf{R}, \textbf{R}' ) \rangle \ll \Delta \textrm{Lk}_\textrm{pr}$, where $\langle ... \rangle$ means ensemble average over all of the DNA conformations]. In terms of mathematical equations, this means that we can simply remove $2 \pi \tau \Delta \textrm{Lk}^\textrm{F} ( \textbf{R}, \textbf{R}' )$ from the exponent of $F_1$ function [see Eq.~\eqref{functions-F1-F2}], increasing $\Delta \textrm{Lk}_\textrm{pr}$ by the amount corresponding to the average value of $\Delta \textrm{Lk}^\textrm{F} ( \textbf{R}, \textbf{R}' )$ at the entry point of the nucleoprotein complex.

As for the remained local DNA twist and linking number change terms in $F_1$ function, $ - \tau \lambda \Delta \textrm{Lk}^\textrm{F} ( \textbf{R}, \textbf{R}' ) + \tau \lambda \Delta \textrm{Tw} ( \textbf{R}, \textbf{R}' ) = - \tau \lambda \Delta \textrm{Wr}^\textrm{F} ( \textbf{R}, \textbf{R}' )$, they were originally introduced into the model in order to shift the boundary between extended and supercoiled DNA conformations back to the experimentally measured position on the DNA phase diagram due to the failure of the Fuller's formula to accurately describe the writhe number of supercoiled DNA plectonemes, see comments before Eq.~\eqref{correction-term-intro} in Section~\ref{sec:energy_terms}. However, since we can safely remove $2 \pi \tau \Delta \textrm{Lk}^\textrm{F} ( \textbf{R}, \textbf{R}' )$ term from $F_1$ function, there is no need in keeping the remaining $\tau \lambda \Delta \textrm{Wr}^\textrm{F} ( \textbf{R}, \textbf{R}' )$ part as the main cause for its existence is eliminated from $F_1$ function. Thus, we can neglect both terms, $\tau (2\pi - \lambda) \Delta \textrm{Lk}^\textrm{F} ( \textbf{R}, \textbf{R}' )$ and $\tau \lambda \Delta \textrm{Tw} ( \textbf{R}, \textbf{R}' )$, in the exponent of $F_1$ function in the case of nucleoprotein complexes formed by DNA-wrapping proteins.

As a result, the elements of $\textbf{S}_{01}$ matrix in the case of DNA-wrapping proteins can be obtained simply by putting $\tau = 0$ in Eq.~\eqref{S01-matrix-elements-1}. Then by taking into account that $I_r (0) = \delta_{r0}$ and $\mathscr{L}^{k'}_0 ( 0 ) = 2 i_{k'} ( 0 ) = 2 \delta_{k'0}$, we get the following formula for the elements of $\textbf{S}_{01}$ matrix:
\begin{multline} \label{S01-matrix-elements-3}
\left( \textbf{S}_{01} \right)_{vv'} =
	\delta_{q0} \times 4\pi^3 \sqrt{ \left( 2s \!+\! 1 \right) \!\left( 2s' \!\!+\! 1 \right) } \,
	\left( -1 \right)^{q'} \! e^{ - a_\textrm{pr} - c_\textrm{pr} }
	\sum_{t,k}
	\left( 2t \!+\! 1 \right) \!\left( 2k \!+\! 1 \right) 
	\, I_{q'} \!\left( c_\textrm{pr} \right)
	D^{\,t}_{0,q'} \!\left( \textbf{A}_\textrm{in} \right) \times \\[-2pt]
	\times 
	\mathscr{L}^t_{q'} \!\left( -a_\textrm{pr} \right) \mathscr{L}^k_0 \!\left( - b_0 f \right) \times 
    \left(\begin{array}{ccc}
		t & \,k & \,s \\
		0 & \,0 & \,0
	\end{array}\right)^{\!\!2}
	\left(\begin{array}{ccc}
		t & \,0 & \,s' \\
		0 & \,0 & \,0
	\end{array}\right)
    \left(\begin{array}{ccc}
		t & \,0 & s' \\
		q' & \,0 & -q'
	\end{array}\right)
\end{multline}

The above expression can be further simplified by noting that Wigner 3-j symbols satisfy the next equation (p. 1058, ref.~\cite{Messiah_1965}):
\begin{equation} \label{Wigner-3j-special}
	\left(\begin{array}{ccc}
		t & \,0 & s' \\
		q' & \,0 & -q'
	\end{array}\right) =	
	\delta_{ts'} \times \frac{ \left( -1 \right)^{s'\!-q'} }{ \sqrt{ 2s' \!\!+\! 1 } }
\end{equation}

Substituting Eq.~\eqref{Wigner-3j-special} into Eq.~\eqref{S01-matrix-elements-3} and taking into account that $\mathscr{L}^k_0 ( - b_0 f ) = 2 \, i_k ( b_0 f )$, we obtain the final formula for the elements of $\textbf{S}_{01}$ matrix in the case of DNA-wrapping proteins:
\begin{multline} \label{S01-matrix-elements-4}
\left( \textbf{S}_{01} \right)_{vv'} =
	\delta_{q0} \times 8\pi^3 \sqrt{ \left( 2s \!+\! 1 \right) \!\left( 2s' \!\!+\! 1 \right) } \,
	e^{ - a_\textrm{pr} - c_\textrm{pr} }
	I_{q'} \!\left( c_\textrm{pr} \right)
	\mathscr{L}^{s'}_{q'} \!\left( -a_\textrm{pr} \right)
	D^{\,s'}_{0,q'} \!\left( \textbf{A}_\textrm{in} \right) \times \\[0pt]
	\times \sum_{k}
	\left( 2k \!+\! 1 \right) 
	i_k \!\left(  b_0 f \right)
    \left(\begin{array}{ccc}
		s' & \,k & \,s \\
		0 & \,0 & \,0
	\end{array}\right)^{\!\!2}
\end{multline}


\subsection{Elements of $\textbf{S}_{30}$ and $\textbf{S}_{31}$ matrices}\label{sec:expansion-S30-S31}

Formulas for the elements of matrices $\textbf{S}_{30}$ and $\textbf{S}_{31}$ are derived in pretty much the same way as in the case of matrix $\textbf{S}_{01}$, which has been discussed in the previous section. The only major difference is that in the case of matrices $\textbf{S}_{30}$ and $\textbf{S}_{31}$ we do not need to perform the integration step. Namely, expressions for the elements of $\textbf{S}_{30}$ and $\textbf{S}_{31}$ matrices can be easily found from the expansion series of $T_{30}$ and $T_{31}$ transfer-functions, which in turn can obtained from the expansion formula of $F_1$ function by multiplying it by $e^{ \frac{ \mu_\textrm{pr} + \mu_\textrm{off} }{ K } }$ prefactor in the case of $T_{30}$ function or $e^{ \frac{ \mu_\textrm{pr} + \mu_\textrm{off} }{ K } + J_\textrm{pr} }$ prefactor in the case of $T_{31}$ (here $K=3$), and making the following parameter substitutions: $a \rightarrow a_\textrm{pr}$, $b \rightarrow b_0$, $c \rightarrow c_\textrm{pr}$, $\lambda \rightarrow \lambda_\textrm{pr}$, $\textbf{A} \rightarrow \textbf{A}_\textrm{out}$ in the case of $T_{30}$ function or $\textbf{A} \rightarrow \textbf{A}_\textrm{ht}$ in the case of $T_{31}$ function, see Eq.~\eqref{all-local-energies-DNA-bending}, \eqref{Transfer-matrix-elements} and \eqref{functions-F1-F2}. Thus, from Eq.~\eqref{F1-function-expansion-2}-\eqref{F1-function-expansion-coeff} we have the following result for the expansion coefficients of $T_{31}$ function:
\begin{align} \label{T31-expansion-coeff}
\left( T_{31} \right)^{p'\!\!,\,q'\!\!,\,s'}_{p,\,q,\,s} =
	\delta_{pp'} \times \pi^2 \sqrt{ \left( 2s \!+\! 1 \right) \!\left( 2s' \!\!+\! 1 \right) } \,
	\left( -1 \right)^{q+q'} \!
	e^{ \frac{ \mu_\textrm{pr} + \mu_\textrm{off} }{ K } + J_\textrm{pr} - a_\textrm{pr} - c_\textrm{pr} 
	+ \tau \lambda_\textrm{pr} \Delta \textrm{Tw} \left( \textbf{I}, \textbf{A}_\textrm{ht} \right) }
	\!\!\sum_{t,k,k'\!,r} \!\!
	\left( 2t \!+\! 1 \right) \!\left( 2k \!+\! 1 \right) \!\left( 2k' \!\!+\! 1 \right)
	i^{-r} \times & \nonumber \\[-2pt]
	\times \, e^{ -i \left( q' \!- r \right) \omega_\textrm{pr} } 
	I_{q'-r} \!\left( c_\textrm{pr} \sqrt{ 1 \!+\! \chi_\textrm{pr}^2 } \right) 
	I_r \!\left( \tau \left[ 1 \!-\! \textstyle{ \frac{ \lambda_\textrm{pr} }{ 2\pi } } \right] \right) 
	D^{\,t}_{q-r,q'\!-r} \!\left( \textbf{A}_\textrm{ht} \right) \!\times 
	\mathscr{L}^t_{q'-r} \!\left( -a_\textrm{pr} \right) \mathscr{L}^k_r \!\left( - b_0 f \right)
	\mathscr{L}^{k'}_r \!\!\left( 0 \right) \times & \nonumber \\[3pt]
	\times
    \left(\begin{array}{ccc}
		t & k & \,s \\
		p\!-\!r & r & \,-p
	\end{array}\right)
	\left(\begin{array}{ccc}
		t & k & \,s \\
		q\!-\!r & r & \,-q
	\end{array}\right)
	\left(\begin{array}{ccc}
		t & k' & \,s' \\
		p\!-\!r & r & \,-p
	\end{array}\right)
    \left(\begin{array}{ccc}
		t & k' & \,s' \\
		q'\!\!-\!r & r & \,-q'
	\end{array}\right) \hspace{6pt} &
\end{align}

\noindent
Where $\chi_\textrm{pr} = \frac{ \tau \lambda_\textrm{pr} }{ 2 \pi c_\textrm{pr} }$ and $\omega_\textrm{pr} = \tan^{-1} ( \chi_\textrm{pr} )$.

As for the expansion coefficients of $T_{30}$ function, they have absolutely the same mathematical form as in the above equation, with the matrix $\textbf{A}_\textrm{ht}$ being replaced by matrix $\textbf{A}_\textrm{out}$ and with $J_\textrm{pr}$ being removed from the exponential function. 
 
Anyway, combining Eq.~\eqref{Transfer-matrix-expansion-coeff}, \eqref{2D-matrix-forms} and \eqref{T31-expansion-coeff}, it is then straightforward to obtain the following formula for the elements of $\textbf{S}_{31}$ matrix:
\begin{align} \label{S31-matrix-elements-1}
\left( \textbf{S}_{31} \right)_{vv'} =
	\pi^2 \sqrt{ \left( 2s \!+\! 1 \right) \!\left( 2s' \!\!+\! 1 \right) } \,
	\left( -1 \right)^{q+q'} \!
	e^{ \frac{ \mu_\textrm{pr} + \mu_\textrm{off} }{ K } + J_\textrm{pr} - a_\textrm{pr} - c_\textrm{pr}
	+ \tau \lambda_\textrm{pr} \Delta \textrm{Tw} \left( \textbf{I}, \textbf{A}_\textrm{ht} \right) }
	\!\!\sum_{t,k,k'\!,r} \!\!
	\left( 2t \!+\! 1 \right) \!\left( 2k \!+\! 1 \right) \!\left( 2k' \!\!+\! 1 \right)
	i^{-r} \times & \nonumber \\[-2pt]
	\times \, e^{ -i \left( q' \!- r \right) \omega_\textrm{pr} } 
	I_{q'-r} \!\left( c_\textrm{pr} \sqrt{ 1 \!+\! \chi_\textrm{pr}^2 } \right) 
	I_r \!\left( \tau \left[ 1 \!-\! \textstyle{ \frac{ \lambda_\textrm{pr} }{ 2\pi } } \right] \right) 
	D^{\,t}_{q-r,q'\!-r} \!\left( \textbf{A}_\textrm{ht} \right) \!\times 
	\mathscr{L}^t_{q'-r} \!\left( -a_\textrm{pr} \right) \mathscr{L}^k_r \!\left( - b_0 f \right)
	\mathscr{L}^{k'}_r \!\!\left( 0 \right) \times & \nonumber \\[3pt]
	\times
    \left(\begin{array}{ccc}
		t & k & \,s \\
		-r & r & \,0
	\end{array}\right)
	\left(\begin{array}{ccc}
		t & k & \,s \\
		q\!-\!r & r & \,-q
	\end{array}\right)
	\left(\begin{array}{ccc}
		t & k' & \,s' \\
		-r & r & \,0
	\end{array}\right)
    \left(\begin{array}{ccc}
		t & k' & \,s' \\
		q'\!\!-\!r & r & \,-q'
	\end{array}\right) \hspace{6pt} &
\end{align}

\noindent
Where $v = q+s(s+1)$ and  $v' = q'+s'(s'+1)$.

In the special case of $\textbf{A}_\textrm{ht} = \textbf{I}$, D-function $D^{\,t}_{q-r,q'\!-r} ( \textbf{A}_\textrm{ht} )$ turns into the Kronecker delta, $\delta_{qq'}$, and the above equation simplifies to:
\begin{align} \label{S31-matrix-elements-2}
\left( \textbf{S}_{31} \right)_{vv'} =
	\delta_{qq'} \times \pi^2 \sqrt{ \left( 2s \!+\! 1 \right) \!\left( 2s' \!\!+\! 1 \right) } \,
	e^{ \frac{ \mu_\textrm{pr} + \mu_\textrm{off} }{ K } + J_\textrm{pr} - a_\textrm{pr} - c_\textrm{pr} }
	\!\!\sum_{t,k,k'\!,r} \!\!
	\left( 2t \!+\! 1 \right) \!\left( 2k \!+\! 1 \right) \!\left( 2k' \!\!+\! 1 \right)
	i^{-r} \, e^{ -i \left( q - r \right) \omega_\textrm{pr} } \times & \nonumber \\[-2pt]
	\times I_{q-r} \!\left( c_\textrm{pr} \sqrt{ 1 \!+\! \chi_\textrm{pr}^2 } \right) 
	I_r \!\left( \tau \left[ 1 \!-\! \textstyle{ \frac{ \lambda_\textrm{pr} }{ 2\pi } } \right] \right) \times 
	\mathscr{L}^t_{q-r} \!\left( -a_\textrm{pr} \right) \mathscr{L}^k_r \!\left( - b_0 f \right)
	\mathscr{L}^{k'}_r \!\!\left( 0 \right) \times & \nonumber \\[3pt]
	\times
    \left(\begin{array}{ccc}
		t & k & \,s \\
		-r & r & \,0
	\end{array}\right)
	\left(\begin{array}{ccc}
		t & k & \,s \\
		q\!-\!r & r & \,-q
	\end{array}\right)
	\left(\begin{array}{ccc}
		t & k' & \,s' \\
		-r & r & \,0
	\end{array}\right)
    \left(\begin{array}{ccc}
		t & k' & \,s' \\
		q\!-\!r & r & \,-q
	\end{array}\right) \hspace{6pt} &
\end{align}

As for the elements of $\textbf{S}_{30}$ matrix, they can be obtained from Eq.~\eqref{S31-matrix-elements-1} simply by replacing $\textbf{A}_\textrm{ht}$ matrix with $\textbf{A}_\textrm{out}$, and removing $J_\textrm{pr}$ from the exponential function. In addition, it should be noted that similarly to the case of $\textbf{S}_{00}$ matrix, the presence of $\delta_{qq'}$ and $\delta_{q0}$ prefactors in Eq.~\eqref{S00-matrix-elements-1} and Eq.~\eqref{S01-matrix-elements-1} leads to nullification of $q'$ index of the matrix $\textbf{S}_{30}$ in the DNA partition function calculations, and as a result we get the following formula for the elements of $\textbf{S}_{30}$ matrix:
\begin{align} \label{S30-matrix-elements-1}
\left( \textbf{S}_{30} \right)_{vv'} =
	\delta_{q'0} \times \pi^2 \sqrt{ \left( 2s \!+\! 1 \right) \!\left( 2s' \!\!+\! 1 \right) } \,
	\left( -1 \right)^q \!
	e^{ \frac{ \mu_\textrm{pr} + \mu_\textrm{off} }{ K } - a_\textrm{pr} - c_\textrm{pr} + \tau \lambda_\textrm{pr}
	\Delta \textrm{Tw} \left( \textbf{I}, \textbf{A}_\textrm{out} \right) }
	\!\!\sum_{t,k,k'\!,r} \!\!
	\left( 2t \!+\! 1 \right) \!\left( 2k \!+\! 1 \right) \!\left( 2k' \!\!+\! 1 \right)
	i^{-r} \times & \nonumber \\[-2pt]
	\times \, e^{ i r \omega_\textrm{pr} } 
	I_r \!\left( c_\textrm{pr} \sqrt{ 1 \!+\! \chi_\textrm{pr}^2 } \right) 
	I_r \!\left( \tau \left[ 1 \!-\! \textstyle{ \frac{ \lambda_\textrm{pr} }{ 2\pi } } \right] \right) 
	D^{\,t}_{q-r,-r} \!\left( \textbf{A}_\textrm{out} \right) \!\times 
	\mathscr{L}^t_r \!\left( -a_\textrm{pr} \right) \mathscr{L}^k_r \!\left( - b_0 f \right)
	\mathscr{L}^{k'}_r \!\!\left( 0 \right) \times & \nonumber \\[3pt]
	\times
    \left(\begin{array}{ccc}
		t & k & \,s \\
		-r & r & \,0
	\end{array}\right)
	\left(\begin{array}{ccc}
		t & k & \,s \\
		q\!-\!r & r & \,-q
	\end{array}\right)
	\left(\begin{array}{ccc}
		t & k' & \,s' \\
		-r & r & \,0
	\end{array}\right)^{\!\!2}
	\hspace{6pt} &
\end{align}

Respectively, for the special case of $\textbf{A}_\textrm{out} = \textbf{I}$ we have:
\begin{align} \label{S30-matrix-elements-2}
\left( \textbf{S}_{30} \right)_{vv'} =
	\delta_{q0} \delta_{q'0} \times \pi^2 \sqrt{ \left( 2s \!+\! 1 \right) \!\left( 2s' \!\!+\! 1 \right) } \,
	e^{ \frac{ \mu_\textrm{pr} + \mu_\textrm{off} }{ K } - a_\textrm{pr} - c_\textrm{pr} }
	\!\!\sum_{t,k,k'\!,r} \!\!
	\left( 2t \!+\! 1 \right) \!\left( 2k \!+\! 1 \right) \!\left( 2k' \!\!+\! 1 \right)
	i^{-r} \times & \nonumber \\[-2pt]
	\times \, e^{ i r \omega_\textrm{pr} } 
	I_r \!\left( c_\textrm{pr} \sqrt{ 1 \!+\! \chi_\textrm{pr}^2 } \right) 
	I_r \!\left( \tau \left[ 1 \!-\! \textstyle{ \frac{ \lambda_\textrm{pr} }{ 2\pi } } \right] \right) \times 
	\mathscr{L}^t_r \!\left( -a_\textrm{pr} \right) \mathscr{L}^k_r \!\left( - b_0 f \right)
	\mathscr{L}^{k'}_r \!\!\left( 0 \right) \times & \nonumber \\[3pt]
	\times
    \left(\begin{array}{ccc}
		t & k & \,s \\
		-r & r & \,0
	\end{array}\right)^{\!\!2}
	\left(\begin{array}{ccc}
		t & k' & \,s' \\
		-r & r & \,0
	\end{array}\right)^{\!\!2}
	\hspace{6pt} &
\end{align}

Finally, using the same reasoning as in the previous section, we can easily obtain the elements of $\textbf{S}_{31}$ and $\textbf{S}_{30}$ matrices for the case of nucleoprotein complexes formed by DNA-wrapping proteins by putting $\tau = 0$ in Eq.~\eqref{S31-matrix-elements-1} and \eqref{S30-matrix-elements-1}, simultaneously replacing $b_0$ with $r_\textrm{pr}/K$ and adding $2 \pi \tau \Delta \textrm{Lk}_\textrm{pr} / K$ term into the exponential function [see Eq.~\eqref{all-local-energies-DNA-wrapping}]. This leads us to the following formulas for DNA-wrapping proteins:
\begin{multline} \label{S31-matrix-elements-3}
\left( \textbf{S}_{31} \right)_{vv'} =
	4 \pi^2 \sqrt{ \left( 2s \!+\! 1 \right) \!\left( 2s' \!\!+\! 1 \right) } \,
	\left( -1 \right)^q \!
	e^{ \frac{ \mu_\textrm{pr} + \mu_\textrm{off} + 2 \pi \tau \Delta \textrm{Lk}_\textrm{pr} }{ K } + J_\textrm{pr}
	- a_\textrm{pr} - c_\textrm{pr} }
	I_{q'} \!\left( c_\textrm{pr} \right)
	D^{\,s'}_{q,q'} \!\left( \textbf{A}_\textrm{ht} \right) 
	\mathscr{L}^{s'}_{q'} \!\left( -a_\textrm{pr} \right) \times \\[0pt]
	\times
	\sum_k 
	\left( 2k \!+\! 1 \right)
	i_k \!\left( r_\textrm{pr} f / K \right)
    \left(\begin{array}{ccc}
		s' & k & \,s \\
		0 & 0 & \,0
	\end{array}\right)
	\left(\begin{array}{ccc}
		s' & \,k & s \\
		q & \,0 & -q
	\end{array}\right)
\end{multline}

\noindent 
and
\begin{multline} \label{S30-matrix-elements-3}
\left( \textbf{S}_{30} \right)_{vv'} =
	\delta_{q'0} \times 8 \pi^2 \sqrt{ \left( 2s \!+\! 1 \right) \!\left( 2s' \!\!+\! 1 \right) } \,
	\left( -1 \right)^q \!
	e^{ \frac{ \mu_\textrm{pr} + \mu_\textrm{off} + 2 \pi \tau \Delta \textrm{Lk}_\textrm{pr} }{ K } - a_\textrm{pr} - 
	c_\textrm{pr} }
	I_0 \!\left( c_\textrm{pr} \right)
	D^{\,s'}_{q,0} \!\left( \textbf{A}_\textrm{out} \right) 
	i_{s'} \!\left( a_\textrm{pr} \right) \times \\[0pt]
	\times
	\sum_k 
	\left( 2k \!+\! 1 \right)
	i_k \!\left( r_\textrm{pr} f / K \right)
    \left(\begin{array}{ccc}
		s' & k & \,s \\
		0 & 0 & \,0
	\end{array}\right)
	\left(\begin{array}{ccc}
		s' & \,k & s \\
		q & \,0 & -q
	\end{array}\right)
\end{multline}

\noindent
Where we have taken into account Eq.~\eqref{Wigner-3j-special}, and that $I_r (0) = \delta_{r0}$, $\mathscr{L}^{k'}_0 ( 0 ) = 2 i_{k'} ( 0 ) = 2 \delta_{k'0}$ and $\mathscr{L}^k_0 ( - r_\textrm{pr} f / K ) = 2 i_k ( r_\textrm{pr} f / K )$.

\subsection{Elements of $\textbf{S}_{12}$ and $\textbf{S}_{23}$ matrices}\label{sec:expansion-S12-S23}

Next, to obtain the elements of $\textbf{S}_{12}$ and $\textbf{S}_{23}$ matrices, we need to find the expansion coefficients of $T_{12}$ and $T_{23}$ transfer-functions, which can be easily acquired from Eq.~\eqref{F2-function-expansion-coeff} by noting that both of these functions have the same mathematical form as $F_2$ function from Eq.~\eqref{functions-F1-F2}. Indeed, it is not hard to show that by substituting $b_0 \rightarrow b$, $\lambda_\textrm{pr} \rightarrow \lambda$, $\textbf{A}_1 \rightarrow \textbf{A}$ or $\textbf{A}_2 \rightarrow \textbf{A}$ into $F_2$ function and multiplying it by $e^{ \frac{ \mu_\textrm{pr} + \mu_\textrm{off} }{ K } }$ (where $K=3$) we eventually come to the formulas for $T_{12}$ and $T_{23}$ transfer-functions, respectively [see Eq.~\eqref{all-local-energies-DNA-bending}, \eqref{Transfer-matrix-elements} and \eqref{functions-F1-F2}]. As a result, by making the same parameter substitutions in Eq.~\eqref{F2-function-expansion-coeff}, we can get the following expression for the expansion coefficients of $T_{12}$ function:
\begin{multline}\label{T12-expansion-coeff}
\!\!\!\!\!\!\!\left( T_{12} \right)^{p'\!\!,\,q'\!\!,\,s'}_{p,\,q,\,s} =
	\delta_{pp'} \times \frac{1}{4} \sqrt{ \left( 2s \!+\! 1 \right) \!\left( 2s' \!\!+\! 1 \right) } \,
	\left( -1 \right)^{q+q'} \!
	e^{ \frac{ \mu_\textrm{pr} + \mu_\textrm{off} }{ K } + 
	\tau \lambda_\textrm{pr} \Delta \textrm{Tw} \left( \textbf{I}, \textbf{A}_1 \right) }
	\!\!\sum_{t,k,k'\!,r} \!\!
	\left( 2t \!+\! 1 \right) \!\left( 2k \!+\! 1 \right) \!\left( 2k' \!\!+\! 1 \right)
	i^{-r} I_r \!\left( \tau \left[ 1 \!-\! \textstyle{ \frac{ \lambda_\textrm{pr} }{ 2\pi } } \right] \right) 
	\times \\[-2pt]
\times D^{\,t}_{q-r,q'\!-r} \!\left( \textbf{A}_1 \right)
	\mathscr{L}^k_r \!\left( - b_0 f \right) \mathscr{L}^{k'}_r \!\!\left( 0 \right) \times
    \left(\begin{array}{ccc}
		t & k & \,s \\
		p\!-\!r & r & \,-p
	\end{array}\right)
	\left(\begin{array}{ccc}
		t & k & \,s \\
		q\!-\!r & r & \,-q
	\end{array}\right)
	\left(\begin{array}{ccc}
		t & k' & \,s' \\
		p\!-\!r & r & \,-p
	\end{array}\right)
    \left(\begin{array}{ccc}
		t & k' & \,s' \\
		q'\!\!-\!r & r & \,-q'
	\end{array}\right)
\end{multline}

From Eq.~\eqref{Transfer-matrix-expansion-coeff}, \eqref{2D-matrix-forms} and \eqref{T12-expansion-coeff}, it is then easy to find the elements of $\textbf{S}_{12}$ matrix:
\begin{multline}\label{S12-matrix-elements-1}
\left( \textbf{S}_{12} \right)_{vv'} =
	\frac{1}{4} \sqrt{ \left( 2s \!+\! 1 \right) \!\left( 2s' \!\!+\! 1 \right) } \,
	\left( -1 \right)^{q+q'} \!
	e^{ \frac{ \mu_\textrm{pr} + \mu_\textrm{off} }{ K } + 
	\tau \lambda_\textrm{pr} \Delta \textrm{Tw} \left( \textbf{I}, \textbf{A}_1 \right) }
	\!\!\sum_{t,k,k'\!,r} \!\!
	\left( 2t \!+\! 1 \right) \!\left( 2k \!+\! 1 \right) \!\left( 2k' \!\!+\! 1 \right)
	i^{-r} I_r \!\left( \tau \left[ 1 \!-\! \textstyle{ \frac{ \lambda_\textrm{pr} }{ 2\pi } } \right] \right) 
	\times \\[-2pt]
\times D^{\,t}_{q-r,q'\!-r} \!\left( \textbf{A}_1 \right)
	\mathscr{L}^k_r \!\left( - b_0 f \right) \mathscr{L}^{k'}_r \!\!\left( 0 \right) \times
    \left(\begin{array}{ccc}
		t & k & \,s \\
		-r & r & \,0
	\end{array}\right)
	\left(\begin{array}{ccc}
		t & k & \,s \\
		q\!-\!r & r & \,-q
	\end{array}\right)
	\left(\begin{array}{ccc}
		t & k' & \,s' \\
		-r & r & \,0
	\end{array}\right)
    \left(\begin{array}{ccc}
		t & k' & \,s' \\
		q'\!\!-\!r & r & \,-q'
	\end{array}\right)
\end{multline}

As for $\textbf{S}_{23}$ matrix, its elements are described by absolutely the same formula as Eq.~\eqref{S12-matrix-elements-1}, where matrix $\textbf{A}_1$ is replaced by the matrix $\textbf{A}_2$.

In the special case of $\textbf{A}_1 = \textbf{I}$, the above expression for the elements of $\textbf{S}_{12}$ matrix can be further simplified with the help of Eq.~\eqref{F2-function-expansion-coeff-special}, and it can be shown that in this case we have:
\begin{equation}\label{S12-matrix-elements-2}
\left( \textbf{S}_{12} \right)_{vv'} =
	\delta_{qq'} \times \sqrt{ \left( 2s \!+\! 1 \right) \!\left( 2s' \!\!+\! 1 \right) } \,
	\left( -1 \right)^q
	e^{ \frac{ \mu_\textrm{pr} + \mu_\textrm{off} }{ K } } \times 
	\sum_k
	\left( 2k \!+\! 1 \right) i_k \!\left( b_0 f \right)
    \left(\begin{array}{ccc}
		s' & k & \,s \\
		0 & 0 & \,0
	\end{array}\right)
	\left(\begin{array}{ccc}
		s' & \,k & s \\
		q & \,0 & -q
	\end{array}\right),
\end{equation}

\noindent
with the identical equation holding for the elements of $\textbf{S}_{23}$ matrix when $\textbf{A}_2 = \textbf{I}$.

Finally, it should be noted that in the case of nucleoprotein complexes formed by DNA-wrapping proteins the elements of matrices $\textbf{S}_{12}$ and $\textbf{S}_{23}$ have the same form as in Eq.~\eqref{S12-matrix-elements-2} with the only difference being that the model parameter $b_0$ is replaced by $r_\textrm{pr} / K$ and the whole formula is multiplied by $e^{ \frac{ 2 \pi \tau \Delta \textrm{Lk}_\textrm{pr} }{ K } }$, see Eq.~\eqref{all-local-energies-DNA-wrapping}, \eqref{Transfer-matrix-elements} and \eqref{functions-F1-F2}. As a result, for DNA-wrapping proteins we get the following formula:
\begin{multline}\label{S12-matrix-elements-3}
\left( \textbf{S}_{12} \right)_{vv'} = \left( \textbf{S}_{23} \right)_{vv'} =
	\delta_{qq'} \times \sqrt{ \left( 2s \!+\! 1 \right) \!\left( 2s' \!\!+\! 1 \right) } \,
	\left( -1 \right)^q
	e^{ \frac{ \mu_\textrm{pr} + \mu_\textrm{off} + 2 \pi \tau \Delta \textrm{Lk}_\textrm{pr} }{ K } } \times \\
	\times \sum_k
	\left( 2k \!+\! 1 \right) i_k \!\left( r_\textrm{pr} f / K \right)
    \left(\begin{array}{ccc}
		s' & k & \,s \\
		0 & 0 & \,0
	\end{array}\right)
	\left(\begin{array}{ccc}
		s' & \,k & s \\
		q & \,0 & -q
	\end{array}\right)
\end{multline}

\subsection{Elements of the boudnary condition matrix, $\textbf{V}_0$}\label{sec:expansion-boundary}

Finally, for the calculations of the DNA paratition function besides $\textbf{S}_{nm}$ matrices we also need to know the matrix $\textbf{V}_0$ describing the boundary conditions imposed on the orientations of the DNA ends. Following our previous work \cite{Efremov_2016}, here we will consider two cases of the boundary condition function, $\xi ( \textbf{R}_N, \textbf{R}_1 )$: 1) $\xi ( \textbf{R}_N, \textbf{R}_1 ) = 1$ that corresponds to the scenario of unconstrained DNA ends' orientations, and 2) $\xi ( \textbf{R}_N, \textbf{R}_1 ) = \delta ( \textbf{R}_N \textbf{z}_0 - \textbf{z}_0 ) \, \delta ( \textbf{R}_1 - \textbf{I} )$, which depicts a DNA whose last segment always stays collinear to the lab $\textbf{z}_0$-axis, but is allowed to freely rotate about it, and the first segment having a fixed orientation corresponding to the global coordinate system, $( \textbf{x}^{}_0,\textbf{y}_0,\textbf{z}^{}_0 )$, -- a setup that is frequently used in \textit{in vitro} experiments. 

As has been previously shown in ref.~\cite{Efremov_2016}, the expansion coefficients of $\sigma_0 ( \textbf{R}_N, \textbf{R}_1 ) = e^{ b_0 f ( \textbf{z}_0 \cdot \textbf{R}_N \textbf{z}_0 ) } \, \xi ( \textbf{R}_N, \textbf{R}_1 )$ function from Eq.~\eqref{Sigma-matrix-expansion-coeff} are described by the following expression in the case of the DNA ends-free orientation boundary condition, $\xi ( \textbf{R}_N, \textbf{R}_1 ) = 1$ (see Eq.~(C15) in ref.~\cite{Efremov_2016}):
\begin{equation} \label{sigma0-unconstrained-expansion-coeff}
\left( \sigma_0 \right)^{\;p_1\!,\,q_1\!,\,s_1}_{p_N \!,\,q_N\!,\,s_N} = 
	\delta_{p_1 0} \delta_{q_1 0} \delta_{s_1 0} \delta_{p_N 0} \delta_{q_N 0} \times 8\pi^2 
	\sqrt{ 2s_N \!+\! 1 } \; i_{s_N} \!\left( b_0 f \right);
\end{equation}

\noindent
whereas, in the case of the DNA ends $\textbf{z}_0$-axis collinear boundary condition, $\xi ( \textbf{R}_N, \textbf{R}_1 ) = \delta ( \textbf{R}_N \textbf{z}_0 - \textbf{z}_0 ) \, \delta ( \textbf{R}_1 - \textbf{I} )$, we have (see Eq.~(C13) in ref.~\cite{Efremov_2016}):
\begin{equation} \label{sigma0-parallel-expansion-coeff}
\left( \sigma_0 \right)^{\;p_1\!,\,q_1\!,\,s_1}_{p_N \!,\,q_N\!,\,s_N} = 
	\delta_{p_N 0} \delta_{q_N 0} \delta_{p_1 q_1} \times \frac{1}{4\pi} \sqrt{ \left( 2s_1 \!+\! 1 \right)
	\!\left( 2s_N \!+\! 1 \right) } \; e^{ b_0 f }
\end{equation}

Substituting the above formulas into Eq.~\eqref{Sigma-matrix-expansion-coeff} and \eqref{2D-matrix-forms}, we finally obtain the elements of $\textbf{V}_0$ matrix, which in the case of the DNA ends-free orientation boundary condition have the following look:
\begin{equation} \label{V0-unconstrained}
\left( \textbf{V}_0 \right)_{v_N v_1} = 
	\delta_{q_1 0} \delta_{s_1 0} \delta_{q_N 0} \times 8\pi^2 
	\sqrt{ 2s_N \!+\! 1 } \; i_{s_N} \!\left( b_0 f \right)
\end{equation}

\noindent
Where $v_1 = q_1 + s_1 ( s_1 + 1 )$ and  $v_N = q_N + s_N ( s_N + 1 )$. 

Alternatively, in the case of the DNA ends $\textbf{z}_0$-axis collinear boundary condition, we have:
\begin{equation} \label{V0-parallel}
\left( \textbf{V}_0 \right)_{v_N v_1} = 
	\delta_{q_1 0} \delta_{q_N 0} \times \frac{1}{4\pi} \sqrt{ \left( 2s_1 \!+\! 1 \right)
	\!\left( 2s_N \!+\! 1 \right) } \; e^{ b_0 f }
\end{equation}

\section{DNA transfer-matrices for different types of DNA-binding proteins}
\label{Appendix-E}

While in the previous Appendix sections we have derived formulas for the DNA partition function and transfer-matrix elements for the special case of DNA-binding proteins that have a fixed binding site size of $K = 3$ DNA segments, it can be seen from the above equations that very similar approach works as well for the case of proteins that have an arbitrarily large binding site on DNA, $K$. As a result, it can be shown that in the general case the DNA partition function can be calculated using the same Eq.~\eqref{DNA-protein-partition-function-10}, where the DNA transfer-matrix, $\textbf{L}$, and the boundary condition matrices, $\textbf{Y}$ and $\textbf{U}$, have the following block-forms:
\begin{equation} \label{Block-matrices-general-1}
\textbf{L} =     
\left( 
\begin{array}{cccccc}
	\textbf{S}_{00} & \textbf{S}_{01} & 0 & 0 & \cdots & 0 \\[2pt]	
	0 & 0 & \textbf{S}_{12} & 0 & \cdots & 0 \\[2pt]
	0 & 0 & 0 & \textbf{S}_{23} & \cdots & 0 \\[-4pt]
	\vdots & \vdots & \vdots & \vdots & \ddots & \vdots \\[2pt]
	0 & 0 & 0 & 0 & \cdots & \textbf{S}_{K-1,K} \\[2pt]
	\textbf{S}_{K0} & \textbf{S}_{K1} & 0 & 0 & \cdots & 0
\end{array}  
\!\right)
\textrm{,} \quad
\textbf{Y} =     
\left(
\begin{array}{c}
\textbf{V}_0 \\[2pt]
0 \\[2pt]
0 \\[2pt]
\vdots \\[2pt]
0 \\[2pt]
\end{array}  
\right)
\quad \textrm{and} \quad
\textbf{U} =     
\left(
\begin{array}{c}
\textbf{I} \\[2pt]
0 \\[2pt]
0 \\[2pt]
\vdots \\[2pt]
0 \\[2pt]
\end{array}  
\right)^{\!\!\textrm{T}}
\end{equation}

\noindent
Here block-matrices $\textbf{S}_{nm}$ and $\textbf{V}_0$ are defined by the same mathematical expressions as in Appendices~\ref{sec:expansion-S00}-\ref{sec:expansion-boundary}.

By performing the following change in the matrices' notation: $\textbf{S}_{00} \rightarrow \textbf{S}_\textrm{B}$, $\textbf{S}_{01} \rightarrow \textbf{S}_\textrm{in}$, $\textbf{S}_{K0} \rightarrow \textbf{S}_\textrm{out}$, $\textbf{S}_{K1} \rightarrow \textbf{S}_\textrm{ht}$, $\textbf{S}_{12} \rightarrow \textbf{S}_{\textrm{pr},1}$, ..., $\textbf{S}_{K-1,K} \rightarrow \textbf{S}_{\textrm{pr},K-1}$ and $\textbf{V}_0 \rightarrow \textbf{V}_\textrm{B}$, the above formulas for the DNA transfer-matrix, $\textbf{L}$, and the boundary condition matrix, $\textbf{Y}$, can be re-written in a more convenient form:
\begin{equation} \label{Block-matrices-general-2}
\textbf{L} =     
\left( 
\begin{array}{cccccc}
	\textbf{S}_\textrm{B} & \textbf{S}_\textrm{in} & 0 & 0 & \cdots & 0 \\[2pt]	
	0 & 0 & \textbf{S}_\textrm{pr,1} & 0 & \cdots & 0 \\[2pt]
	0 & 0 & 0 & \textbf{S}_\textrm{pr,2} & \cdots & 0 \\[-4pt]
	\vdots & \vdots & \vdots & \vdots & \ddots & \vdots \\[2pt]
	0 & 0 & 0 & 0 & \cdots & \textbf{S}_\textrm{pr,K-1} \\[2pt]
	\textbf{S}_\textrm{out} & \textbf{S}_\textrm{ht} & 0 & 0 & \cdots & 0
\end{array}  
\!\right)
\quad \textrm{and} \quad
\textbf{Y} =     
\left(
\begin{array}{c}
\textbf{V}_\textrm{B} \\[2pt]
0 \\[2pt]
0 \\[2pt]
\vdots \\[2pt]
0 \\[2pt]
\end{array}  
\right)
\end{equation}

\noindent
Where matrix $\textbf{S}_\textrm{B}$ describes the contribution of bare DNA segments being in B-DNA state to the resulting transfer-matrix, $\textbf{L}$; whereas, matrices $\textbf{S}_\textrm{in}$ and $\textbf{S}_\textrm{out}$ correspond to the DNA segments sitting next to the entry and exit points of nucleoprotein complexes. As for matrices $\textbf{S}_\textrm{pr,j}$, they depict contributions of the DNA segments residing inside the nucleoprotein complexes. Finally, matrix $\textbf{S}_\textrm{ht}$ represents DNA segments located at the interface between nucleoprotein complexes that occupy neighbouring DNA sites in a head-to-tail configuration. 

Eq.~\eqref{Block-matrices-general-2} can be further generalized by taking into account that protein-unbound DNA segments can transit between a number of alternative structural states such as B-, L- and P-DNA. In this case, by including all of the energy terms from Eq.~\eqref{bare-DNA-sum-intro} into the DNA partition function calculations, it can be shown that the DNA transfer-matrix, $\textbf{L}$, and boundary condition matrices, $\textbf{Y}$ and $\textbf{U}$, assume the followin forms (for additional details see ref.~\cite{Efremov_2016}):
\begin{equation} \label{Block-matrices-general-3}
\textbf{L} =     
\left(
\begin{array}{cccccccc}
	\textbf{S}_\textrm{P} & \textbf{S}_\textrm{P} e^{-J} & \textbf{S}_\textrm{P} e^{-J} & 0	& 0 & 0 & \cdots & 0 \\[2pt]
	\textbf{S}_\textrm{L} e^{-J} & \textbf{S}_\textrm{L} & \textbf{S}_\textrm{L} e^{-J} & 0	& 0 & 0 & \cdots & 0 \\[2pt]
	\textbf{S}_\textrm{B} e^{-J} & \textbf{S}_\textrm{B} e^{-J} & \textbf{S}_\textrm{B} & \textbf{S}_\textrm{in} & 0 
	& 0 & \cdots & 0 \\[2pt]	
	0 & 0 & 0 & 0 & \textbf{S}_{\textrm{pr},1} & 0 & \cdots & 0 \\[2pt]
	0 & 0 & 0 & 0 & 0 & \textbf{S}_{\textrm{pr},2} & \cdots & 0 \\[-4pt]
	\vdots & \vdots & \vdots & \vdots & \vdots & \vdots & \ddots & \vdots \\[2pt]
	0 & 0 & 0 & 0 & 0 & 0 & \cdots & \textbf{S}_{\textrm{pr},K-1} \\[2pt]
	0 & 0 & \textbf{S}_\textrm{out} & \textbf{S}_\textrm{ht} & 0 & 0 & \cdots & 0
\end{array}  
\!\right)
\textrm{,} \quad
\textbf{Y} =     
\left( 
\begin{array}{c}
\textbf{V}_\textrm{P} \\[2pt]
\textbf{V}_\textrm{L} \\[2pt]
\textbf{V}_\textrm{B} \\[2pt]
0 \\[2pt]
0 \\[2pt]
\vdots \\[2pt]
0 \\[2pt]
\end{array}  
\right)
\quad \textrm{and} \quad
\textbf{U} =     
\left( 
\begin{array}{c}
\textbf{I} \\[2pt]
\textbf{I} \\[2pt]
\textbf{I} \\[2pt]
0 \\[2pt]
0 \\[2pt]
\vdots \\[2pt]
0 \\[2pt]
\end{array}  
\right)^{\!\!\textrm{T}}
\end{equation}

\noindent
In the above formula, $J$ is the domain wall penalty that accounts for the cooperativity of the DNA structural transitions, describing the molecule preference for structural uniformity \cite{Sarkar_2001, Oberstrass_2012}. The elements of matrices $\textbf{S}_\textrm{B}$, $\textbf{S}_\textrm{L}$ and $\textbf{S}_\textrm{P}$ corresponding to bare DNA segments being in B-, L- and P-DNA states, respectively, are described by Eq.~\eqref{S-block-matrix} shown below, which is very similar to Eq.~(10) and (15) derived in ref.~\cite{Efremov_2016}; likewise, matrices $\textbf{V}_\textrm{B}$, $\textbf{V}_\textrm{L}$ and $\textbf{V}_\textrm{P}$ representing the boundary conditions for different structural forms of the DNA end segments are described by Eq.~\eqref{V-matrices}, which is basically a combination of Eq.~(11), (12) and (17) from ref.~\cite{Efremov_2016}. 

While Eq.~\eqref{Block-matrices-general-3} can be used in the general case to estimate the partition function of DNA interacting with proteins, it should be noted that in the case of large nucleoprotein complexes (like nucleosomes, which bind to $\sim 147$ bp of DNA) matrix $\textbf{L}$ will be of a very big size. This may result in considerable slowdown of the DNA partition function calculations as the computational complexity of the matrices product increases as $S^n$, where $S$ is the matrix size and $n$ typically has a value in the range of $2 < n < 3$. However, this problem can be easily circumvented by reducing the size of the DNA transfer-matrix by several times via slight coarse-graining of the polygonal chain representing DNA. 

Namely, to accurately describe local DNA deformations, in all of the formulas for the elements of matrices $\textbf{S}_\textrm{B}$, $\textbf{S}_\textrm{L}$ and $\textbf{S}_\textrm{P}$, we need to select the DNA segment size, $b_n$, to be much smaller than the DNA bending and twisting persistence lengths, $A_n$ and $C_n$, in the respective DNA state (i.e., $b_n \ll A_n$ and $C_n$, where $n = -2$, $-1$ and $0$). As a result, in all our calculations the DNA segment size was set equal to $q = 1.5$ bp. Hence, the length of bare B-DNA segments was $b_0 = 0.5$ nm; whereas, the lengths of L- and P-DNA segments were $b_{-1} = 0.675$ nm and $b_{-2} = 0.85$ nm, respectively. 

Yet to accurately predict the DNA transitions between different states, it is not necessary to model DNA at such a high detalization level. Instead, it is possible to use more coarse-grained representation of DNA by dividing all of the DNA segments in multiples of some number $Q$, assigning to each of the resulting groups a new matrix, $\widetilde{\textbf{S}}_u$, that equals to the product of matrices corresponding to the DNA segments from this group. For example, for $Q$ consecutive B-DNA segments we assign a new matrix $\widetilde{\textbf{S}}_\textrm{B} = \textbf{S}^Q_\textrm{B}$. Likewise, for the first $Q$ DNA segments in a nucleoprotein complex we assign matrix $\widetilde{\textbf{S}}_{\textrm{pr},1}$ that equals to $\prod_{j=1}^Q \textbf{S}_{\textrm{pr},j}$, etc. As a result, we obtain a new DNA transfer-matrix, whose number of rows and columns are both decreased by $\sim Q$ times:
\begin{equation} \label{Block-matrices-general-4}
\textbf{L} =     
\left(
\begin{array}{cccccccc}
	\widetilde{\textbf{S}}_\textrm{P} & \widetilde{\textbf{S}}_\textrm{P} e^{-J} & 
	\widetilde{\textbf{S}}_\textrm{P} e^{-J} & \widetilde{\textbf{S}}_\textrm{in,P} e^{-J} & 0 & 0 & \cdots & 0 \\[2pt]
	\widetilde{\textbf{S}}_\textrm{L} e^{-J} & \widetilde{\textbf{S}}_\textrm{L} & 
	\widetilde{\textbf{S}}_\textrm{L} e^{-J} & \widetilde{\textbf{S}}_\textrm{in,L} e^{-J} & 0 & 0 & \cdots & 0 \\[2pt]
	\widetilde{\textbf{S}}_\textrm{B} e^{-J} & \widetilde{\textbf{S}}_\textrm{B} e^{-J} & 
	\widetilde{\textbf{S}}_\textrm{B} & \widetilde{\textbf{S}}_\textrm{in,B} & 0 & 0 & \cdots & 0 \\[2pt]	
	0 & 0 & 0 & 0 & \widetilde{\textbf{S}}_{\textrm{pr},1} & 0 & \cdots & 0 \\[2pt]
	0 & 0 & 0 & 0 & 0 & \widetilde{\textbf{S}}_{\textrm{pr},2} & \cdots & 0 \\[-4pt]
	\vdots & \vdots & \vdots & \vdots & \vdots & \vdots & \ddots & \vdots \\[2pt]
	0 & 0 & 0 & 0 & 0 & 0 & \cdots & \widetilde{\textbf{S}}_{\textrm{pr},K/Q-1} \\[2pt]
	\widetilde{\textbf{S}}_\textrm{out} e^{-J} & \widetilde{\textbf{S}}_\textrm{out} e^{-J} & 
	\widetilde{\textbf{S}}_\textrm{out} & \widetilde{\textbf{S}}_\textrm{ht} & 0 & 0 & \cdots & 0
\end{array}  
\!\right)
\end{equation}

\noindent
Here it is assumed that the protein binding size, $K$, is a multiple of $Q$. Matrices $\widetilde{\textbf{S}}_u$ in the above equation are defined by the following formulas: 1) $\widetilde{\textbf{S}}_u = \textbf{S}^Q_u$ and $\widetilde{\textbf{S}}_\textrm{in,u} = \textbf{S}^{Q-1}_u \textbf{S}_\textrm{in}$ for $u$ = B, L or P; 2) $\widetilde{\textbf{S}}_{\textrm{pr},n} = \prod_{j=(n-1)Q+1}^{nQ} \textbf{S}_{\textrm{pr},j}$ for $n = 1,...,\frac{K}{Q}\!-\!1$; 3) $\widetilde{\textbf{S}}_\textrm{out} = \prod_{j=K-Q+1}^{K-1} \textbf{S}_{\textrm{pr},j} \times \textbf{S}_\textrm{out}$; and 4) $\widetilde{\textbf{S}}_\textrm{ht} = \prod_{j=K-Q+1}^{K-1} \textbf{S}_{\textrm{pr},j} \times \textbf{S}_\textrm{ht}$.

As for the boundary condition matrices, $\textbf{Y}$ and $\textbf{U}$, they still have the same forms as in Eq.~\eqref{Block-matrices-general-3} with the number of zero rows being decreased by the respective amount of times after the coarse-graining procedure. Combining the above results together, it can be shown that the number of matrix blocks comprising the matrix $\textbf{L}$ reduces to $(\frac{K}{Q}+3) \times (\frac{K}{Q}+3)$; and the sizes of matrices $\textbf{Y}$ and $\textbf{U}$ become $(\frac{K}{Q}+3) \times 1$ and $1 \times (\frac{K}{Q}+3)$ blocks, respectively, where each block is a square $( s_\textrm{max} \!+\! 1 )^2 \times ( s_\textrm{max} \!+\! 1 )^2$ matrix with $s_\textrm{max}$ being the index of the highest DNA bending / twisting harmonic considered in the DNA partition function calculations, see comments after Eq.~\eqref{2D-matrix-forms}.

Using the coarse-grained DNA transfer-matrix, $\textbf{L}$, and boundary condition matrices, $\textbf{Y}$ and $\textbf{U}$, the DNA partition function, $Z_{f,\tau}$, can be found as:
\begin{equation} \label{DNA-protein-partition-function-11}
Z_{f,\tau} = \textrm{Tr} \!\left[ \textbf{U} \textbf{L}^{ \left( N\!-\!1 \right) / Q } \textbf{Y} \right]
\end{equation}

\noindent
Here it is assumed that $N\!-\!1$ is a multiple of $Q$, where $N$ is the total number of DNA segments in the polygonal chain representing DNA.

For the purpose of demonstration, in the next three sections we describe the DNA transfer-matrices for the three major types of DNA-binding proteins studied in this work, taking into account that in all of the calculations reported in the main text the value of $Q$ was set to $Q  = 12$ DNA segments.

\subsection{Transfer-matrix of DNA interacting with DNA-stiffening proteins}\label{sec:DNA-stiffening}

First, we start with the simplest type of DNA-binding proteins that form straight nucleoprotein filaments along DNA, as in this case matrices $\textbf{A}_\textrm{in}$ and $\textbf{A}_\textrm{out}$ denoting the equilibrium orientations of DNA segments at the entry and exit points of nucleoprotein complexes as well as matrices $\textbf{A}_j$ depicting the relative orientations of neighbouring DNA segments inside nucleoprotein complexes are all equal to the unit Euler matrix, $\textbf{I}$. 

By using the coarse-graining approach described above, it is not hard to show that for the case of DNA-stiffening proteins with the binding site size of $K = 12$ DNA segments discussed in the main text, the DNA transfer-matrix, $\textbf{L}$, and boundary condition matrices, $\textbf{Y}$ and $\textbf{U}$, take the following forms:
\begin{equation} \label{Block-matrices-DNA-stiffening}
\textbf{L} =     
\left(
\begin{array}{cccc}
	\textbf{S}^{12}_\textrm{P} & \textbf{S}^{12}_\textrm{P} e^{-J} & 
	\textbf{S}^{12}_\textrm{P} e^{-J} & \textbf{S}^{11}_\textrm{P} \textbf{S}_\textrm{in} e^{-J} \\[2pt]
	\textbf{S}^{12}_\textrm{L} e^{-J} & \textbf{S}^{12}_\textrm{L} & 
	\textbf{S}^{12}_\textrm{L} e^{-J} & \textbf{S}^{11}_\textrm{L} \textbf{S}_\textrm{in} e^{-J} \\[2pt]
	\textbf{S}^{12}_\textrm{B} e^{-J} & \textbf{S}^{12}_\textrm{B} e^{-J} & 
	\textbf{S}^{12}_\textrm{B} & \textbf{S}^{11}_\textrm{B} \textbf{S}_\textrm{in} \\[2pt]	
	\textbf{S}^{11}_\textrm{pr} \textbf{S}_\textrm{out} e^{-J} & 
	\textbf{S}^{11}_\textrm{pr} \textbf{S}_\textrm{out} e^{-J} & 
	\textbf{S}^{11}_\textrm{pr} \textbf{S}_\textrm{out} & \textbf{S}^{11}_\textrm{pr} \textbf{S}_\textrm{ht}
\end{array}  
\!\right)
\textrm{,} \quad
\textbf{Y} =     
\left( 
\begin{array}{c}
\textbf{V}_\textrm{P} \\[2pt]
\textbf{V}_\textrm{L} \\[2pt]
\textbf{V}_\textrm{B} \\[2pt]
0
\end{array}  
\right)
\quad \textrm{and} \quad
\textbf{U} =     
\left( 
\begin{array}{c}
\textbf{I} \\[2pt]
\textbf{I} \\[2pt]
\textbf{I} \\[2pt]
0
\end{array}  
\right)^{\!\!\textrm{T}}
\end{equation}

\noindent
Where matrices $\textbf{S}_\textrm{in}$, $\textbf{S}_\textrm{out}$, $\textbf{S}_\textrm{ht}$ and $\textbf{S}_\textrm{pr}$ are defined by Eq.~\eqref{S01-matrix-elements-2}, \eqref{S30-matrix-elements-2}, \eqref{S31-matrix-elements-2} and \eqref{S12-matrix-elements-2}, respectively. As for the remaining matrices, $\textbf{S}_u$ and $\textbf{V}_u$, where $u = $ P, L or B, their elements can be found using Eq.~(10) and (15), (11), (12) and (17) from ref.~\cite{Efremov_2016}:
\begin{multline} \label{S-block-matrix}
\left( \textbf{S}_u \right)_{vv'} =
	\delta_{q0} \delta_{q'0} \times 
	\pi^2 \sqrt{ \left( 2s \!+\! 1 \right) \left( 2s' \!\!+\! 1 \right) } \, 
	e^{ - a_u - c_u - q [ \mu_u - 2\pi\tau \Delta lk_0^{(u)} ] } 
	\!\sum_{t,k,k'\!,r} \!\!
	\left( 2t \!+\! 1 \right) \!\left( 2k \!+\! 1 \right) \!\left( 2k' \!\!+\! 1 \right)
	i^{-r} e^{ i r \omega_u } \times \\[-2pt]
	\times \, I_r \!\left( c_u \sqrt{ 1 \!+\! \chi_u^2 } \right) 
	I_r \!\left( \tau \left[ 1 \!-\! \textstyle{ \frac{ \lambda_u }{ 2\pi } } \right] \right) 
	\!\times 
	\mathscr{L}^t_r \!\left( -a_u \right) \mathscr{L}^k_r \!\left( - b_u f \right)
	\mathscr{L}^{k'}_r \!\!\left( 0 \right) \times
	\left(\!\begin{array}{ccc}
		t & k & \,s \\
	   -r & r & \,0
	\end{array}\right)^{\!2}
	\left(\!\begin{array}{ccc}
		t & k' & \,s' \\
	   -r & r & \,0
	\end{array}\right)^{\!2}
\end{multline}

and 
\begin{equation} \label{V-matrices}
\begin{cases}
\left( \textbf{V}_u \right)_{vv'} = 
	\delta_{q0} \delta_{q'0} \delta_{s'0} \times 8\pi^2 
	\sqrt{ 2s \!+\! 1 } \; i_s \!\left( b_u f \right) e^{ - q [ \mu_u - 2\pi\tau \Delta lk_0^{(u)} ] }
	& \textrm{-- the DNA ends-free orientation boundary condition} \\[5pt]
	\left( \textbf{V}_u \right)_{vv'} = 
	\delta_{q0} \delta_{q'0} \times \frac{1}{4\pi} \sqrt{ \left( 2s \!+\! 1 \right)
	\!\left( 2s' \!\!+\! 1 \right) } \; e^{ b_u f - q [ \mu_u - 2\pi\tau \Delta lk_0^{(u)} ] }
	& \textrm{-- the DNA ends } \textbf{z}_0 \textrm{-axis collinear boundary condition}
\end{cases}
\end{equation} 

\noindent
Here $v = q + s ( s + 1 )$ and  $v' = q' + s' ( s' + 1 )$; $a_\textrm{P} = a_{-2}$, $a_\textrm{L} = a_{-1}$ and $a_\textrm{B} = a_0$; $b_\textrm{P} = b_{-2}$, $b_\textrm{L} = b_{-1}$ and $b_\textrm{B} = b_0$, etc., where parameters $a_u$, $b_u$, $c_u$, $\lambda_u$, $\chi_u$, $\omega_u$, $\mu_u$ and $\Delta lk_0^{(u)}$ take the values corresponding to the respective DNA state $u = $ P, L or B.

\subsection{Transfer-matrix of DNA interacting with DNA-bending proteins}\label{sec:DNA-bending}

In the case of the DNA-bending protein shown on Figure~\ref{fig1}(d) that has a binding site of $K = 24$ DNA segments, the relative orientations of DNA segments inside the corresponding nucleoprotein complexes are described by two rotation matrices: the unit matrix $\textbf{A}_1 = \textbf{I}$ that represents the relative orientations of the DNA segments closer to the entry and exit points of the complex, and by the matrix $\textbf{A}_2 = \textbf{A}_2 ( \pi, 0.2, \pi)$ that corresponds to the DNA segments residing in the middle part of the complex. As a result, the coarse-grained transfer-matrix of DNA interacting with such DNA-bending proteins takes the following form: 
\begin{equation} \label{Block-matrices-DNA-bending}
\textbf{L} =     
\left(
\begin{array}{ccccc}
	\textbf{S}^{12}_\textrm{P} & \textbf{S}^{12}_\textrm{P} e^{-J} & 
	\textbf{S}^{12}_\textrm{P} e^{-J} & \textbf{S}^{11}_\textrm{P} \textbf{S}_\textrm{in} e^{-J} & 0 \\[2pt]
	\textbf{S}^{12}_\textrm{L} e^{-J} & \textbf{S}^{12}_\textrm{L} & 
	\textbf{S}^{12}_\textrm{L} e^{-J} & \textbf{S}^{11}_\textrm{L} \textbf{S}_\textrm{in} e^{-J} & 0 \\[2pt]
	\textbf{S}^{12}_\textrm{B} e^{-J} & \textbf{S}^{12}_\textrm{B} e^{-J} & 
	\textbf{S}^{12}_\textrm{B} & \textbf{S}^{11}_\textrm{B} \textbf{S}_\textrm{in} & 0 \\[2pt]	
	0 & 0 & 0 & 0 & \textbf{S}^{5}_\textrm{pr,1} \textbf{S}^{7}_\textrm{pr,2} \\[2pt]
	\textbf{S}^{6}_\textrm{pr,2} \textbf{S}^{5}_\textrm{pr,1} \textbf{S}_\textrm{out} e^{-J} & 
	\textbf{S}^{6}_\textrm{pr,2} \textbf{S}^{5}_\textrm{pr,1} \textbf{S}_\textrm{out} e^{-J} & 
	\textbf{S}^{6}_\textrm{pr,2} \textbf{S}^{5}_\textrm{pr,1} \textbf{S}_\textrm{out} & 0 & 0
\end{array}  
\!\right)
\end{equation}

\noindent
Here the elements of $\textbf{S}_\textrm{pr,1}$ and $\textbf{S}_\textrm{pr,2}$ blocks are defined by Eq.~\eqref{S12-matrix-elements-2} and \eqref{S12-matrix-elements-1}, respectively, where in Eq.~\eqref{S12-matrix-elements-1} matrix $\textbf{A}_1$ is replaced by the matrix $\textbf{A}_2$. As for the rest of the matrix blocks, they have absolutely the same forms as in the case of the DNA-stiffening protein described in the previous section. 

It should be noted that in the above formula it is assumed that $\textbf{S}_\textrm{ht} = 0$ as the existing crystallographic data seem to indicate that IHF proteins do not bind to neighbouring DNA sites and thus do not form extended nucleoprotein filaments in a head-to-tail configuration \cite{Rice_1996}. Indeed, by setting $\textbf{S}_\textrm{ht} = 0$, it is possible to reproduce such a volume exclusion effect, which is observed for IHF-DNA complexes, and from the panels shown in Figure~\ref{figS5}(b) it can be seen that in this case the DNA occupancy fraction by nucleoprotein complexes never goes above $\sim 60\%$.

As for the boundary condition matrices, $\textbf{Y}$ and $\textbf{U}$, their mathematical forms are very similar to those in Eq.~\eqref{Block-matrices-DNA-stiffening}, with the only difference being that the total number of zero matrix blocks in both $\textbf{Y}$ and $\textbf{U}$ matrices is equal to $2$ instead of $1$: 
\begin{equation} \label{Block-matrices-DNA-bending-2}
\textbf{Y} =     
\left( 
\begin{array}{c}
\textbf{V}_\textrm{P} \\[2pt]
\textbf{V}_\textrm{L} \\[2pt]
\textbf{V}_\textrm{B} \\[2pt]
0 \\[2pt]
0
\end{array}  
\right)
\quad \textrm{and} \quad
\textbf{U} =     
\left( 
\begin{array}{c}
\textbf{I} \\[2pt]
\textbf{I} \\[2pt]
\textbf{I} \\[2pt]
0 \\[2pt]
0
\end{array}  
\right)^{\!\!\textrm{T}}
\end{equation}

\noindent
Where the elements of block-matrices $\textbf{V}_u$, $u =$ P, L or B, are defined by Eq.~\eqref{V-matrices}.

\subsection{Transfer-matrix of DNA interacting with DNA-wrapping proteins}\label{sec:DNA-wrapping}

The last group of architectural proteins considered in this study are DNA-wrapping proteins that include histone tetramers and octamers that upon binding to DNA form tetrasome and nucleosome complexes, respectively. 

Recalling that all of the DNA segments residing inside this type of nucleoprotein complexes are represented by straight intervals aligned along the line connecting the entry and exit points of the DNA, it can be seen that the block-matrices $\textbf{S}_{\textrm{pr},j}$ ($j = 1,...,K\!-\!1$) from Eq.~\eqref{Block-matrices-general-3} corresponding to such DNA segments are defined by Eq.~\eqref{S12-matrix-elements-3}. As a result, after the coarse-graining procedure, we get the following transfer-matrix for DNA interacting with histone octamers, which have the binding site size of $K = 96$ DNA segments:
\begin{equation} \label{Block-matrices-nucleosomes}
\textbf{L} =     
\left(
\begin{array}{cccccccc}
	\textbf{S}^{12}_\textrm{P} & \textbf{S}^{12}_\textrm{P} e^{-J} & \textbf{S}^{12}_\textrm{P} e^{-J} &
	\textbf{S}^{11}_\textrm{P} \textbf{S}_\textrm{in} e^{-J} & 0 & 0 & \cdots & 0 \\[2pt]
	\textbf{S}^{12}_\textrm{L} e^{-J} & \textbf{S}^{12}_\textrm{L} & \textbf{S}^{12}_\textrm{L} e^{-J} &
	\textbf{S}^{11}_\textrm{L} \textbf{S}_\textrm{in} e^{-J} & 0 & 0 & \cdots & 0 \\[2pt]
	\textbf{S}^{12}_\textrm{B} e^{-J} & \textbf{S}^{12}_\textrm{B} e^{-J} & \textbf{S}^{12}_\textrm{B} &
	\textbf{S}^{11}_\textrm{B} \textbf{S}_\textrm{in} & 0 & 0 & \cdots & 0 \\[2pt]	
	0 & 0 & 0 & 0 &\textbf{S}^{12}_\textrm{pr} & 0 & \cdots & 0 \\[2pt]
	0 & 0 & 0 & 0 & 0 & \textbf{S}^{12}_\textrm{pr} & \cdots & 0 \\[-4pt]
	\vdots & \vdots & \vdots & \vdots & \vdots & \vdots & \ddots & \vdots \\[2pt]
	0 & 0 & 0 & 0 & 0 & 0 & \cdots & \textbf{S}^{12}_\textrm{pr} \\[2pt]
	\textbf{S}^{11}_\textrm{pr} \textbf{S}_\textrm{out} e^{-J} & 
	\textbf{S}^{11}_\textrm{pr} \textbf{S}_\textrm{out} e^{-J} & 
	\textbf{S}^{11}_\textrm{pr} \textbf{S}_\textrm{out} & 0 & 0 & 0 & \cdots & 0
\end{array}  
\!\right)
\end{equation}

\noindent
In the above formula, the matrix $\textbf{S}_\textrm{pr}$ is defined by Eq.~\eqref{S12-matrix-elements-3}; and matrices $\textbf{S}_\textrm{in}$ and $\textbf{S}_\textrm{out}$ are described by Eq.~\eqref{S01-matrix-elements-4} and \eqref{S30-matrix-elements-3}, respectively, where $\textbf{A}_\textrm{in} = \textbf{A}_\textrm{in} (0, 2.12, -0.79)$ and $\textbf{A}_\textrm{out} = \textbf{A}_\textrm{out} (-0.79, 2.12, 0)$, see Table~\ref{tab:Protein-parameters}. As for the remaining blocks, $\textbf{S}_u$, where $u =$ P, L or B, they have absolutely the same form as in the case of DNA-stiffening proteins, see Eq.~\eqref{S-block-matrix}. The total size of the DNA transfer-matrix, $\textbf{L}$, in Eq.~\eqref{Block-matrices-nucleosomes} equals to $11 \times 11$ block-matrices.

As for the boundary condition matrices, $\textbf{Y}$ and $\textbf{U}$, they are described by formulas similar to Eq.~\eqref{Block-matrices-DNA-bending-2}, where the total number of zero blocks is increased to $8$ for both matrices, $\textbf{Y}$ and $\textbf{U}$.

In contrast to nucleosomes, which assume only left-handed helicity, existing experimental data show that tetrasomes can transit between the left- and right-handed nucleoprotein complex conformations, see Section~\ref{sec:results-DNA-wrap}. To accurately reflect this experimental fact in the transfer-matrix calculations, it is thus necessary to include both left- and right-handed tetrasome complexes into the DNA transfer-matrix, with each tetrasome structure being described by its own entry and exit block-matrices, $\textbf{S}_\textrm{in,L}$ and $\textbf{S}_\textrm{out,L}$, and $\textbf{S}_\textrm{in,R}$ and $\textbf{S}_\textrm{out,R}$, respectively. As a result, it can be shown that in the case of DNA interaction with histone tetramers, the DNA transfer-matrix has the following form:
\begin{equation} \label{Block-matrices-tetrasomes}
\textbf{L} =     
\left(
\begin{array}{ccccccccccc}
	\textbf{S}^{12}_\textrm{P} & \textbf{S}^{12}_\textrm{P} e^{-J} & 
	\textbf{S}^{12}_\textrm{P} e^{-J} & \textbf{S}^{11}_\textrm{P} \textbf{S}_\textrm{in,L} e^{-J} & 0 & 0 & 0 & 
	\textbf{S}^{11}_\textrm{P} \textbf{S}_\textrm{in,R} e^{-J} & 0 & 0 & 0 \\[2pt]
	\textbf{S}^{12}_\textrm{L} e^{-J} & \textbf{S}^{12}_\textrm{L} & 
	\textbf{S}^{12}_\textrm{L} e^{-J} & \textbf{S}^{11}_\textrm{L} \textbf{S}_\textrm{in,L} e^{-J} & 0 & 0 & 0 &
	\textbf{S}^{11}_\textrm{L} \textbf{S}_\textrm{in,R} e^{-J} & 0 & 0 & 0 \\[2pt]
	\textbf{S}^{12}_\textrm{B} e^{-J} & \textbf{S}^{12}_\textrm{B} e^{-J} & 
	\textbf{S}^{12}_\textrm{B} & \textbf{S}^{11}_\textrm{B} \textbf{S}_\textrm{in,L} & 0 & 0 & 0 & 
	\textbf{S}^{11}_\textrm{B} \textbf{S}_\textrm{in,R} & 0 & 0 & 0 \\[2pt]	
	0 & 0 & 0 & 0 & \textbf{S}^{12}_\textrm{pr} & 0 & 0 & 0 & 0 & 0 & 0 \\[2pt]
	0 & 0 & 0 & 0 & 0 & \textbf{S}^{12}_\textrm{pr} & 0 & 0 & 0 & 0 & 0 \\[2pt]
	0 & 0 & 0 & 0 & 0 & 0 & \textbf{S}^{12}_\textrm{pr} & 0 & 0 & 0 & 0 \\[2pt]
	\textbf{S}^{11}_\textrm{pr} \textbf{S}_\textrm{out,L} e^{-J} & 
	\textbf{S}^{11}_\textrm{pr} \textbf{S}_\textrm{out,L} e^{-J} & 
	\textbf{S}^{11}_\textrm{pr} \textbf{S}_\textrm{out,L} & 0 & 0 & 0 & 0 & 0 & 0 & 0 & 0 \\[2pt]
	0 & 0 & 0 & 0 & 0 & 0 & 0 & 0 & \textbf{S}^{12}_\textrm{pr} & 0 & 0 \\[2pt]
	0 & 0 & 0 & 0 & 0 & 0 & 0 & 0 & 0 & \textbf{S}^{12}_\textrm{pr} & 0 \\[2pt]
	0 & 0 & 0 & 0 & 0 & 0 & 0 & 0 & 0 & 0 & \textbf{S}^{12}_\textrm{pr} \\[2pt]
	\textbf{S}^{11}_\textrm{pr} \textbf{S}_\textrm{out,R} e^{-J} & 
	\textbf{S}^{11}_\textrm{pr} \textbf{S}_\textrm{out,R} e^{-J} & 
	\textbf{S}^{11}_\textrm{pr} \textbf{S}_\textrm{out,R} & 0 & 0 & 0 & 0 & 0 & 0 & 0 & 0
\end{array}  
\!\right)
\end{equation}

\noindent
Here the matrix block $\textbf{S}_\textrm{pr}$ is defined by Eq.~\eqref{S12-matrix-elements-3}; matrices $\textbf{S}_\textrm{in,L}$ and $\textbf{S}_\textrm{in,R}$ are described by Eq.~\eqref{S01-matrix-elements-4} with matrices $\textbf{A}_\textrm{in,L}$ and $\textbf{A}_\textrm{in,R}$ being equal to $\textbf{A}_\textrm{in,L} = \textbf{A}_\textrm{in,L} (0, 2.26, -1.11)$ and $\textbf{A}_\textrm{in,R} = \textbf{A}_\textrm{in,R} (0, 2.26, 1.11)$, respectively (see Table~\ref{tab:Protein-parameters}); and matrices $\textbf{S}_\textrm{out,L}$ and $\textbf{S}_\textrm{out,R}$ are determined by Eq.~\eqref{S30-matrix-elements-3}, where  $\textbf{A}_\textrm{out,L} = \textbf{A}_\textrm{out,L} (-1.11, 2.26, 0)$ and $\textbf{A}_\textrm{out,R} = \textbf{A}_\textrm{out,R} (1.11, 2.26, 0)$, accordingly. Finally, the remaining blocks, $\textbf{S}_u$, where $u =$ P, L or B, have the same forms as in the case of DNA-stiffening proteins, see Eq.~\eqref{S-block-matrix}.

Lastly, in the case of tetrasomes, the boundary condition matrices $\textbf{Y}$ and $\textbf{U}$ are described by the same Eq.~\eqref{Block-matrices-DNA-bending-2}, where the total number of zero blocks is increased to $8$ for both matrices, $\textbf{Y}$ and $\textbf{U}$.

\section{General algorithm for the DNA transfer-matrix calculations}
\label{Appendix-F}

Using the formulas derived in the above appendix sections, it is not very hard now to build a general algorithm for finding the partition function of DNA and estimation of the observable parameters, such as the DNA extension, superhelical density and occupancy fraction by DNA-binding proteins, that characterize the conformational state of DNA. Namely, the algorithm includes the following steps:

\begin{enumerate}[label={\arabic*)}]
\item Calculate the elements $( \textbf{S}_u )_{vv'}$ and $( \textbf{V}_u )_{vv'}$ of square matrices $\textbf{S}_u$ and $\textbf{V}_u$, where $u =$ P, L or B, by using Eq.~\eqref{S-block-matrix} and \eqref{V-matrices}. The size of these matrices, $( s_\textrm{max} \!+\! 1 )^2 \times ( s_\textrm{max} \!+\! 1 )^2$, is determined by the index $s_\textrm{max}$ of the highest DNA bending / twisting harmonic considered in the DNA partition function calculations, such that $0 \leq v = q + s ( s + 1 ) \leq ( s_\textrm{max} \!+\!1)^2 \!-\! 1$ and  $0 \leq v' = q' + s' ( s' + 1 ) \leq ( s_\textrm{max} \!+\! 1)^2 \!-\! 1$, where $-s \leq q \leq s$, $-s' \leq q' \leq s'$ and $0 \leq s,s' \leq s_\textrm{max}$.

\item Using Eq.~\eqref{S01-matrix-elements-1}-\eqref{S12-matrix-elements-3} find the rest of the matrices, $\textbf{S}_\textrm{in}$, $\textbf{S}_\textrm{out}$, $\textbf{S}_\textrm{ht}$ and $\textbf{S}_{\textrm{pr},j}$, characterizing the physical and geometric properties of nucleoprotein complexes that may form on the DNA. Here index $j = 1,...,K\!-\!1$, where $K$ is the protein binding site size on DNA (i.e., number of DNA segments bound to a single protein). All of these matrices are of $( s_\textrm{max} \!+\! 1)^2 \times ( s_\textrm{max} \!+\! 1 )^2$ size as well.

\item Following Eq.~\eqref{Block-matrices-general-3} form the DNA transfer-matrix and boundary condition matrices, $\textbf{L}$, $\textbf{Y}$ and $\textbf{U}$, required for the calculation of the DNA partition function. In the case of big nucleoprotein complexes, which bind to a large number of DNA segments, use Eq.~\eqref{Block-matrices-general-4} instead to form coarse-grained DNA transfer-matrix, reducing the number of zero matrix blocks in matrices $\textbf{Y}$ and $\textbf{U}$ accordingly. 

\item Apply Eq.~\eqref{DNA-protein-partition-function-10} [or Eq.~\eqref{DNA-protein-partition-function-11} in the case of	the coarse-grained DNA transfer-matrix] to obtain the value of the DNA partition function, $Z_{f,\tau}$, at given force ($f$) and torque ($\tau$) constraints.

\item Using the above steps 1-4, find the DNA extension ($z$), linking number change ($\Delta \textrm{Lk}$), and the total number of protein-bound ($N_\textrm{pr}$) and bare ($N_u$) DNA segments in each of the states, $u =$ L or P, by calculating the following derivatives of the DNA partition function:
\begin{align} \label{observables}
& z \!\left( f, \tau \right) = \frac{ \partial \ln Z_{f,\tau} }{ \partial f }
\quad \textrm{and} \quad
\Delta \textrm{Lk} \!\left( f, \tau \right) = \frac{1}{2\pi} \frac{ \partial \ln Z_{f,\tau} }{ \partial \tau } \bigg\rvert \!\! \vcenter{\vspace{4pt} \hbox{ $ \substack{ \tau \lambda_u = \textrm{const} \\ \!\!\! \chi_u = \textrm{const} \\ \, u = \textrm{B,L,P,pr} } $ } }
\nonumber \\[5pt]
& N_\textrm{pr} \!\left( f, \tau \right) = K \frac{ \partial \ln Z_{f,\tau} }{ \partial \mu_\textrm{pr} }
\quad \textrm{and} \quad
N_u \!\left( f, \tau \right) = -\frac{1}{q} \frac{ \partial \ln Z_{f,\tau} }{ \partial \mu_u }
\end{align}

\noindent 
Where $\mu_\textrm{L} = \mu_{-1}$ and $\mu_\textrm{P} = \mu_{-2}$ are the base-pairing energies in the respective DNA states; and $\mu_\textrm{pr}$ is the protein binding energy to DNA, see Section~\ref{sec:energy_terms}. In the above formula for the DNA linking number change, parameters $\chi_u$ ($u =$ B, L, P and pr) as well as all products $\tau \lambda_u$, which are used in computations of the elements of matrices $\textbf{S}_u$, $\textbf{S}_\textrm{in}$, $\textbf{S}_\textrm{out}$, $\textbf{S}_\textrm{ht}$ and $\textbf{S}_{\textrm{pr},j}$ [see Eq.~\eqref{S-block-matrix} and Eq.~\eqref{S01-matrix-elements-1}-\eqref{S12-matrix-elements-3}], are treated as constants during the differentiation process. 

\item By fixing the value of the applied force ($f = f_0$) or torque ($\tau = \tau_0$) in the above equations, plot the DNA force-extension and torque-extension curves [$z (f) |_{\tau = \tau_0} = z (f, \tau_0)$ and $z (\tau) |_{f = f_0} = z (f_0, \tau)$] as well as force-superhelical density and torque-superhelical density curves [$\sigma (f) |_{\tau = \tau_0} = \Delta \textrm{Lk} (f, \tau_0) / \textrm{Lk}_0$ and $\sigma (\tau) |_{f = f_0} = \Delta \textrm{Lk} (f_0, \tau) / \textrm{Lk}_0$] to obtain insights into the global DNA conformation under various mechanical constraints imposed on the DNA. Here $\textrm{Lk}_0$ is the linking number of a torsionally relaxed DNA being in B-DNA state, see Section~\ref{sec:theory_outline}.
\end{enumerate}

Finally, we would like to note that in order to obtain accurate estimation of the DNA partition function in the above algorithm it is necessary to know the exact value of the model parameter $\mu_\textrm{off}$, which in contrast to other model parameters that can be measured in experiments, has to be determined numerically. Namely, by performing several iterative DNA transfer-matrix calculations, parameter $\mu_\textrm{off}$ must be changed in such a way until the occupancy fraction, $O (f,\tau)$, of a relaxed DNA ($f = 0$ pN and $\tau = 0$ pN$\cdot$nm) obeys the classical exponential relation, $O (0,0) = e^{\mu_\textrm{pr}}$, at sufficiently large negative values of the protein binding energy to DNA ($\mu_\textrm{pr} < - 5$). The reason why $\mu_\textrm{pr}$ has to be negative in these calculations is to keep the amount of protein-bound DNA segments at a low level, since otherwise behaviour of the DNA occupancy fraction will strongly deviate from the simple exponential law described here.

In this study, parameter $\mu_\textrm{off}$ for each of the DNA-binding proteins was estimated by setting $\mu_\textrm{pr} = - \ln 10^3$ and adjusting $\mu_\textrm{off}$ until the occupancy fraction of a relaxed DNA reached the level of $O (0,0) = 0.1 \%$. The final values obtained for $\mu_\textrm{off}$ parameter for different types of DNA-binding proteins explored in this work were: 1) $\mu_\textrm{off} = -55.0$ $k_\textrm{B} T$ for the DNA-stiffening protein described in Section~\ref{sec:results-DNA-stiff}; 2) $\mu_\textrm{off} = -108.7$ $k_\textrm{B} T$ for the DNA-bending protein described in Section~\ref{sec:results-DNA-bend}; 3) $\mu_\textrm{off} = -217.0$ $k_\textrm{B} T$ for left- and right-handed histone tetramers, and $\mu_\textrm{off} = -432.0$ $k_\textrm{B} T$ for histone octamers described in Section~\ref{sec:results-DNA-wrap}.

Lastly, it should be noted that while in this study we considered the simplest scenario when upon binding to DNA proteins form completely folded nucleoprotein complexes, the constructed transfer-matrix formalism can be easily generalized to describe nucleoprotein complexes that can be partially unfolded by mechanical forces applied to the DNA. Namely, by replacing zero matrix blocks in the third row of the DNA transfer-matrix, $\textbf{L}$, defined by Eq.~\eqref{Block-matrices-general-3} with matrices similar to $\textbf{S}_\textrm{in}$, and zero matrix blocks in the third column with matrices similar to $\textbf{S}_\textrm{out}$, we will immediately get a new DNA transfer-matrix, which not only depicts completely folded, but also partially unfolded nucleoprotein complexes as well. Such approach may prove to be useful in future studies for detailed investigation of the protein-DNA binding energy landscape based on single-molecule experiments aimed at exploration of the nucleoprotein complexes' unfolding upon mechanical stretching of DNA.


%

\pagebreak

\newpage
~

\newpage
~

\newpage
~

\newpage
~

\newpage
~

\newpage
~

\newpage
~

\newpage
~

\renewcommand\thefigure{S1}
\begin{figure}[!htb]
\includegraphics[width=0.9\textwidth]{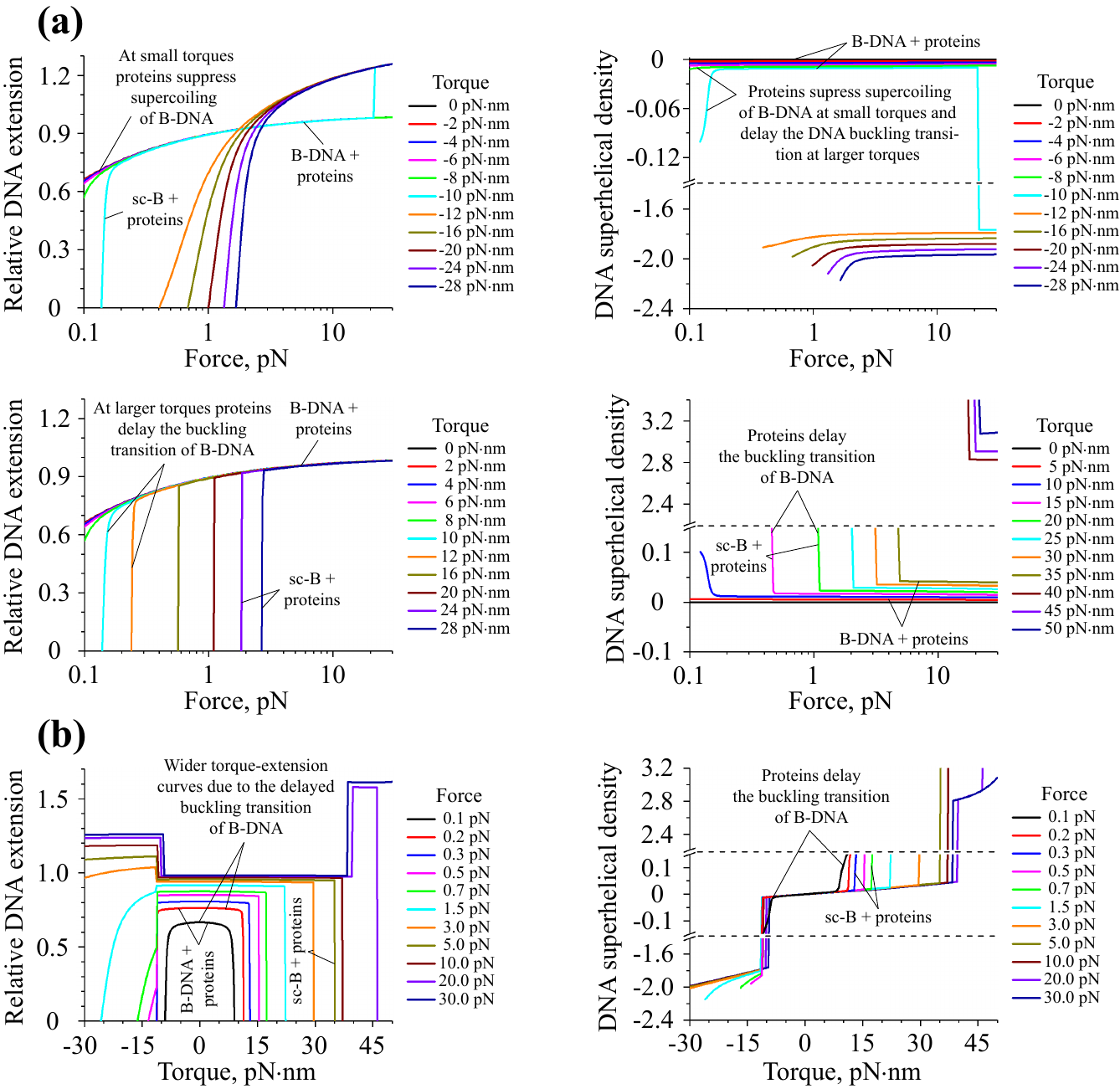}
\caption{
\textbf{Mechanical response of DNA interacting with DNA-stiffening proteins to the applied force and torque constraints, large scale view.} The panels show a wider view of Figure~\ref{fig3}, demonstrating \textbf{(a)} force-extension and force-superhelical density curves of DNA obtained at different values of the torque, $\tau$, as well as \textbf{(b)} torque-extension and torque-superhelical density curves of DNA obtained at different values of the force, $f$, in the presence of DNA-stiffening proteins in solution. In all panels, the DNA extension is normalized to the total contour length of DNA in B-form. Abbreviation sc-B is used to indicate a supercoiled state of B-DNA.
}
\label{figS1}
\end{figure}

\newpage
~

\renewcommand\thefigure{S2}
\begin{figure}[!htb]
\includegraphics[width=0.9\textwidth]{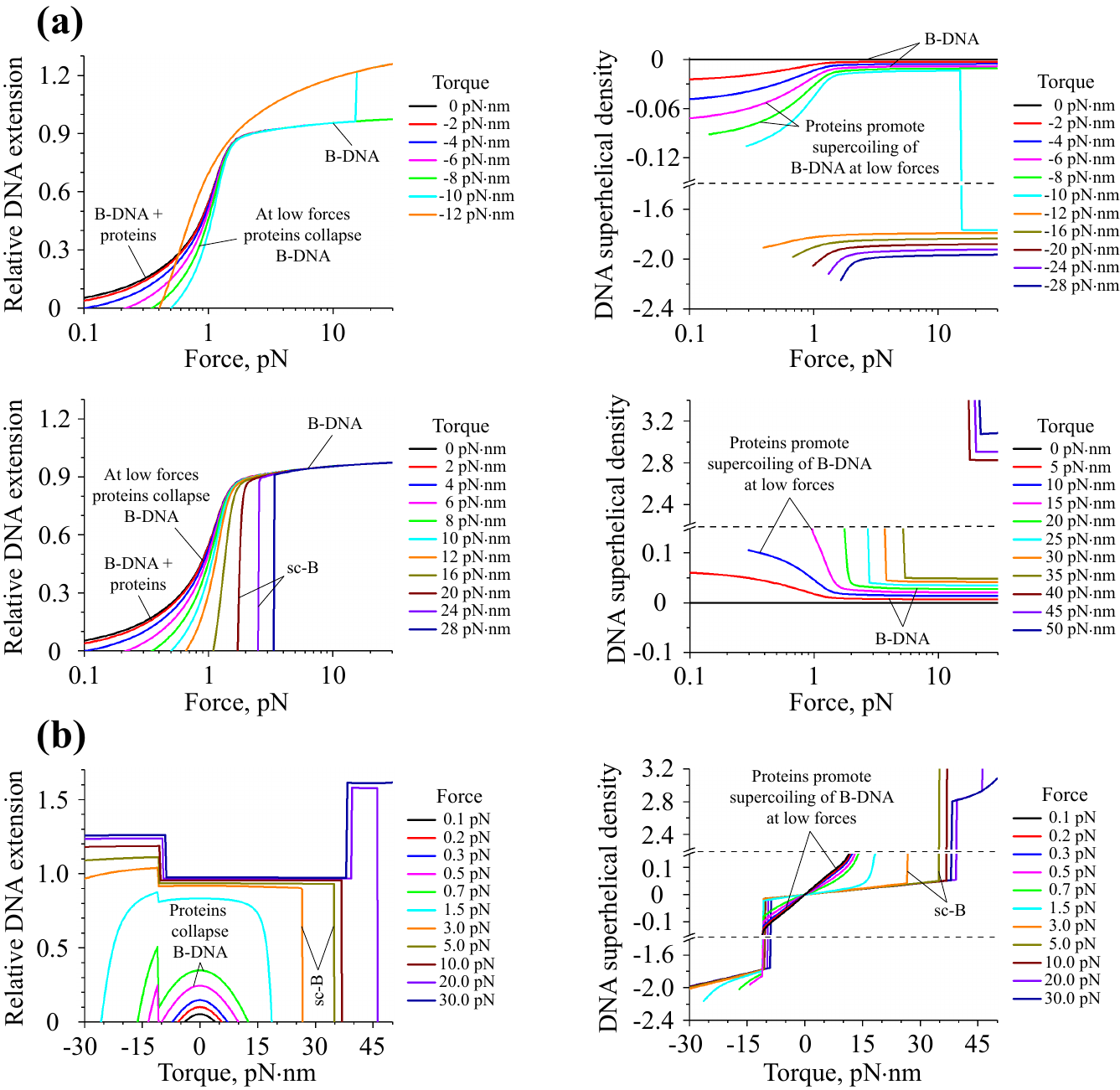}
\caption{
\textbf{Mechanical response of DNA interacting with DNA-bending proteins to the applied force and torque constraints, large scale view.} The panels show a wider view of Figure~\ref{fig4}, demonstrating \textbf{(a)} force-extension and force-superhelical density curves of DNA obtained at different values of the torque, $\tau$, as well as \textbf{(b)} torque-extension and torque-superhelical density curves of DNA obtained at different values of the force, $f$, in the presence of DNA-bending proteins in solution. In all panels, the DNA extension is normalized to the total contour length of DNA in B-form. Abbreviation sc-B is used to indicate a supercoiled state of B-DNA.
}
\label{figS2}
\end{figure}

\newpage
~

\renewcommand\thefigure{S3}
\begin{figure}[!htb]
\includegraphics[width=0.9\textwidth]{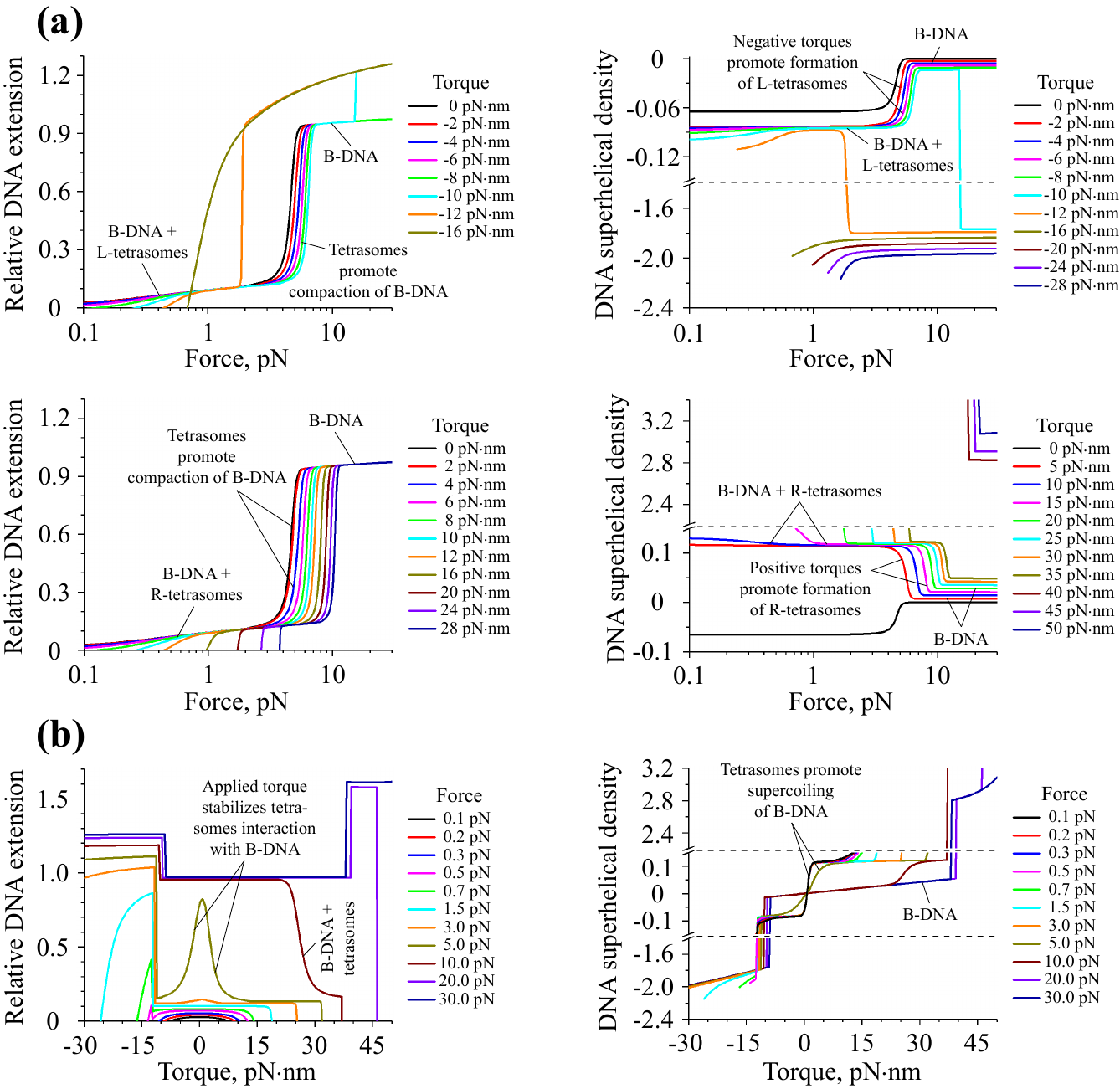}
\caption{
\textbf{Mechanical response of DNA interacting with histone tetramers to the applied force and torque constraints, large scale view.} The panels show a wider view of Figure~\ref{fig5}, demonstrating \textbf{(a)} force-extension and force-superhelical density curves of DNA obtained at different values of the torque, $\tau$, as well as \textbf{(b)} torque-extension and torque-superhelical density curves of DNA obtained at different values of the force, $f$, in the presence of histone tetramers in solution. In all panels, the DNA extension is normalized to the total contour length of DNA in B-form. Abbreviations L-tetrasomes and R-tetrasomes are used to indicate left- and right-handed tetrasome complexes, respectively.
}
\label{figS3}
\end{figure}

\newpage
~

\renewcommand\thefigure{S4}
\begin{figure}[!htb]
\includegraphics[width=0.9\textwidth]{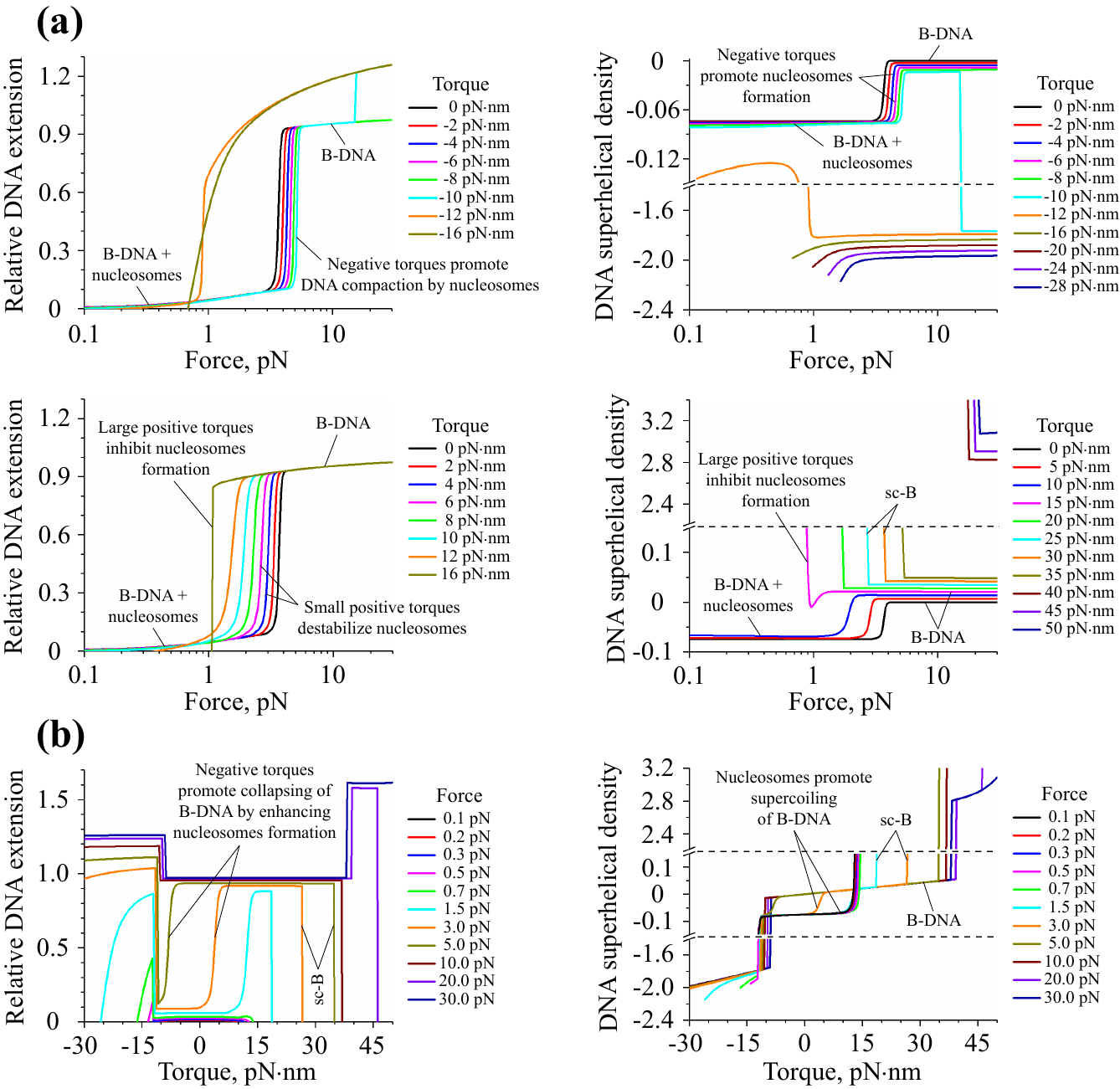}
\caption{
\textbf{Mechanical response of DNA interacting with histone octamers to the applied force and torque constraints, large scale view.} The panels show a wider view of Figure~\ref{fig6}, demonstrating \textbf{(a)} force-extension and force-superhelical density curves of DNA obtained at different values of the torque, $\tau$, as well as \textbf{(b)} torque-extension and torque-superhelical density curves of DNA obtained at different values of the force, $f$, in the presence of histone octamers in solution that upon binding to DNA form nucleosome complexes. In all panels, the DNA extension is normalized to the total contour length of DNA in B-form. Abbreviation sc-B is used to indicate a supercoiled state of B-DNA.
}
\label{figS4}
\end{figure}

\newpage
~

\renewcommand\thefigure{S5} 
\begin{figure}[!htb]
\includegraphics[width=\textwidth]{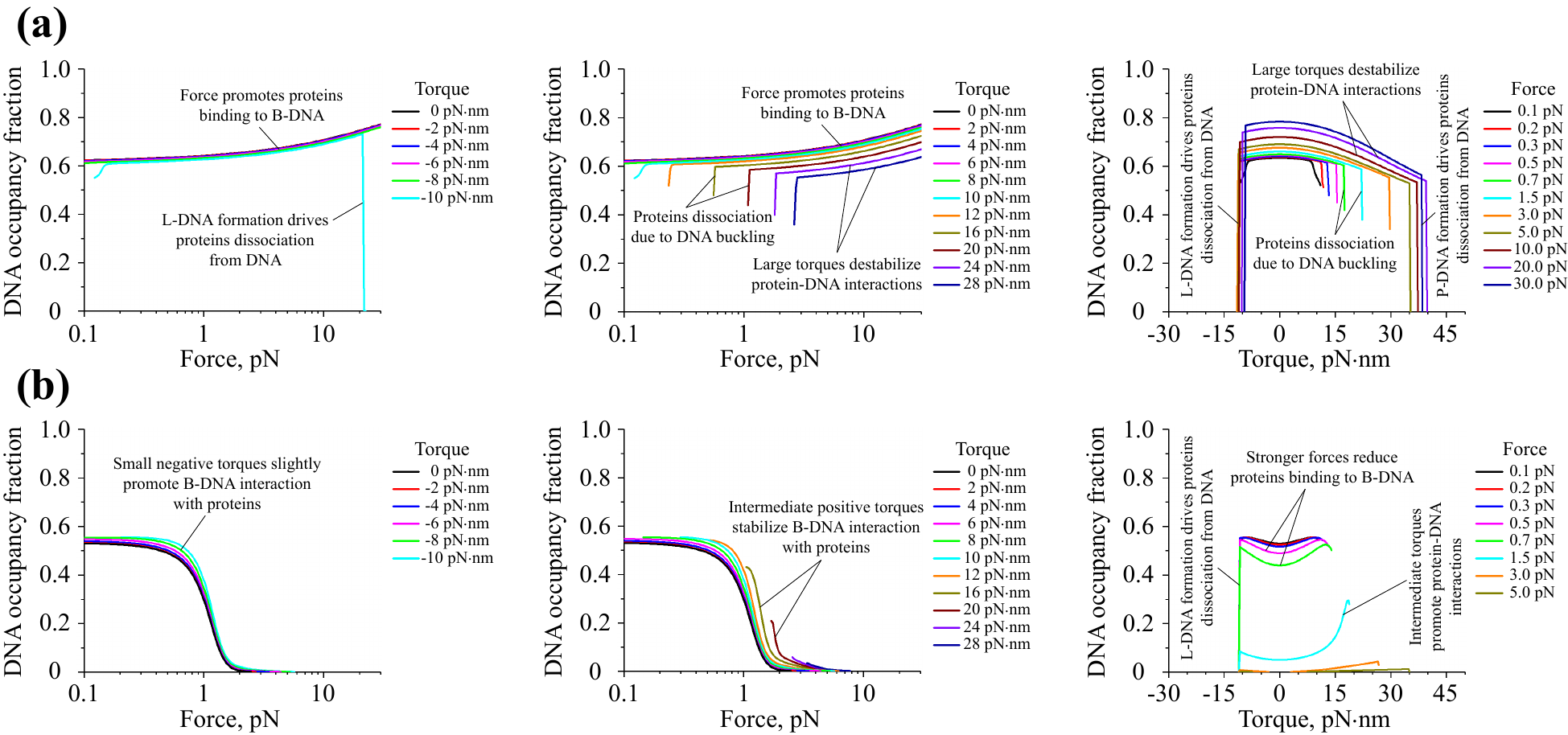}
\caption{
\textbf{DNA occupancy fractions by DNA-stiffening (a) and DNA-bending (b) proteins as functions of the applied force and torque constraints.} From the graphs shown on panel \textbf{(a)}, it can be seen that the DNA occupancy fraction by DNA-stiffening proteins is quite sensitive to the mechanical constraints imposed on the DNA. While strong stretching forces cause $\sim 30 \%$ increase in the number of protein-bound DNA segments, the applied torque has an opposite effect on the DNA-binding affinity of DNA-stiffening proteins, decreasing the DNA coating by nucleoprotein complexes. In contrast, the binding affinity of DNA-bending proteins is considerably increased in the presence of torque (of either sign) exerted to the DNA, as can be seen from the right graph on panel \textbf{(b)}. Furthermore, from the rest of the plots shown on panel \textbf{(b)}, it is clear that mechanical stretching of DNA results in destabilization of the nucleoprotein complexes formed by DNA-bending proteins at forces $f \geq 1.0$ pN. Thus, DNA-stiffening and DNA-bending proteins demonstrate completely different response to force and torque constraints applied to the DNA. On both panels \textbf{(a)} and \textbf{(b)}, the calculated curves are shown only for the working range of the Fuller's formula, which was used to compute the DNA linking number change in the transfer-matrix calculations, see Eq.~\eqref{delta-Lk-intro-1}-\eqref{delta-Lk-intro-3}.
}
\label{figS5}
\end{figure}

\newpage
~

\renewcommand\thefigure{S6} 
\begin{figure}[!htb]
\includegraphics[width=\textwidth]{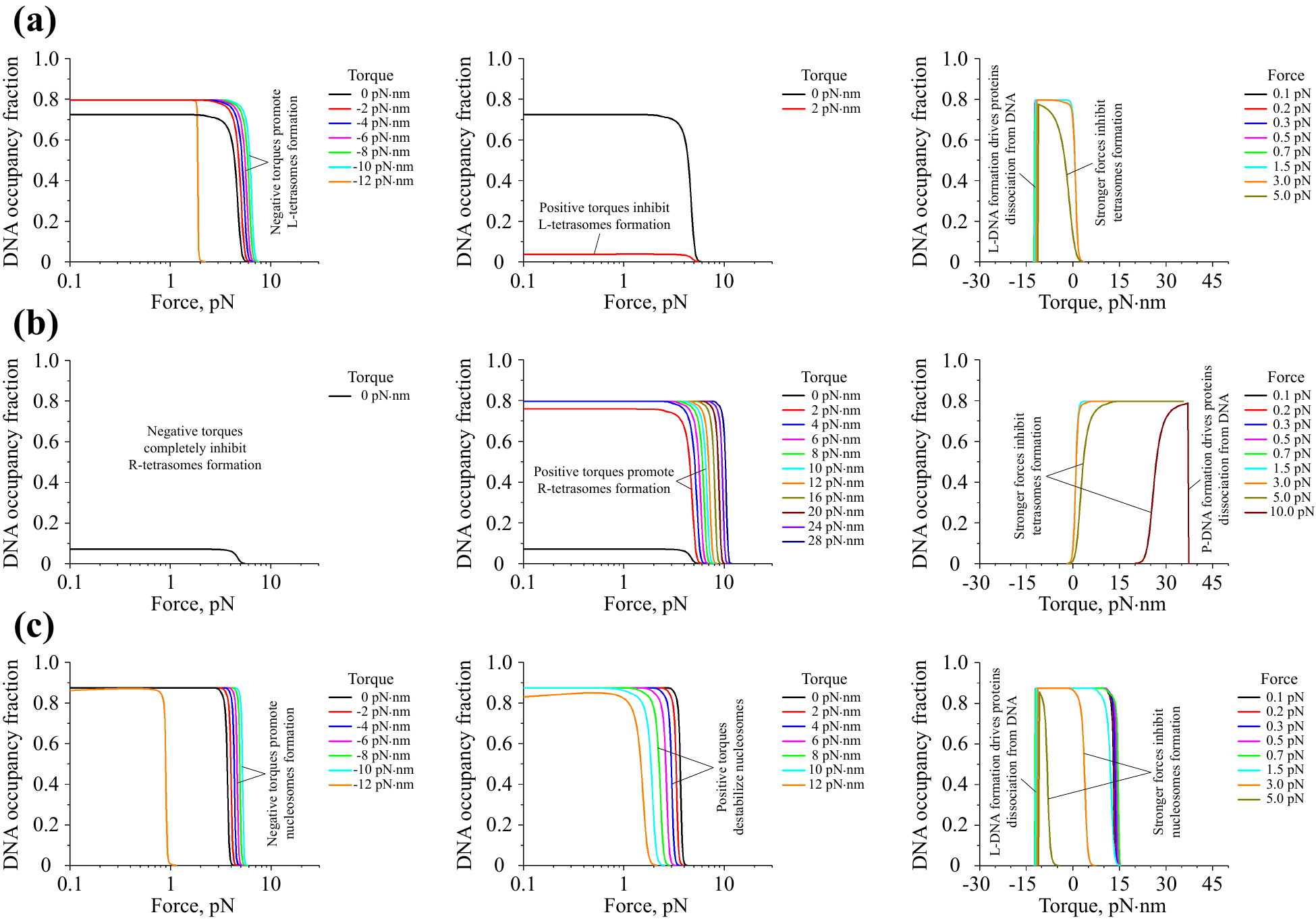}
\caption{
\textbf{DNA occupancy fractions by left-handed (a) and right-handed (b) histone tetrasomes, and nucleosome complexes (c) as functions of the applied force and torque constraints.} From panels \textbf{(a)} and \textbf{(b)}, it can be seen that chirality of tetrasome complexes is highly sensitive to the direction (i.e., sign) of the torque applied to the DNA. While at negative torques tetrasomes assume the left-handed conformation, at positive torques they flip to the right-handed structure. Furthermore, the graphs plotted on panels \textbf{(a)} and \textbf{(b)} indicate that negative and positive torques exerted to DNA not only cause changes in the tetrasome architecture, but also lead to enhancement of the DNA-binding affinities of histone tetramers that have left- and right-handed chiralities, respectively, resulting in increased value of the stretching force required for their dissociation from the DNA. As for nucleosomes, from the graphs shown on the panel \textbf{(c)} it can be seen that these nucleoprotein complexes behave in very much the same way as left-handed tetrasomes with the only difference being that nucleosomes do not have the capability to change their chirality to the right-handed one at positive torques. As a result, positive torques applied to DNA strongly destabilize nucleosome complexes, and already at moderate torques of $\tau \geq 15$ pN$\cdot$nm nucleosomes practically do not assemble on DNA, giving a way to formation of supercoiled bare DNA structures.
}
\label{figS6}
\end{figure}

\end{document}